\documentclass{article}

\usepackage{amsmath}
\usepackage{amssymb}
\usepackage{graphicx}
\usepackage{authblk}
\usepackage{hyperref}
\usepackage{setspace}
\usepackage{fullpage}

\title{Hypergraph Discretization of the Cauchy Problem in General Relativity via Wolfram Model Evolution}
\author{Jonathan Gorard\footnote{\url{jg865@cam.ac.uk}}}

\begin{document}
\maketitle

\begin{abstract}
Although the traditional form of the Einstein field equations is intrinsically four-dimensional, the field of numerical general relativity focuses on the reformulation of these equations as a ${3 + 1}$-dimensional Cauchy problem, in which Cauchy initial data is specified on a three-dimensional spatial hypersurface, and then evolved forwards in time. The Wolfram model offers an inherently discrete formulation of the Einstein field equations as an a priori Cauchy problem, in which Cauchy initial data is specified on a single spatial hypergraph, and then evolved by means of hypergraph substitution rules, with the resulting causal network corresponding to the conformal structure of spacetime. This article introduces a new numerical general relativity code based upon the conformal and covariant Z4 (CCZ4) formulation with constraint-violation damping, with the option to reduce to the standard BSSN formalism if desired, with Cauchy data defined over hypergraphs; the code incorporates an unstructured generalization of the adaptive mesh refinement technique proposed by Berger and Colella, in which the topology of the hypergraph is refined or coarsened based upon local conformal curvature terms. We validate this code numerically against a variety of standard spacetimes, including Schwarzschild black holes, Kerr black holes, maximally extended Schwarzschild black holes, and binary black hole mergers (both rotating and non-rotating), and explicitly illustrate the relationship between the discrete hypergraph topology and the continuous Riemannian geometry that is being approximated. Finally, we compare the results produced by this code to the results obtained by means of pure Wolfram model evolution (without the underlying PDE system), using a hypergraph substitution rule that provably satisfies the Einstein field equations in the continuum limit, and show that the two sets of discrete spacetimes converge to the same limiting geometries.
\end{abstract}

\tableofcontents

\newpage

\section{Introduction}

The Einstein field equations are notoriously difficult to solve; indeed, the only analytical solutions that are currently known (such as the Schwarzschild solution and the Kerr solution) are ones which depend inherently upon a very high degree of a priori symmetry of the spacetime, so as to allow one to reduce the ten independent components of the spacetime metric tensor down to just a handful. Moreover, since the equations are highly nonlinear, perturbative approaches tend to fail whenever the regimes are strongly relativistic, and even in symmetrical cases, questions of non-perturbative stability (such as in the case of the exterior Kerr solution) still remain unanswered. The field of modern numerical general relativity, which arguably began in a serious way with the work of Pretorius\cite{pretorius}, Baker et al.\cite{baker} and Campanelli et al.\cite{campanelli} on the numerical simulation of binary black hole inspirals\cite{chu}\cite{centrella}, seeks to overcome these difficulties by formulating general relativity as a Cauchy problem, in which Cauchy initial data is specified on a three-dimensional spatial hypersurface which is then evolved forwards (or backwards) in time using standard numerical methods such as finite element, finite difference or finite volume schemes. In this way, one effectively reformulates the four-dimensional Einstein field equations over spacetime as a ${3 + 1}$-dimensional explicit time evolution scheme, in which the evolution equations relate the data on pairs of neighboring spacelike hypersurfaces.

On the other hand, the Wolfram model\cite{wolfram}\cite{wolfram2} is an intrinsically discrete spacetime formalism based on hypergraph transformation dynamics\cite{gorard}\cite{gorard2}, in which a causal graph representing the conformal structure of spacetime is generated algorithmically by means of an abstract rewriting system defined over hypergraphs; much work has already been done in developing the underlying mathematical formulations of both general relativity and quantum mechanics in the context of the Wolfram model, and in investigating connections to related formalisms such as causal set theory\cite{gorard3} and categorical quantum mechanics\cite{gorard4}. However, one notable feature of this formalism for our present purposes is that the natural formulation of the Einstein field equations in Wolfram model systems is as a discrete Cauchy problem, in which the initial hypergraph defines the Cauchy initial data, and the hypergraph transformation rules define the time evolution dynamics; it is already known that one can use the Ollivier-Ricci constructions\cite{ollivier} of both scalar and sectional curvatures in Wolfram model hypergraphs so as to recover a discrete form of the Einstein-Hilbert action which subsequently reduces to the standard Benincasa-Dowker action over causal sets\cite{benincasa} for particular classes of rules (specifically those satisfying a weak ergodicity hypothesis, as well as an asymptotic dimension preservation condition).

The main objective of the present article is to demonstrate the applicability of the Wolfram model as a practical tool for solving the Einstein field equations numerically. We begin by introducing a new numerical general relativity code based on a canonical decomposition of the Einstein field equations via the conformal and covariant Z4 (CCZ4) formulation with constraint-violation damping due to Alic et al.\cite{alic} and Bona et al.\cite{bona}, and with the option to reduce to the standard BSSN evolution scheme of Baumgarte, Shapiro, Shibata and Nakamura\cite{baumgarte}\cite{shibata}, if desired. The Cauchy initial data is defined over a spatial hypergraph, and the code incorporates an implementation of a modified local adaptive mesh refinement (AMR) algorithm based on the work of Berger and Colella, appropriately generalized so as to deal with topologically unstructured meshes defined by arbitrary hypergraphs; in this implementation, the topology of the hypergraph is either refined or coarsened on the basis of the behavior of the local (spatial) conformal curvature tensor. We proceed to validate this code numerically against a wide variety of standard spacetimes, as well as comparing the computed sequences of hypergraphs against the sequences of hypergraphs obtained by pure Wolfram model evolution (in the absence of the underlying PDE system) using hypergraph transformation rules that provably satisfy the Einstein field equations in the continuum limit, and we show that the two sets of spacetimes converge to the same sets of Lorentzian manifolds in the continuum limit.

We start by introducing the Z4 formulation of the Einstein field equations in Section \ref{sec:Section1}, which expresses the equations in a strongly hyperbolic form by performing the standard ${3 + 1}$ (ADM\cite{arnowitt}) decomposition of the spacetime metric, as in the conventional BSSN equations, but in which the ADM Hamiltonian and momentum constraints are treated instead as dynamical variables, thus allowing modes which violate these constraints to propagate freely throughout the computational domain. We show how the Z4 system may be augmented with additional terms that have the effect of exponentially suppressing these constraint-violating modes, with a resultant increase in numerical stability. In Section \ref{sec:Section2}, we proceed to show how the Z4 system can be written in a fully covariant form (just as with the BSSN equations), as demonstrated by Brown\cite{brown}, Baumgarte et al.\cite{baumgarte2} and Sanchis-Gual et al.\cite{sanchisgual}, and can also be written in terms of a purely conformally-invariant spatial metric tensor ${\tilde{\gamma}_{i j}}$, thus yielding the conformal and covariant Z4 (CCZ4) formulation. We also introduce the various gauge conditions that are employed for the simulations presented in this article, particularly the \textit{1 + log} slicing and gamma driver conditions that are used for the stable simulation of dynamical spacetimes involving binary black holes, and discuss how the moving puncture gauge conditions allow us effectively to manage the presence of black hole singularities without excision.

In Section \ref{sec:Section3}, we describe the implementation details of our new numerical general relativity code, with a particular emphasis on the explicit fourth-order Runge-Kutta time evolution scheme used for the evolution of the equations, the structure of the fourth-order finite difference stencils chosen (for cases with and without advection terms), and how they can be applied in a hypergraph with no a priori coordinate system and a totally unstructured topology. We also emphasize the significance of the Kreiss-Oliger term\cite{kreiss} for the dissipation of spurious high-frequency modes, the implementation of a modification of the local AMR algorithm of Berger and Colella\cite{berger} for the purpose of refining or coarsening the hypergraph topology based on the local conformal curvature tensor, and the role of our higher-order weighted essentially non-oscillatory (WENO) scheme implementation for the extrapolation of boundary states in the refinement algorithm. In Section \ref{sec:Section4}, we present the results of standard numerical test cases to illustrate the convergence of our implementation, including static Schwarzschild black hole spacetimes, rapidly rotating Kerr black hole spacetimes (in Boyer-Lindquist coordinates) with various rotational parameters, maximally extended Schwarzschild black hole spacetimes (in Kruskal-Szekeres coordinates), head-on collision tests for static Schwarschild black holes (with Brill-Lindquist initial data), and head-on collision tests for rapidly rotating Kerr black holes (also with Brill-Lindquist initial data). In each of these test cases, we present the convergence properties of the numerical scheme with respect to the ${L_1}$, ${L_2}$ and ${L_{\infty}}$ norms, examining the stability properties of the ADM Hamiltonian, linear momentum and angular momentum constraints; for the case of the head-on black hole collision tests, we also compute the complex radial Weyl scalar ${r \Psi_4}$ and use it to extract gravitational radiation information during both the collision and ringdown phases of the merger, demonstrating the convergence properties of the extracted gravitational waves also. We find approximately fourth-order (stable) convergence throughout.

Finally, in Section \ref{sec:Section5}, we illustrate how the relevant dynamical quantities in the CCZ4 equations (such as the spatial extrinsic curvature tensor, the spatial conformal metric tensor, etc.) can be computed purely on the basis of the hypergraph topology using the standard Ollivier-Ricci constructions and their various generalizations. Therefore, we can compare our previous numerical results against results obtained by pure Wolfram model evolution (without knowledge of the underlying PDE system), by selecting a hypergraph transformation rule that provably satisfies the Einstein field equations in the continuum limit, which can in turn be done by locating rules which satisfy weak ergodicity and asymptotic dimension preservation (which have previously been shown to yield extremization of the discrete Einstein-Hilbert action). These comparisons show that the discrete spacetimes produced by numerical solution of the CCZ4 equations and the discrete spacetimes produced by pure Wolfram model evolution are asymptotically equivalent (i.e. they yield the same sets of Lorentzian manifolds in the continuum limit), with the pure Wolfram model evolution demonstrating approximately second-order (stable) convergence throughout. In Section \ref{sec:Section6}, we make some concluding remarks regarding these results and discuss potential directions for future research.

Note that all of the code required to reproduce the simulations and visualizations presented throughout this article is freely available through the \textit{Wolfram Function Repository} (e.g. \url{https://resources.wolframcloud.com/FunctionRepository/resources/IntrinsicCurvedManifoldToGraph/} and \url{https://resources.wolframcloud.com/FunctionRepository/resources/ExtrinsicCurvedManifoldToGraph/} for setting up initial Cauchy data, \url{https://resources.wolframcloud.com/FunctionRepository/resources/CurvedSpacetimeRegionSprinkling} for producing the characteristic structure of evolutions, \url{https://resources.wolframcloud.com/FunctionRepository/resources/WolframRicciCurvatureScalar} and \url{https://resources.wolframcloud.com/FunctionRepository/resources/WolframRicciCurvatureTensor} for extracting curvature invariants from the intermediate geometries, etc.)

\section{The Z4 Formulation with Constraint Damping}
\label{sec:Section1}

Since general relativity is a generally covariant theory, meaning that physical solutions of the Einstein field equations are given by equivalence classes defined in terms of gauge transformations whose gauge group (the diffeomorphism group) is that of smooth coordinate transformations of the form:

\begin{equation}
y^{\mu} = f^{\mu} \left( x^{\nu} \right),
\end{equation}
it follows that the field equations alone:

\begin{equation}
R_{\mu \nu} = 8 \pi \left( T_{\mu \nu} - \frac{1}{2} T g_{\mu \nu} \right),
\end{equation}
do not, in general, provide sufficient information to determine the values of the ten independent coefficients of the metric tensor ${g_{\mu \nu}}$. Here, $T$ denotes the usual trace of the energy-momentum tensor:

\begin{equation}
T = g_{\mu \nu} T^{\mu \nu}.
\end{equation}
The Z4 formulation, originally introduced by Bona, Ledvinka, Palenzuela and \u{Z}\'a\u{c}ek\cite{bona2}, is an extension of the conventional Einstein field equations that is explicitly covariant, since it is derived (in the vacuum case) from the following covariant Lagrangian density\cite{bona3}:

\begin{equation}
\mathcal{L} = \sqrt{g} g^{\mu \nu} \left[ R_{\mu \nu} + 2 \nabla_{\mu} Z_{\nu} \right],
\end{equation}
with the elliptic energy and momentum constraints of the ADM formalism replaced by simple algebraic constraints on the newly-introduced 4-vector ${Z_{\mu}}$. More specifically, by introducing a generic action of the form:

\begin{equation}
S = \int \mathcal{L} d^4 x,
\end{equation}
writing out the Ricci curvature tensor ${R_{\mu \nu}}$ explicitly in terms of coefficients of the Levi-Civita connection (with symmetrization of indices denoted by round brackets):

\begin{equation}
R_{\mu \nu} = \partial_{\rho} \Gamma_{\mu \nu}^{\rho} - \partial_{\left( \mu \right.} \Gamma_{\left. \nu \right) \rho}^{\rho} + \Gamma_{\rho \sigma}^{\rho} \Gamma_{\mu \nu}^{\sigma} - \Gamma_{\sigma \mu}^{\rho} \Gamma_{\rho \nu}^{\sigma},
\end{equation}
and performing a Palatini variation, in which the metric density:

\begin{equation}
h^{\mu \nu} = \sqrt{g} g^{\mu \nu},
\end{equation}
the coefficients of the connection ${\Gamma_{\mu \nu}^{\rho}}$ and the 4-vector ${Z_{\mu}}$ are each varied independently, the variations of the metric density ${h^{\mu \nu}}$ in particular yield the following system of vacuum Z4 field equations:

\begin{equation}
R_{\mu \nu} + \nabla_{\mu} Z_{\nu} + \nabla_{\nu} Z_{\mu} = 0.
\end{equation}
Incorporating the energy-momentum contributions ${T_{\mu \nu}}$ yields:

\begin{equation}
R_{\mu \nu} + \nabla_{\mu} Z_{\nu} + \nabla_{\nu} Z_{\mu} = 8 \pi \left( T_{\mu \nu} - \frac{1}{2} T g_{\mu \nu} \right),
\end{equation}
with the ADM energy and momentum constraints simply reducing to:

\begin{equation}
Z_{\mu} = 0,
\end{equation}
as we shall describe in more detail shortly.

Brodbeck, Frittelli, H\"ubner and Reula\cite{brodbeck} proposed a very general formalism for evolving systems of equations with constraints, known as the ${\lambda}$-system, in which the constraint surface forms a set of attractor solutions for the system. This is achieved by supplementing the system with an auxiliary variable ${\lambda}$ for each constraint, with the time derivatives of the ${\lambda}$ variables designating the constraint variables themselves (i.e. ${\lambda}$ denotes a kind of ``time integral'' of a constraint variable), in such a way as to preserve and/or introduce symmetric hyperbolicity of the underlying system. Then, for the case of the four energy-momentum constraints of the Einstein field equations, the evolution equations can be supplemented with damping terms of the general form:

\begin{equation}
\frac{\partial \lambda}{\partial t} = \alpha_0 \mathcal{C} - \beta_0 \lambda, \qquad \frac{\partial \lambda^i}{\partial t} = \alpha_1 \mathcal{C}^i - \beta_1 \lambda^i,
\end{equation}

\begin{equation}
\frac{\partial \lambda^{i j}_{k}}{\partial t} = \alpha_3 \mathcal{C}^{i j}_k - \beta_3 \lambda^{i j}_{k}, \qquad \frac{\partial \lambda^{i j}_{k l}}{\partial t} = \alpha_4 \mathcal{C}^{i j}_{k l} - \beta_4 \lambda^{i j}_{k l},
\end{equation}
for constraint quantities ${\mathcal{C}}$, ${\mathcal{C}^i}$, ${\mathcal{C}^{i j}_{k}}$ and ${\mathcal{C}^{i j}_{k l}}$, constants ${\alpha_i \neq 0}$ and ${\beta_i > 0}$, and where the tensor-valued ${\lambda}$ variables are assumed to inherit all of the symmetries of the corresponding constraint variables ${\mathcal{C}}$. The lower-order nature of these equations ensures that they do not alter the hyperbolicity properties of the system, whilst still having the effect of damping the ${\lambda}$ variables (and hence also the constraint variables). As demonstrated by Gundlach, Mart\'in-Garc\'ia, Calabrese and Hinder\cite{gundlach}, the ${Z_{\mu}}$ variables of the Z4 formulation may be considered to be ${\lambda}$ variables, with damping terms that can be introduced in covariant notation by replacing the vacuum Z4 equations:

\begin{equation}
R_{\mu \nu} + \nabla_{\mu} Z_{\nu} + \nabla_{\nu} Z_{\mu} = 0,
\end{equation}
with the damped equations:

\begin{equation}
R_{\mu \nu} + \nabla_{\mu} Z_{\nu} + \nabla_{\nu} Z_{\mu} - \kappa_1 \left[ t_{\mu} Z_{\nu} + t_{\nu} Z_{\mu} - \left( 1 + \kappa_2 \right) g_{\mu \nu} t^{\sigma} Z_{\sigma} \right] = 0,
\end{equation}
or, in non-vacuum form:

\begin{equation}
R_{\mu \nu} + \nabla_{\mu} Z_{\nu} + \nabla_{\nu} Z_{\mu} - \kappa_1 \left[ t_{\mu} Z_{\nu} + t_{\nu} Z_{\mu} - \left( 1 + \kappa_2 \right) g_{\mu \nu} t^{\sigma} Z_{\sigma} \right] = 8 \pi \left( T_{\mu \nu} - \frac{1}{2} T g_{\mu \nu} \right),
\end{equation}
for real constants ${\kappa_1 \geq 0}$ and ${\kappa_2}$, and for a non-vanishing timelike vector field ${t^{\mu}}$. In traced-reversed form, the vacuum equations thus become:

\begin{equation}
G_{\mu \nu} + \nabla_{\mu} Z_{\nu} + \nabla_{\nu} Z_{\mu} - g_{\mu \nu} \nabla_{\sigma} Z_{\sigma} - \kappa_1 \left( t_{\mu} Z_{\nu} + t_{\nu} Z_{\mu} + \kappa_2 g_{\mu \nu} t^{\sigma} Z_{\sigma} \right) = 0.
\end{equation}
By taking the divergence of this trace-reversed system and applying the contracted Bianchi identities:

\begin{equation}
\nabla_{\rho} R^{\rho \mu} = \frac{1}{2} \nabla_{\mu} R, \qquad \implies \nabla^{\mu} G_{\mu \nu} = 0,
\end{equation}
we obtain a second-order wave equation for the 4-vector ${Z_{\mu}}$:

\begin{equation}
\nabla_{\mu} \nabla^{\mu} Z_{\nu} + R_{\mu \nu} Z^{\mu} - \kappa_1 \nabla^{\mu} \left( t_{\mu} Z_{\nu} + t_{\nu} Z_{\mu} + \kappa_2 g_{\mu \nu} t^{\sigma} Z_{\sigma} \right) = 0,
\end{equation}
or alternatively:

\begin{equation}
\Box Z_{\nu} + R_{\mu \nu} Z^{\mu} - \kappa_1 \nabla^{\mu} \left( t_{\mu} Z_{\nu} + t_{\nu} Z_{\mu} + \kappa_2 g_{\mu \nu} t^{\sigma} Z_{\sigma} \right) = 0,
\end{equation}
using the covariant wave operator:

\begin{equation}
\Box = \nabla_{\mu} \nabla^{\mu}.
\end{equation}
If we consider now a ${3 + 1}$ decomposition of the metric tensor (as described in more detail subsequently), then we can simply take ${t^{\mu}}$ to be the future-pointing unit normal vector to a given constant-time slice, i.e.:

\begin{equation}
t^{\mu} = n^{\mu},
\end{equation}
such that the wave equation for ${Z_{\mu}}$ can be rewritten (after rearrangement) as:

\begin{equation}
\nabla^{\nu}_{\nu} Z_{\mu} + R_{\mu \nu} Z^{\nu} = - \kappa_1 \nabla^{\nu} \left[ n_{\mu} Z_{\nu} + n_{\nu} Z_{\mu} + \kappa_2 g_{\mu \nu} n_{\sigma} Z^{\sigma} \right].
\end{equation}

Following the techniques of Gundlach et al.\cite{gundlach}, we construct a linearized form of the constraint equations by perturbing a background solution ${g_{\mu \nu}^{\left( 0 \right)}}$ obeying the Einstein field equations with:

\begin{equation}
R_{\mu \nu}^{\left( 0 \right)} = 0, \qquad \text{ and } \qquad Z_{\mu}^{\left( 0 \right)} = 0,
\end{equation}
to obtain:

\begin{equation}
g_{\mu \nu} = g_{\mu \nu}^{\left( 0 \right)} + \epsilon g_{\mu \nu}^{\left( 1 \right)}.
\end{equation}
Up to first-order in ${\epsilon}$, this constraint violation will then obey the following linearized vector wave equation over the background spacetime:

\begin{equation}
\Box^{\left( 0 \right)} Z_{\nu}^{\left( 1 \right)} - \kappa_1 \nabla^{\mu \left( 0 \right)} \left( t_{\mu}^{\left( 0 \right)} Z_{\nu}^{\left( 1 \right)} + t_{\nu}^{\left( 0 \right)} Z_{\mu}^{\left( 1 \right)} + \kappa_2 g_{\mu \nu}^{\left( 0 \right)} t^{\sigma \left( 0 \right)} Z_{\sigma}^{\left( 1 \right)} \right) = 0.
\end{equation}
In the high-frequency limit, i.e. the limit in which the wavelength of ${Z{\mu}^{\left( 1 \right)}}$ is negligible compared to the curvature scale of the background spacetime, the background metric ${g_{\mu \nu}^{\left( 0 \right)}}$ can therefore be approximated as the flat Minkowski metric, such that the covariant wave operator reduces to the Minkowski wave operator:

\begin{equation}
\Box = - \partial_{t}^{2} + \partial_i \partial^i,
\end{equation}
where we have adopted the metric convention that:

\begin{equation}
t_{\mu} t^{\mu} = -1.
\end{equation}
Shifting to the frame in which:

\begin{equation}
t^{\mu} = \left( 1, 0, 0, 0 \right),
\end{equation}
our linearized wave equations will therefore reduce to:

\begin{equation}
\Box Z_{0}^{\left( 1 \right)} - \kappa_1 \left[ \left( 2 + \kappa_2 \right) \partial_{t}^{\left( 0 \right)} Z_{0}^{\left( 1 \right)} - \partial^{i \left( 0 \right)} Z_{i}^{\left( 1 \right)} \right] = 0,
\end{equation}
for the time components, and:

\begin{equation}
\Box Z_{i}^{\left( 1 \right)} - \kappa_1 \left( \partial_{t}^{\left( 0 \right)} Z_{i}^{\left( 1 \right)} + \kappa_2 \partial_{i}^{\left( 0 \right)} Z_{0}^{\left( 1 \right)} \right) = 0,
\end{equation}
for the space components.

Making the following ansatz for a sinusoidal plane-wave solution:

\begin{equation}
Z_{\mu} \left( t, x^{i} \right) = \exp \left( s t + i \omega_i x^i \right) \hat{Z}^{\mu},
\end{equation}
where ${\omega_i}$ are real angular frequencies, $s$ is an arbitrary complex number and ${\hat{Z}_{\mu}}$ is a real amplitude, and moreover denoting the component of ${Z_i}$ in the direction of the angular frequency ${\omega_i}$ by ${Z_n}$ and the projection of ${Z_i}$ in the direction normal to the angular frequency ${\omega_i}$ by ${Z_A}$, we obtain the following eigenvalue problem:

\begin{equation}
\begin{bmatrix}
-s^2 - \omega^2 - \kappa_1 \left( 2 + \kappa_2 \right) s & \kappa_1 i \omega & 0\\
- \kappa_1 \kappa_2 i \omega & -s^2 - \omega^2 - \kappa_1 s & 0\\
0 & 0 & -s^2 - \omega^2 - \kappa_1 s
\end{bmatrix} \begin{bmatrix}
\hat{Z}_0\\
\hat{Z}_n\\
\hat{Z}_A
\end{bmatrix} = 0.
\end{equation}
Whereas the four eigenvalues $s$ arising from the ${Z_A}$ equation are independent of the choice of ${\kappa_2}$, namely:

\begin{equation}
s = - \frac{\kappa_1}{2} \pm \sqrt{\left( \frac{\kappa_1}{2} \right)^2 - \omega^2},
\end{equation}
each with algebraic multiplicity two, the remaining four eigenvalues $s$ are not so straightforward. For the case ${\kappa_2 = -1}$, they become:

\begin{equation}
s = - \kappa_1 \pm i \omega, \qquad \text{ and } \qquad s = \pm i \omega,
\end{equation}
yielding undamped modes for all ${\omega_i}$, although the eigenstructure is such that all constraint-related modes are damped whenever ${\kappa_2 > -1}$ and ${\kappa_1 > 0}$, such that the constraint surface becomes an attractor, as required. For instance, for the case ${\kappa_2 = 0}$, the eigenvalues $s$ have the form:

\begin{equation}
s = - \kappa_1 \pm \sqrt{\kappa_{1}^{2} - \omega^2},
\end{equation}
each with algebraic multiplicity two, yielding damping for all ${\omega_i \neq 0}$.

If we now perform the canonical ${3 + 1}$ decomposition of spacetime, using the standard ADM line element\cite{arnowitt}:

\begin{equation}
d s^2 = - \alpha^2 d t^2 + \gamma_{i j} \left( d x^i + \beta^i d t \right) \left( d x^j + \beta^j d t \right),
\end{equation}
with the lapse function ${\alpha}$ and shift vector ${\beta^i}$ being the ADM gauge variables, and denoting the induced spatial metric tensor on the constant-time slices by ${\gamma_{i j}}$, we obtain the conventional ADM system of evolution equations for the spatial metric tensor ${\gamma_{i j}}$:

\begin{equation}
\partial_t \gamma_{i j} = -2 \alpha K_{i j} + \nabla_i \beta_j + \nabla_j \beta_i,
\end{equation}
and the extrinsic curvature tensor ${K_{i j}}$:

\begin{multline}
\partial_t K_{i j} = \alpha \left( R_{i j}^{\left( 3 \right)} - 2 K_{i k} K_{j}^{k} + K_{i j} K \right) - \nabla_i \nabla_j \alpha - 8 \pi G \alpha \left[ S_{i j} - \frac{1}{2} \gamma_{i j} \left( S - \tau \right) \right]\\
+ \beta^k \partial_k K_{i j} + K_{i k} \partial_j \beta^k + K_{k j} \partial_i \beta^k,
\end{multline}
where ${\tau}$, ${S_{i j}}$ and $S$ are simple functions of the energy-momentum tensor ${T_{\mu \nu}}$:

\begin{equation}
\tau = T_{\mu \nu} n^{\mu} n^{\nu} , \qquad S_{i j} = T_{\mu \nu} \gamma_{i}^{\mu} \gamma_{j}^{\nu}, \qquad S = \gamma^{i j} S_{i j},
\end{equation}
and where $K$ denotes the trace of the extrinsic curvature tensor ${K_{i j}}$ (i.e. the mean curvature):

\begin{equation}
K = K_{i j} \gamma^{i j}.
\end{equation}
Note that the extrinsic curvature tensor ${K_{\mu \nu}}$ may be defined abstractly in terms of the Lie derivative ${\mathcal{L}}$ as:

\begin{equation}
K_{\mu \nu} = - \frac{1}{2} \mathcal{L}_{\mathbf{n}} \gamma_{\mu \nu} = - \gamma_{\mu}^{\tau} \gamma_{\nu}^{\sigma} \nabla_{\tau}^{\left( 4 \right)} n_{\sigma} = - \nabla_{\nu}^{\left( 3 \right)} n_{\mu},
\end{equation}
where ${\nabla^{\left( 4 \right)}}$ denotes the covariant derivative over spacetime and ${\nabla^{\left( 3 \right)}}$ denotes the covariant derivative on a given spacelike hypersurface, and ${\mathcal{L}_{\mathbf{n}}}$ indicates the Lie derivative along the normal vector ${\mathbf{n}}$. As referenced previously, the ADM Hamiltonian (i.e. energy) constraint is then given by:

\begin{equation}
H = R^{\left( 3 \right)} - K_{i j} K^{i j} + K^2,
\end{equation}
and the three ADM momentum constraints are given in turn by:

\begin{equation}
M_i = \gamma^{j l} \left( \partial_l K_{i j} - \partial_i K_{j l} - \Gamma_{j l}^{m} K_{m i} + \Gamma_{j i}^{m} K_{m l} \right),
\end{equation}
with spatial Ricci scalar ${R^{\left( 3 \right)}}$.

Considering the ADM equations to be a system of first-order evolution equations in ${\mathbb{R}^{n + 1}}$ over a set of tensor fields ${\tilde{\phi}^{\alpha}}$ of the form:

\begin{equation}
\partial_t \tilde{\phi}^{\alpha} = A_{\beta}^{i \alpha} \partial_i \tilde{\phi}^{\beta} + B_{\beta}^{\alpha} \tilde{\phi}^{\beta},
\end{equation}
for tensor fields ${A_{\beta}^{i \alpha}}$ and ${B_{\beta}^{\alpha}}$ over ${\mathbb{R}^{n + 1}}$, the system is only weakly hyperbolic, in the sense that for any covector ${\omega_i}$ in ${\mathbb{R}^n}$, the matrix ${A_{\beta}^{i \alpha} \omega_i}$ has purely real eigenvalues, but is not (necessarily) diagonalizable\cite{reula}\cite{york}. However, by replacing the ADM energy and momentum constraints with evolution equations for the ${Z_{\mu}}$ 4-vector, one forces the system to be strongly hyperbolic, such that ${A_{\beta}^{i \alpha} \omega_i}$ is also diagonalizable. The leads to a ${3 + 1}$ decomposition of the Z4 formulation, including damping terms, consisting of the usual evolution equations for the spatial metric tensor ${\gamma_{i j}}$:

\begin{equation}
\left( \partial_t - \mathcal{L}_{\boldsymbol{\beta}} \right) \gamma_{i j} = -2 \alpha K_{i j},
\end{equation}
and the extrinsic curvature tensor ${K_{i j}}$:

\begin{multline}
\left( \partial_t - \mathcal{L}_{\boldsymbol{\beta}} \right) K_{i j} = - \nabla_i \alpha_j + \alpha \left[ R_{i j} + \nabla_i Z_j + \nabla_j Z_i - 2 K_{i}^{l} K_{l j} + \left( K - 2 \Theta \right) K_{i j} - \kappa_1 \left( 1 + \kappa_2 \right) \Theta \gamma_{i j} \right]\\
- 8 \pi \alpha \left[ S_{i j} - \frac{1}{2} \left( S - \tau \right) \gamma_{i j} \right],
\end{multline}
as well as an evolution equation for the projection ${\Theta}$ of the 4-vector ${Z_{\mu}}$ in the normal direction:

\begin{equation}
\Theta = n_{\mu} Z^{\mu} = \alpha Z^0,
\end{equation}
namely:

\begin{equation}
\left( \partial_t - \mathcal{L}_{\boldsymbol{\beta}} \right) \Theta = \frac{\alpha}{2} \left[ R + 2 \nabla_j Z^j + \left( K - 2 \Theta \right) K - K^{i j} K_{i j} - 2 \frac{Z^j \alpha_j}{\alpha} - 2 \kappa_1 \left( 2 + \kappa_2 \right) \Theta - 16 \pi \tau \right],
\end{equation}
and finally an evolution equation for the 4-vector ${Z_{\mu}}$ itself:

\begin{equation}
\left( \partial_t - \mathcal{L}_{\boldsymbol{\beta}} \right) Z_i = \alpha \left[ \nabla_j \left( K_{i}^{j} - \delta_{i}^{j} K \right) + \partial_i \Theta - 2 K_{i}^{j} Z_j - \Theta \frac{\alpha_i}{\alpha} - \kappa_1 Z_i - 8 \pi S_i \right],
\end{equation}
where, in the above, we have used ${\tau}$, ${S_i}$ and ${S_{i j}}$ to denote the standard functions of the energy-momentum tensor ${T_{\mu \nu}}$:

\begin{equation}
\tau = n_{\mu} n_{\nu} T^{\mu \nu}, \qquad S_i = n_{\nu} T_{i}^{\nu}, \qquad S_{i j} = T_{i j}.
\end{equation}

\section{The Conformal and Covariant Z4 Formulation with Gauge Conditions}
\label{sec:Section2}

By beginning from this fully covariant variation of the Z4 equations and proceeding to perform a conformal decomposition of the individual terms\cite{bernuzzi}\cite{hilditch}, we are able to derive a conformal and covariant formulation of Z4 (elsewhere known as CCZ4), which is the formulation that we shall use for all of the numerical simulations presented in the remainder of this article. We can begin by rewriting the spatial metric tensor ${\gamma_{i j}}$ as a conformal spatial metric tensor ${\tilde{\gamma}_{i j}}$:

\begin{equation}
\tilde{\gamma}_{i j} = \phi^2 \gamma_{i j},
\end{equation}
with conformal factor ${\phi}$, chosen in such a way as to ensure a unit determinant for ${\tilde{\gamma}_{i j}}$:

\begin{equation}
\phi = \left( \mathrm{det} \left( \gamma_{i j} \right) \right)^{-\frac{1}{6}}.
\end{equation}
Next, we can decompose the total extrinsic curvature tensor ${K_{i j}}$ into a sum of its trace $K$:

\begin{equation}
K = K_{i j} \gamma^{i j},
\end{equation}
and its trace-free components ${\hat{A}_{i j}}$:

\begin{equation}
\hat{A}_{i j} = \phi^2 \left( K_{i j} - \frac{1}{3} K \gamma_{i j} \right).
\end{equation}
Finally, we can decompose the spatial Ricci curvature tensor ${R_{i j}^{\left( 3 \right)}}$ into a sum:

\begin{equation}
R_{i j}^{\left( 3 \right)} = \tilde{R}_{i j}^{\left( 3 \right)} + \tilde{R}_{i j}^{\phi \left( 3 \right)},
\end{equation}
of a tensor ${\tilde{R}_{i j}^{\left( 3 \right)}}$ containing only spatial derivatives of the conformal spatial metric tensor ${\tilde{\gamma}_{i j}}$:

\begin{equation}
\tilde{R}_{i j}^{\left( 3 \right)} = - \frac{1}{2} \hat{\gamma}^{l m} \partial_l \partial_m \tilde{\gamma}_{i j} + \tilde{\gamma}_{k \left( i \right.} \partial_{\left. j \right)} \tilde{\Gamma}^{k} + \tilde{\Gamma}^{k} \tilde{\Gamma}_{\left( i j \right) k} + \tilde{\gamma}^{l m} \left[ 2 \tilde{\Gamma}_{l \left( i \right.}^{k} \tilde{\Gamma}_{\left. j \right) k m} + \tilde{\Gamma}_{i m}^{k} \tilde{\Gamma}{k j l} \right],
\end{equation}
and another tensor ${\tilde{R}_{i j}^{\phi \left( 3 \right)}}$ containing only conformal terms:

\begin{equation}
\tilde{R}_{i j}^{\phi \left( 3 \right)} = \frac{1}{\phi^2} \left[ \phi \left( \tilde{\nabla}_i \tilde{\nabla}_j \phi + \tilde{\gamma}_{i j} \tilde{\nabla}^l \tilde{\nabla}_l \phi \right) - 2 \tilde{\gamma}_{i j} \tilde{\nabla}^l \phi \tilde{\nabla}_l \phi \right],
\end{equation}
where we have introduced:

\begin{equation}
\tilde{\Gamma}^i = \tilde{\gamma}^{j k} \tilde{\Gamma}_{j k}^{i} = \tilde{\gamma}_{i j} \tilde{\gamma}^{k l} \partial_l \tilde{\gamma}_{j k},
\end{equation}
and:

\begin{equation}
\tilde{\Gamma}_{j k}^{i} = \frac{1}{2} \tilde{\gamma}^{i l} \left( \partial_j \tilde{\gamma}_{k l} + \partial_k \tilde{\gamma}_{j l} - \partial_l \tilde{\gamma}_{j k} \right),
\end{equation}
as Christoffel symbols associated with the conformal spatial metric tensor ${\tilde{\gamma}_{i j}}$.

Thus, we can express the fully conformal and covariant variation of the Z4 formulation as a system of evolution equations for the conformal spatial metric tensor ${\tilde{\gamma}_{i j}}$:

\begin{equation}
\partial_t \tilde{\gamma}_{i j} = -2 \alpha \tilde{A}_{i j} + 2 \tilde{\gamma}_{k \left( i \right.} \partial_{\left. j \right)} \beta^k - \frac{2}{3} \tilde{\gamma}_{i j} \partial_k \beta^k + \beta^k \partial_k \tilde{\gamma}_{i j},
\end{equation}
the trace-free components of the extrinsic curvature tensor ${\tilde{A}_{i j}}$:

\begin{multline}
\partial_t \tilde{A}_{i j} = \phi^2 \left[ - \nabla_i \nabla_j \alpha + \alpha \left( R_{i j} + \nabla_i Z_j + \nabla_j Z_i - 8 \pi S_{i j} \right) \right]^{TF} + \alpha \tilde{A}_{i j} \left( K - 2 \Theta \right) - 2 \alpha \tilde{A}_{i l} \tilde{A}_{j}^{l}\\
+ 2 \tilde{A}_{k \left( i \right.} \partial_{\left. j \right)} \beta^k - \frac{2}{3} \tilde{A}_{i j} \partial_k \beta^k + \beta^k \partial_k \tilde{A}_{i j},
\end{multline}
the trace of the extrinsic curvature tensor $K$:

\begin{equation}
\partial_t K = - \nabla^i \nabla_i \alpha + \alpha \left( R + 2 \nabla_i Z^i + K^2 - 2 \Theta K \right) + \beta^j \partial_j K - 3 \alpha \kappa_1 \left( 1 + \kappa_2 \right) \Theta + 4 \pi \alpha \left( S - 3 \tau \right),
\end{equation}
the conformal factor ${\phi}$:

\begin{equation}
\partial_t \phi = \frac{1}{3} \alpha \phi K - \frac{1}{3} \phi \partial_k \beta^k + \beta^k \partial_k \phi,
\end{equation}
the projection ${\Theta}$ of the ${Z_n}$ 4-vector in the normal direction:

\begin{equation}
\partial_t \Theta = \frac{1}{2} \alpha \left( R + 2 \nabla_i Z^i - \tilde{A}_{i j} \tilde{A}^{i j} + \frac{2}{3} K^2 - 2 \Theta K \right) - Z^i \partial_i \alpha + \beta^k \partial_k \Theta - \alpha \kappa_1 \left( 2 + \kappa_2 \right) \Theta - 8 \pi \alpha \tau,
\end{equation}
and finally an evolution equation for the newly-introduced vector quantity ${\hat{\Gamma}^i}$, defined in terms of the Christoffel symbol ${\tilde{\Gamma}^i}$ for the conformal spatial metric tensor:

\begin{equation}
\hat{\Gamma}^i = \tilde{\Gamma}^i + 2 \tilde{\gamma}^{i j} Z_j,
\end{equation}
namely:

\begin{multline}
\partial_t \hat{\Gamma}^i = 2 \alpha \left( \tilde{\Gamma}_{j k}^{i} \tilde{A}^{j k} - 3 \tilde{A}^{j k} \frac{\partial_j \phi}{\phi} - \frac{2}{3} \tilde{\gamma}^{i j} \partial_j K \right) + 2 \tilde{\gamma}^{k i} \left( \alpha \partial_k \Theta - \Theta \partial_k \alpha - \frac{2}{3} \alpha K Z_k \right) - 2 \tilde{A}^{i j} \partial_j \alpha\\
+ \tilde{\gamma}^{k l} \partial_k \partial_l \beta^i + \frac{1}{3} \tilde{\gamma}^{i k} \partial_k \partial_l \beta^l + \frac{2}{3} \tilde{\Gamma}^i \partial_k \beta^k - \tilde{\Gamma}^k \partial_k \beta^i + 2 \kappa_3 \left( \frac{2}{3} \tilde{\gamma}^{i j} Z_j \partial_k \beta^k - \tilde{\gamma}^{j k} Z_j \partial_k \beta^i \right) + \beta^k \partial_k \hat{\Gamma}^i\\
-2 \alpha \kappa_1 \tilde{\gamma}^{i j} Z_j - 16 \pi \alpha \tilde{\gamma}^{i j} S_j,
\end{multline}
where the additional parameter ${\kappa_3}$ has been introduced to prevent instabilities from being introduced via the quadratic terms in the evolution equation for the Christoffel symbol ${\hat{\Gamma}^i}$, which in turn corresponds to the evolution of the vector ${Z_i}$. Choosing ${\kappa_3 = 1}$ yields the fully covariant formulation of the conformal Z4 system, which is unstable for spacetimes containing black holes (in which case the covariance of the conformal decomposition can be broken, for instance by choosing ${\kappa_3 = \frac{1}{2}}$, in order to restore numerical stability). However, these numerical instabilities are a consequence of the fact that the damping coefficient ${\kappa_1}$ is multiplied by the lapse function ${\alpha}$, and the product ${\alpha \kappa_1}$ clearly approaches zero close to a black hole singularity. Thus, as proposed by Alic, Kastaun and Rezzolla\cite{alic2}, if we replace the damping coefficient ${\kappa_1}$ with a lapse-dependent variant:

\begin{equation}
\kappa_1 \to \frac{\kappa_1}{\alpha}, \qquad \text{ or } \qquad \kappa_1 \to \frac{\kappa_1}{2} \left( \alpha + \frac{1}{\alpha} \right),
\end{equation}
then one retains covariance of the Z4 system (strictly speaking, the terms containing ${\kappa_1}$ are only spatially covariant, since they now depend on details of the ${3 + 1}$ decomposition, but all other evolution equations remain fully covariant). For the black hole spacetimes simulated within this article, we use the simpler ${\kappa_1 \to \frac{\kappa_1}{\alpha}}$ damping choice.

All that remains is to define an appropriate set of gauge conditions that will determine the ``slicing'' of the  spacetime by specifying the requisite \textit{driving conditions} for the lapse function ${\alpha}$ and the shift vector ${\beta^i}$\cite{alcubierre}, and where the optimal choice is ultimately dependent on the detailed physics of the phenomenon being simulated. For the purposes of this article, we shall principally choose the general \textit{alpha-driver}\cite{balakrishna}\cite{bona4} condition for the lapse function, which can be written as a first-order differential equation of the form:

\begin{equation}
\partial_t \alpha = - \mu_{\alpha_1} \alpha^{\mu_{\alpha_2}} \left( K - 2 \Theta \right) + \mu_{\alpha_3} \beta^k \partial_k \alpha,
\end{equation}
for arbitrary scalar parameters ${\mu_{\alpha_1}}$, ${\mu_{\alpha_2}}$ and ${\mu_{\alpha_3}}$, such that the standard \textit{1 + log} slicing condition (used in this article for the stable simulation of black hole inspirals), namely:

\begin{equation}
\partial_t \alpha = -2 \alpha \left( K - 2 \Theta \right) + \beta^k \partial_k \alpha,
\end{equation}
corresponds to the case ${\mu_{\alpha_1} = 2}$, ${\mu_{\alpha_2} = 1}$ and ${\mu_{\alpha_3} = 1}$\cite{alcubierre2}. However, for problems involving spherical symmetry, we opt instead to use the much simpler \textit{maximal slicing} condition\cite{alcubierre3}\cite{alcubierre4}, defined by a second-order differential equation of the form:

\begin{equation}
\nabla^2 \alpha = \alpha K_{i j} K^{i j},
\end{equation}
which trivially preserves the extrinsic curvature quantities:

\begin{equation}
K - 2 \Theta = 0, \qquad \text{ and } \qquad \partial_t \left( K - 2 \Theta \right) = 0,
\end{equation}
across all slices. On the other hand, we use the general \textit{gamma-driver} conditions for the shift vector\cite{anninos}\cite{smarr}, which can be written in terms of an auxiliary vector field ${B^i}$ and arbitrary scalar parameters ${\eta_1}$, ${\eta_2}$, ${\mu_{\beta_1}}$ and ${\mu_{\beta_2}}$ as:

\begin{equation}
\partial_t \beta^i = \eta_1 B^i, \qquad \text{ with } \qquad \partial_t B^i = \mu_{\beta_1} \alpha^{\mu_{\beta_2}} \partial_t \hat{\Gamma}^i - \eta_2 B^i,
\end{equation}
in the simplified case (without advection terms), or as:

\begin{equation}
\partial_t \beta^i = \eta_1 B^i + \beta^k \partial_k \beta^i, \qquad \text{ with } \qquad \partial_t B^i = \mu_{\beta_1} \alpha^{\mu_{\beta_2}} \partial_t \hat{\Gamma}^i - \beta^k \partial_k \hat{\Gamma}^i + \beta^k \partial_k B^i - \eta_2 B^i,
\end{equation}
in the more general case (with advection terms). The usual hyperbolic driver condition involves choosing the parameters ${\eta_1 = \frac{3}{4}}$, ${\mu_{\beta_1} = 1}$, ${\mu_{\beta_2} = 0}$ and ${\eta_2 = 1}$, although in some cases the choice ${\eta_1 = 1}$ is more robust, since the ${\eta_1 = \frac{3}{4}}$ may result in weak hyperbolicity of the system when the lapse function ${\alpha}$ approaches 1. For the most part, the tests shown in this article use the choice ${\eta_1 = \frac{3}{4}}$, since it makes manifest the connection between the CCZ4 formulation and the celebrated BSSN formulation of Baumgarte, Shapiro, Shibata and Nakamura\cite{nakamura}.

In order to see how the standard BSSN equations can be recovered as a limiting case of the CCZ4 formulation, it is first helpful to recall how the BSSN equations are derived from the canonical ${3 + 1}$ ADM decomposition\cite{york}, consisting of evolution equations for the spatial metric tensor ${\gamma_{i j}}$:

\begin{equation}
\left( \partial_t - \mathcal{L}_{\beta} \right) \gamma_{i j} = -2 \alpha K_{i j},
\end{equation}
and the extrinsic curvature tensor ${K_{i j}}$:

\begin{equation}
\left( \partial_t - \mathcal{L}_{\beta} \right) K_{i j} = - \nabla_i \nabla_j \alpha + \alpha \left( R_{i j} + K K_{i j} - 2 K_{i k} K^{k}_{j} \right),
\end{equation}
with Hamiltonian constraint (in the vacuum case):

\begin{equation}
H = R + K^2 - K_{i j} K^{i j} = 0,
\end{equation}
and momentum constraint (in the vacuum case):

\begin{equation}
M^i = \nabla_j \left( K^{i j} - \gamma^{i j} K \right) = 0,
\end{equation}
where ${\nabla_i}$ denotes, as usual, the covariant derivative operator on a given spacelike hypersurface, and ${R_{i j}}$ denotes the spatial Ricci curvature tensor. Specifically, one constructs a conformal spatial metric tensor ${\tilde{\gamma}_{i j}}$ in the usual way (i.e. by applying a conformal transformation with scale factor ${\psi}$):

\begin{equation}
\gamma_{i j} = \psi^4 \tilde{\gamma}_{i j},
\end{equation}
with the factor ${\psi}$ chosen so as to ensure that the determinant of the conformal metric tensor ${\tilde{\gamma}_{i j}}$ is identically 1, i.e:

\begin{equation}
\psi = \mathrm{det} \left( \gamma_{i j} \right)^{\frac{1}{12}}, \qquad \tilde{\gamma}_{i j} = \psi^{-4} \gamma_{i j} = \mathrm{det} \left( \gamma_{i j} \right)^{- \frac{1}{3}} \gamma_{i j},
\end{equation}
such that ${\mathrm{det} \left( \gamma_{i j} \right) = 1}$. Now, instead of evolving the spatial metric tensor ${\gamma_{i j}}$ and the extrinsic curvature tensor ${\kappa_{i j}}$, we can involve instead the conformal spatial metric tensor ${\tilde{\gamma}_{i j}}$ (plus a scalar function ${\phi}$ of the conformal factor ${\psi}$):

\begin{equation}
\tilde{\gamma}_{i j} = e^{-4 \phi} \gamma_{i j}, \qquad \text{ with } \qquad \phi = \ln \left( \psi \right) = \frac{1}{2} \ln \left( \mathrm{det} \left( \gamma_{i j} \right) \right),
\end{equation}
and the conformal version of the trace-free part of the extrinsic curvature tensor ${\tilde{A}_{i j}}$ (plus the extrinsic curvature scalar $K$):

\begin{equation}
\tilde{A}_{i j} = e^{-4 \phi} A_{i j}, \qquad \text{ with } \qquad K = \gamma_{i j} K^{i j},
\end{equation}
where, as usual, ${A_{i j}}$ is just the ordinary trace-free part:

\begin{equation}
A_{i j} = K_{i j} - \frac{1}{3} \gamma_{i j} K,
\end{equation}
such that ${\mathrm{det} \left( \tilde{\gamma}_{i j} \right) = 1}$ and ${\mathrm{tr} \left( \tilde{A}_{i j} \right) = 0}$. We also use the standard conformal connection functions ${\tilde{\Gamma}^i}$, as introduced previously:

\begin{equation}
\tilde{\Gamma}^i = \tilde{\gamma}^{j k} \tilde{\Gamma}_{j k}^{i} = - \partial_j \tilde{\gamma}^{i j},
\end{equation}
with ${\tilde{\Gamma}^{i}_{j k}}$ being the Christoffel symbols of the conformal metric, and with the latter equality contingent upon the fact that ${\mathrm{det} \left( \tilde{\gamma}_{i j} \right) = 1}$.

In terms of the BSSN variables ${\left\lbrace \phi, K, \tilde{\gamma}_{i j}, \tilde{A}_{i j}, \tilde{\Gamma}^i \right\rbrace}$, the ADM evolution equation for the spatial metric tensor ${\gamma_{i j}}$ now decomposes into a pair of evolution equations for the conformal spatial metric tensor ${\tilde{\gamma}_{i j}}$ and the scalar function ${\phi}$ of the conformal factor ${\psi}$, namely:

\begin{equation}
\left( \partial_t - \mathcal{L}_{\beta} \right) \tilde{\gamma}_{i j} = -2 \alpha \tilde{A}_{i j}, \qquad \text{ and } \qquad \left( \partial_t - \mathcal{L}_{\beta} \right) \phi = - \frac{1}{6} \alpha K,
\end{equation}
respectively. Next, the ADM evolution equation for the extrinsic curvature tensor ${K_{i j}}$ decomposes into a pair of evolution equations for the conformal version of the trace-free part of the extrinsic curvature tensor ${\tilde{A}_{i j}}$ and the extrinsic curvature scalar $K$, namely:

\begin{equation}
\left( \partial_t - \mathcal{L}_{\beta} \right) \tilde{A}_{i j} = e^{-4 \phi} \left[ - \nabla_i \nabla_j \alpha + \alpha R_{i j} \right]^{TF} + \alpha \left( K \tilde{A}_{i j} - 2 \tilde{A}_{i k} \tilde{A}_{j}^{k} \right),
\end{equation}
and:

\begin{equation}
\left( \partial_t - \mathcal{L}_{\beta} \right) K = - \nabla^i \nabla_i \alpha + \alpha \left( \tilde{A}_{i j} \tilde{A}^{i j} + \frac{1}{3} K^2 \right),
\end{equation}
respectively, where ${\left[ \cdot \right]^{TF}}$ denotes, as usual, the trace-free part of the expression with respect to the spatial metric tensor ${\gamma_{i j}}$, and where we have used the vacuum Hamiltonian constraint:

\begin{equation}
H = R + K^2 - K_{i j} K^{i j} = 0,
\end{equation}
as a means of eliminating the Ricci scalar term $R$ from the right-hand side of the evolution equation for ${\tilde{A}_{i j}}$, i.e:

\begin{equation}
R = K_{i j} K^{i j} - K^2 = \tilde{A}_{i j} \tilde{A}^{i j} - \frac{2}{3} K^2.
\end{equation}
The ADM momentum constraint yields the following constraint equation for the conformal version of the trace-free part of the extrinsic curvature tensor ${\tilde{A}_{i j}}$:

\begin{equation}
\partial_j \tilde{A}^{i j} = - \tilde{\Gamma}_{j k}^{i} \tilde{A}^{j k} - 6 \tilde{A}^{i j} \partial_j \phi + \frac{2}{3} \tilde{\gamma}^{i j} \partial_j K,
\end{equation}
whilst an evolution equation for the conformal connection functions ${\tilde{\Gamma}^i}$ can be derived as a consequence of the evolution equation for the conformal spatial metric tensor ${\tilde{\gamma}_{i j}}$, namely:

\begin{equation}
\partial_t \tilde{\Gamma}^i = -2 \left( \alpha \partial_j \tilde{A}^{i j} + \tilde{A}^{i j} \partial_j \alpha \right) - \partial_j \left( \mathcal{L}_{\beta} \tilde{\gamma}^{i j} \right).
\end{equation}

Thus, we see that the BSSN formalism requires the Hamiltonian constraint to be satisfied exactly, in order for the Ricci scalar $R$ to be eliminated from the right-hand side of the evolution equation for the conformal version of the trace-free part of the extrinsic curvature tensor ${\tilde{A}_{i j}}$, whereas the CCZ4 system derives the evolution equation for ${\tilde{A}_{i j}}$ as an immediate corollary from the trace-free part of the evolution equation for the extrinsic curvature tensor ${K_{i j}}$ in the ordinary ADM decomposition, at least when combined with additional terms dependent upon the 4-vector ${Z_i}$ and its normal projection ${\Theta}$. Moreover, the trace of the extrinsic curvature tensor in the BSSN formalism, henceforth denoted ${K^{BSSN}}$, is related to the extrinsic curvature scalar $K$ in the CCZ4 formulation by:

\begin{equation}
K^{BSSN} = K - 2 \Theta,
\end{equation}
for much the same reason (i.e. it follows from using the Hamiltonian constraint to remove the dependency on the Ricci scalar $R$ in BSSN, as opposed to using the trace-free part of the evolution equation for the extrinsic curvature tensor ${K_{i j}}$ in the ADM decomposition plus ${Z_i}$ and ${\Theta}$ terms). Therefore, we can see that we immediately recover the standard BSSN equations in the special case where ${\Theta = 0}$ and ${Z^i = 0}$, and where the Hamiltonian constraint is used to remove the explicit dependence on the Ricci scalar term from the evolution equation for the extrinsic curvature scalar $K$ in the CCZ4 formulation:

\begin{equation}
\partial_t K = - \nabla^i \nabla_i \alpha + \alpha \left( R + 2 \nabla_i Z^i + K^2 - 2 \Theta K \right) + \beta^j \beta_j K - 3 \alpha \kappa_1 \left( 1 + \kappa_2 \right) \Theta + 4 \pi \alpha \left( S - 3 \tau \right),
\end{equation}
as required.

As mentioned above, the two additional scalar fields that are present in the CCZ4 system but absent in the Z4 system, namely ${\mathrm{det} \left( \tilde{\gamma}_{i j} \right)}$ and ${\mathrm{tr} \left( \tilde{A}_{i j} \right)}$, thus introduce an additional pair of consistency constraints:

\begin{equation}
\mathrm{det} \left( \tilde{\gamma}_{i j} \right) = 1, \qquad \text{ and } \qquad \mathrm{tr} \left( \tilde{A}_{i j} \right) = 0,
\end{equation}
which must now be enforced. It is tempting simply to relax these constraints, enforcing them only for the data defined on the initial hypersurface, thus introducing a pair of additional dynamical modes\cite{alcubierre5}. Since the traces of the evolution equations for the conformal spatial metric tensor ${\tilde{\gamma}_{i j}}$:

\begin{equation}
\partial_t \tilde{\gamma}_{i j} = -2 \alpha \tilde{A}_{i j} +2 \tilde{\gamma}_{k \left( i \right.} \partial_{\left. j \right)} \beta^k - \frac{2}{3} \tilde{\gamma}_{i j} \partial_k \beta^k + \beta^k \partial_k \tilde{\gamma}_{i j},
\end{equation}
and the trace-free components of the extrinsic curvature tensor ${\tilde{A}_{i j}}$:

\begin{multline}
\partial_t \tilde{A}_{i j} = \phi^2 \left[ - \nabla_i \nabla_j \alpha + \alpha \left( R_{i j} + \nabla_i Z_j + \nabla_j Z_i - 8 \pi S_{i j} \right) \right]^{TF} + \alpha \tilde{A}_{i j} \left( K - 2 \Theta \right) - 2 \alpha \tilde{A}_{i l} \tilde{A}_{j}^{l}\\
+ 2 \tilde{A}_{k \left( i \right.} \partial_{\left. j \right)} \beta^k - \frac{2}{3} \tilde{A}_{i j} \partial_k \beta^k + \beta^k \partial_k \tilde{A}_{i j},
\end{multline}
are both trivial, these new dynamical modes will always propagate along characteristic lines that are normal to the spacelike hypersurfaces. However, in this case the trace of the ${-2 \alpha \tilde{A}_{i j}}$ term in the first evolution equation must be kept under control in order to prevent spurious numerical oscillations, by instead evolving an equation of the form:

\begin{equation}
\partial_t \tilde{\gamma}_{i j} = -2 \alpha \left( \tilde{A}_{i j} - \frac{1}{3} \tilde{\gamma}_{i j} \tilde{A}_{k l} \tilde{\gamma}^{k l} \right) + 2 \tilde{\gamma}_{k \left( i \right.} \partial_{\left. j \right)} \beta^k - \frac{2}{3} \tilde{\gamma}_{i j} \partial_k \beta^k + \beta^k \partial_k \tilde{\gamma}_{i j}.
\end{equation}
Similarly, the trace of ${\tilde{A}_{i j}}$ in the second evolution equation must also be kept under control, for instance by introducing a damping term that is proportional to ${\tilde{\gamma}_{i j} \mathrm{tr} \left( \tilde{A}_{i j} \right)}$. Consequently, it is in many respects simpler just to enforce the constraints explicitly as part of the numerical integration algorithm by simply removing the trace of ${\tilde{A}_{i j}}$ and rescaling the conformal spatial metric tensor ${\tilde{\gamma}_{i j}}$ accordingly, such that the characteristic structure of the dynamical modes remains the same as in the original Z4 system. This is the standard approach adopted in the context of the BSSN formalism, and it is the approach that we opt for in this article.

For the case of simulations involving black holes, we choose to deal with the singularities using the \textit{moving punctures method}\cite{brandt}, based on the work of Misner and Wheeler\cite{misner} (and later Brill and Lindquist\cite{brill}), in which the metric is effectively ``factorized'' into an analytical part containing the singularity, and a numerical part that remains singularity-free. More specifically, having rescaled the spatial metric tensor ${\gamma_{i j}}$ and the extrinsic curvature tensor ${K_{i j}}$ by the Brill-Lindquist conformal factor ${\psi_{BL}}$ in the usual way:

\begin{equation}
\gamma_{i j} = \psi_{BL}^{4} \tilde{\gamma}_{i j}, \qquad \text{ and } \qquad K_{i j} = \psi_{BL}^{-2} \tilde{K}_{i j},
\end{equation}
the Hamiltonian constraint reduces to an elliptic differential equation for the scalar field ${\psi_{BL}}$:

\begin{equation}
\nabla^2 \psi_{BL} + \frac{1}{8} \tilde{K}^{i j} \tilde{K}_{i j} \psi_{BL}^{-7} = 0,
\end{equation}
whilst the momentum constraint reduces to a simple divergence-free condition on the extrinsic curvature:

\begin{equation}
\nabla_i \tilde{K}^{i j} = 0.
\end{equation}
Henceforth, we shall assume a black hole with linear momentum ${P^i}$ and angular momentum/spin ${S^i}$, described by the following solution ${\tilde{K}_{PS}^{i j}}$ to the momentum constraint:

\begin{equation}
\tilde{K}_{PS}^{i j} = \frac{3}{2 r^2} \left( P^i n^j + P^j n^i - \left( \tilde{\gamma}^{i j} - n^a n^b \right) P^k n_k \right) + \frac{3}{r^3} \left( \epsilon^{i k l} S_k n_l n^j + \epsilon^{j k l} S_k n_l n^i \right),
\end{equation}
with radial normal vector ${n^i}$, and with the standard relationship between the spherical and Cartesian coordinate systems:

\begin{equation}
r = \sqrt{x^2 + y^2 + z^2}, \qquad \text{ and } \qquad n^i = \frac{x^i}{r},
\end{equation}
due to the conformal flatness of the spatial metric tensor ${\tilde{\gamma}_{i j}}$. By time symmetry, we can write the general Schwarzschild solution to the constraint equations for a set of $N$ black holes as:

\begin{equation}
\psi_{BL} = 1 + \sum_{i = 1}^{N} \frac{m_{\left( i \right)}}{2 \left\lvert \mathbf{r} - \mathbf{r}_{\left( i \right)} \right\rvert}, \qquad \text{ with } \qquad \tilde{K}_{i j} = 0,
\end{equation}
where ${m_{\left( i \right)}}$ denotes the mass of the $i$th black hole, such that the total ADM mass is simply ${\sum\limits_{i = 1}^{N} m_{\left( i \right)}}$, and where ${\mathbf{r}_{\left( i \right)}}$ denotes the coordinate position of the $i$th black hole. Topologically, we interpret this Misner-Wheeler-Brill-Lindquist (MWBL) solution to be homeomorphic to ${\mathbb{R}^3}$ with the points ${\mathbf{r} = \mathbf{r}_{\left( i \right)}}$ (known as \textit{punctures}) removed, thus preserving regularity of the Brill-Lindquist conformal factor ${\psi_{BL}}$ at the expense of isometry of the Schwarzschild spacetime.

Now, following the approach of Brandt and Br\"ugmann\cite{brandt}, we can solve the momentum constraint equation by using the following form of the conformal extrinsic curvature tensor ${\tilde{K}^{j k}}$:

\begin{equation}
\tilde{K}^{j k} = \sum_{i = 1}^{N} \tilde{K}^{j k}_{PS \left( i \right)},
\end{equation}
with each black hole having its own origin point ${\mathbf{r}_{\left( i \right)}}$, its own linear momentum ${\mathbf{P}_{\left( i \right)}}$ and its own angular momentum/spin ${\mathbf{S}_{\left( i \right)}}$, with these quantities yielding the standard ADM parameters in the limit of infinite black hole separation. Next, we can use the solve the Hamiltonian constraint equation using the known value of the conformal extrinsic curvature tensor ${\tilde{K}_{i j}}$ by rewriting the Brill-Lindquist conformal factor ${\psi_{BL}}$ in terms of scalar functions $u$ and ${\alpha}$ of the form:

\begin{equation}
\psi_{BL} = \frac{1}{\alpha} + u, \qquad \text{ with } \qquad \frac{1}{\alpha} = \sum_{i = 1}^{N} \frac{m_{\left( i \right)}}{2 \left\lvert \mathbf{r} - \mathbf{r}_{\left( i \right)} \right\rvert}.
\end{equation}
On a hypersurface that is homeomorphic to ${\mathbb{R}^3}$ with punctures, the Laplacian of the ${\frac{1}{\alpha}}$ term is identically zero, such that the Hamiltonian constraint reduces to:

\begin{equation}
\nabla^2 u + \beta \left( 1 + \alpha u \right)^{-7} = 0, \qquad \text{ with } \qquad \beta = \frac{1}{8} \alpha^7 \tilde{K}^{i j} K_{i j},
\end{equation}
with the following boundary condition imposed at a sufficiently large distance from the punctures:

\begin{equation}
u - 1 = O \left( r^{-1} \right),
\end{equation}
so as to guarantee asymptotic flatness at infinity. The ``factorization'' of the spatial metric tensor ${\gamma_{i j}}$ is therefore given simply by:

\begin{equation}
\gamma_{i j} = \left( \psi_{BL} + u \right)^4 \delta_{i j}.
\end{equation}
First, we show that the conformal spatial metric becomes flat as we approach the punctures, by applying the following coordinate inversion through a sphere enclosing the $i$th puncture:

\begin{equation}
r_{inv} = \frac{a^2}{r},
\end{equation}
for some parameter $a$, in order to show that the $i$th puncture is a point lying at infinity within some alternative asymptotically flat spacetime. The metric line element transforms as follows:

\begin{equation}
ds^2 = \psi^3 \left( dr^2 + r^2 d \Omega^2 \right) = \psi_{inv}^{4} \left( d r_{inv} + r_{inv}^{2} d \Omega^2 \right),
\end{equation}
under this choice of coordinate transformation, where we have defined a non-isometric transformation of the conformal factor: 

\begin{equation}
\psi_{inv} = \psi \frac{r}{a}.
\end{equation}
Since we chose above the following form for the Brill-Lindquist conformal factor ${\psi_{BL}}$:

\begin{equation}
\psi_{BL} = \sum_{i = 1}^{N} \frac{m_{\left( i \right)}}{2 \left\lvert \mathbf{r} - \mathbf{r}_{\left( i \right)} \right\rvert} + u,
\end{equation}
it follows that, under the choice:

\begin{equation}
a = \frac{m_{\left( i \right)}}{2},
\end{equation}
for a puncture of bare mass ${m_{\left( i \right)}}$, we obtain, after coordinate inversion:

\begin{equation}
\psi_{inv} = 1 + \frac{m_{\left( i \right) inv}}{2 r_{inv}} + O \left( \frac{1}{r_{inv}^{2}} \right),
\end{equation}
where:

\begin{equation}
m_{\left( i \right) inv} = m_{\left( i \right)} \left( u \left( \mathbf{r}_{\left( i \right)} \right) + \sum_{j \neq i} \frac{m_{\left( j \right)}}{2 \left\lvert \mathbf{r}_{\left( i \right)} - \mathbf{r}_{\left( j \right)} \right\rvert} \right),
\end{equation}
demonstrating that the spatial metric becomes conformally flat in the vicinity of the punctures, as required.

All that remains is to prove that the metric is also asymptotically flat \textit{at} the punctures, i.e. that the conformal extrinsic curvature tensor after coordinate inversion ${\tilde{K}_{i j}^{inv}}$ is of the form:

\begin{equation}
\tilde{K}_{i j}^{inv} = O \left( r_{inv}^{-2} \right).
\end{equation}
With respect to the original physical (non-conformal) variables, the coordinate transformation is trivial, namely:

\begin{equation}
r_{inv} = \frac{a^2}{r}, \qquad \text{ and } \qquad \Lambda_{k}^{j} = \frac{\partial x^j}{\partial x_{inv}^{k}} = \left( \frac{a^2}{r_{inv}^{2}} \right) L_{k}^{j},
\end{equation}
where we have introduced the following tensor quantity:

\begin{equation}
L_{k}^{j} = \delta_{k}^{j} - 2 n^j n_k.
\end{equation}
However, with respect to the newly-introduced conformal variables, one must also account for the transformation of the Briller-Lindquist conformal factor ${\psi_{BL}}$ by using:

\begin{equation}
\tilde{K}_{i j} = \psi_{inv}^{2} K_{i j}^{inv}, \qquad \text{ where } \qquad K_{i j}^{inv} = \left( \Lambda K \right)_{i j},
\end{equation}
to deduce that:

\begin{equation}
K_{i j}^{inv} = \left( \frac{a}{r_{inv}} \right)^6 \left( L K \right)_{i j}.
\end{equation}
Consequently, when applied to our explicit solution to the momentum constraint equation for a single black hole with linear momentum ${P^i}$ and angular momentum/spin ${S^i}$, namely:

\begin{equation}
\tilde{K}_{PS}^{i j} = \frac{3}{2 r^2} \left( P^i n^j + P^j n^i - \left( \tilde{\gamma}^{i j} - n^i n^j \right) P^k n_k \right) + \frac{3}{r^3} \left( \epsilon^{i k l} S_k n_l n^j + \epsilon^{j k l} S_k n_l n^i \right),
\end{equation}
we see that the coordinate inversion maps the linear momentum terms of order ${r^{-2}}$ to corresponding terms of order ${r_{inv}^{-4}}$, and the angular momentum/spin terms of order ${r^{-3}}$ to corresponding terms of order ${r_{inv}^{-3}}$. This analysis can be extended to the more general case of multiple black holes using our previous prescription:

\begin{equation}
K^{j k} = \sum_{i = 1}^{N} K_{PS \left( i \right)}^{j k},
\end{equation}
thus demonstrating asymptotic flatness at the punctures, as required. For this reason, the moving punctures strategy (combined with the \textit{1 + log} slicing condition for the lapse function ${\alpha}$ and the \textit{gamma-driver} gauge condition for the shift vector ${\beta^i}$) is used for dealing with all black hole singularities described in the present article, with the positivity of the lapse function ${\alpha > 0}$ being explicitly hard-coded.

\section{Numerical Integration and Refinement Algorithms}
\label{sec:Section3}

The set of fields on any given spacelike hypersurface forms a state vector ${\Phi \left( \mathbf{x}, t \right)}$ (assuming local spatial coordinates ${\mathbf{x}}$ and time coordinate $t$) of the general form:

\begin{equation}
\Phi \left( \mathbf{x}, t \right) = \left( \phi_1 \left( \mathbf{x}, t \right), \phi_2 \left( \mathbf{x}, t \right), \phi_3 \left( \mathbf{x}, t \right), \dots \right),
\end{equation}
which can, in turn, be evolved in time by solving a general set of equations of motion:

\begin{equation}
\frac{\partial \Phi \left( \mathbf{x}, t \right)}{\partial t} = \mathcal{F} \left( \Phi \left( \mathbf{x}, t \right) \right),
\end{equation}
for some non-linear operator on ${\mathcal{F}}$ acting on ${\Phi}$ (with the non-linearity of the operator ${\mathcal{F}}$ following as a consequence of the non-linearity of the Einstein field equations themselves). A solution is obtained by discretizing the space and time coordinates ${\left( \mathbf{x}, t \right)}$ and then applying a finite difference method in order to achieve explicit time stepping ${t \to t + \Delta t}$, with the fluxes at each node in the coordinate grid being computed anew at each time step\cite{calabrese}. More specifically, such a system can then be solved as a simple initial value problem of the form:

\begin{equation}
\frac{d \Phi \left( \mathbf{x}, t \right)}{d t} = F \left( t, \Phi \left( \mathbf{x}, t \right) \right), \qquad \text{ with } \qquad \Phi \left( \mathbf{x}, t_0 \right) = \Phi_0,
\end{equation}
using for instance a standard fourth-order Runge-Kutta explicit numerical method\cite{kreiss2}\cite{levy}, by selecting a finite step-size ${\Delta t > 0}$ satisfying the Courant condition, and introducing the following discretized variables:

\begin{equation}
\Phi \left( \mathbf{x}, t_{n + 1} \right) = \Phi \left( \mathbf{x}, t_n \right) + \frac{1}{6} \Delta t \left( k_1 + 2 k_2 + 2 k_3 + k_4 \right), \qquad \text{ and } \qquad t_{n + 1} = t_n + \Delta t,
\end{equation}
with:

\begin{equation}
k_1 = F \left( t_n, \Phi \left( \mathbf{x}, t_n \right) \right), \qquad k_4 = F \left( t_n + \Delta t, \Phi \left( \mathbf{x}, t_n \right) + \Delta k_3 \right),
\end{equation}
and:

\begin{equation}
k_2 = F \left( t_n + \frac{\Delta t}{2}, \Phi \left( \mathbf{x}, t_n \right) + \Delta t \frac{k_1}{2} \right), \qquad k_3 = F \left( t_n + \frac{\Delta t}{2}, \Phi \left( \mathbf{x}, t_n \right) + \Delta t \frac{k_2}{2} \right).
\end{equation}

By discretizing the spatial coordinates ${\mathbf{x}}$ across three coordinate axes indexed by $i$, $j$ and $k$, one therefore obtains a discrete form of the state vector ${\Phi_{i, j, k}^{n}}$ at time ${t_n}$, with ${\Delta x = x_{i + 1} - x_i}$, ${\Delta y = y_{j + 1} - y_j}$, ${\Delta z = z_{k + 1} - z_k}$, for which we use the following standard fourth-order centered finite difference stencils (as proposed by Zlochower, Baker Campanelli and Lousto\cite{zlochower}) for the first derivatives (in $x$):

\begin{equation}
\partial_x F \left( t_n, \Phi_{i, j, k}^{n} \right) = \frac{1}{12 \Delta x} \left( F \left( t_n, \Phi_{i - 2, j, k}^{n} \right) - 8 F \left( t_n, \Phi_{i - 1, j, k}^{n} \right) + 8 F \left( t_n, \Phi_{i + 1, j, k}^{n} \right) - F \left( t_n, \Phi_{i + 2, j, k}^{n} \right) \right),
\end{equation}
the second derivatives (in $x$):

\begin{multline}
\partial_{x x} F \left( t_n, \Phi_{i, j, k}^{n} \right) = \frac{1}{12 \left( \Delta x \right)^2} \left( - F \left( t_n, \Phi_{i + 2, j, k}^{n} \right) + 16 F \left( t_n, \Phi_{i + 1, j, k}^{n} \right) - 30 F \left( t_n, \Phi_{i, j, k}^{n} \right) \right.\\
\left. + 16 F \left( t_n, \Phi_{i - 1, j, k}^{n} \right) - F \left( t_n, \Phi_{i - 2, j, k}^{n} \right) \right),
\end{multline}
and mixed derivatives (in $x$ and $y$):

\begin{multline}
\partial_{x y} F \left( t_n, \Phi_{i, j, k}^{n} \right) = \frac{1}{144 \Delta x \Delta y} \left[ F \left( t_n, \Phi_{i - 2, j - 2, k}^{n} \right) - 8 F \left( t_n, \Phi_{i - 1, j - 2, k}^{n} \right) + 8 F \left( t_n, \Phi_{i + 1, j - 2, k}^{n} \right) \right.\\
\left. - F \left( t_n, \Phi_{i + 2, j - 2, k}^{n} \right) - 8 \left( F \left( t_n, \Phi_{i - 2, j - 1, k}^{n} \right) - 8 F \left( t_n, \Phi_{i - 1, j - 1, k}^{n} \right) + 8 F \left( t_n, \Phi_{i + 1, j - 1, k}^{n} \right) \right. \right.\\
\left. \left. - F \left( t_n, \Phi_{i + 2, j - 1, k}^{n} \right) \right) + 8 \left( F \left( t_n, \Phi_{i - 2, j + 1, k}^{n} \right) - 8 F \left( t_n, \Phi_{i - 1, j + 1, k}^{n} \right) + 8 F \left( t_n, \Phi_{i + 1, j + 1, k}^{n} \right) \right. \right.\\
\left. \left. - F \left( t_n, \Phi_{i + 2, j + 1, k}^{n} \right) \right) - \left( F \left( t_n, \Phi_{i - 2, j + 2, k}^{n} \right) - 8 F \left( t_n, \Phi_{i - 1, j + 2, k}^{n} \right) + 8 F \left( t_n, \Phi_{i + 1, j + 2, k}^{n} \right) \right. \right.\\
\left. \left. - F \left( t_n, \Phi_{i + 2, j + 2, k}^{n} \right) \right) \right],
\end{multline}
where the mixed derivative form is simply deduced by sequentially applying the $x$ and $y$ derivatives in any order. However, when advection terms of the form ${\beta^i \partial_i F \left( t_n, \Phi_{i, j, k}^{n} \right)}$ are present, we opt instead to use the following fourth-order upwinded finite difference stencil for the first derivatives (in $x$):

\begin{multline}
\partial_x F \left( t_n, \Phi_{i, j, k}^{n} \right) = \frac{1}{12 \Delta x} \left( - F \left( t_n, \Phi_{i - 3, j, k}^{n} \right) + 6 F \left( t_n, \Phi_{i - 2, j, k}^{n} \right) - 18 F \left( t_n, \Phi_{i - 1, j, k}^{n} \right) \right.\\
\left. + 10 F \left( t_n, \Phi_{i, j, k}^{n} \right) + 3 F \left( t_n, \Phi_{i + 1, j, k}^{n} \right) \right),
\end{multline}
whenever ${\beta^x < 0}$, and:

\begin{multline}
\partial_x F \left( t_n, \Phi_{i, j, k}^{n} \right) = \frac{1}{12 \Delta x} \left( F \left( t_n, \Phi_{i + 3, j, k}^{n} \right) - 6 F \left( t_n, \Phi_{i + 2, j, k}^{n} \right) + 18 F \left( t_n, \Phi_{i + 1, j, k}^{n} \right) \right.\\
\left. - 10 F \left( t_n, \Phi_{i, j, k}^{n} \right) -3 F \left( t_n, \Phi_{i - 1, j, k}^{n} \right) \right),
\end{multline}
whenever ${\beta^x > 0}$, and likewise for all other directional derivatives (and combinations thereof).

\begin{figure}[ht]
\centering
\includegraphics[width=0.495\textwidth]{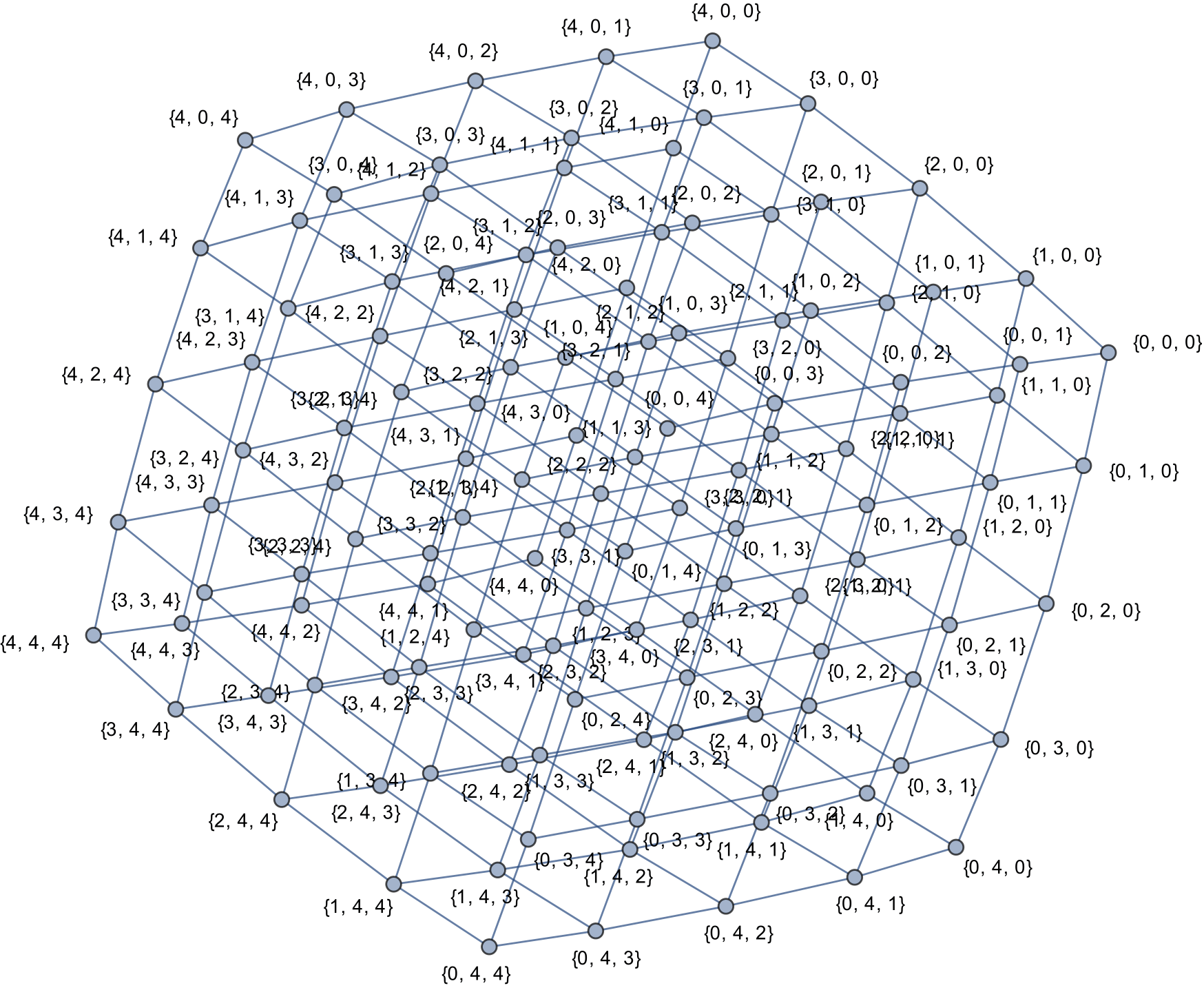}
\caption{A coordinate scheme for vertices in a three-dimensional grid-like spatial hypergraph.}
\label{fig:Figure1}
\end{figure}

\begin{figure}[ht]
\centering
\includegraphics[width=0.495\textwidth]{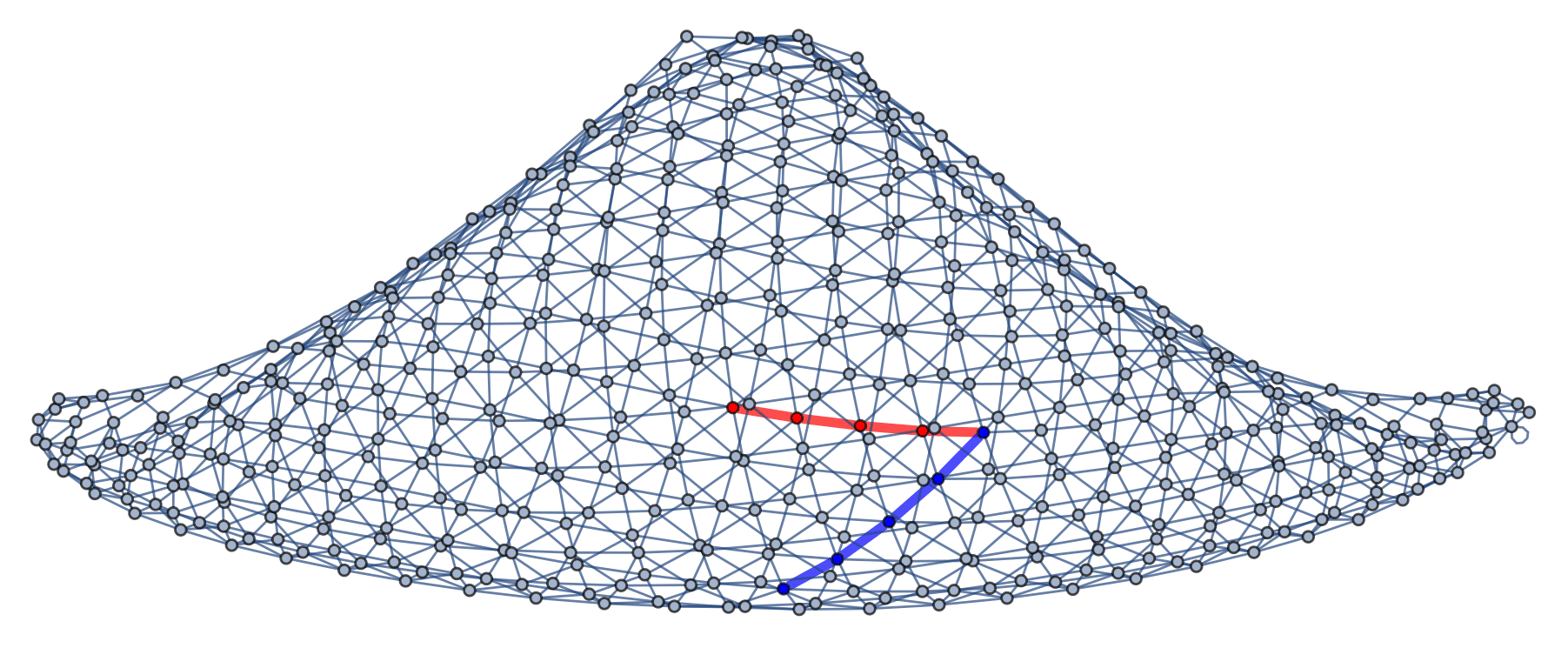}
\includegraphics[width=0.495\textwidth]{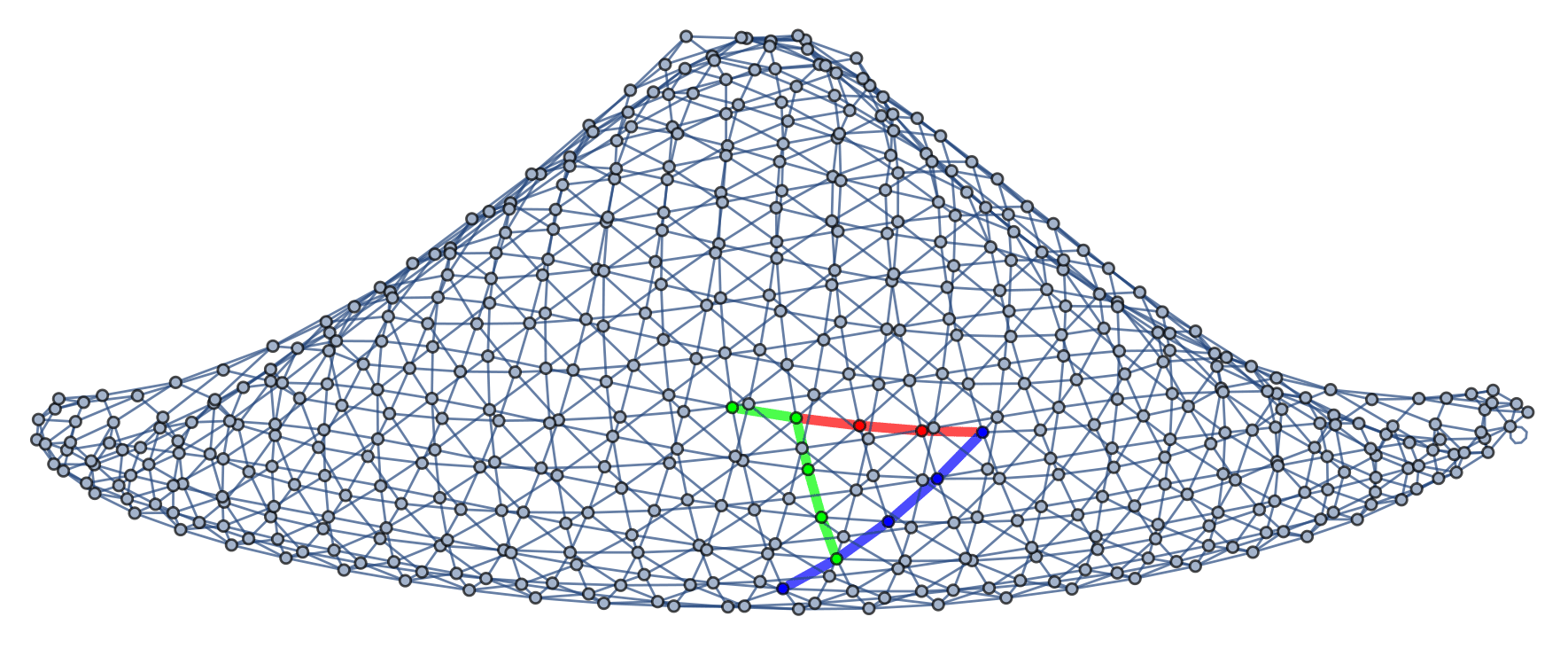}
\caption{On the left, a pair of geodesics in a spatial hypergraph with a two-dimensional Riemannian manifold-like limiting structure, as generated by the set substitution system ${\left\lbrace \left\lbrace x, x, y \right\rbrace, \left\lbrace x, z, w \right\rbrace \right\rbrace \to \left\lbrace \left\lbrace w, w, v \right\rbrace, \left\lbrace v, w, y \right\rbrace, \left\lbrace z, y, v \right\rbrace \right\rbrace}$. On the right, a ``perpendicular'' geodesic being ``dropped'' from the red geodesic onto the blue geodesic, corresponding to a value for the inner product of the red and blue geodesics of ${\frac{3}{4}}$.}
\label{fig:Figure2}
\end{figure}

For a spatial hypergraph with a regular three-dimensional grid-like structure, such as that depicted in Figure \ref{fig:Figure1}, introducing an appropriate coordinate scheme for the vertices in such a way that the fourth-order finite difference scheme described above can be applied is trivial. However, for more general spatial hypergraphs with non-trivial geometries and topologies, the problem of constructing a coordinate scheme becomes considerably more complex. Nevertheless, given a pair of graph geodesics ${g_1}$ and ${g_2}$ in an arbitrary spatial hypergraph, we can compute an \textit{inner product} ${\left\langle g_1, g_2 \right\rangle}$ by minimizing over the lengths of all geodesics ${g_n}$ from the endpoint of ${g_1}$ to any point on ${g_2}$; the construction of this minimum-length geodesic corresponds to a graph-theoretic analog of the operation of \textit{dropping a perpendicular} in elementary Euclidean geometry, thus allowing us to compute a \textit{projection} of ${g_1}$ onto ${g_2}$, hence equipping the hypergraph with an inner product structure. This procedure is demonstrated in Figure \ref{fig:Figure2}, in which the inner product between the red and green geodesics is shown (by means of the green geodesic) to be equal to ${\frac{3}{4}}$. This inner product structure is trivially positive-definite, and will satisfy the requisite axioms of linearity and conjugate symmetry if and only if the hypergraph has a reasonable interpretation as a continuous manifold in the continuum limit. The inner product structure consequently allows us to define a local set of orthonormal coordinate axes in the hypergraph, such that the finite difference scheme shown above becomes readily applicable.

\begin{figure}[ht]
\centering
\includegraphics[width=0.495\textwidth]{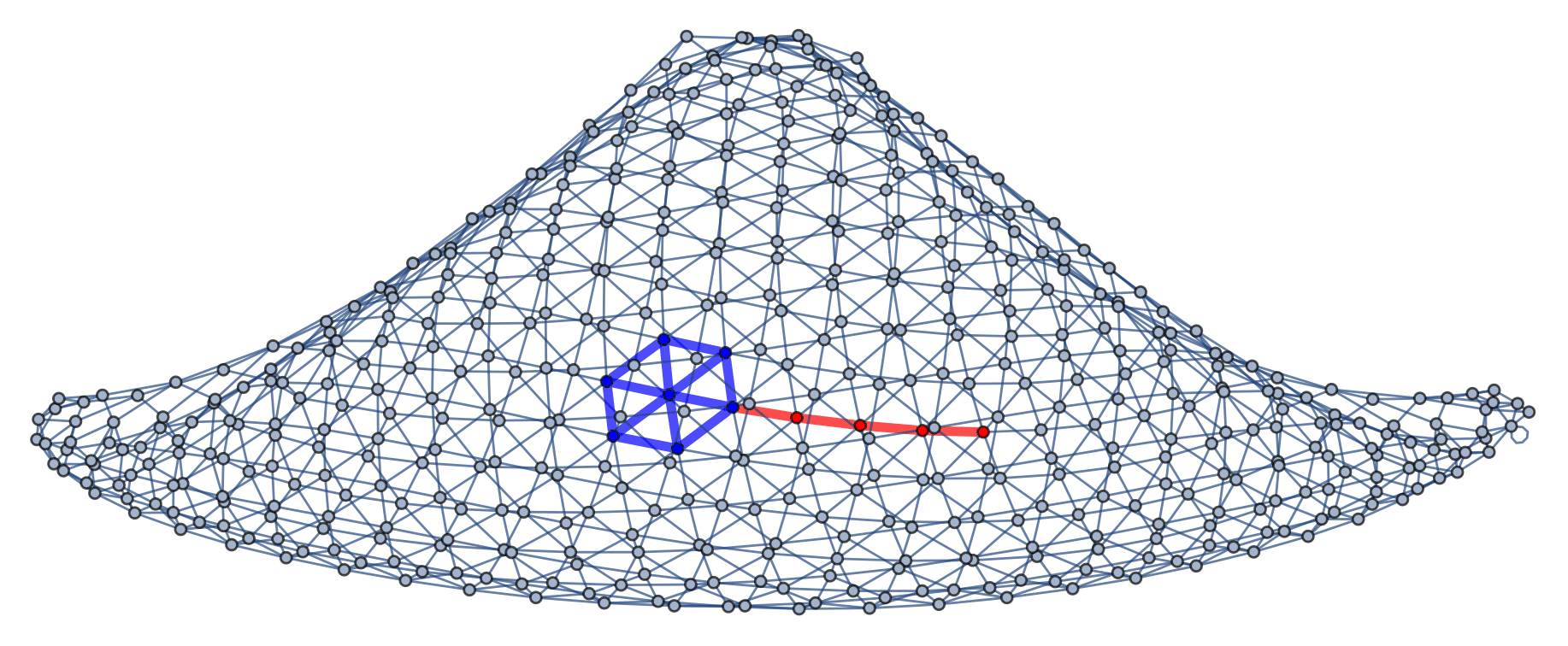}
\caption{Interpolating a value at the endpoint of the red geodesic using a generalized bicubic interpolation scheme over the 7 vertices contained within the blue neighborhood.}
\label{fig:Figure3}
\end{figure}

Since such a coordinate scheme depends upon geodesics of arbitrary extent, it no longer suffices (for reasons of continuity) to treat the endpoint of each geodesic as a single vertex for the purposes of the finite difference stencils - rather, one must interpolate a value at the endpoint using some finite neighborhood in the hypergraph. Depending upon the limiting dimension of the hypergraph, this can be done using some appropriate generalization of a bicubic:

\begin{equation}
f \left( x, y \right) = \sum_{i = 0}^{3} \sum_{j = 0}^{3} a_{i j} x^i y^j,
\end{equation}
or tricubic:

\begin{equation}
f \left( x, y, z \right) = \sum_{i = 0}^{3} \sum_{j = 0}^{3} \sum_{k = 0}^{3} a_{i j k} x^i y^j z^k,
\end{equation}
interpolation scheme\cite{lekien}, with ${2^{2 n}}$ coefficients $a$ (where $n$ is the limiting dimension of the hypergraph), as shown in Figure \ref{fig:Figure3}, in which a value at the endpoint of the red geodesic is interpolated using values at the 7 vertices contained within the blue neighborhood.

Due to the possibility of the appearance of spurious high-frequency modes in solutions produced using finite difference schemes of this type, and the resultant possibility of the formation of numerical instabilities, we employ the numerical dissipation scheme of Kreiss and Oliger\cite{kreiss}, in which the right-hand sides of the time evolution equations are modified to incorporate a Kreiss-Oliger dissipation term, i.e:

\begin{equation}
\partial_t \Phi \left( \mathbf{x}, t \right) \to \partial_t \Phi \left( \mathbf{x}, t \right) + Q \Phi \left( \mathbf{x}, t \right),
\end{equation}
with the general form of the Kreiss-Oliger dissipation operator $Q$ being:

\begin{equation}
Q = \frac{1}{2^{2 r}} \sigma \sum_{i} \left( -1 \right)^r \Delta x_{i}^{2r - 1} \left( D_{i +} \right)^r \rho \left( D_{i -} \right)^r.
\end{equation}
In the above, the dissipation operator is of the order ${2 r}$, which is appropriate for a finite difference scheme of order ${2 r - 2}$ accuracy, ${\sigma}$ is a parameter denoting the strength of dissipation (here, we take ${\sigma}$ to be on the order of ${10^{-2}}$), ${\rho}$ is a weighting function (within this article, ${\rho = 1}$ everywhere except the boundary of the spacetime, where ${\rho = 0}$), ${\Delta x_i}$ denotes the grid spacing in the $i$th direction and ${D_{i +}}$ and ${D_{i -}}$ are the forward and backward finite differencing operators in the $i$th direction, respectively. Thus, for the fourth-order accurate finite difference scheme presented here, the modified time evolution equations (including Kreiss-Oliger dissipation term) have the specific form:

\begin{multline}
\partial_t \Phi_{i, j k} \to \partial_t \Phi_{i, j, k} + \frac{\sigma}{64 \Delta x} \left( \Phi_{i + 3, j, k} - 6 \Phi_{i + 2, j, k} + \right.\\
\left. 15 \Phi_{i + 1, j, k} - 20 \Phi_{i, j, k} +15 \Phi_{i - 1, j, k} - 6 \Phi_{i - 2, j, k} + \Phi_{i - 3, j, k} \right),
\end{multline}
in the $x$-direction, and likewise for all other directional derivatives.

Finally, we employ a natural graph-theoretic generalization of the local adaptive mesh refinement (AMR) algorithm proposed by Berger and Colella\cite{berger}, building upon previous work of Berger and Oliger\cite{berger2} and Gropp\cite{gropp}, in which (presented in two dimensions, for simplicity) a PDE is discretized over a rectangular domain $D$ using a quadrilateral mesh with a sequence of refinement levels ${l = 1, 2, \dots, l_{max}}$:

\begin{equation}
G_l = \bigcup_k G_{l, k},
\end{equation}
where ${G_{l, k}}$ denotes a regular grid with mesh spacing ${\Delta x_l}$, with ${l = 1}$ corresponding to the coarsest such grid, such that:

\begin{equation}
D = G_1 = \bigcup_{k} G_{1, k},
\end{equation}
where we have used a slight abuse of notation to equate the domain $D$ with the grid ${G_1}$ that covers it. Each grid at level 1 must be a subset of the regular rectangular discretization of the entire domain $D$, so as to ensure that all of the grids correctly align. Overlapping grids $j$ and $k$ at level $l$:

\begin{equation}
G_{l, j} \cap G_{l, k} \neq \emptyset,
\end{equation}
where ${j \neq k}$, must also be configured in such a way that the finite difference approximation does not depend upon the details of the decomposition of $D$. As a consequence, a single point ${\left( x, y \right) \in D}$ may lie within multiple distinct grids, in which case the solution ${\Phi \left( x, y, t \right)}$ will be computed using (one of) the finest grid(s) containing ${\left( x, y \right)}$.

The hierarchical nesting of grids needs only to obey two axioms - the first being that fine grids must always be contained within the boundaries of a cell in the next coarser grid, and the second being that a cell (not at the domain boundary) at refinement level $l$ must be separated from a cell at refinement level ${l - 2}$ by at least one cell at refinement level ${l - 1}$, in any direction in the immediate von Neumann neighborhood. The grids are then refined in both time and space, using the same refinement ratio $r$:

\begin{equation}
r = \frac{\Delta x_{l - 1}}{\Delta x_l},
\end{equation}
such that one has:

\begin{equation}
\frac{\Delta t_l}{\Delta x_l} = \frac{\Delta t_{l - 1}}{\Delta x_{l - 1}} = \cdots = \frac{\Delta t_1}{\Delta x_1}.
\end{equation}
This is done so as to ensure numerical stability of the explicit finite difference scheme, since each finer mesh has a smaller Courant number, which must therefore be accounted for using:

\begin{equation}
\Delta t_{l + 1} = \frac{\Delta t^l}{n^l}, \qquad \text{ with } \qquad n^l = \frac{1}{r} = \frac{\Delta t^{l + 1}}{\Delta t^l}.
\end{equation}

\begin{figure}[ht]
\centering
\includegraphics[width=0.345\textwidth]{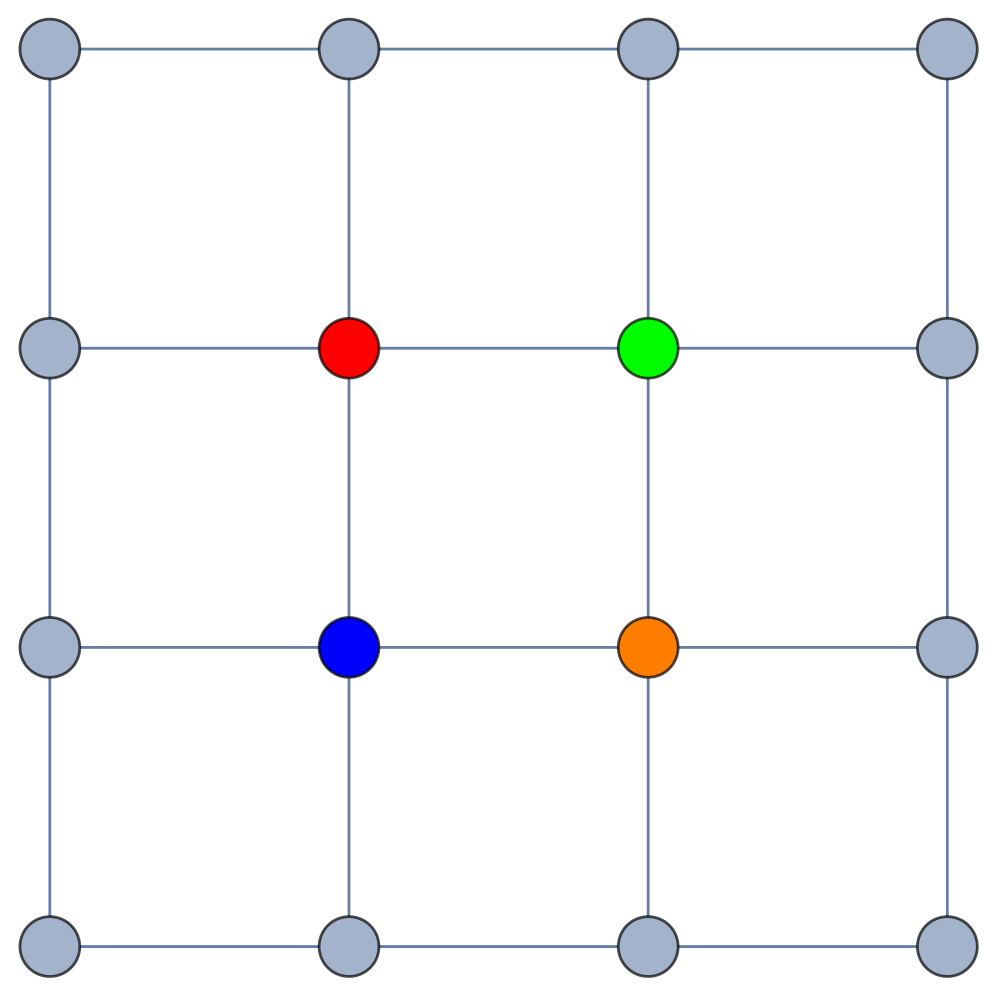}\hspace{0.1\textwidth}
\includegraphics[width=0.395\textwidth]{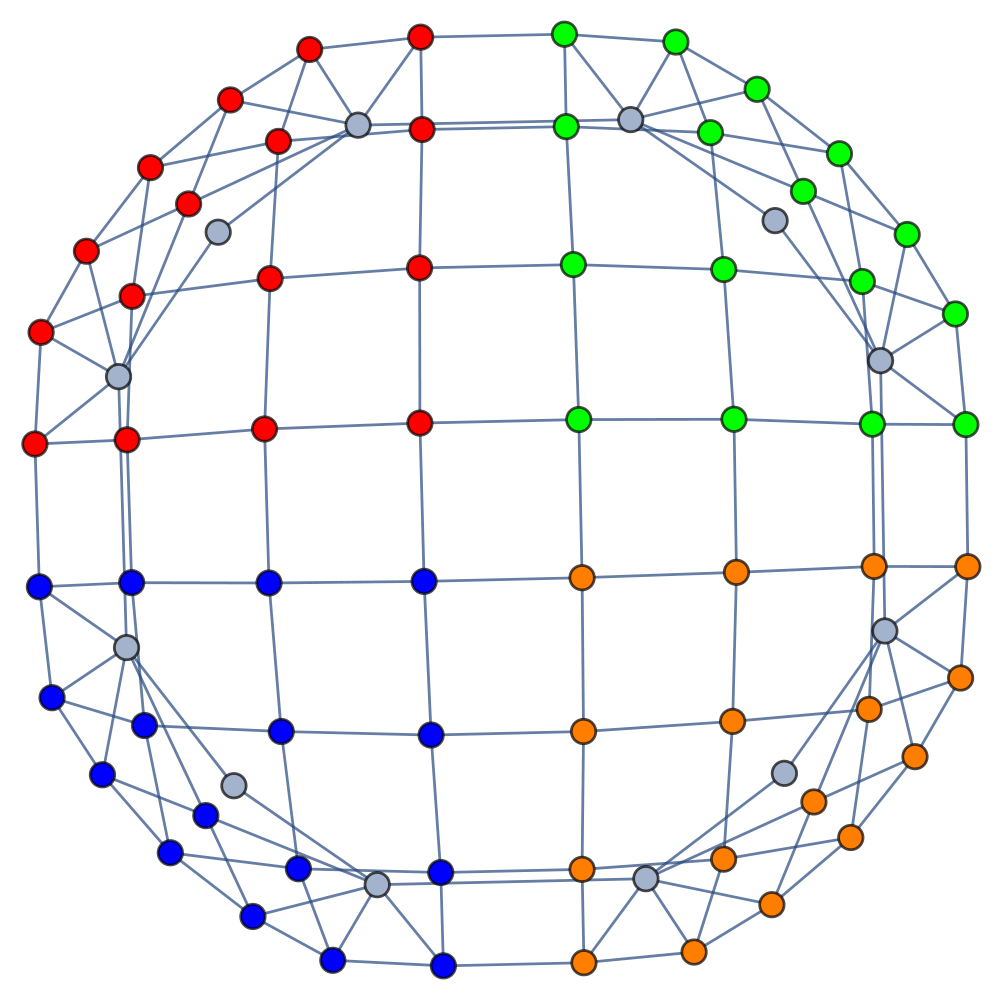}
\caption{On the left, a two-dimensional grid-like spatial hypergraph representing the neighborhood structure of a coarse (unrefined) two-dimensional quadrilateral mesh, with the red, green, blue and orange vertices marked for refinement. On the right, the refined spatial hypergraph with the red, green, blue and orange vertices replaced with finer (4-by-4) two-dimensional grids.}
\label{fig:Figure4}
\end{figure}

\begin{figure}[ht]
\centering
\includegraphics[width=0.345\textwidth]{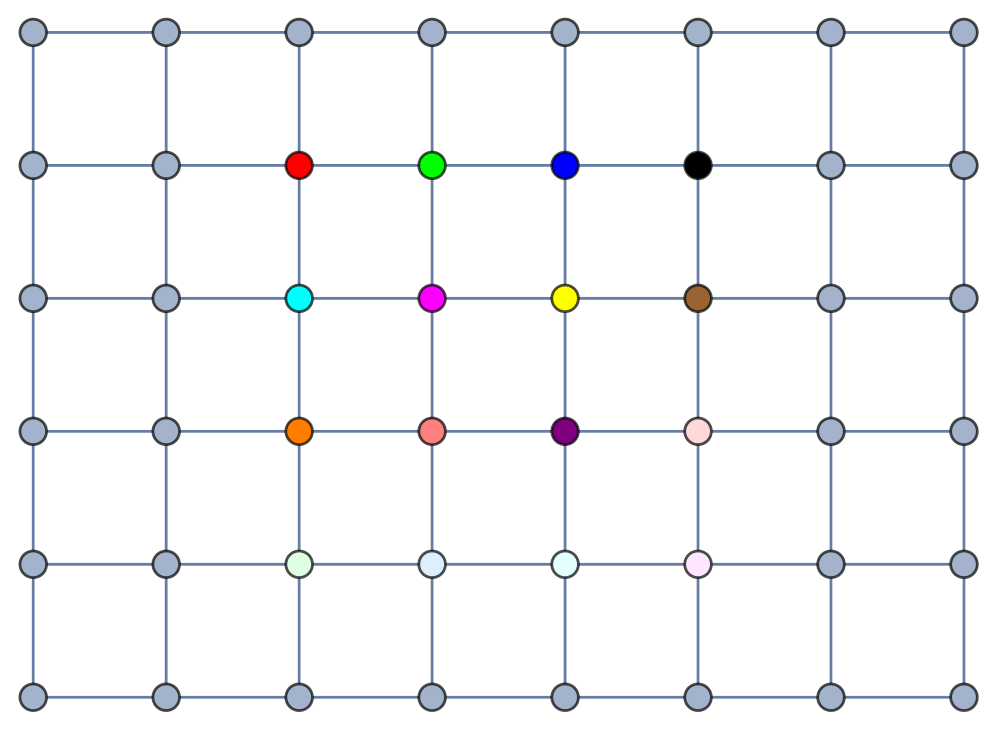}\hspace{0.1\textwidth}
\includegraphics[width=0.395\textwidth]{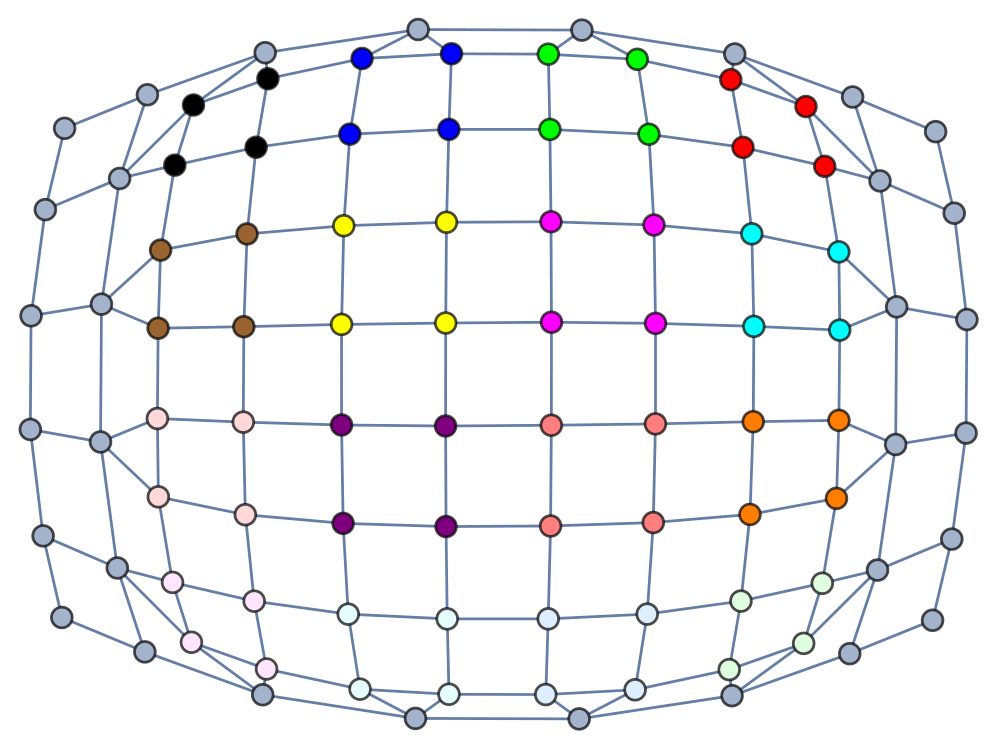}
\caption{On the left, a two-dimensional grid-like spatial hypergraph representing the neighborhood structure of a coarse (unrefined) two-dimensional quadrilateral mesh, with the twelve colored vertices marked for refinement. On the right, the refined spatial hypergraph with the colored vertices replaced with finer (2-by-2) two-dimensional grids.}
\label{fig:Figure5}
\end{figure}

This entire procedure may now be described purely graph-theoretically, using vertices as a means of storing cell information, with the neighborhood structure betweens cells being encoded by edges between vertices, and with (potentially overlapping) grids being represented as (potentially overlapping) subgraphs. For instance, in Figure \ref{fig:Figure4}, four colored vertices (red, green, blue and orange) in a two-dimensional grid-like spatial hypergraph are refined by replacing them each with 4-by-4 two-dimensional grids, whilst in Figure \ref{fig:Figure5}, twelve colored vertices are refined by replacing them each with 2-by-2 two-dimensional grids, whilst still preserving all information regarding the grid hierarchy and the neighborhood structure in both cases.

From here, we proceed on the assumption that our underlying finite difference scheme can be described as an explicitly conservative update scheme of the form (assuming, for instance, a three-dimensional spatial dimension):

\begin{multline}
\Phi_{i, j, k}^{n + 1} = \Phi_{i, j, k}^{n} - \frac{\Delta t}{\Delta x} \left( F \left( \Phi_{i + \frac{1}{2}, j, k}^{n} \right) - F \left( \Phi_{i - \frac{1}{2}, j, k}^{n} \right) \right) - \frac{\Delta t}{\Delta y} \left( G \left( \Phi_{i, j + \frac{1}{2}, k}^{n} \right) - G \left( \Phi_{i, j - \frac{1}{2}, k}^{n} \right) \right)\\
- \frac{\Delta t}{\Delta z} \left( H \left( \Phi_{i, j, k + \frac{1}{2}}^{n} \right) - H \left( \Phi_{i, j, k - \frac{1}{2}}^{n} \right) \right),
\end{multline}
where $F$, $G$ and $H$ denote the inter-cell flux functions in the $x$, $y$ and $z$ directions, respectively, and ${\Phi_{i \pm \frac{1}{2}, j, k}}$, ${\Phi_{i, j \pm \frac{1}{2}, k}}$, ${\Phi_{i, j, k \pm \frac{1}{2}}}$ denote the boundary-extrapolated values of the state vector ${\Phi}$. Each grid within the domain has its own independently-defined state vector, all of which (modulo boundary conditions) are also integrated independently. This procedure is recursive, such that when integrating a grid up to time ${t + \Delta t}$, all finer grids must be integrated up to time $t$. The explicit conservative update scheme presented above must be modified in the cases where a given coarse cell (with refinement level ${l - 1}$) is either overlayed by a refined grid with refinement level $l$, or is adjacent to an interface with a fine grid with refinement level $l$. In the former case, values at the coarse grid level (i.e. ${l - 1}$) are defined in terms of conservative averages of the values at the fine grid level (i.e. $l$), namely:

\begin{equation}
\frac{1}{r^2} \sum_{p = 0}^{r - 1} \sum_{q = 0}^{r - 1} \sum_{s = 0}^{r - 1} \Phi_{k + p, m + q, n + s}^{fine} \to \Phi_{i, j, k}^{coarse},
\end{equation}
where ${\Phi^{coarse}}$ denotes the state vector at the coarse level ${l - 1}$, and ${\Phi^{fine}}$ denotes the state vector at the fine level $l$, such that, at each subsequent integration step, the values at the coarse grid level are computed via this same conservative average. In the above, $i$, $j$ and $k$ denote the discrete spatial coordinate indices of the coarse cell in the coarse grid, whilst the same cell in the fine grid spans the discrete intervals ${\left[ k, k + r - 1 \right]}$, ${\left[ m, m + r - 1 \right]}$ and ${\left[ n, n + r - 1 \right]}$ in the fine grid.

\begin{figure}[ht]
\centering
\includegraphics[width=0.495\textwidth]{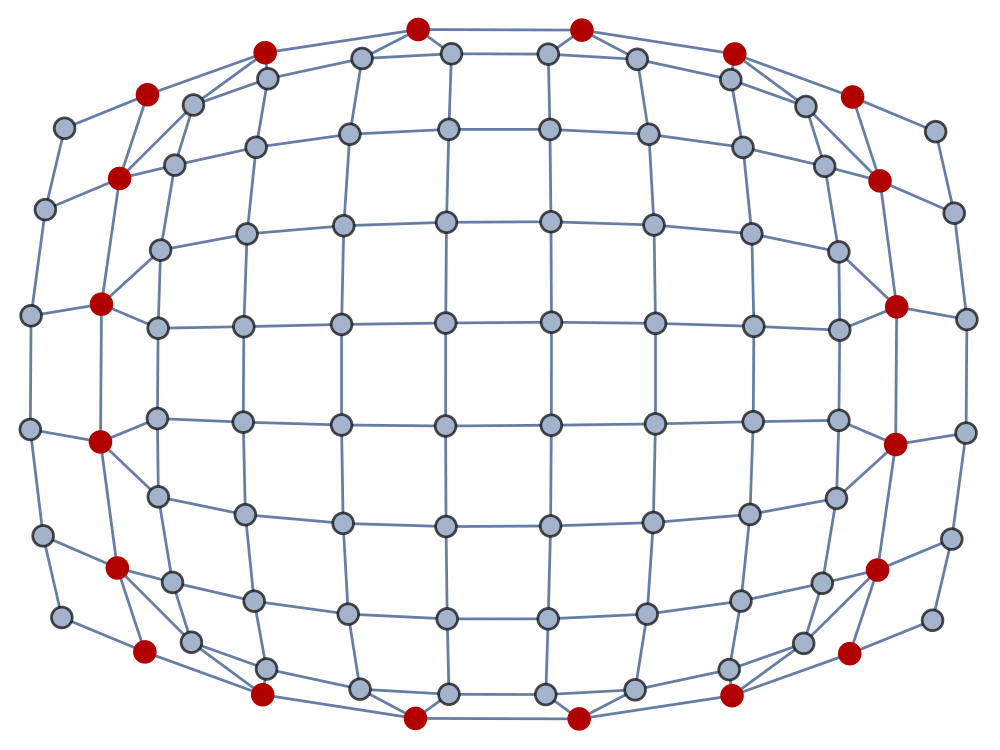}
\caption{A refined two-dimensional spatial hypergraph, with twelve vertices replaced with finer (2-by-2) two-dimensional grids, and with the twenty coarse vertices that are adjacent to interfaces with fine grids (and for which the finite difference scheme must therefore be modified) highlighted in dark red.}
\label{fig:Figure6}
\end{figure}

The latter case is slightly more subtle, since in order to guarantee conservation of the finite difference scheme within a grid hierarchy in which coarse cells are adjacent to fine cells, incoming fluxes to the fine grid (across the boundary between the coarse and fine grids) must equal outgoing fluxes of the coarse cell. The configuration of coarse cells for which the finite difference scheme must be modified, for the case of a refined two-dimensional spatial hypergraph in which twelve vertices have been replaced by 2-by-2 two-dimensional grids, is shown in Figure \ref{fig:Figure6}. This means that the difference scheme must be modified so as to be of the general form (for the case in which the coarse-to-fine grid flux is in the $x$ direction):

\begin{multline}
\Phi_{i, j, k} \left( t + \Delta t_{coarse} \right) = \Phi_{i, j, k} \left( t \right) - \frac{\Delta t_{coarse}}{\Delta x} \left[ F \left( \Phi_{i + \frac{1}{2}, j, k} \left( t \right) \right) \right.\\
\left. - \frac{1}{r^2} \sum_{q = 0}^{r - 1} \sum_{p = 0}^{r - 1} \sum_{s = 0}^{r - 1} F \left( \Phi_{k + \frac{1}{2}, m + p, n + s} \left( t + q \Delta t_{fine} \right) \right) \right] - \frac{\Delta t_{coarse}}{\Delta y} \left[ G \left( \Phi_{i, j + \frac{1}{2}, k} \left( t \right) \right) - G \left( \Phi_{i, j - \frac{1}{2}, k} \left( t \right) \right) \right]\\
- \frac{\Delta t_{coarse}}{\Delta z} \left[ H \left( \Phi_{i, j, k + \frac{1}{2}} \left( t \right) \right) - H \left( \Phi_{i, j, k - \frac{1}{2}} \left( t \right) \right) \right],
\end{multline}
and likewise for the other coordinate directions. In the above, ${\Delta t_{coarse}}$ denotes the stable time step at the coarse level ${l - 1}$ and ${\Delta t_{fine}}$ denotes the stable time step at the fine level $l$, with ${\Delta x}$, ${\Delta y}$ and ${\Delta z}$ signifying the spatial sizes of the coarse grid, and, as previously, the coarse cell has indices $i$, $j$ and $k$, with the fine grid spanning ${\left[ k, k + r - 1 \right]}$, ${\left[ m, m + r - 1 \right]}$ and ${\left[ n, n + r - 1 \right]}$.

Pragmatically, we implement this modified scheme as a correction step to the na\"ive (coarse) flux computation, in which this flux is subtracted from ${\Phi_{i, j, k}^{coarse} \left( t + \Delta t_{coarse} \right)}$ and the fine fluxes are added instead, as detailed above. We begin by constructing a tensor ${\delta F}$ of fluxes at those coarse grid boundaries which also correspond to outer boundaries of fine grids:

\begin{equation}
\delta F^{old} \left( \Phi_{i + \frac{1}{2}, j, k} \right) = - F^{coarse} \left( \Phi_{i + \frac{1}{2}, j, k} \right),
\end{equation}
etc. We next add a sum of all fine grid fluxes through the boundary at ${\left( i + \frac{1}{2}, j, k \right)}$ after each stable time step ${\Delta t_{fine}}$ for the fine grid:

\begin{equation}
\delta F^{new} \left( \Phi_{i + \frac{1}{2}, j, k} \right) = \delta F^{old} \left( \Phi_{i + \frac{1}{2}, j, k} \right) + \frac{1}{r^2} \sum_{p = 0}^{r - 1} \sum_{q = 0}^{r - 1} F \left( \Phi_{k + \frac{1}{2}, m + p, n + q} \right),
\end{equation}
and then proceed, once $r$ such time steps have elapsed, to use the tensor ${\delta F^{new} \left( \Phi_{i + \frac{1}{2}, j, k} \right)}$ to correct the solutions over the coarse grid such that they match those computed using the modified update formula. For instance, for a coarse cell with coordinate indices ${\left( i + 1, j, k \right)}$, the correction would be of the form:

\begin{equation}
\Phi_{i + 1, j, k}^{coarse (new)} = \Phi_{i + 1, j, k}^{coarse (old)} + \frac{\Delta t_{coarse}}{\Delta x_{coarse}} \delta F^{new} \left( \Phi_{i + \frac{1}{2}, j, k} \right).
\end{equation}
If the coarse cell had coordinate indices ${\left( i + 2, j, k \right)}$ then we would instead make two corrections:

\begin{equation}
\Phi_{i + 1, j, k}^{coarse (new)} = \Phi_{i + 1, j, k}^{coarse (old)} + \frac{\Delta t_{coarse}}{\Delta x_{coarse}} \delta F^{new} \left( \Phi_{i + \frac{1}{2}, j, k} \right) - \frac{\Delta t_{coarse}}{\Delta x_{coarse}} \delta F^{new} \left( \Phi_{i + \frac{3}{2}, j, k} \right),
\end{equation}
and so on, and likewise for fluxes in the other coordinate directions.

\begin{figure}[ht]
\centering
\includegraphics[width=0.495\textwidth]{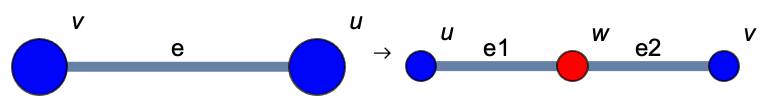}
\includegraphics[width=0.495\textwidth]{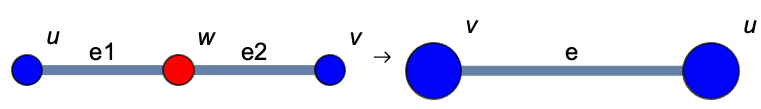}
\caption{On the left, a subdivision of an edge ${e = \left\lbrace u, v \right\rbrace}$ into a pair of edges ${e_1 = \left\lbrace u, w \right\rbrace}$, ${e_2 = \left\lbrace w, v \right\rbrace}$ with a newly-generated vertex $w$. On the right, the smoothing of a pair of edges ${e_1 = \left\lbrace u, w \right\rbrace}$, ${e_2 = \left\lbrace w, v \right\rbrace}$ incident on a common vertex $w$ into a single edge ${e = \left\lbrace u, v \right\rbrace}$.}
\label{fig:Figure7}
\end{figure}

In the purely graph-theoretic setup, this refinement procedure can be performed in a much more unstructured way, by first partitioning the overall graph into subgraphs and then computing the \textit{signatures} ${X \left( x \right)}$, ${Y \left( y \right)}$, ${Z \left( z  \right)}$ of the tagging function ${f \left( x, y, z \right)}$ within each such subgraph (assuming three spatial dimensions):

\begin{equation}
X \left( x \right) = \int f \left( x, y z \right) dy dz, \qquad Y \left( y \right) = \int f \left( x, y, z \right) dx dz, \qquad Z \left( z \right) = \int f \left( x, y, z \right) dx dy,
\end{equation}
with $f$ defined such that ${f \left( x, y, z \right) = 1}$ for any vertex or region that is tagged for refinement, and ${f \left( x, y, z \right) = 0}$ otherwise. Then, by computing the Laplacians ${\partial_{x}^{2} X \left( x \right)}$, ${\partial_{y}^{2} Y \left( y \right)}$ and ${\partial_{z}^{2} Z \left( z \right)}$ of these signatures, we are able to determine the local coordinate direction which maximizes ${\Delta \left( \partial_{i}^{2} X_i \right)}$ (by determining the appropriate inflection points), which then becomes the axis separating the tagged and untagged vertices in the direction that is orthogonal to the signature function (the axis of partition). If there exists a vertex ${x_i}$ with zero signature ${X_i \left( x_i \right) = 0}$ in any direction, then that direction is chosen as the axis of partition instead. If the ratio of tagged vertices to total vertices within a partitioned subgraph is greater than some-predefined ${\epsilon}$ (where ${0 < \epsilon < 1}$), and if the subgraph satisfies the nesting axioms of the grid hierarchy discussed above, then the partitioning procedure ends; otherwise, we recursively subdivide the subgraph into further subgraphs and continue the procedure. The general form of the tagging function ${f \left( x, y, z \right)}$ that we select for this article will set the refinement criterion to be the ${L_2}$ norm of the change in a pre-determined scalar field ${\phi}$ across a vertex exceeding a pre-defined threshold ${\sigma \left( \phi \right)}$, i.e:

\begin{equation}
f \left( x, y, z \right) = \begin{cases}
1, \qquad &\text{ if } \sqrt{\sum\limits_{i = 1}^{3} \left( \phi \left( \mathbf{x} + \Delta x \hat{\mathbf{x}}_i \right) - \phi \left( \mathbf{x} - \Delta x \hat{\mathbf{x}}_i \right) \right)^2} > \sigma \left( \phi \right),\\
0, \qquad &\text{ otherwise}.
\end{cases}
\end{equation}
For the examples shown subsequently, the scalar field ${\phi}$ will invariably be chosen to be a scalar curvature invariant from Riemannian geometry (or a quantity derived from such invariants). A vertex $u$ or $v$ can be refined or coarsened using a procedure based solely on graph homeomorphism, i.e. if either $u$ or $v$ along edge ${e = \left\lbrace u, v \right\rbrace}$ is tagged for refinement, then the edge $e$ can be subdivided to yield a pair of edges ${e_1 = \left\lbrace u, w \right\rbrace}$ and ${e_2 = \left\lbrace w, v \right\rbrace}$, with some newly-generated vertex $w$, whereas if a vertex $w$ that is incident to both vertices $u$ and $v$ on edges ${e_1 = \left\lbrace u, w \right\rbrace}$, ${e_2 = \left\lbrace w, v \right\rbrace}$ is tagged for coarsening, then the edges ${e_1}$ and ${e_2}$ can be smoothed to yield a single edge ${e = \left\lbrace u, v \right\rbrace}$, as shown in Figure \ref{fig:Figure7}.

All that remains is to illustrate how the boundary-extrapolated values of the state vector ${\Phi}$, i.e. ${\Phi_{i \pm \frac{1}{2}, j, k}}$, ${\Phi_{i, j \pm \frac{1}{2}, k}}$ and ${\Phi_{i, j, k \pm \frac{1}{2}}}$, and hence also the inter-cell fluxes ${F \left( \Phi_{i \pm \frac{1}{2}, j, k} \right)}$, ${G \left( \Phi_{i, j \frac{1}{2}, k} \right)}$ and ${H \left( \Phi_{i, j, k \pm \frac{1}{2}} \right)}$, are computed. In a conventional second-order numerical method based on finite volume approximations (such as SLIC or MUSCL-Hancock\cite{sweby}\cite{toro}), the boundary extrapolation step is computed using a simple linearized scheme, namely:

\begin{equation}
\Phi_{i + \frac{1}{2}, j, k}^{L} = \Phi_{i, j, k}^{n} + \frac{1}{2} \phi_{i - \frac{1}{2}, j, k}^{+} \left( \Phi_{i, j, k}^n - \Phi_{i - 1, j, k}^{n} \right),
\end{equation}
for the left boundary-extrapolated value (in the $x$ direction), and:

\begin{equation}
\Phi_{i + \frac{1}{2}, j, k}^{R} = \Phi_{i + 1, j, k}^{n} - \frac{1}{2} \phi_{i + \frac{3}{2}, j, k}^{-} \left( \Phi_{i + 2, j, k}^{n} - \Phi_{i + 1, j, k}^{n} \right),
\end{equation}
for the right boundary-extrapolated value (in the $x$ direction), and likewise for the other coordinate directions. In the above, the diagonal tensors ${\phi^{\pm}}$ are \textit{limiter functions}, i.e. they are functions of the ratio vectors ${\mathbf{r}^{\pm}}$:

\begin{equation}
\phi_{i - \frac{1}{2}, j, k}^{+} = \phi \left( \mathbf{r}_{i -\frac{1}{2}, j, k}^{+} \right), \qquad \text{ and } \qquad \phi_{i + \frac{3}{2}, j, k}^{-} = \phi \left( \mathbf{r}_{i + \frac{3}{2}, j}^{-} \right),
\end{equation}
where the ratio vectors are simply ratios of consecutive variations of the state vector ${\Phi}$:

\begin{equation}
\mathbf{r}_{i - \frac{1}{2}, j, k}^{+} = \frac{\Phi_{i + 1, j, k}^{n} - \Phi_{i, j, k}^{n}}{\Phi_{i, j, k}^{n} - \Phi_{i - 1, j, k}^{n}}, \qquad \text{ and } \qquad \mathbf{r}_{i + \frac{3}{2}, j, k}^{-} = \frac{\Phi_{i + 1, j, k}^{n} - \Phi_{i, j, k}^{n}}{\Phi_{i + 2, j, k}^{n} - \Phi_{i + 1, j, k}^{n}}.
\end{equation}
The diagonal limiter functions ${\phi^{\pm}}$ are specifically chosen so as to guarantee that the numerical scheme satisfies the \textit{total variation diminishing} constraint in each spatial direction, namely that:

\begin{equation}
\sum_{i} \left\lvert \Phi_{i + 1, j, k}^{n + 1} - \Phi_{i, j, k}^{n + 1} \right\rvert \leq \sum_{i} \left\lvert \Phi_{i + 1, j, k}^{n} - \Phi_{i, j, k}^{n} \right\rvert, \qquad \sum_{j} \left\lvert \Phi_{i, j + 1, k}^{n + 1} - \Phi_{i, j, k}^{n + 1} \right\rvert \leq \sum_{j} \left\lvert \Phi_{i, j + 1, k}^{n} - \Phi_{i, j, k}^{n} \right\rvert,
\end{equation}
and:

\begin{equation}
\sum_{k} \left\lvert \Phi_{i, j, k + 1}^{n + 1} - \Phi_{i, j, k}^{n + 1} \right\rvert \leq \sum_{k} \left\lvert \Phi_{i, j, k + 1}^{n} - \Phi_{i, j, k}^{n} \right\rvert,
\end{equation}
which ensures that the scheme preserves monotonicity (and therefore eliminates spurious oscillations from appearing within the solution).

However, by choosing to use a linearized (first-order) spatial reconstruction scheme for performing the boundary-extrapolation step in this way, one effectively limits the accuracy of the underlying numerical method to second-order at best. In order to achieve the desired fourth-order accuracy, we opt instead to use a higher-order \textit{weighted, essentially non-oscillatory} (WENO) method for the spatial reconstruction\cite{jiang}\cite{jiang2}, in which the first-order linear interpolation is replaced with an interpolation using a polynomial function of higher degree. We begin by choosing either a \textit{modal} basis or a \textit{nodal} basis of underlying polynomials, where the basis functions have been rescaled so as to fit a reference unit interval ${\left[ 0, 1 \right]}$ using the following transformation from the spatial coordinates ${\left( x, y, z \right)}$ to the rescaled coordinates ${\left( \xi, \eta, \zeta \right)}$:

\begin{equation}
\xi = \left( x, i \right) = \frac{1}{\Delta x_i} \left( x - x_{i - \frac{1}{2}} \right), \qquad \eta = \left( y, j \right) = \frac{1}{\Delta y_j} \left( y - y_{j - \frac{1}{2}} \right), \qquad \zeta = \zeta \left( z, k \right) = \frac{1}{\Delta z_k} \left( z - z_{k - \frac{1}{2}} \right).
\end{equation}
In this context, the phrase \textit{modal basis} refers to a set of ${M + 1}$ linearly independent polynomials, with degrees ranging from 0 to $M$ inclusive - typically the Legendre polynomials ${P_n \left( x \right)}$, defined by the generating function:

\begin{equation}
\sum_{n = 0}^{\infty} P_n \left( x \right) t^n = \frac{1}{\sqrt{1 - 2 x t + t^2}}.
\end{equation}
On the other hand, the phrase \textit{nodal basis} refers to a set of ${M + 1}$ linearly independent polynomials, all of degree exactly $M$. For the purposes of the simulations presented in this article, we choose to use a nodal basis of polynomials, since there exist explicit numerical experiments\cite{hidalgo} indicating marginally increased efficiency (when using standard Gauss-Legendre quadrature nodes, etc.) over modal basis polynomials.

Henceforth, we shall denote the ${M + 1}$ linearly independent polynomials in the nodal basis by ${\left\lbrace \psi_l \right\rbrace_{l = 1}^{M + 1}}$, and they are taken to be Lagrange polynomials, i.e. polynomials of the form ${\ell_j \left( x \right)}$:

\begin{equation}
\ell_j \left( x \right) = \prod_{1 \leq m \leq M + 1, m \neq j} \frac{x - x_m}{x_j - x_j} = \frac{\left( x - x_1 \right)}{\left( x_j - x_1 \right)} \cdots \frac{\left( x - x_{ j - 1} \right)}{\left( x_j - x_{j - 1} \right)} \frac{ \left( x - x_{j + 1} \right)}{\left( x_j - x_{j + 1} \right)} \cdots \frac{\left( x - x_{M + 1} \right)}{\left( x_j - x_{M + 1} \right)},
\end{equation}
where ${1 \leq j \leq M + 1}$, which interpolate between a set of ${M + 1}$ nodal points ${\left( x_j, y_j \right)}$, from ${\left( x_1, y_1 \right)}$ up to ${\left( x_{M + 1}, y_{M + 1} \right)}$, such that no two ${x_j}$ are the same. Henceforth, we denote these ${M + 1}$ nodal points by ${\left\lbrace x_k \right\rbrace_{k = 1}^{M + 1}}$, with the property that:

\begin{equation}
\psi_l \left( x_k \right) = \delta_{l k}, \qquad \text{ where } \qquad l, k = 1, 2, \dots, M + 1.
\end{equation}
With basis polynomials of this form, the spatial reconstruction within each cell (i.e. vertex) ${I_{i, j, k}}$ can be performed using the following numerical stencils ${\mathcal{S}_{i, j, k}}$ in each of three Cartesian directions:

\begin{equation}
\mathcal{S}_{i, j, k}^{s, x} \bigcup_{e = i - L}^{i + R} I_{e, j, k}, \qquad \mathcal{S}_{i, j, k}^{s, y} \bigcup_{e = j - L}^{j + R} I_{i, e, k}, \qquad \mathcal{S}_{i, j, k}^{s, z} \bigcup_{e = k - L}^{k + R} I_{i, j, e},
\end{equation}
where the parameters $L$ and $R$ denote the spatial extent of the numerical stencil to the left and to the right, respectively (these parameters will thus depend both upon the particular choice of stencil, and the chosen order of accuracy of the spatial reconstruction). More specifically, the number of cells contained within a given stencil will be equal to ${M + 1}$, which is in turn equal to the order of accuracy of the spatial reconstruction (with $M$ being the degree of the basis polynomials). In general, for a spatial reconstruction with an odd order of accuracy, and thus with an even degree of basis polynomial $M$, three stencils are used: one which is centered (with ${s = 1}$, ${L = R = \frac{M}{2}}$), one which is left-sided (with ${s = 2}$, ${L = M}$, ${R = 0}$) and one which is right-sided (with ${s = 3}$, ${L = 0}$, ${R = M}$). Likewise, for a spatial reconstruction with an even order of accuracy, and thus with an odd degree of basis polynomial $M$, four stencils are used: two which are centered (with ${s = 0}$, ${L = \left\lfloor \frac{M}{2} \right\rfloor + 1}$, ${R = \left\lfloor \frac{M}{2} \right\rfloor}$, and ${s = 1}$, ${L = \left\lfloor \frac{M}{2} \right\rfloor}$, ${R = \left\lfloor \frac{M}{2} \right\rfloor + 1}$, respectively), one which is left-sided (as previously) and one which is right-sided (as previously).

To illustrate how this spatial reconstruction scheme works, consider the case of reconstruction in the $x$ direction (the $y$ and $z$ directions are directly analogous); first, use the nodal basis polynomials ${\Psi_{l} \left( \xi \right)}$ to expand the spatial reconstruction polynomial ${\mathbf{w}_{h}^{s, x} \left( x, t^n \right)}$ as follows:

\begin{equation}
\mathbf{w}_{h}^{s, x} \left( x, t^n \right) = \sum_{p = 0}^{M} \Psi_p \left( \xi \right) \hat{\mathbf{w}}_{i, j, k, p}^{n, s},
\end{equation}
which we can write more succinctly using the Einstein summation convention:

\begin{equation}
\sum_{p = 0}^{M} \Psi_p \left( \xi \right) \hat{\mathbf{w}}_{i, j, k, p}^{n, s} = \Psi_p \left( \xi \right) \hat{\mathbf{w}}_{i, j, k, p}^{n, s}.
\end{equation}
Next, if we impose the weak (integral) form of a conservation law across every cell within the numerical stencil ${\mathcal{S}_{i, j, k}^{s, x}}$, then we obtain the following linear system of equations:

\begin{equation}
\forall I_{e, j, k} \in \mathcal{S}_{i, j, k}^{s, x}, \qquad \frac{1}{\Delta x_e} \int_{x_{e - \frac{1}{2}}}^{x_{e + \frac{1}{2}}} \Psi_p \left( \xi \left( x \right) \right) \hat{\mathbf{w}}_{i, j, k, p}^{n, s} dx = \bar{\Phi}_{e, j, k}^{n},
\end{equation}
where we have introduced the notation ${\bar{\Phi}_{i, j, k}^{n}}$ as a convenient shorthand for the (spatial) integral average of the solution vector ${\Phi}$ within the cell ${I_{i, j, k}}$ at time ${t^n}$, hence allowing us to determine the reconstruction coefficients ${\hat{\mathbf{w}}_{i, j, k, p}^{n, s}}$ simply by solving the resulting linear system. Then, with the reconstruction polynomials within each stencil thus obtained, by taking a nonlinear combination of the polynomials in each individual stencil (where the choice of weights is dependent upon the initial data) of the general form:

\begin{equation}
\mathbf{w}_{h}^{x} \left( x, t^n \right) = \Psi_p \left( \xi \right) \hat{\mathbf{w}}_{i, j, k, p}^{n}, \qquad \text{ where } \qquad \hat{\mathbf{w}}_{i, j, k, p}^{n} = \sum_{s = 1}^{N_s} \omega_s \hat{\mathbf{w}}_{i, j, k, p}^{n, s},
\end{equation}
we are able now to produce a reconstruction for the overall computational cell. In the above, ${N_s}$ is equal either to 3 or 4, depending upon whether the polynomial degree $M$ is even or odd, respectively, and the data-dependent nonlinear weights ${\omega_s}$ are given by:

\begin{equation}
\omega_s = \frac{\tilde{\omega}_s}{\sum\limits_q \tilde{\omega}_q}, \qquad \text{ where } \qquad \tilde{\omega}_s = \frac{\lambda_s}{\left( \sigma_s + \epsilon \right)^r}.
\end{equation}
In order to reduce the impact of spurious oscillations, thus making the scheme essentially non-oscillatory, we have introduced a dependence upon an \textit{oscillator indicator function} ${\sigma_s}$:

\begin{equation}
\sigma_s = \Sigma_{l m} \tilde{\mathbf{w}}_{l}^{n, s} \tilde{\mathbf{w}}_{m}^{n, s},
\end{equation}
which we compute by means of an \textit{oscillation indicator matrix} ${\Sigma_{l m}}$\cite{dumbser}:

\begin{equation}
\Sigma_{l m} = \sum_{\alpha = 1}^{M} \int_{0}^{1} \frac{\partial^{\alpha} \Psi_l \left( \xi \right)}{\partial \xi^{\alpha}} \frac{\partial^{\alpha} \Psi_m \left( \xi \right)}{\partial \xi^{\alpha}} d \xi.
\end{equation}
For the remainder of this article, we choose to use a spatial reconstruction scheme with third-order accuracy (and therefore in which ${M = 2}$), with parameter choices ${\lambda_s = 1}$ and ${\lambda_s = 10^5}$ for one-sided and centered stencils, respectively (as determined by explicit numerical experiment), and where we set ${\epsilon = 10^{-4}}$ so as to prevent division by zero for the case of smooth solutions, and ${r = 8}$.

\section{Numerical Convergence Tests}
\label{sec:Section4}

\subsection{Static Schwarzschild Black Hole}

We begin by evolving a static Schwarzschild black hole, i.e. a spacetime defined by the usual line element (assuming henceforth that ${G = c = 1}$)\cite{misner2}:

\begin{equation}
g = ds^2 = - \left( 1 - \frac{r_s}{r} \right) dt^2 + \left( 1 - \frac{r_s}{r} \right)^{-1} dr^2 + r^2 d \Omega^2,
\end{equation}
in the Schwarzschild coordinate system ${\left( t, r, \theta, \phi \right)}$, where ${d \Omega^2}$ denotes the induced Riemannian spatial metric on the 2-sphere ${S^2 \subset E^3}$:

\begin{equation}
d \Omega^2 = d \theta^2 + \sin^2 \left( \theta \right) d \phi^2,
\end{equation}
with $r$ denoting the radial coordinate, ${\Omega}$ denoting an arbitrary point on the 2-sphere ${S^2}$, ${\theta}$ denoting the colatitude of point ${\theta}$, ${\phi}$ denoting the longitude of point ${\phi}$, and ${r_s}$ denoting the Schwarzschild radius for a black hole of mass $M$, i.e. ${r_s = 2M}$. However, since we choose to evolve the black hole with an isotropic gauge (i.e. keeping the metric conformally flat, with the lapse function set to ${\alpha = 0}$ everywhere), we use isotropic coordinates that are adapted to the family of nested spheres; specifically, we use the spacetime line element:

\begin{equation}
g = ds^2 = - \frac{\left( 1 - \frac{r_s}{4R} \right)^2}{\left( 1 + \frac{r_s}{4R} \right)^2} dt^2 + \left( 1 + \frac{r_s}{4R} \right)^4 \left( dx^2 + dy^2 + dz^2 \right),
\end{equation}
with modified radial coordinate $R$:

\begin{equation}
R = \sqrt{x^2 + y^2 + z^2},
\end{equation}
for all points outside the horizon, i.e. with ${R > \frac{r_s}{4}}$. The conformal factor ${\psi}$ has the following very elementary form in terms of Schwarzschild coordinates:

\begin{equation}
\psi = \left( 1 + \frac{M}{2r} \right)^{-2}.
\end{equation}

\begin{figure}[ht]
\centering
\includegraphics[width=0.325\textwidth]{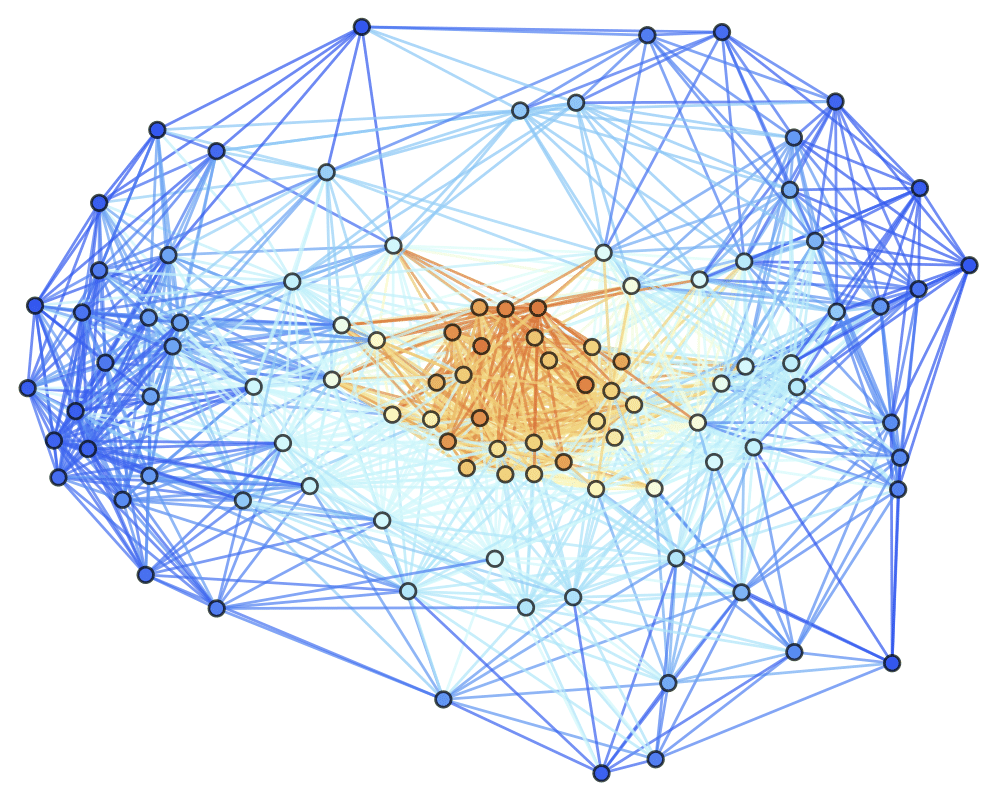}
\includegraphics[width=0.325\textwidth]{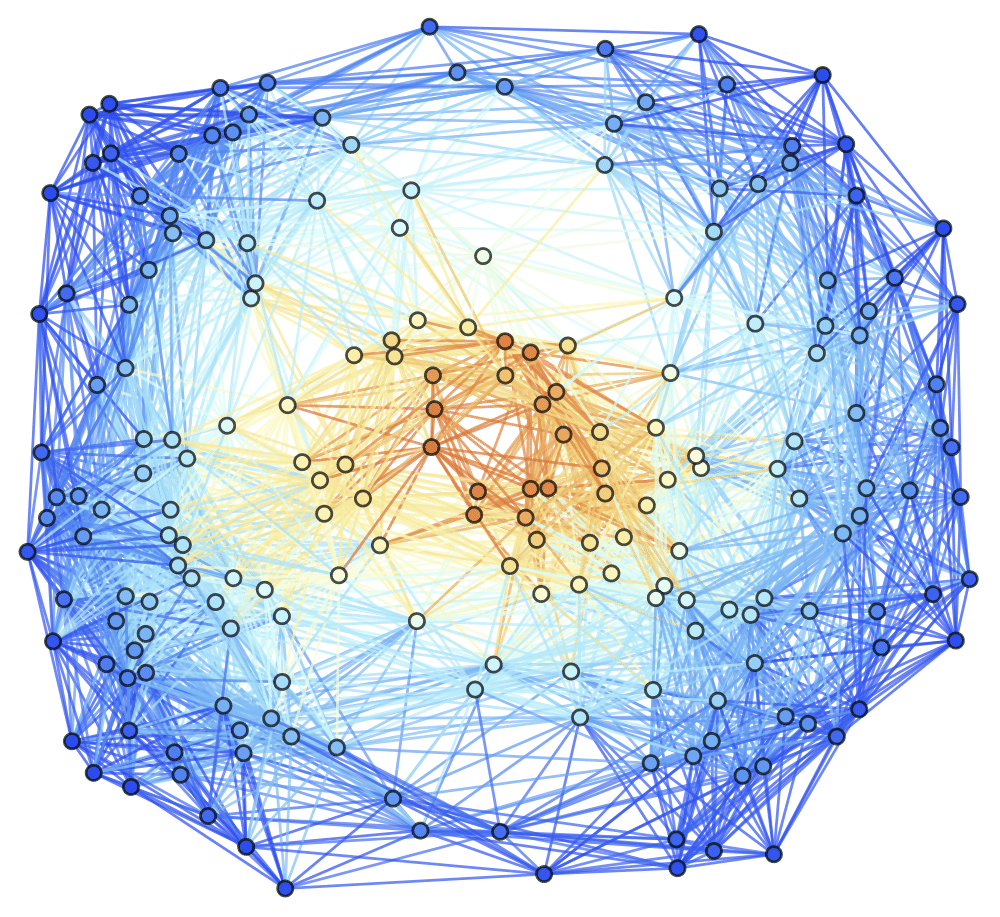}
\includegraphics[width=0.325\textwidth]{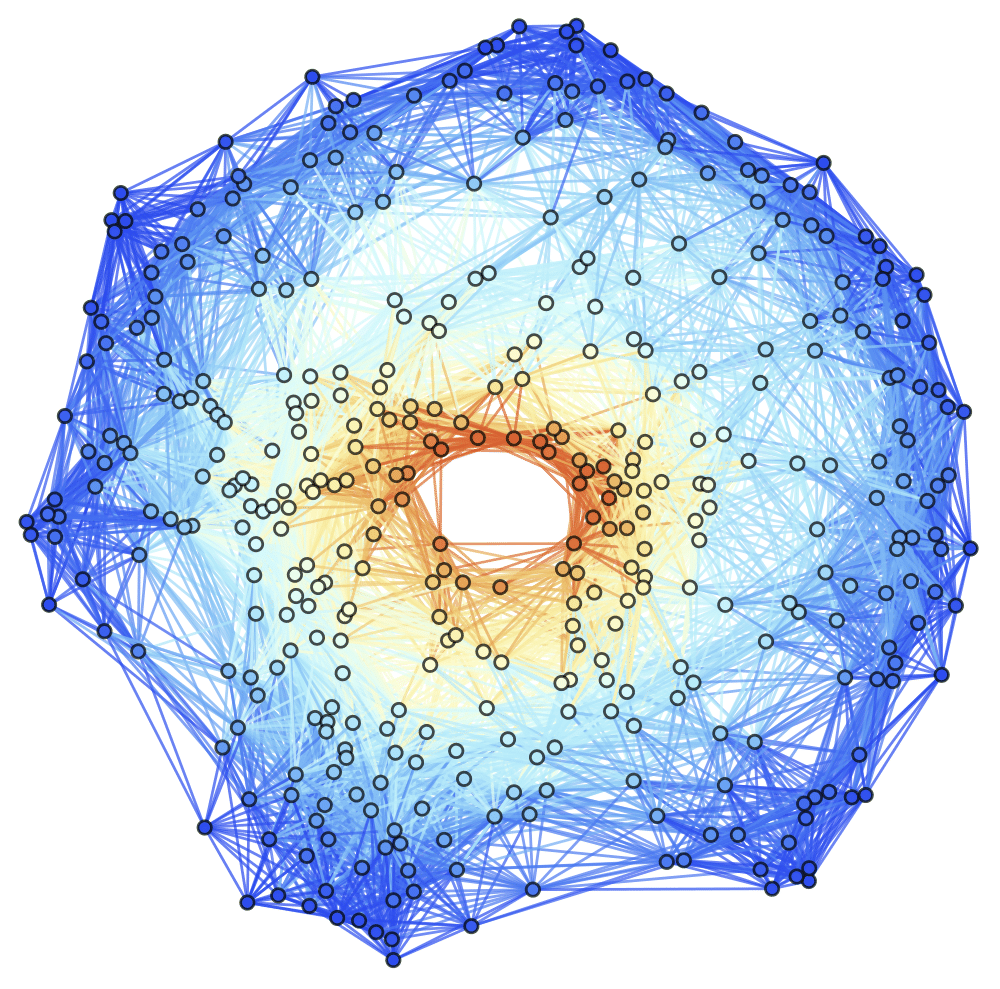}
\caption{Spatial hypergraphs corresponding to the initial hypersurface configuration of the static Schwarzschild black hole test at time ${t = 0 M}$, with resolutions of 100, 200 and 400 vertices, respectively. The hypergraphs have been adapted and colored using the local curvature in the Schwarzschild conformal factor ${\psi}$.}
\label{fig:Figure8}
\end{figure}

\begin{figure}[ht]
\centering
\includegraphics[width=0.325\textwidth]{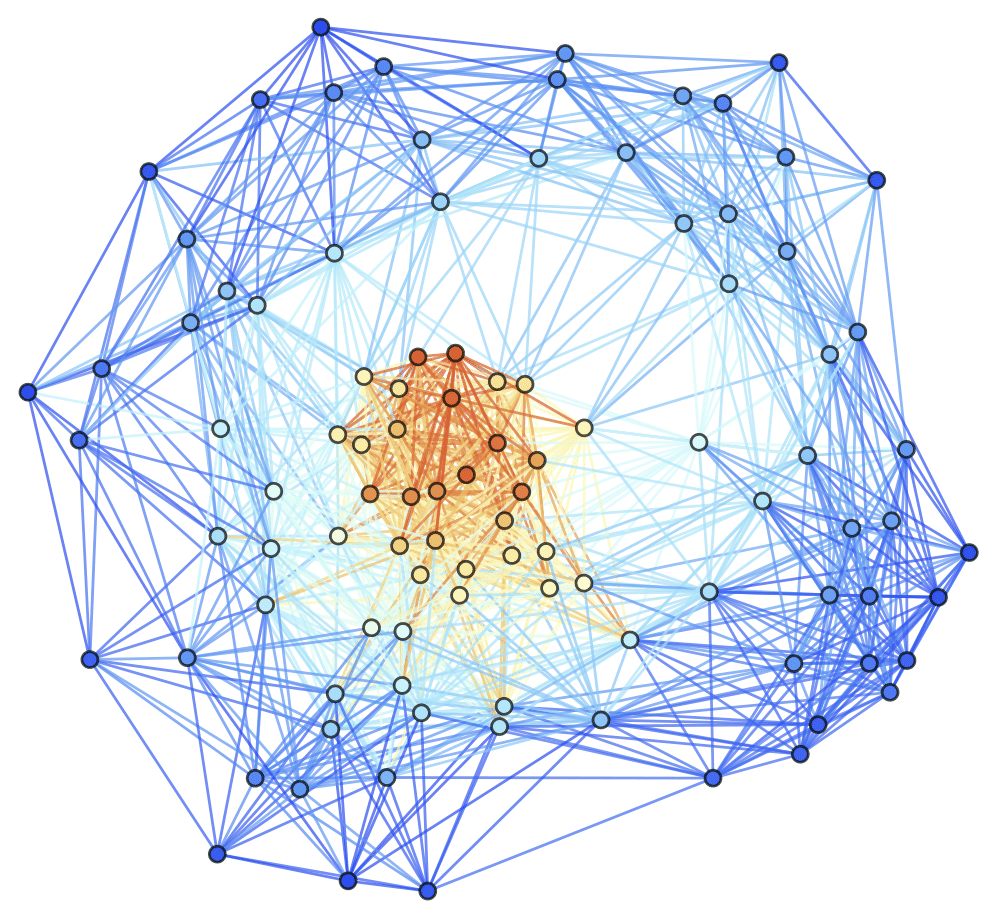}
\includegraphics[width=0.325\textwidth]{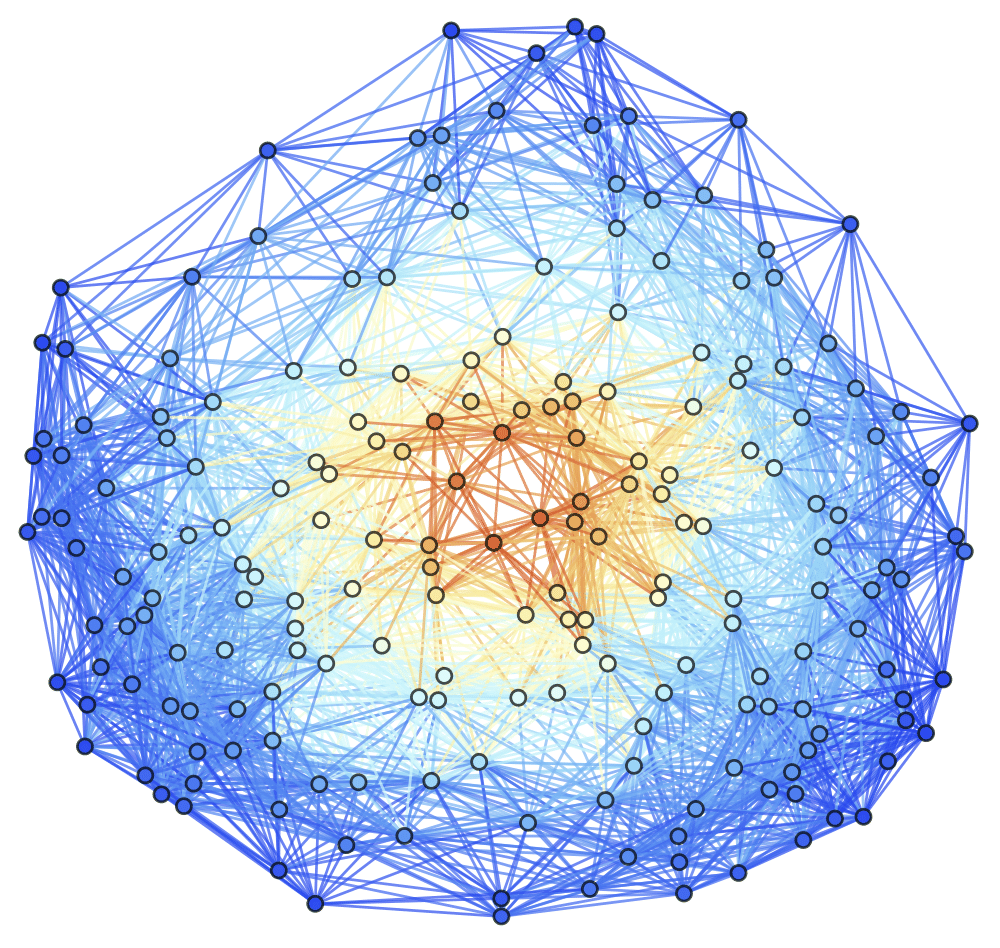}
\includegraphics[width=0.325\textwidth]{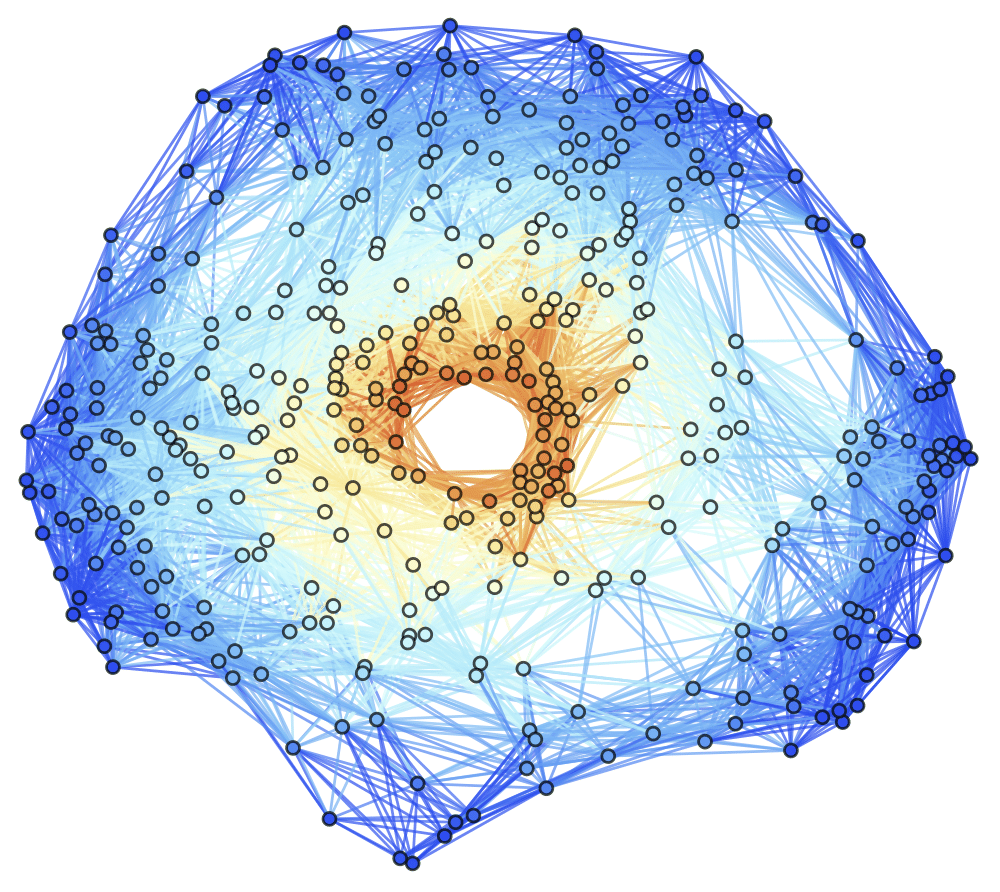}
\caption{Spatial hypergraphs corresponding to the final hypersurface configuration of the static Schwarzschild black hole test at time ${t = 1000 M}$, with resolutions of 100, 200 and 400 vertices, respectively. The hypergraphs have been adapted and colored using the local curvature in the Schwarzschild conformal factor ${\psi}$.}
\label{fig:Figure9}
\end{figure}

\begin{figure}[ht]
\centering
\includegraphics[width=0.325\textwidth]{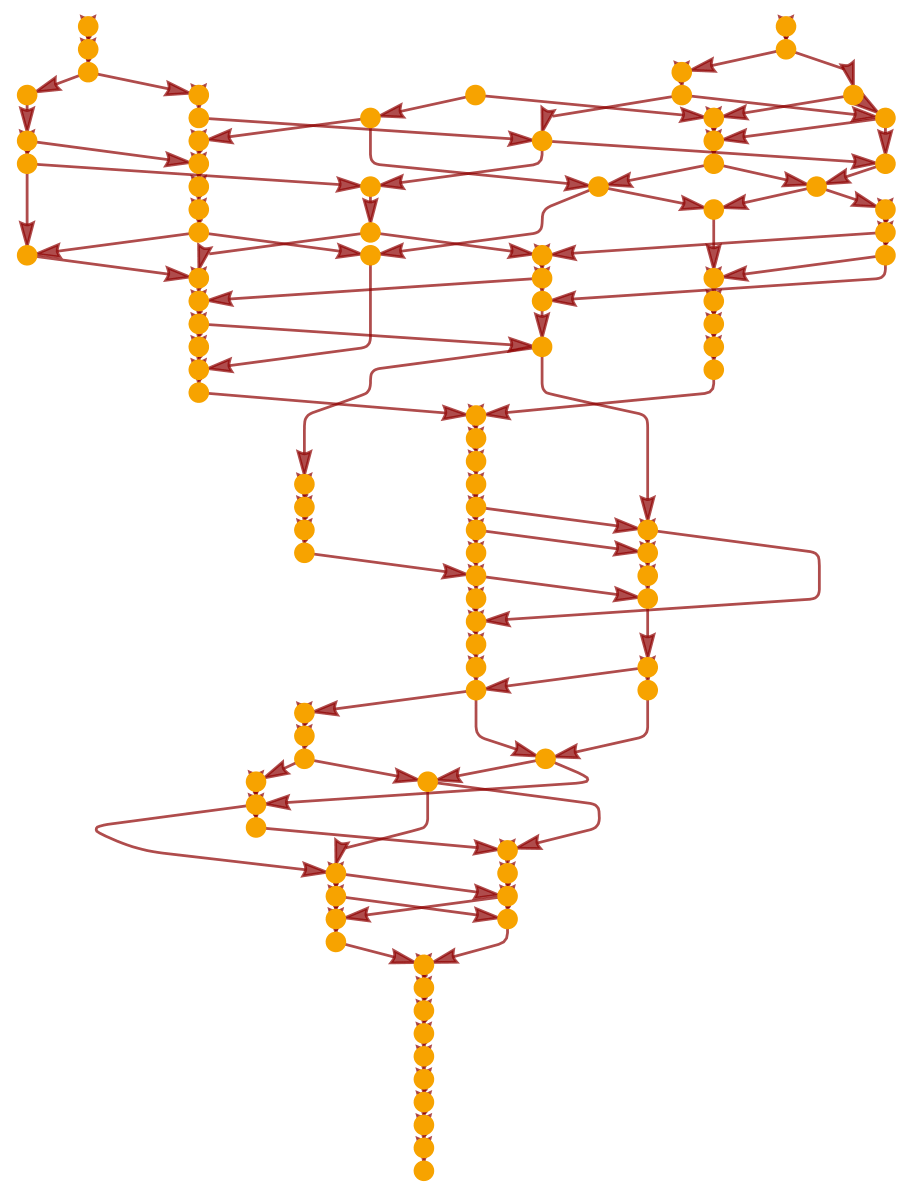}
\includegraphics[width=0.325\textwidth]{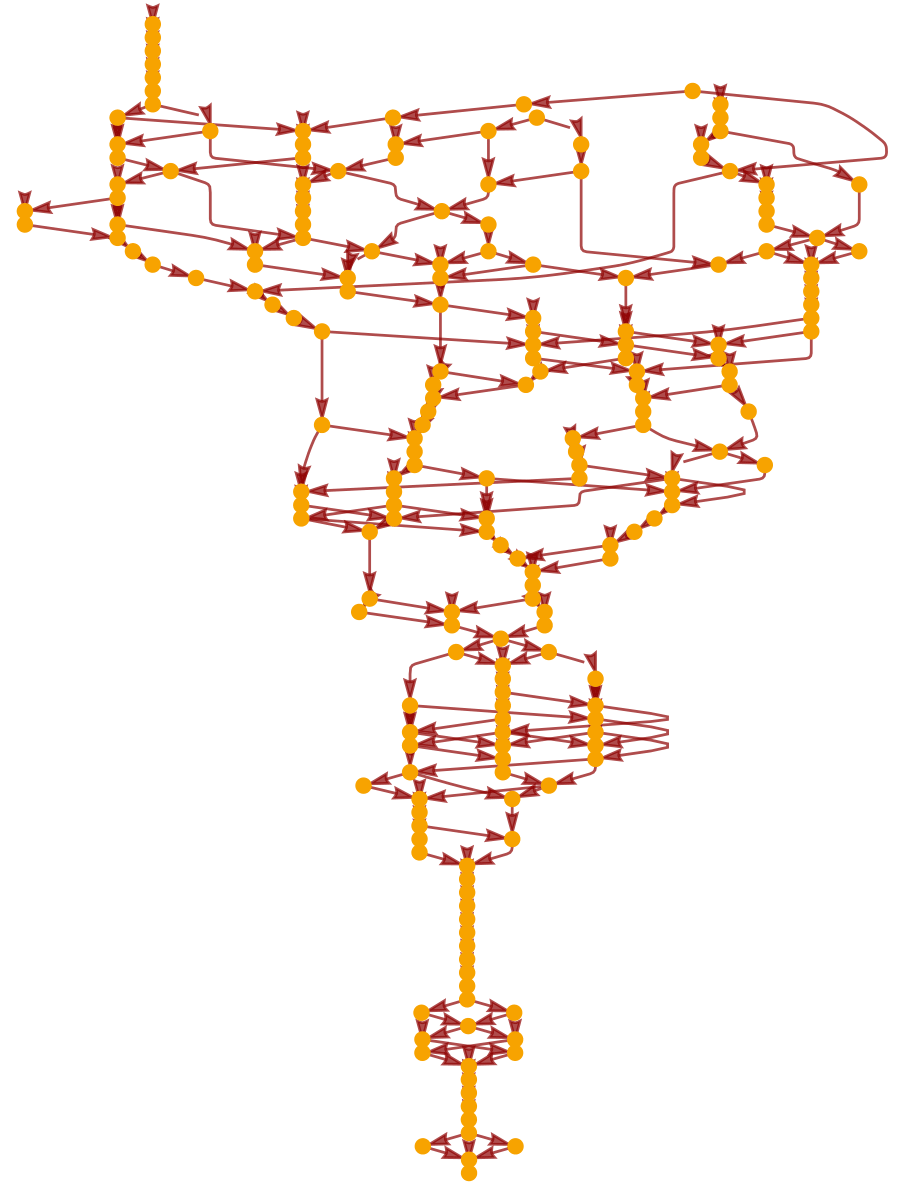}
\includegraphics[width=0.325\textwidth]{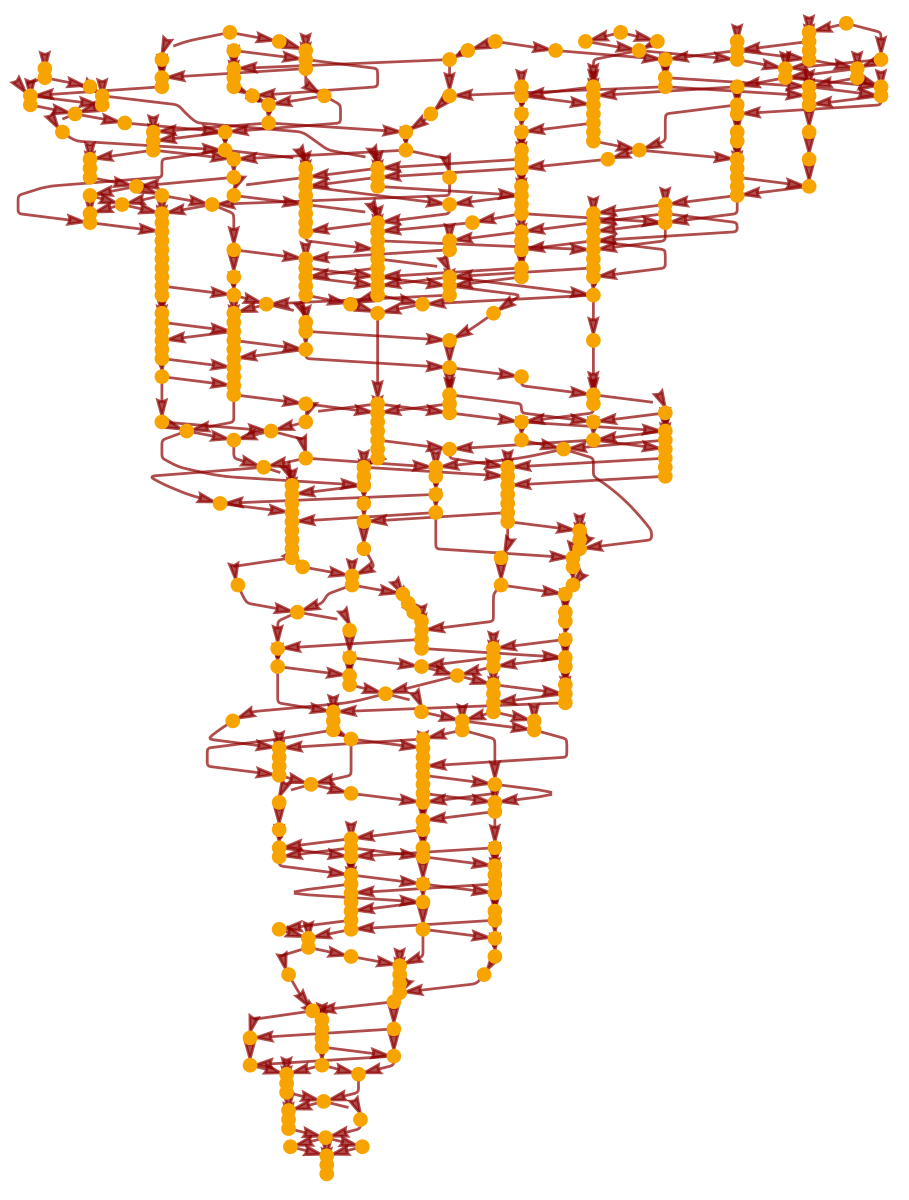}
\caption{Causal graphs corresponding to the discrete characteristic structure of the static Schwarzschild black hole test after time ${t = 1000 M}$, with resolutions of 100, 200 and 400 hypergraph vertices, respectively.}
\label{fig:Figure10}
\end{figure}

\begin{figure}[ht]
\centering
\includegraphics[width=0.325\textwidth]{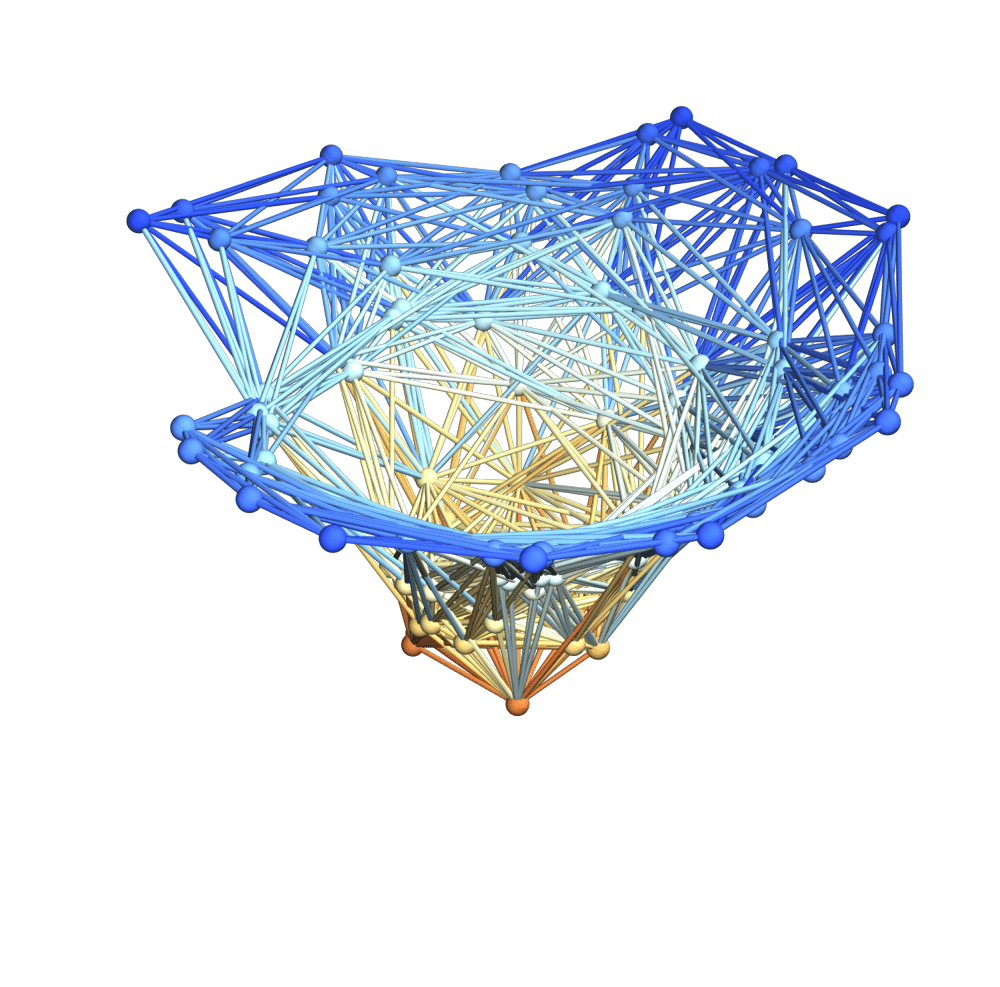}
\includegraphics[width=0.325\textwidth]{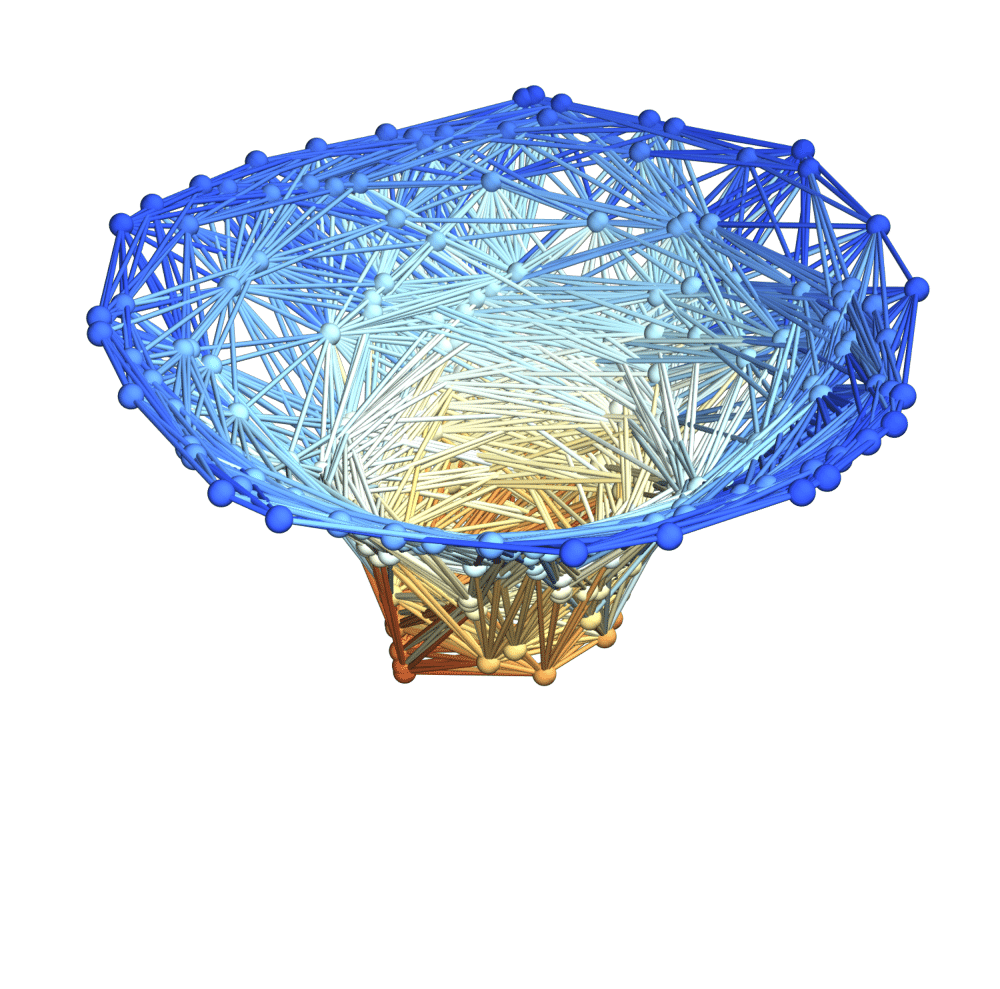}
\includegraphics[width=0.325\textwidth]{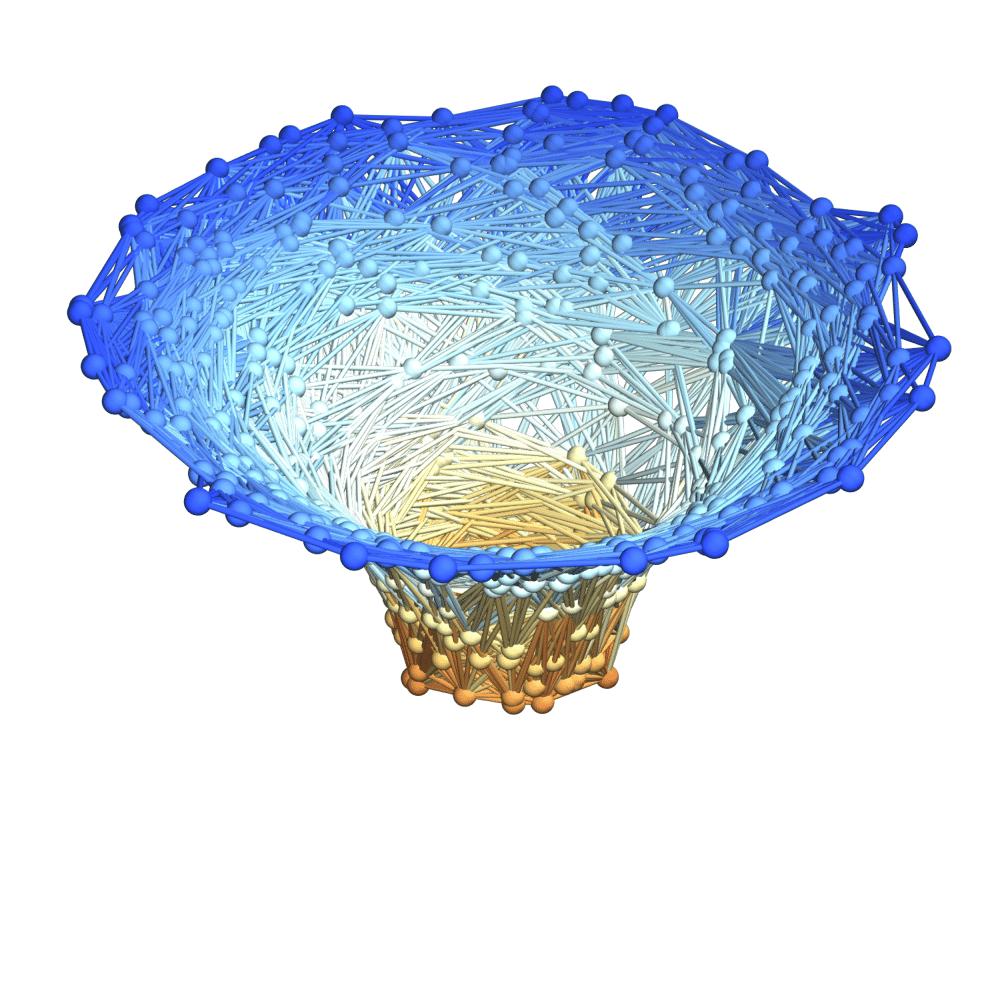}
\caption{Spatial hypergraphs corresponding to projections along the $z$-axis of the initial hypersurface configuration of the static Schwarzschild black hole test at time ${t = 0 M}$, with resolutions of 100, 200 and 400 vertices, respectively. The vertices have been assigned spatial coordinates according to the profile of the Schwarzschild conformal factor ${\psi}$ through a spatial slice perpendicular to the $z$-axis, and the hypergraphs have been adapted and colored using the local curvature in ${\psi}$.}
\label{fig:Figure11}
\end{figure}

\begin{figure}[ht]
\centering
\includegraphics[width=0.325\textwidth]{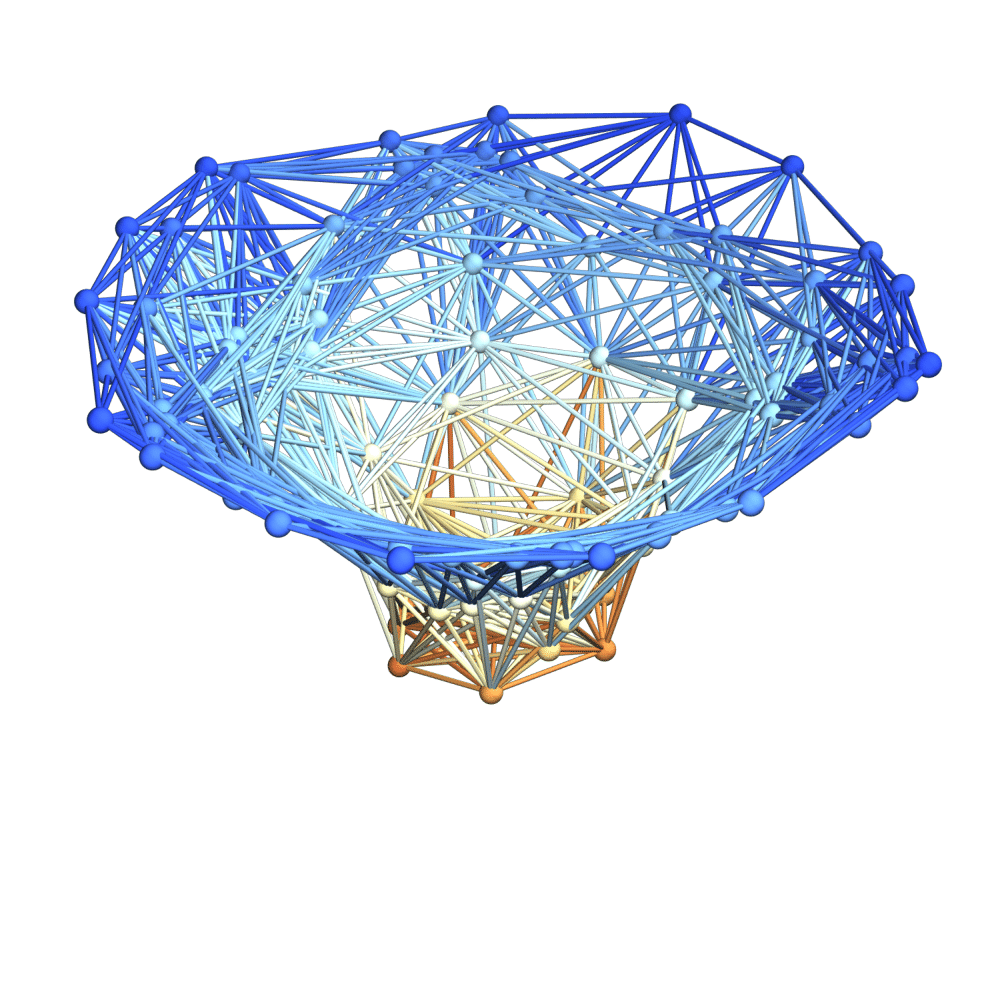}
\includegraphics[width=0.325\textwidth]{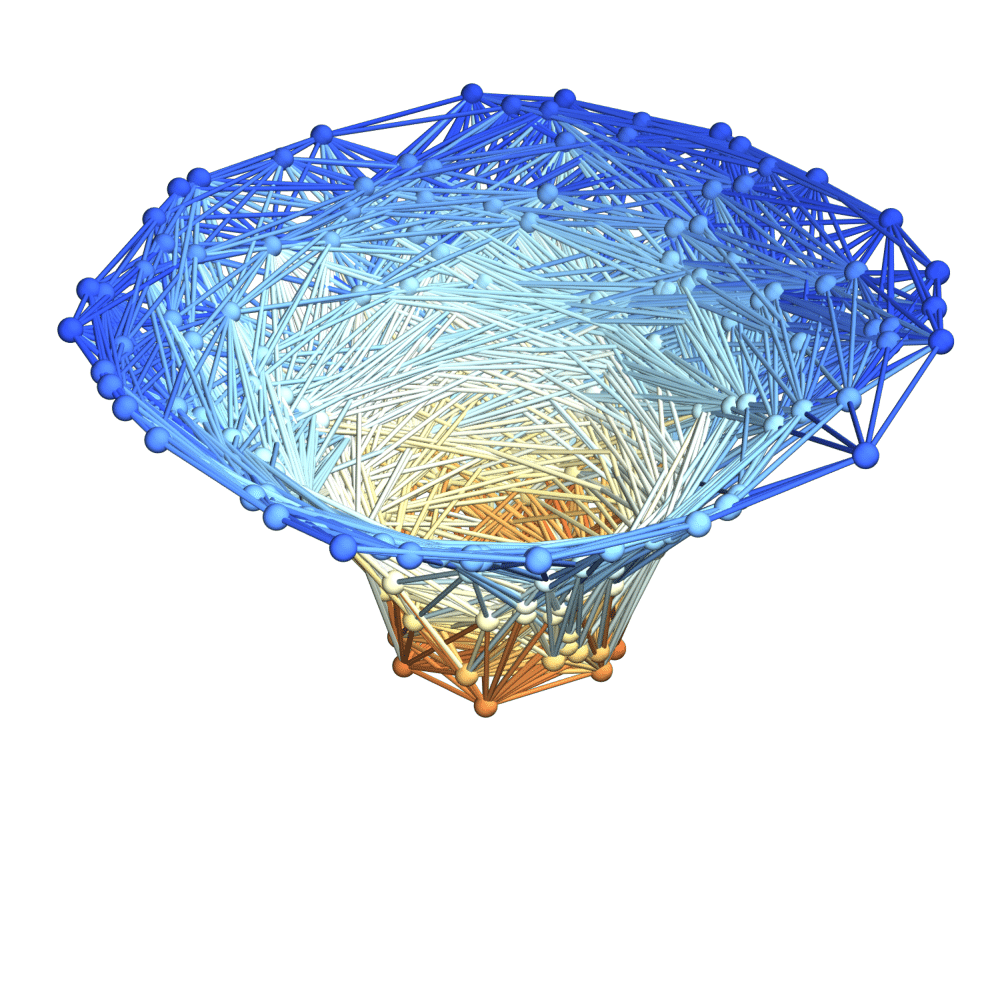}
\includegraphics[width=0.325\textwidth]{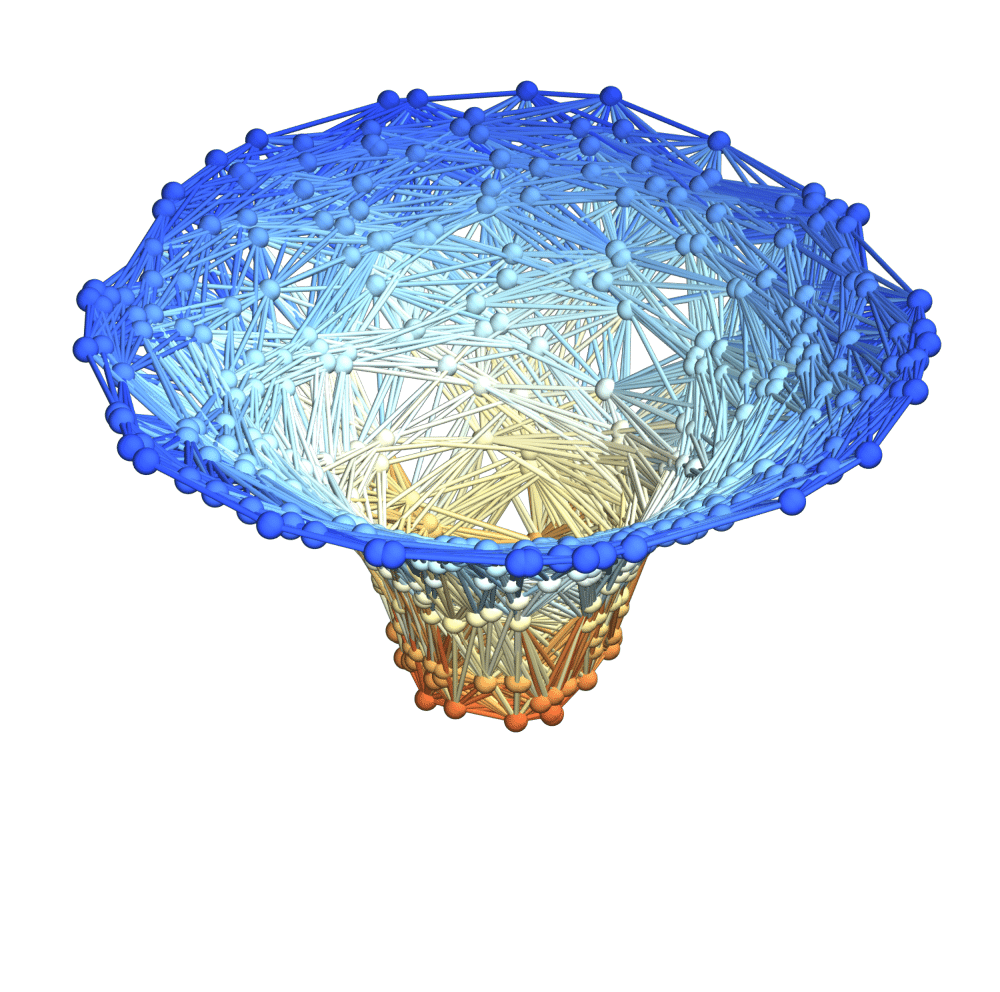}
\caption{Spatial hypergraphs corresponding to projections along the $z$-axis of the final hypersurface configuration of the static Schwarzschild black hole test at time ${t = 1000 M}$, with resolutions of 100, 200 and 400 vertices, respectively. The vertices have been assigned spatial coordinates according to the profile of the Schwarzschild conformal factor ${\psi}$ through a spatial slice perpendicular to the $z$-axis, and the hypergraphs have been adapted and colored using the local curvature in ${\psi}$.}
\label{fig:Figure12}
\end{figure}

Up to redundancies due to symmetry, the Schwarzschild metric exhibits four non-zero components of the Riemann curvature tensor ${R_{i j k l}}$, namely:

\begin{equation}
R_{r r t}^{t} = 2 R_{r \theta r}^{\theta} = 2 R_{r \phi r}^{\phi} = \frac{r_s}{r^2 \left( r_s - r \right)}, \qquad R_{t r t}^{r} = -2 R_{t \theta t}^{\theta} = -2 R_{t \phi t}^{\phi} = \frac{r_s \left( r_s - r \right)}{r^4},
\end{equation}
and:

\begin{equation}
2 R_{\theta \theta t}^{t} = 2 R_{\theta \theta r}^{r} = R_{\theta \phi \theta}^{\phi} = \frac{r_s}{r}, \qquad 2 R_{\phi \phi t}^{t} = 2 R_{\phi \phi r}^{r} = - R_{\phi \phi \theta}^{\theta} = \frac{r_s \sin^2 \left( \theta \right)}{r}.
\end{equation}
Any one of these components, or indeed the conformal factor ${\psi}$, may be used as the scalar field quantity ${\phi}$ in the computation of the refinement criterion for the hypergraph; henceforth, we opt to use the conformal factor ${\phi = \psi}$. We place the outer boundary of the computational domain at ${60 M}$, enforcing Sommerfeld boundary conditions:

\begin{equation}
\frac{\partial f \left( x_i, t \right)}{\partial t} = - \frac{v x_i}{r} \frac{\partial f \left( x_i, t \right)}{\partial x_i} - v \frac{f \left( x_i, t \right) - f_0 \left( x_i, t \right)}{r},
\end{equation}
with $f$ being an arbitrary scalar field, ${r = \sqrt{x_{1}^{2} + x_{2}^{2} + x_{3}^{2}}}$ being the radial distance from the center of the domain, ${f_0}$ being the spacetime boundary field values (for the purposes of the tests presented in this article, this is typically Minkowski space), and $v$ being the radiative velocity (typically ${v = 1}$). The solution is evolved until a final time of ${t = 1000 M}$; the initial and final hypersurface configurations are shown in Figures \ref{fig:Figure8} and \ref{fig:Figure9}, respectively, with resolutions of 100, 200 and 400 vertices, and with the hypergraphs adapted and colored using the Schwarzschild conformal factor ${\psi}$. Figure \ref{fig:Figure10} shows the discrete characteristic structure of these solutions after time ${t = 1000 M}$ (using directed acyclic causal graphs to show discrete characteristic lines), and Figures \ref{fig:Figure11} and \ref{fig:Figure12} show projections along the $z$-axis of the initial and final hypersurface configurations, respectively, with vertices assigned spatial coordinates according to the profile of the Schwarzschild conformal factor ${\psi}$.

We confirm that the ADM mass of the black hole (computed by integrating over a surface surrounding the boundary of asymptotic flatness) remains approximately constant, and the linear and angular momenta of the black hole remain approximately zero, as expected. Here, and henceforth, we compute the ${L_1}$, ${L_2}$ and ${L_{\infty}}$ norms of the Hamiltonian constraint:

\begin{equation}
H = R - K_{i j} K^{i j} + K^2,
\end{equation}
using:

\begin{equation}
\left\lVert H \right\rVert_{1} = \sum_{i} m_i \left\lvert H_{i} \right\rvert, \qquad \left\lVert H \right\rVert_{2} = \sqrt{\sum_{i} m_i \left\lvert H_{i}^{2} \right\rvert}, \qquad \left\lVert H \right\rVert_{\infty} = \max_{i} \left[ m_i \left\lvert H_i \right\rvert \right],
\end{equation}
respectively, where ${m_i}$ denotes the fraction of the total volume of the domain occupied by the $i$th partitioned subgraph, i.e:

\begin{equation}
m_i = \frac{V_i}{V_{tot}}.
\end{equation}
In this, as in all subsequent black hole simulations, the spacetime region in which the lapse ${\alpha < 0.3}$ (corresponding to the approximate location of an event horizon in the moving puncture gauge) is effectively excised by setting $H$ to zero; the standard alpha-driver and gamma-driver conditions are used here. The convergence rates for the Hamiltonian constraint after time ${t = 1000 M}$, with respect to the ${L_1}$, ${L_2}$ and ${L_{\infty}}$ norms, illustrating approximately fourth-order convergence of the finite difference scheme, are shown in Table \ref{tab:Table1}.

\begin{table}[ht]
\centering
\begin{tabular}{|c|c|c|c|c|c|c|}
\hline
Vertices & ${\epsilon \left( L_1 \right)}$ & ${\epsilon \left( L_2 \right)}$ & ${\epsilon \left( L_{\infty} \right)}$ & ${\mathcal{O} \left( L_1 \right)}$ & ${\mathcal{O} \left( L_2 \right)}$ & ${\mathcal{O} \left( L_{\infty} \right)}$\\
\hline\hline
100 & ${6.72 \times 10^{-3}}$ & ${7.73 \times 10^{-3}}$ & ${2.80 \times 10^{-3}}$ & - & - & -\\
\hline
200 & ${3.05 \times 10^{-4}}$ & ${3.31 \times 10^{-4}}$ & ${1.81 \times 10^{-4}}$ & 4.46 & 4.55 & 3.95\\
\hline
400 & ${2.47 \times 10^{-5}}$ & ${1.55 \times 10^{-5}}$ & ${7.97 \times 10^{-6}}$ & 3.63 & 4.41 & 4.51\\
\hline
800 & ${2.16 \times 10^{-6}}$ & ${1.77 \times 10^{-6}}$ & ${3.74 \times 10^{-7}}$ & 3.51 & 3.10 & 4.41\\
\hline
1600 & ${1.02 \times 10^{-7}}$ & ${1.07 \times 10^{-7}}$ & ${1.86 \times 10^{-8}}$ & 4.41 & 4.06 & 4.33\\
\hline
\end{tabular}
\caption{Convergence rates for the static Schwarzschild black hole test with respect to the ${L_1}$, ${L_2}$ and ${L_{\infty}}$ norms for the Hamiltonian constraint $H$ after time ${t = 1000 M}$, showing approximately fourth-order convergence.}
\label{tab:Table1}
\end{table}

\clearpage

\subsection{Rapidly Rotating Kerr Black Hole}

Next, we evolve a rapidly rotating Kerr black hole, i.e. a spacetime defined by the following line element in Boyer-Lindquist coordinates\cite{landau}\cite{visser}\cite{boyer}:

\begin{multline}
g = ds^2 = - \left( 1 - \frac{r_s r}{\Sigma} \right) dt^2 + \frac{\Sigma}{\Delta} dr^2 + \Sigma d \theta^2\\
+ \left( r^2 + a^2 + \frac{r_s r a^2}{\Sigma} \sin^2 \left( \theta \right) \right) \sin^2 \left( \theta \right) d \phi^2 - \frac{2 r_s r a \sin^2 \left( \theta \right)}{\Sigma} dt d \phi,
\end{multline}
where ${\left( r, \theta, \phi \right)}$ denotes the standard three-dimensional orthogonal coordinate system on an oblate spheroid, related to the Cartesian coordinates ${\left( x, y, z \right)}$ by:

\begin{equation}
x = \sqrt{r^2 + a^2} \sin \left( \theta \right) \cos \left( \phi \right), \qquad y = \sqrt{r^2 + a^2} \sin \left( \theta \right) \sin \left( \phi \right), \qquad z = r \cos \left( \theta \right),
\end{equation}
${r_s = 2 M}$ is the usual Schwarzschild radius, $a$ is the ratio of the black hole's angular momentum/spin to its ADM mass:

\begin{equation}
a = \frac{J}{M},
\end{equation}
and ${\Sigma}$ and ${\Delta}$ are useful characteristic length scales, namely:

\begin{equation}
\Sigma = r^2 + a^2 \cos^2 \left( \theta \right), \qquad \text{ and } \qquad \Delta = r^2 - r_s r + a^2.
\end{equation}
However, for the convenience of constructing the initial data for our simulation, we use Kerr-Schild coordinates instead\cite{debney}\cite{balasin}, whose metric tensor ${g_{\mu \nu}}$ is defined by:

\begin{equation}
g_{\mu \nu} = \eta_{\mu \nu} + f k_{\mu} k_{\nu},
\end{equation}
since these allow us to define the initial hypergraph configuration as a perturbation of a flat (Minkowski) metric. In the above, ${\eta_{\mu \nu}}$ is the metric tensor of flat (Minkowski) spacetime, $f$ is a scalar field defined by:

\begin{equation}
f = \frac{2 M r^3}{r^4 + a^2 z^2},
\end{equation}
${\mathbf{k}}$ is a spacetime unit vector of the form:

\begin{equation}
\mathbf{k} = \left( k_0, k_x, k_y, k_z \right) = \left( 1, \frac{rx + ay}{r^2 + a^2}, \frac{ry - ax}{r^2 + a^2}, \frac{z}{r} \right),
\end{equation}
${\left( x, y, z \right)}$ are standard Cartesian coordinates, and $a$ is the angular momentum parameter introduced above (assuming that the angular momentum vector ${\mathbf{J}}$ is defined relative to the positive $z$-axis). It is worth noting that the quantity $r$ in Kerr-Schild coordinates is actually a distorted form of the usual radius $R$, defined implicitly by:

\begin{equation}
\frac{x^2 + y^2}{r^2 + a^2} + \frac{z^2}{r^2} = 1,
\end{equation}
and only becomes equal to $R$ in the limit of a non-rotating black hole (i.e. as $a$ approaches zero):

\begin{equation}
\lim_{a \to 0} \left[ r \right] = R = \sqrt{x^2 + y^2 + z^2}.
\end{equation}
The advantage of using Kerr-Schild coordinates for constructing the initial data is that the determinant of the metric tensor ${\mathrm{det} \left( g_{i j} \right) = -1}$ is constant across all of spacetime except for the curvature singularity at ${r = 0}$ (including points that are arbitrarily close to the curvature singularity).

In the non-relativistic limit (i.e. as $M$ and/or ${r_s}$ approach zero), the Kerr metric in Boyer-Lindquist coordinates reduces to the standard orthogonal metric on an oblate spheroid:

\begin{equation}
\lim_{M, r_s \to 0} \left[ ds^2 \right] = -dt^2 + \frac{\Sigma}{r^2 + a^2} dr^2 + \Sigma d \theta^2 + \left( r^2 + a^2 \right) \sin^2 \left( \theta \right) d \phi^2.
\end{equation}
On the other hand, in the limit of infinite distance (i.e. as $R$ approaches infinity), the Kerr metric in Kerr-Schild coordinates reduces to the Schwarzschild metric in Eddington-Finkelstein coordinates, i.e:

\begin{equation}
\lim_{R \to \infty} \left[ ds^2 \right] = - \left( 1 - \frac{2M}{r} \right) dv^2 + 2 dv dr + r^2 d \Omega^2,
\end{equation}
where, as usual, ${d \Omega^2}$ represents the induced Riemannian spatial metric on the 2-sphere ${S^2 \subset E^3}$:

\begin{equation}
d \Omega^2 = d \theta^2 + \sin^2 \left( \theta \right) d \phi^2,
\end{equation}
and where ${v = t + r^{*}}$ represents the Eddington-Finkelstein time coordinate, defined in terms of the tortoise coordinate ${r^{*}}$:

\begin{equation}
r^{*} = r + 2 M \log \left( \left\lvert \frac{R}{2M} - 1 \right\rvert \right),
\end{equation}
such that the differential equation:

\begin{equation}
\frac{dr^{*}}{dr} = \left( 1 - \frac{2M}{r} \right)^{-1},
\end{equation}
is satisfied, and such that ${r^{*}\to -\infty}$ as ${r \to r_s = 2M}$. The metric also exhibits two surfaces on which the metric appears to be singular, which will both play an important role in our forthcoming numerical simulations. The first is a standard Schwarzschild event horizon, occurring when the purely radial component of the metric tensor ${g_{r r}}$ becomes infinite, such that:

\begin{equation}
\frac{1}{g_{r r}} = 0, \qquad \implies r_I = \frac{r_s \pm \sqrt{r_{s}^{2} - 4 a^2}}{2},
\end{equation}
which we denote using ${r_I}$ for \textit{interior} horizon. The second is a horizon at which the temporal component of the metric tensor ${g_{t t}}$ undergoes a positive-to-negative sign change, such that:

\begin{equation}
g_{t t} = 0, \qquad \implies r_E = \frac{r_s \pm \sqrt{r_{s}^{2} - 4 a^2 \cos^2 \left( \theta \right)}}{2},
\end{equation}
which we denote using ${r_E}$ for \textit{exterior} horizon, with the space between the horizons forming the \textit{ergosphere} of the Kerr black hole. These are both pure coordinate singularities, and the spacetime can be smoothly continued across them.

By applying the following transformation to the Boyer-Lindquist coordinates:

\begin{equation}
\tilde{r} = \frac{\alpha^2 \cos^2 \left( \frac{\lambda}{2} \right) + \delta^2 \sin^2 \left( \frac{\lambda}{2} \right)}{\alpha \sin \left( \lambda \right)} + M,
\end{equation}
with ${0 \leq \lambda \leq \pi}$, an arbitrary positive constant ${\alpha}$ and:

\begin{equation}
\delta = \sqrt{M^2 - a^2},
\end{equation}
we are effectively composing a transformation to the quasi-isotropic radial coordinate ${\bar{r}}$ and a stereographic projection, i.e:

\begin{equation}
\tilde{r} + \bar{r} + M + \frac{\delta^2}{4 \bar{r}}, \qquad \text{ where } \qquad \bar{r} = \frac{\alpha \cos \left( \frac{\lambda}{2} \right)}{2 \sin \left( \frac{\lambda}{2} \right)},
\end{equation}
yielding a metric that is defined for all ${\tilde{r} > M + \delta}$, and singular at ${\tilde{r} = M + \delta}$. The conformal factor ${\psi}$ then becomes elementary in terms of this new coordinate system:

\begin{equation}
\psi = \frac{\Sigma^{\frac{1}{4}}}{\sqrt{\sin \left( \lambda \right)}}.
\end{equation}

On the other hand, if we introduce a new coordinate ${\chi}$ to the Boyer-Lindquist coordinate system using:

\begin{equation}
\chi = \cos \left( \theta \right), \qquad \text{ such that } \qquad \chi \in \left[ -1, 1 \right],
\end{equation}
then one has:

\begin{equation}
\sin \left( \theta \right) = \sqrt{1 - \chi^2}, \qquad \implies d \chi = \sin \left( \theta \right) d \theta, \qquad \implies d \theta = \frac{d \chi}{\sqrt{1 - \chi^2}},
\end{equation}
thus allowing us to replace the ${\theta}$ coordinate in such a way that, in terms of this new coordinate system ${\left( t, r, \chi, \phi \right)}$, the Boyer-Lindquist form of the Kerr metric becomes:

\begin{multline}
ds^2 = - \left( 1 - \frac{2 M r}{r^2 + a^2 \chi^2} \right) dt^2 - \frac{4 a M r \left( 1 - \chi^2 \right)}{r^2 + a^2 \chi^2} d \phi dt + \frac{r^2 + a^2 \chi^2}{r^2 - 2 M r + a^2} dr^2 + \left( r^2 + a^2 \chi^2 \right) \frac{d \chi^2}{1 - \chi^2}\\
+ \left( 1 - \chi^2 \right) \left[ r^2 + a^2 + \frac{2 M a^2 r \left( 1 - \chi^2 \right)}{r^2 + a^2 \chi^2} \right] d \phi^2,
\end{multline}
with the property that all dependencies upon trigonometric functions have been removed, and all metric components can now be written entirely as rational polynomials of the coordinates. The only non-trivial quadratic curvature invariant ${R^{i j k l} R_{i j k l}}$ then becomes:

\begin{equation}
R^{i j k l} R_{i j k l} = \frac{48 \left( r^2 - a^2 \chi^2 \right) \left[ \left( r^2 + a^2 \chi^2 \right)^2 - 16 r^2 a^2 \chi^2 \right]}{\left( r^2 + a^2 \chi^2 \right)^6}.
\end{equation}
In the orthonormal basis defined by the following tetrad:

\begin{equation}
e_{A}^{a} = \begin{bmatrix}
e_{0}^{a}\\
e_{1}^{a}\\
e_{2}^{a}\\
e_{3}^{a}
\end{bmatrix} = \begin{bmatrix}
\frac{r^2 + a^2}{\sqrt{\left( r^2 - 2 M r + a^2 \right) \left( r^2 + a^2 \chi^2 \right)}} & 0 & 0 & \frac{a}{\sqrt{\left( r^2 - 2 M r + a^2 \right) \left( r^2 + a^2 \chi^2 \right)}}\\
0 & \sqrt{\frac{r^2 + a^2 \chi^2}{r^2 - 2 M r + a^2}} & 0 & 0\\
0 & 0 & \sqrt{\frac{r^2 + a^2 \chi^2}{1 - \chi^2}} & 0\\
a \sqrt{\frac{1 - \chi^2}{r^2 + a^2}} & 0 & 0 & \frac{1}{\sqrt{\left( 1 - \chi^2 \right) \left( r^2 + a^2 \right)}}
\end{bmatrix},
\end{equation}
the only non-zero components of the Riemann curvature tensor ${R_{i j k l}}$, up to redundancies due to symmetry, are:

\begin{equation}
R_{0 1 0 1} = -2 R_{0 2 0 2} = -2 R_{0 3 0 3} = 2 R_{1 2 1 2} = 2 R_{1 3 1 3} = - R_{2 3 2 3} = - \frac{2 M r \left( r^2 - 3 a^2 \chi^2 \right)}{\left( r^2 + a^2 \chi^2 \right)^3},
\end{equation}
and:

\begin{equation}
R_{0 1 2 3} = R_{0 2 1 3} = - R_{0 3 1 2} = \frac{2 M a \chi \left( r^2 - a^2 \chi^2 \right)}{\left( r^2 + a^2 \chi^2 \right)^3}.
\end{equation}
Along the equator ${\chi = 0}$, the only non-zero curvature components are now the same as they would be for the Schwarzschild metric, namely:

\begin{equation}
R_{0 1 0 1} = -2 R_{0 2 0 2} = -2 R_{0 3 0 3} = 2 R_{1 2 1 2} = 2 R_{1 3 1 3} = - R_{2 3 2 3} = - \frac{2M}{r^3},
\end{equation}
whereas along the axis of rotation ${\chi = \pm 1}$, they are instead:

\begin{equation}
R_{0 1 0 1} = - 2 R_{0 2 0 2} = - 2 R_{0 3 0 3} = 2 R_{1 2 1 2} = 2 R_{1 3 1 3} = - R_{2 3 2 3} = - \frac{2 M r \left( r^2 - 3 a^2 \right)}{\left( r^2 + a^2 \right)^3},
\end{equation}
and:

\begin{equation}
R_{0 1 2 3} = R_{0 2 1 3} = - R_{0 3 1 2} = \pm \frac{2 M a \left( 3 r^2 - a^2 \right)}{\left( r^2 + a^2 \right)^3}.
\end{equation}
As in the Schwarzschild case, any one of these components may be used as the scalar field quantity ${\phi}$ in the refinement algorithm in place of the conformal factor ${\psi}$, but we choose ${\phi = \psi}$ for the purposes of consistency. The size of the computational domain is chosen to be ${\left( 32 M \right)^3}$, with periodic boundary conditions imposed for the sake of simplicity (although the resultant boundary effects will limit the stable duration of the simulation).

\begin{figure}[ht]
\centering
\includegraphics[width=0.325\textwidth]{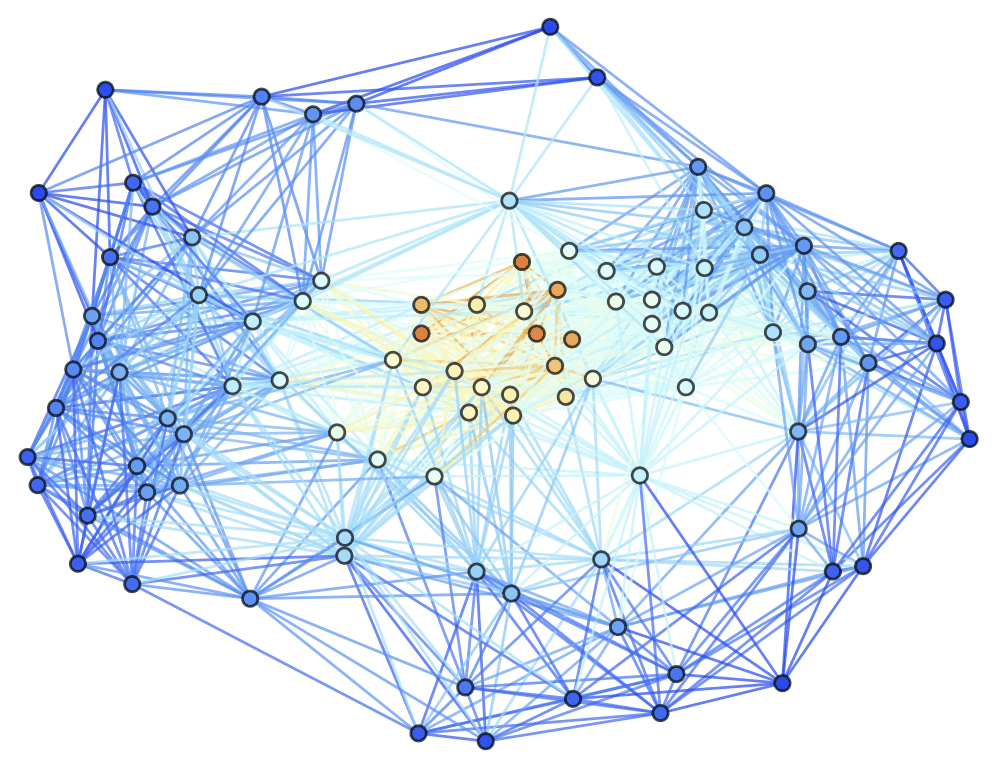}
\includegraphics[width=0.325\textwidth]{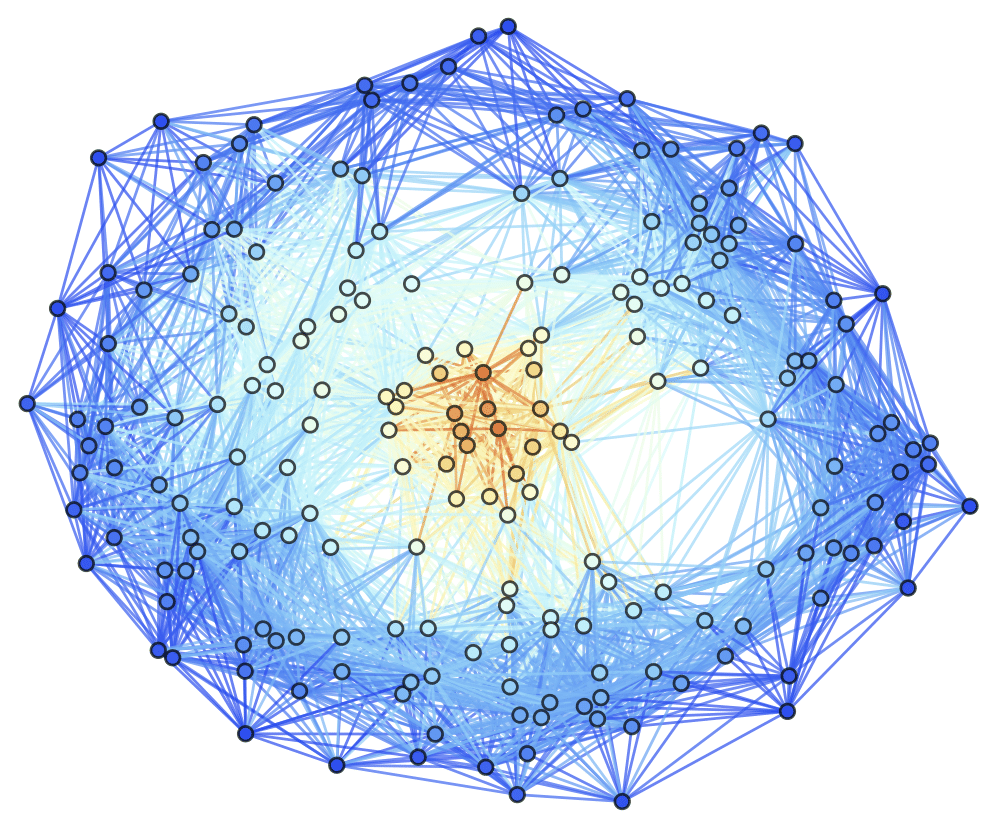}
\includegraphics[width=0.325\textwidth]{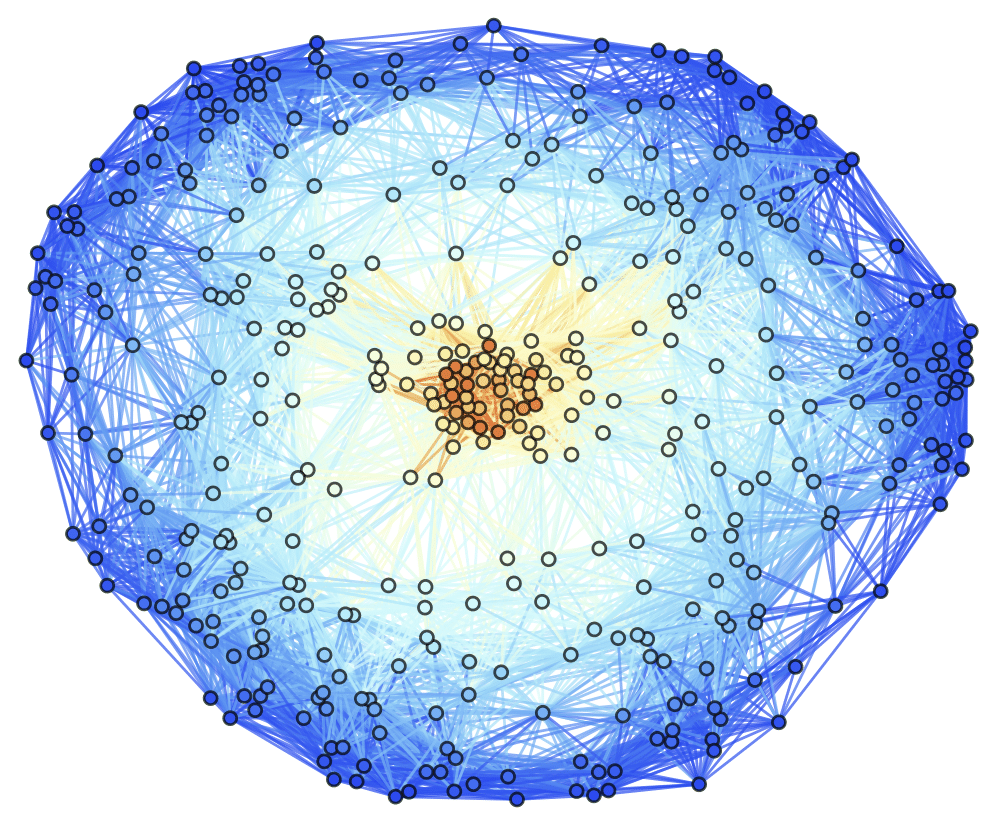}
\caption{Spatial hypergraphs corresponding to the final hypersurface configuration of the rapidly rotating Kerr black hole test with ${a = 0.3}$ at time ${t = 20 M}$, with resolutions of 100, 200 and 400 vertices, respectively. The hypergraphs have been adapted and colored using the local curvature in the Boyer-Lindquist conformal factor ${\psi}$.}
\label{fig:Figure13}
\end{figure}

\begin{figure}[ht]
\centering
\includegraphics[width=0.325\textwidth]{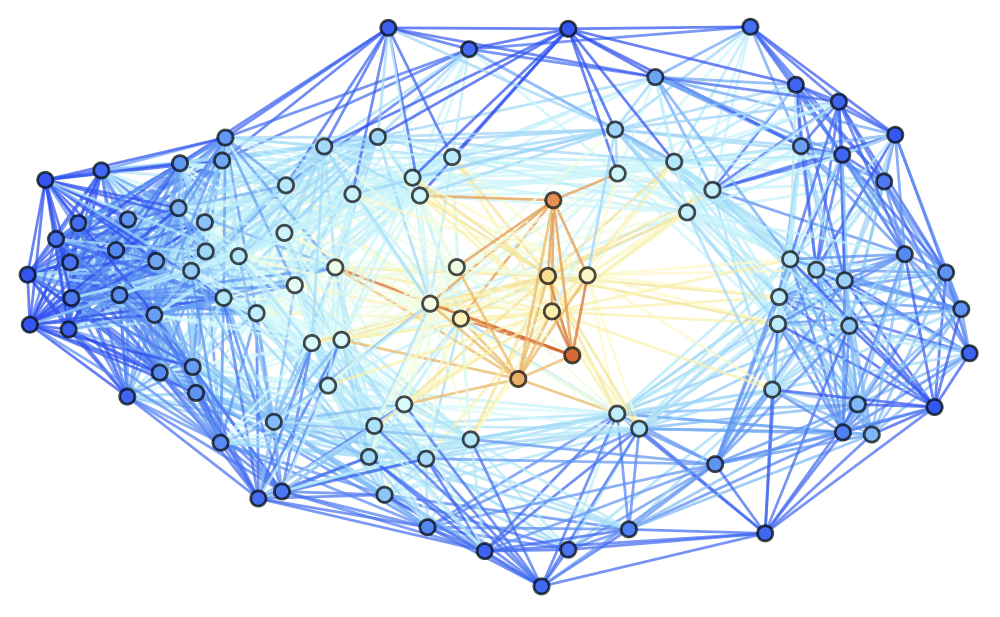}
\includegraphics[width=0.325\textwidth]{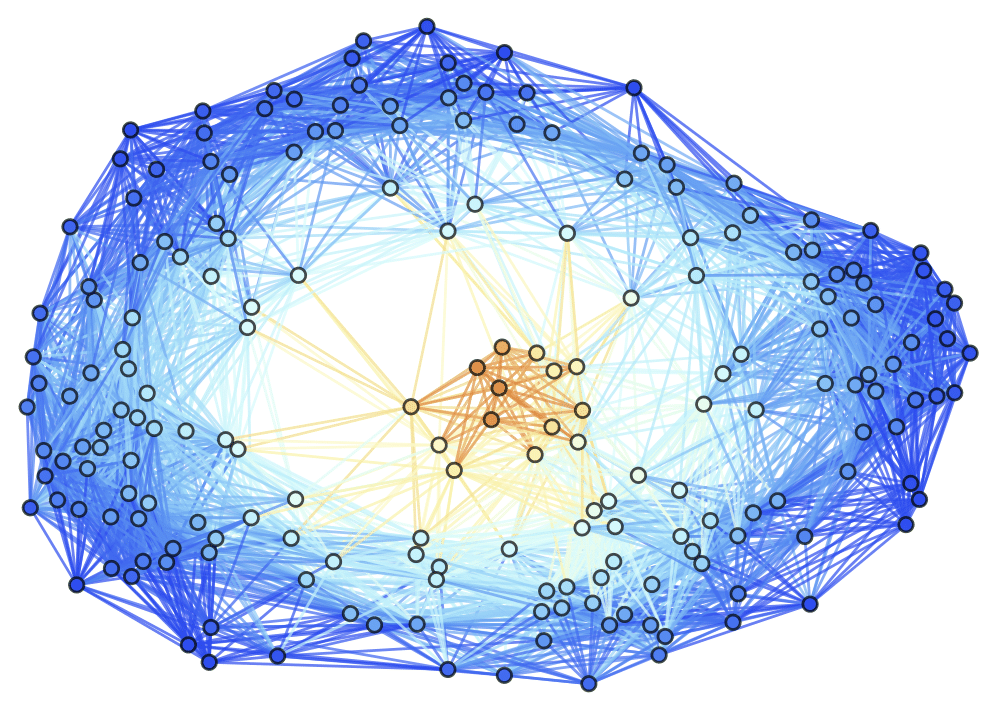}
\includegraphics[width=0.325\textwidth]{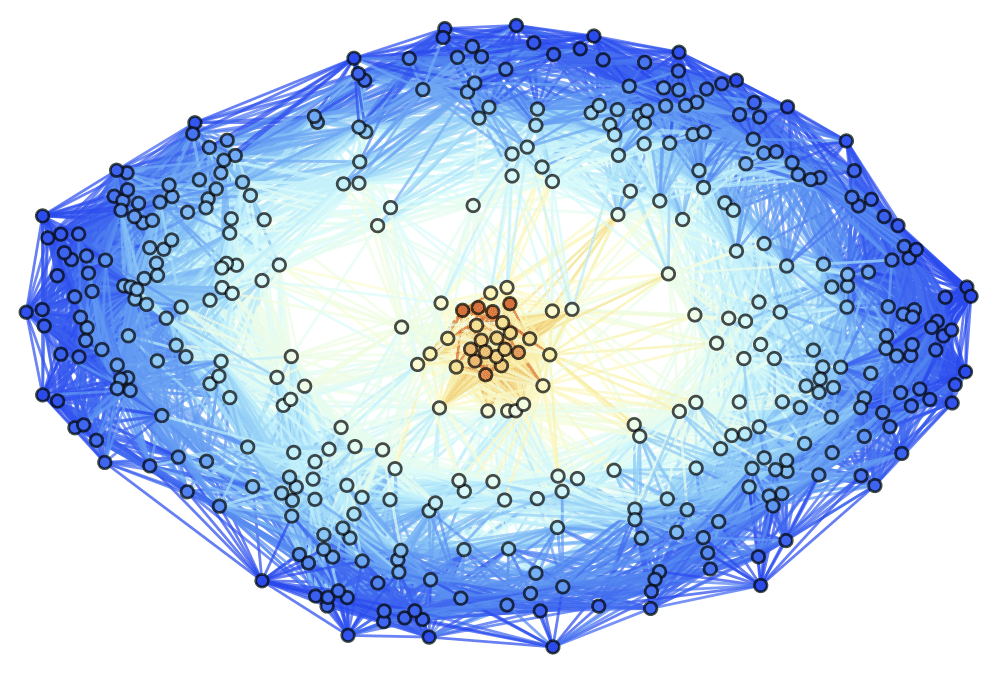}
\caption{Spatial hypergraphs corresponding to the final hypersurface configuration of the rapidly rotating Kerr black hole test with ${a = 0.6}$ at time ${t = 20 M}$, with resolutions of 100, 200 and 400 vertices, respectively. The hypergraphs have been adapted and colored using the local curvature in the Boyer-Lindquist conformal factor ${\psi}$.}
\label{fig:Figure14}
\end{figure}

\begin{figure}[ht]
\centering
\includegraphics[width=0.325\textwidth]{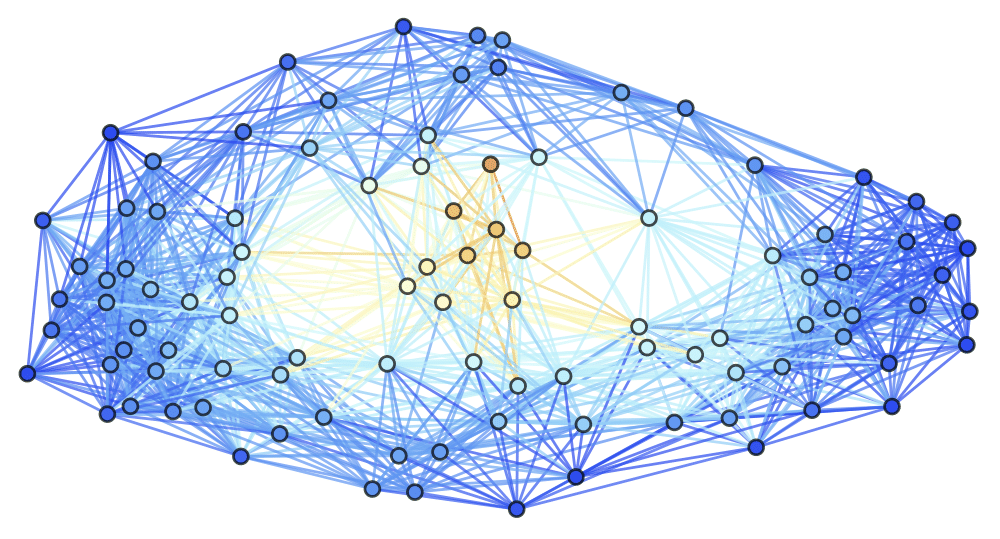}
\includegraphics[width=0.325\textwidth]{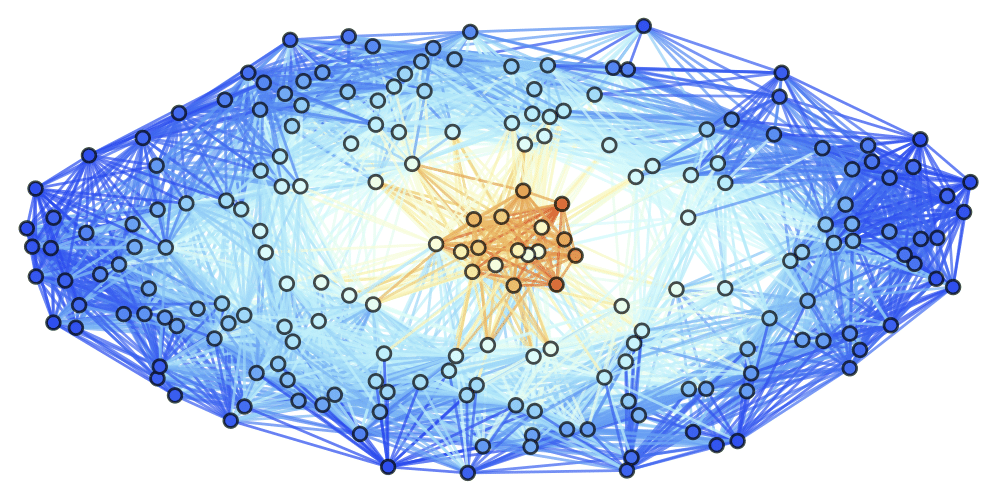}
\includegraphics[width=0.325\textwidth]{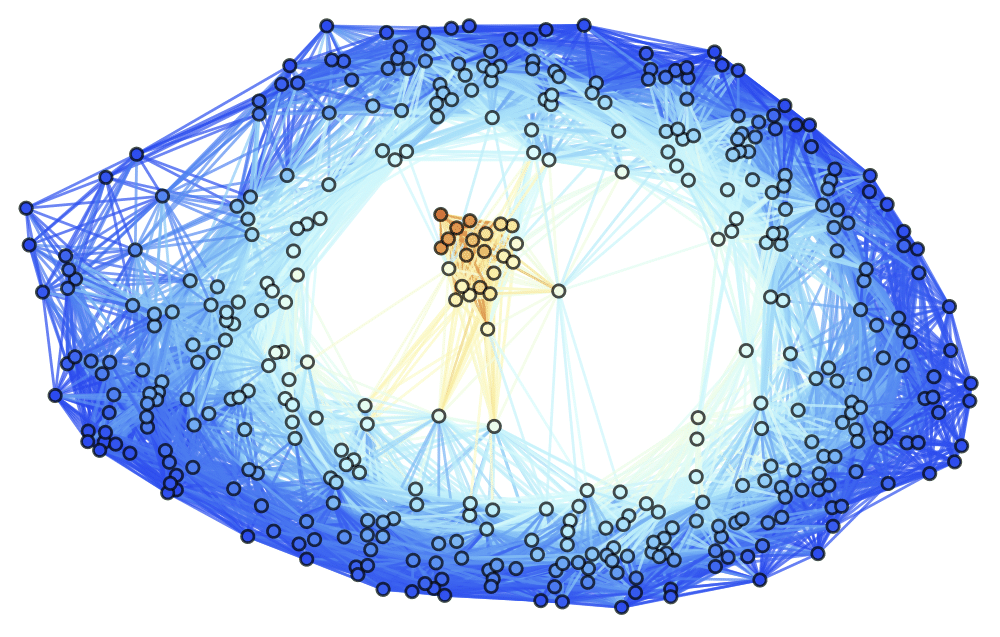}
\caption{Spatial hypergraphs corresponding to final hypersurface configurations of the rapidly rotating Kerr black hole test with ${a = 0.9}$ at time ${t = 20 M}$, with resolutions of 100, 200 and 400 vertices, respectively. The hypergraphs have been adapted and colored using the local curvature in the Boyer-Lindquist conformal factor ${\psi}$.}
\label{fig:Figure15}
\end{figure}

\begin{figure}[ht]
\centering
\includegraphics[width=0.395\textwidth]{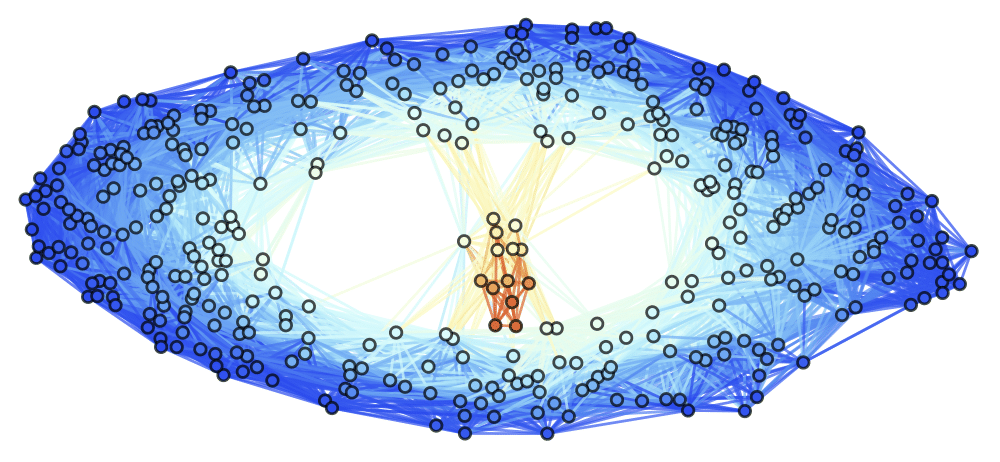}\hspace{0.1\textwidth}
\includegraphics[width=0.495\textwidth]{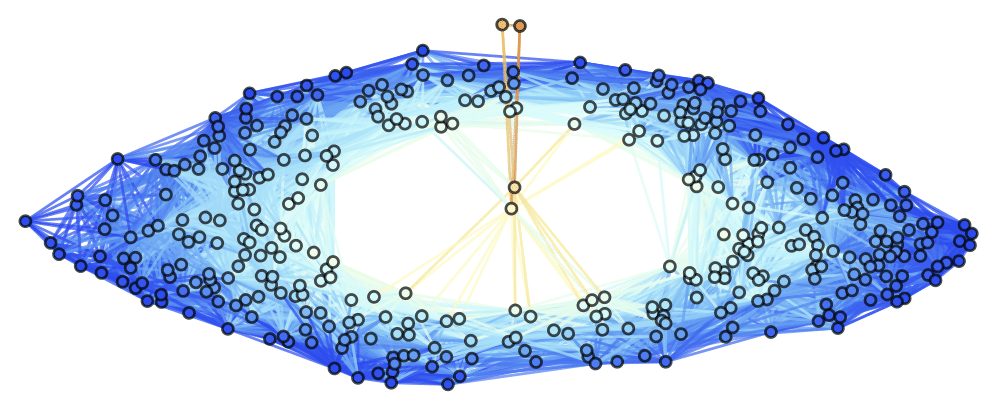}
\caption{Spatial hypergraphs corresponding to final hypersurface configurations of the rapidly rotating Kerr black hole test with a resolution of 400 vertices at time ${t = 20 M}$, for ${a = 0.95}$ and ${a = 0.99}$, respectively, showing extreme separation of the interior and exterior event horizons. The hypergraphs have been adapted and colored using the local curvature in the Boyer-Lindquist conformal factor ${\psi}$.}
\label{fig:Figure16}
\end{figure}

\begin{figure}[ht]
\centering
\includegraphics[width=0.325\textwidth]{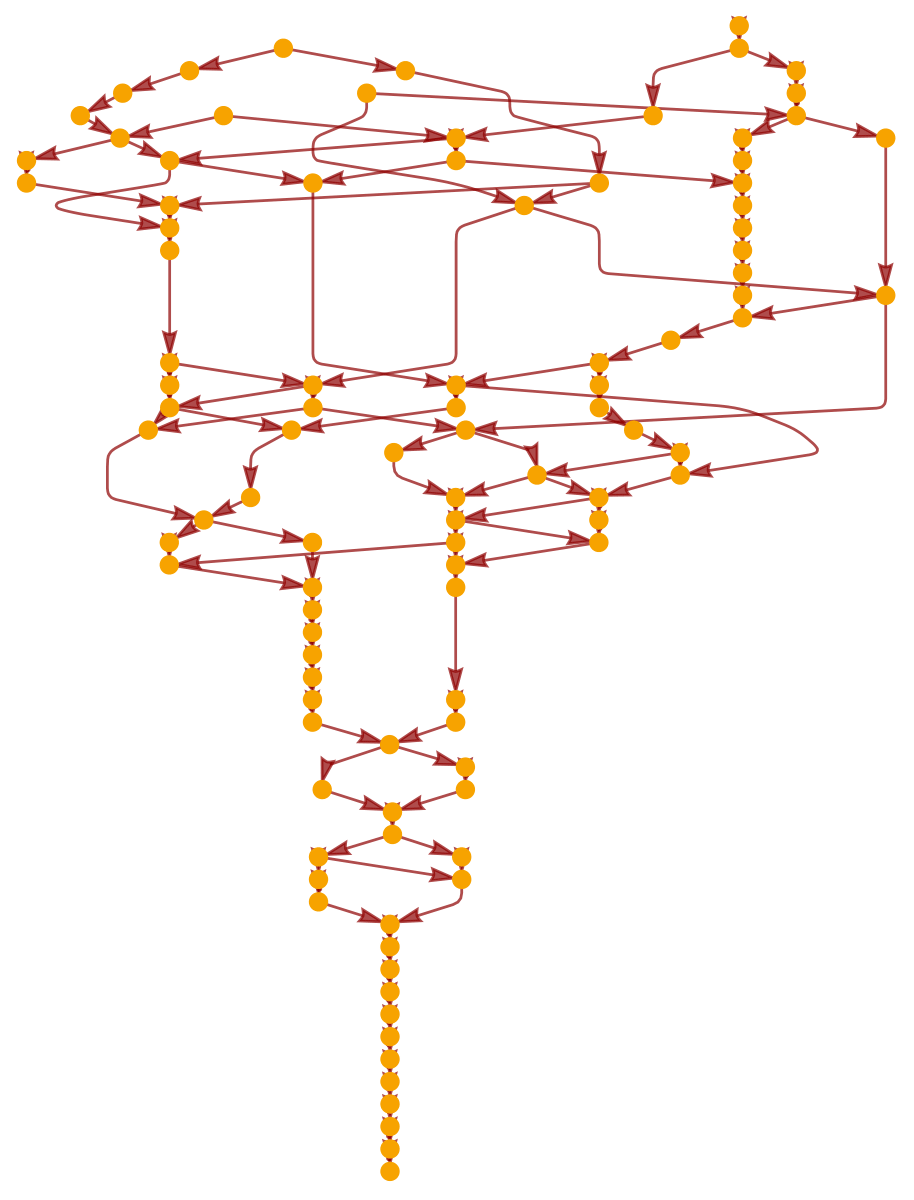}
\includegraphics[width=0.325\textwidth]{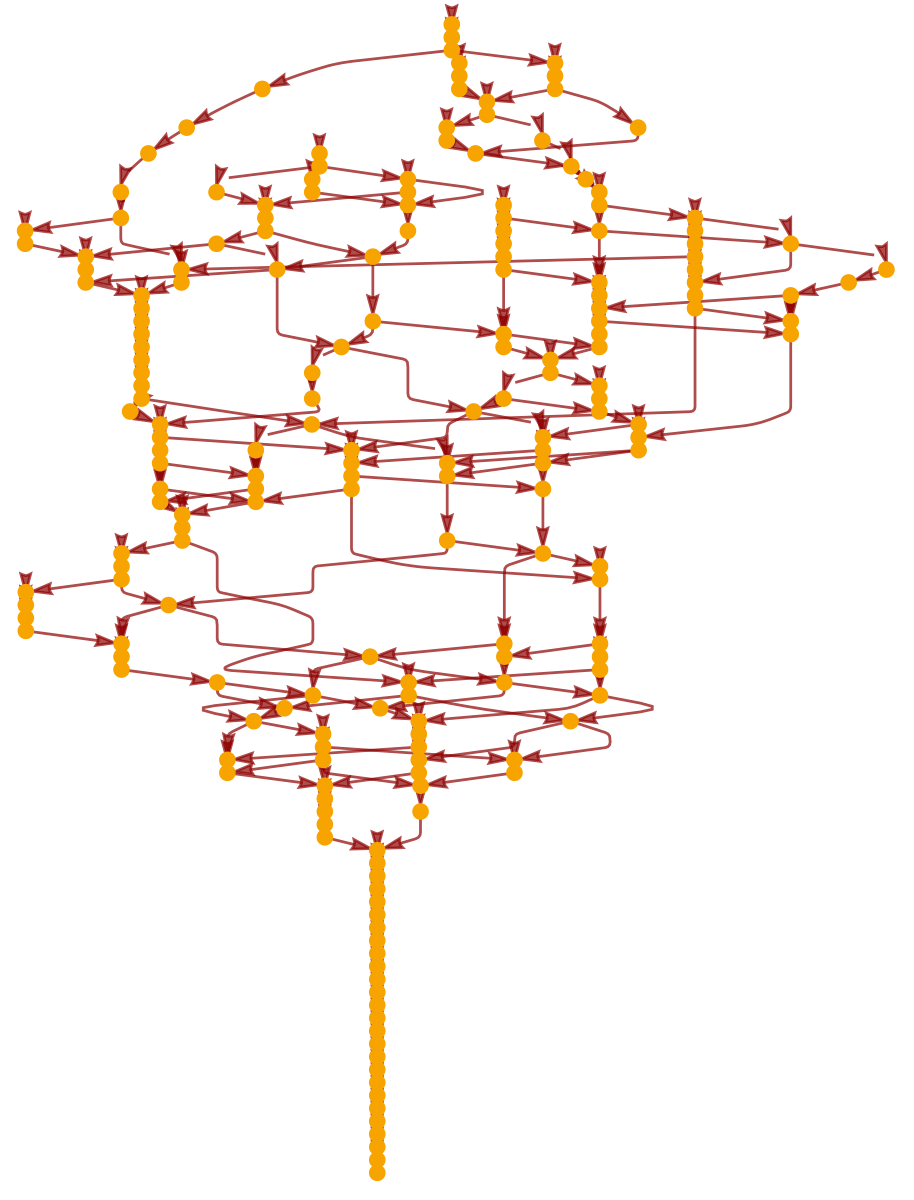}
\includegraphics[width=0.325\textwidth]{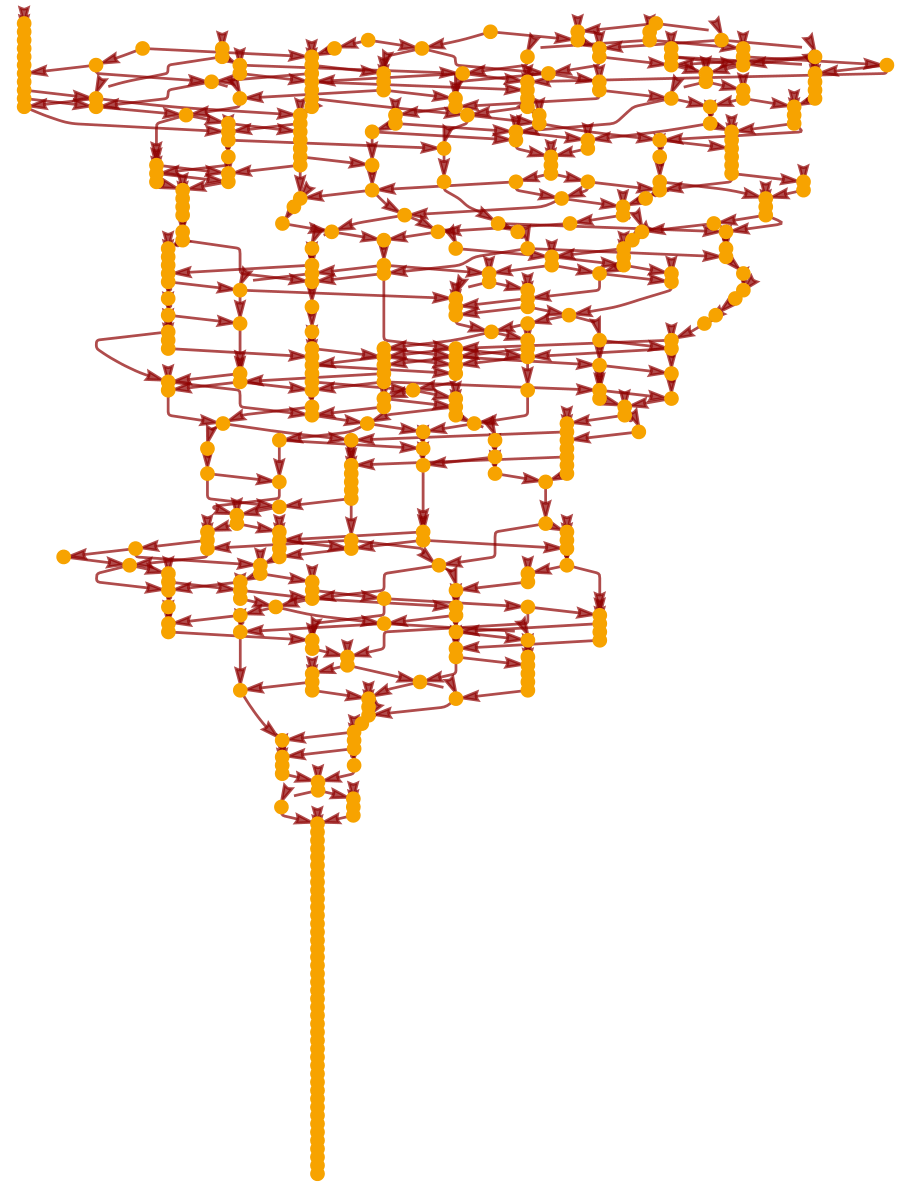}
\caption{Causal graphs corresponding to the discrete characteristic structure of the rapidly rotating Kerr black hole test with ${a = 0.3}$ at time ${t = 20 M}$, with resolutions of 100, 200 and 400 hypergraph vertices, respectively.}
\label{fig:Figure17}
\end{figure}

\begin{figure}[ht]
\centering
\includegraphics[width=0.325\textwidth]{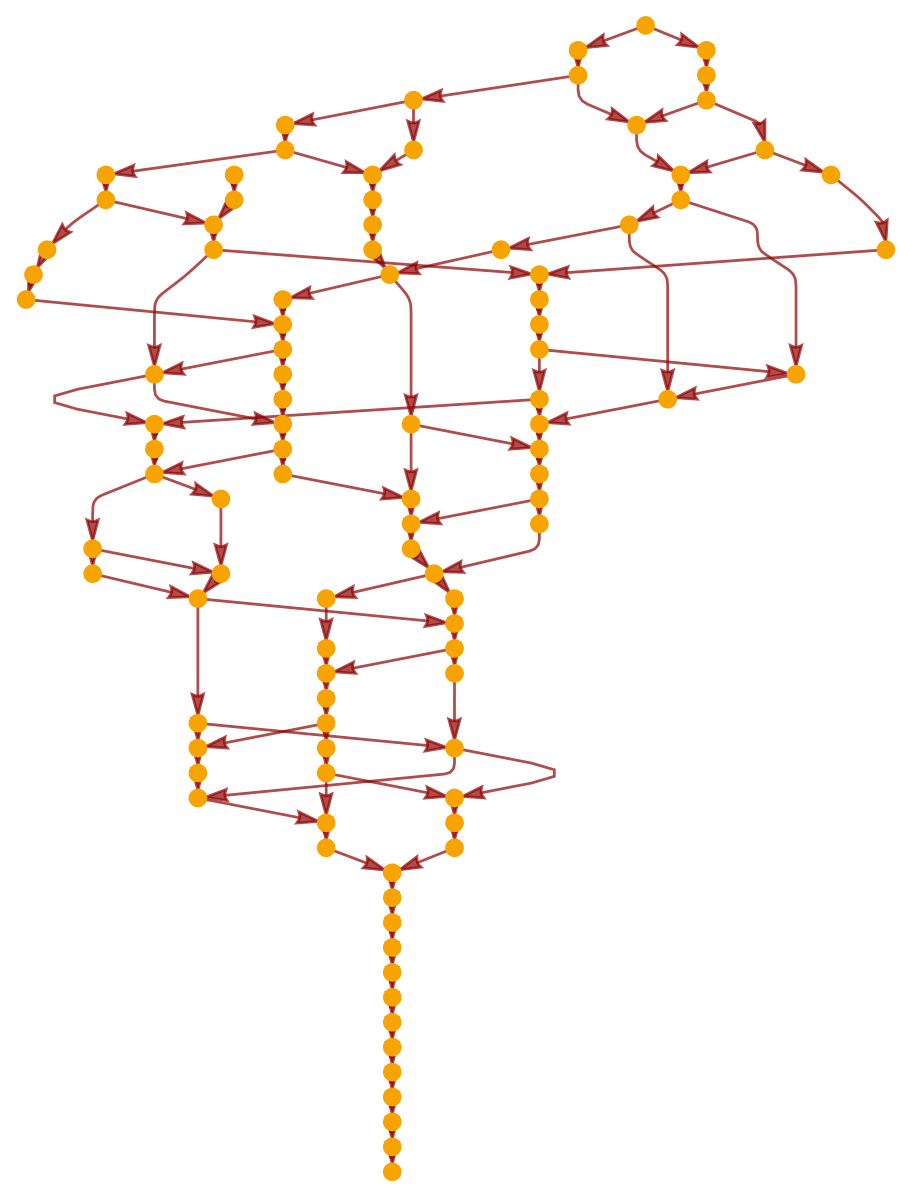}
\includegraphics[width=0.325\textwidth]{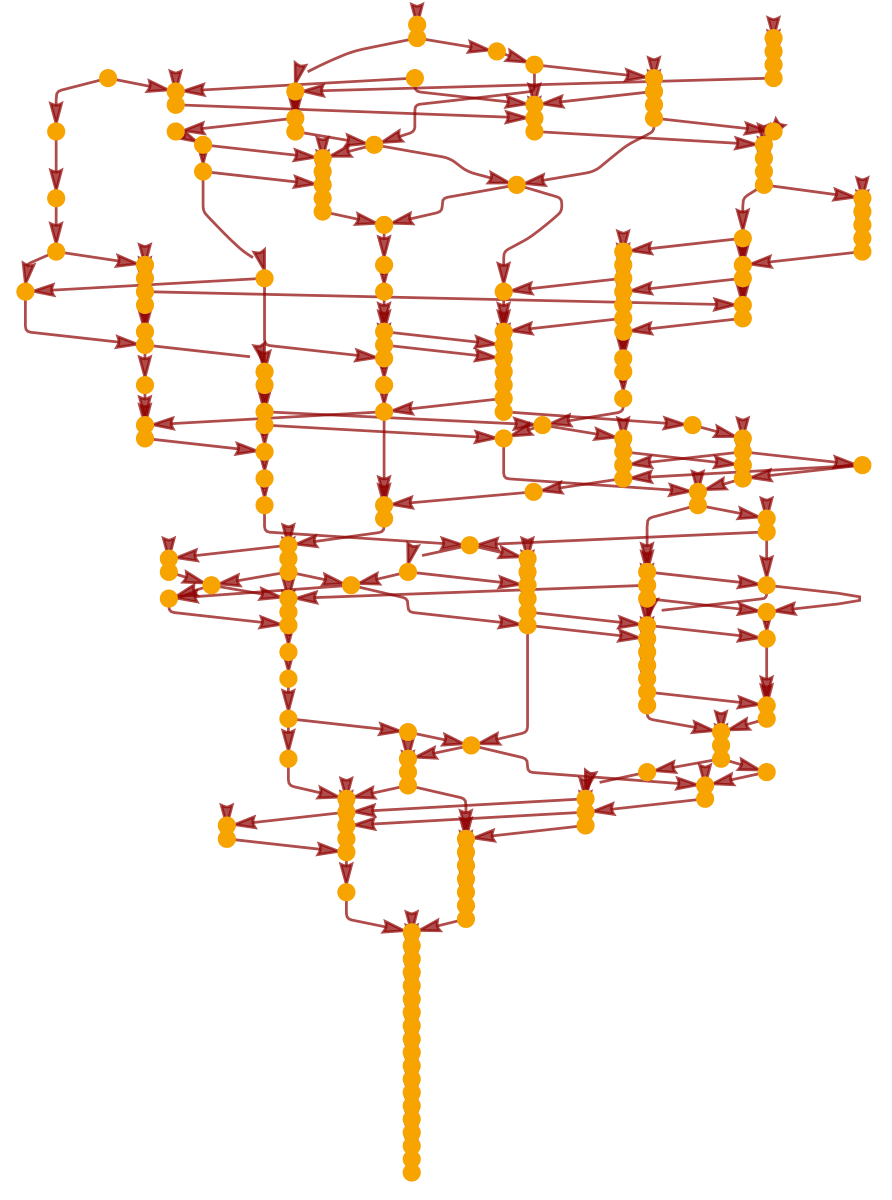}
\includegraphics[width=0.325\textwidth]{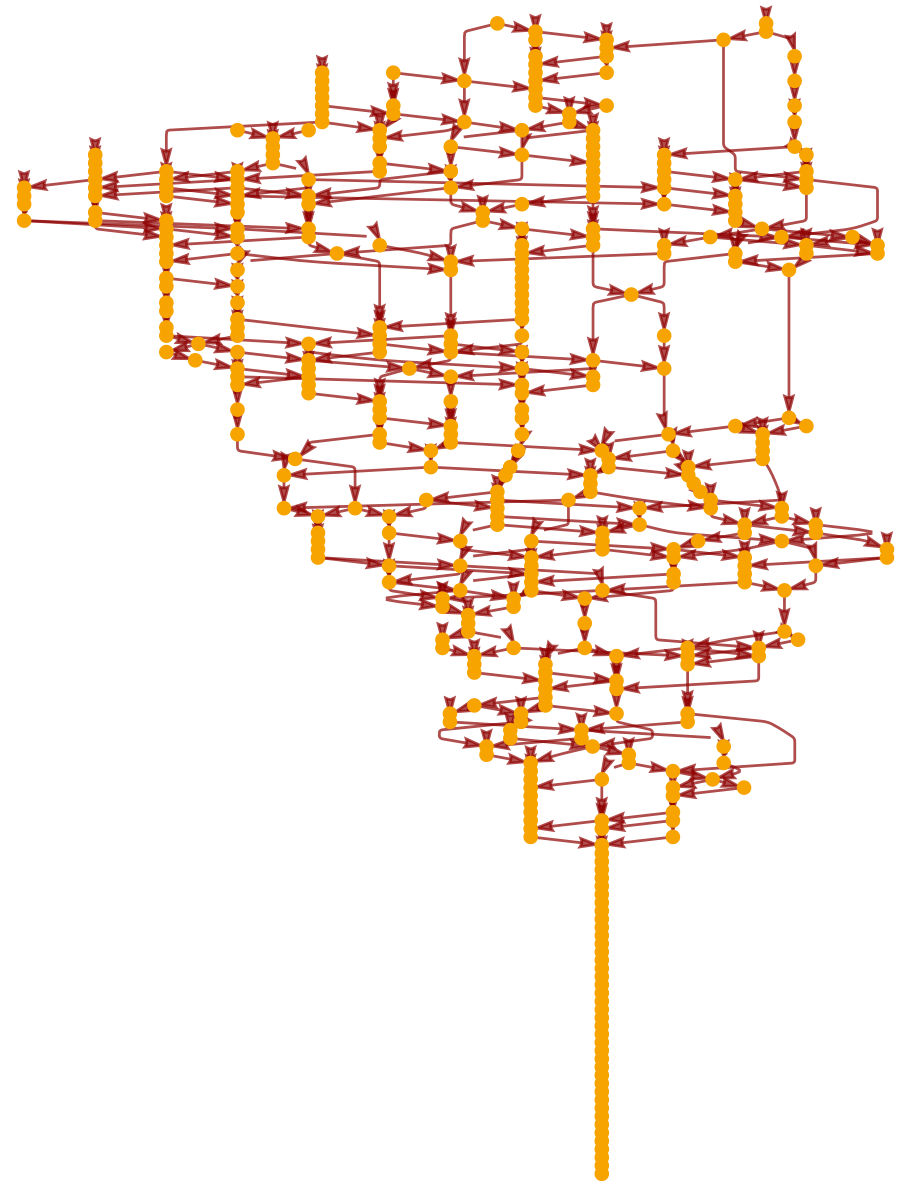}
\caption{Causal graphs corresponding to the discrete characteristic structure of the rapidly rotating Kerr black hole test with ${a = 0.6}$ at time ${t = 20 M}$, with resolutions of 100, 200 and 400 hypergraph vertices, respectively.}
\label{fig:Figure18}
\end{figure}

\begin{figure}[ht]
\centering
\includegraphics[width=0.325\textwidth]{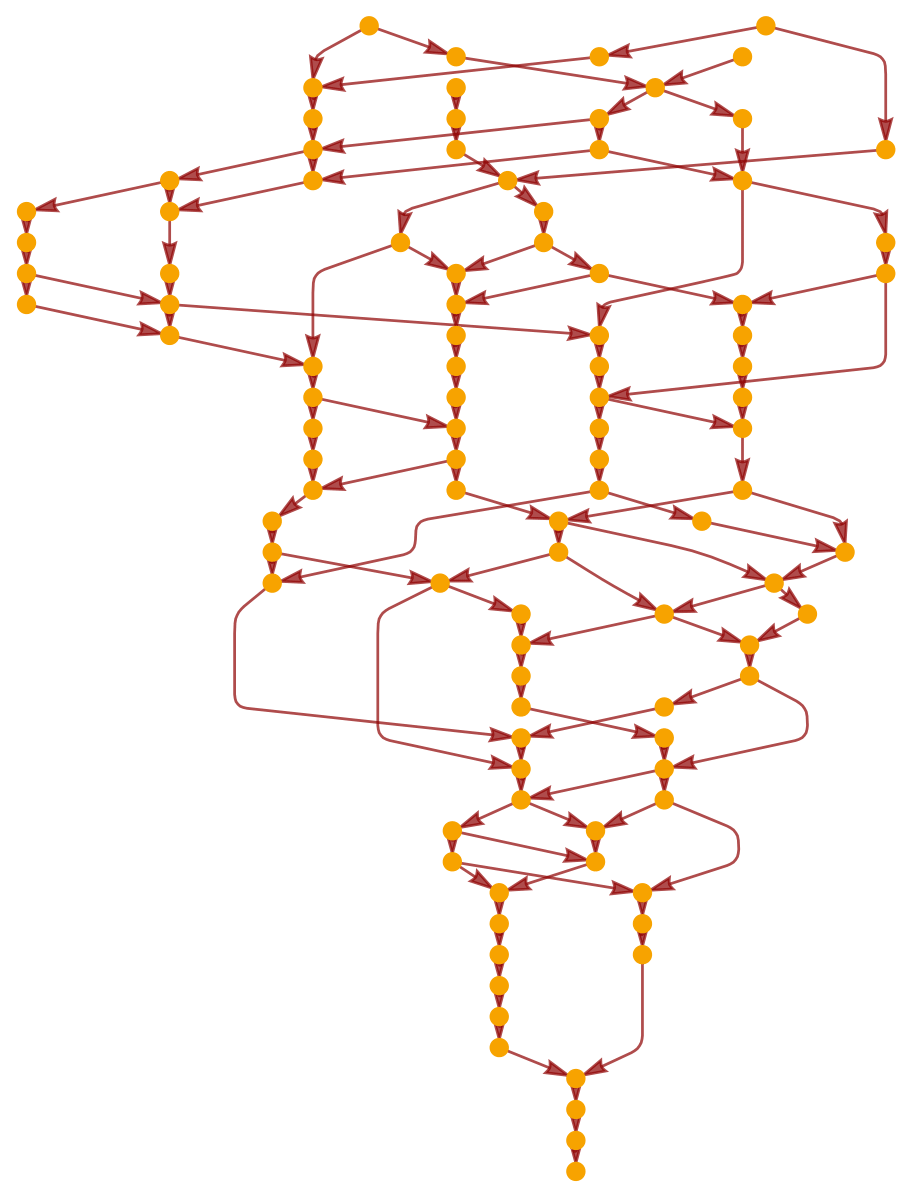}
\includegraphics[width=0.325\textwidth]{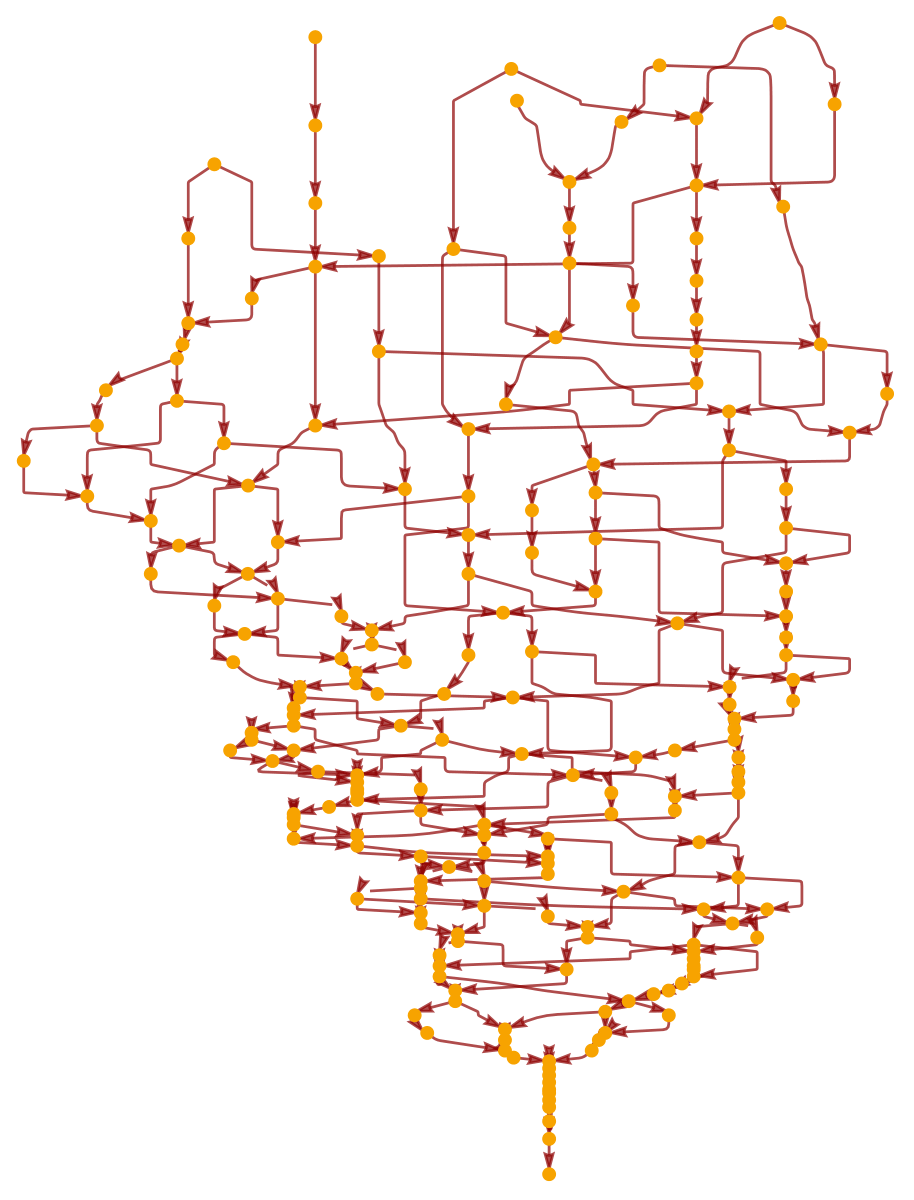}
\includegraphics[width=0.325\textwidth]{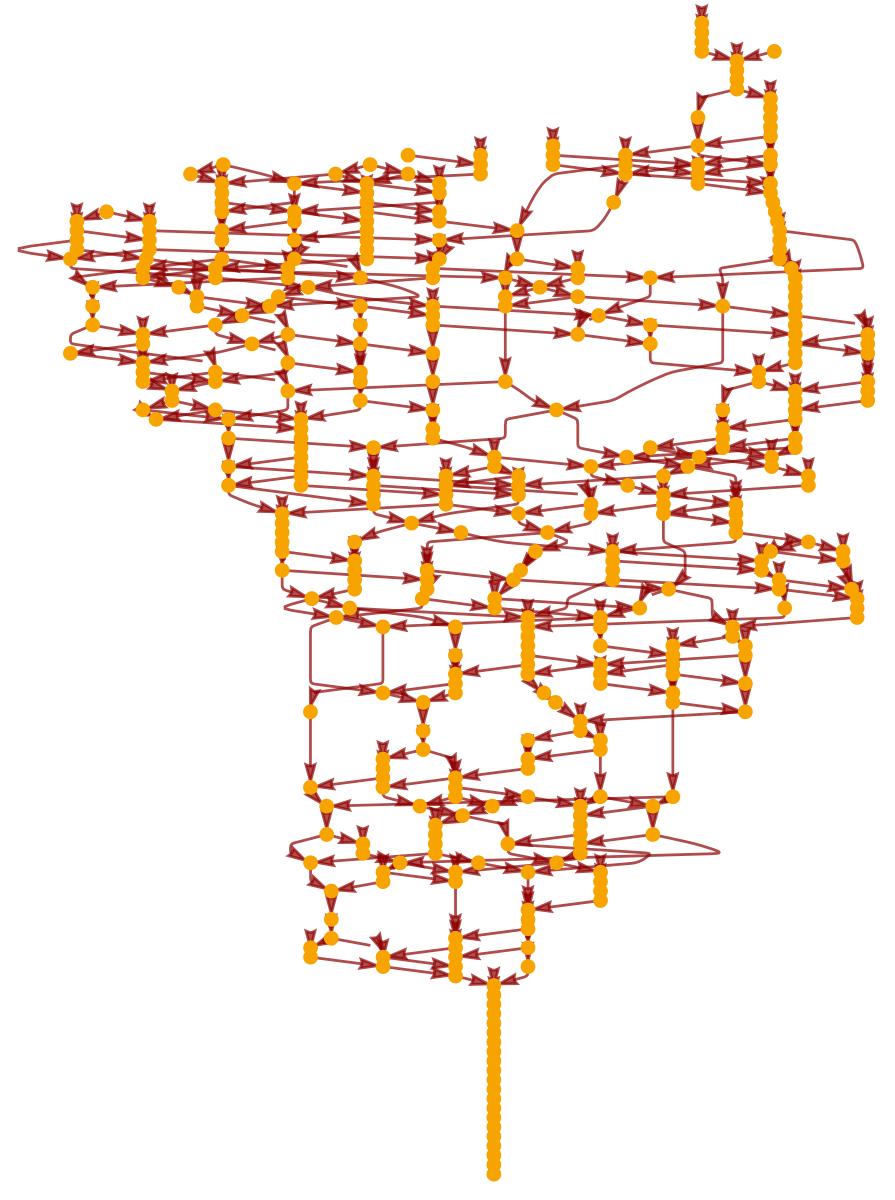}
\caption{Causal graphs corresponding to the discrete characteristic structure of the rapidly rotating Kerr black hole test with ${a = 0.9}$ at time ${t = 20 M}$, with resolutions of 100, 200 and 400 hypergraph vertices, respectively.}
\label{fig:Figure19}
\end{figure}

\begin{figure}[ht]
\centering
\includegraphics[width=0.325\textwidth]{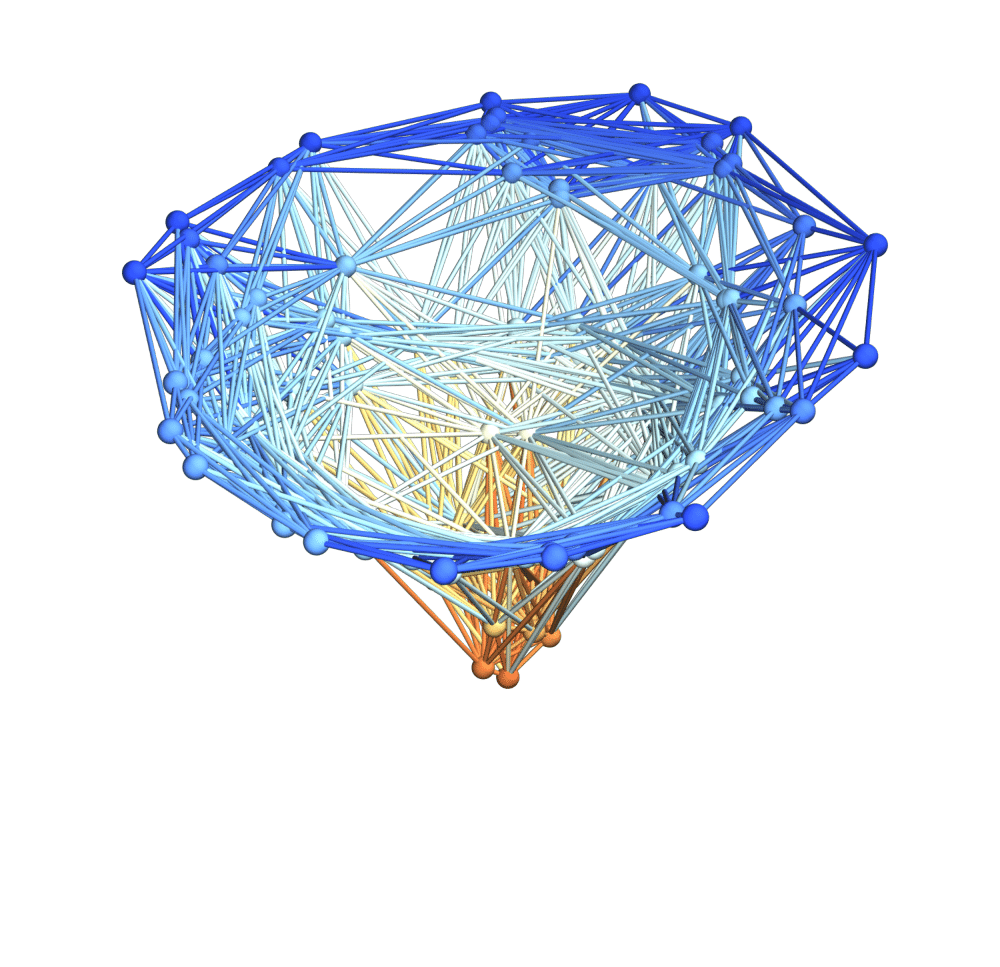}
\includegraphics[width=0.325\textwidth]{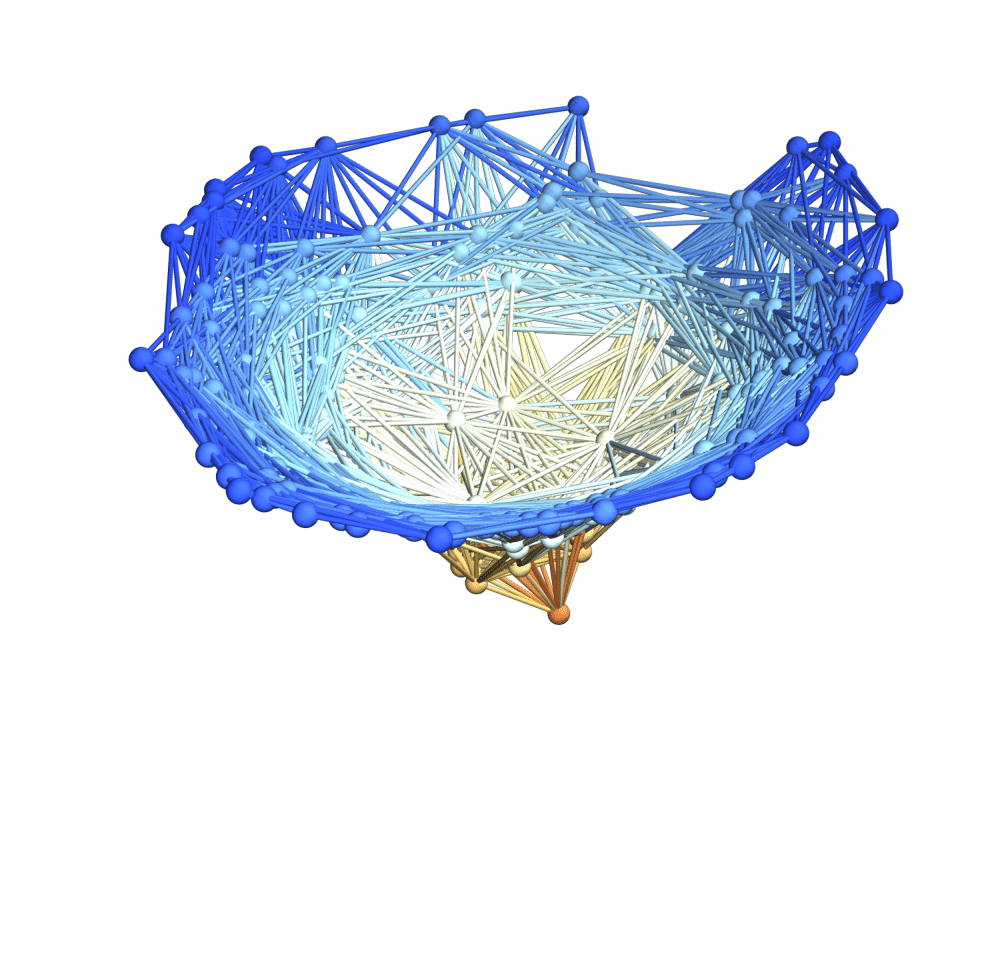}
\includegraphics[width=0.325\textwidth]{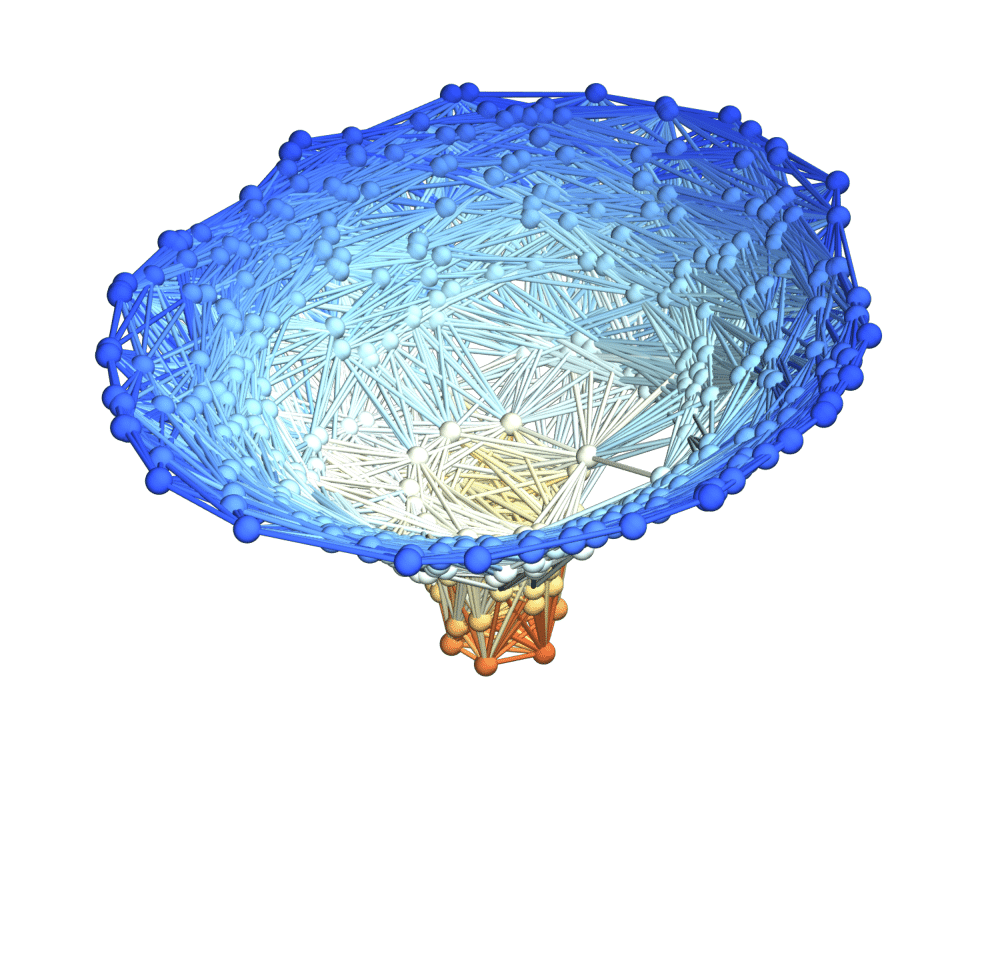}
\caption{Spatial hypergraphs corresponding to projections along the $z$-axis of the final hypersurface configuration of the rapidly rotating Kerr black hole test with ${a = 0.3}$ at time ${t = 20 M}$, with resolutions of 100, 200 and 400 vertices, respectively. The vertices have been assigned spatial coordinates according to the profile of the Boyer-Lindquist conformal factor ${\psi}$ through a spatial slice perpendicular to the $z$-axis, and the hypergraphs have been adapted and colored using the local curvature in ${\psi}$.}
\label{fig:Figure20}
\end{figure}

\begin{figure}[ht]
\centering
\includegraphics[width=0.325\textwidth]{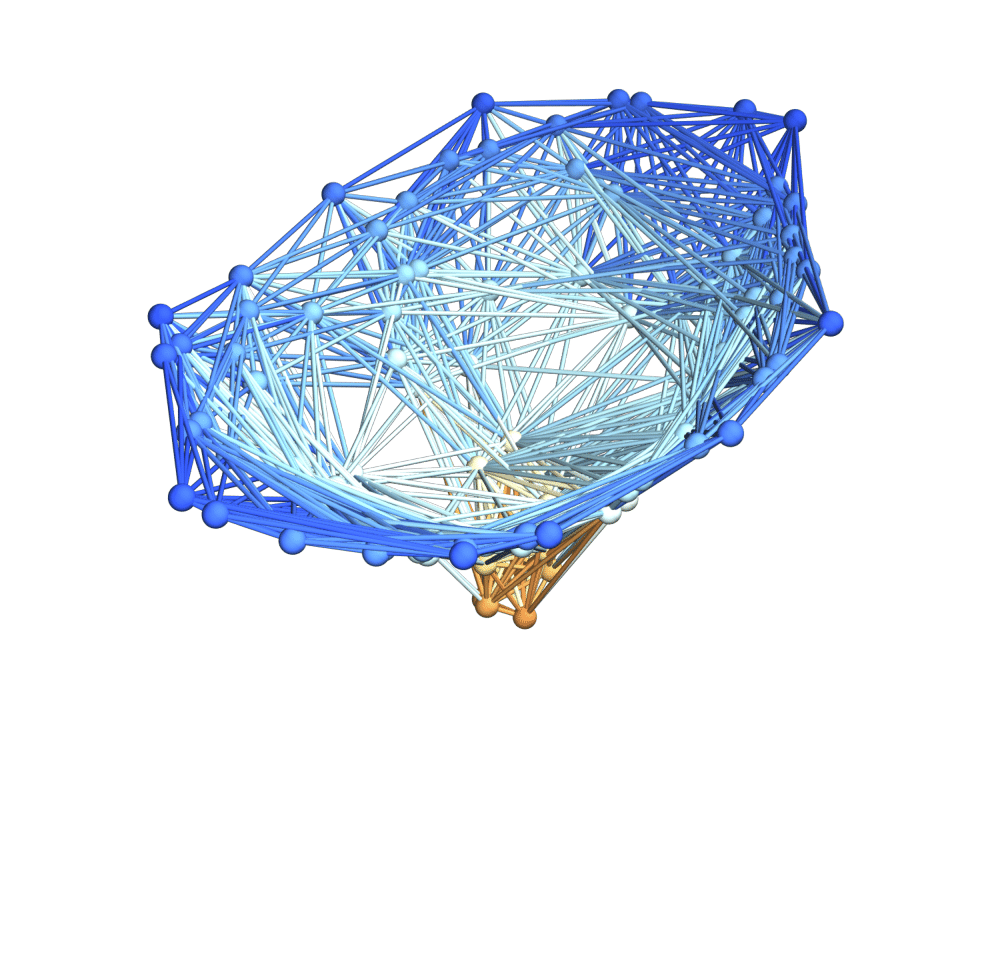}
\includegraphics[width=0.325\textwidth]{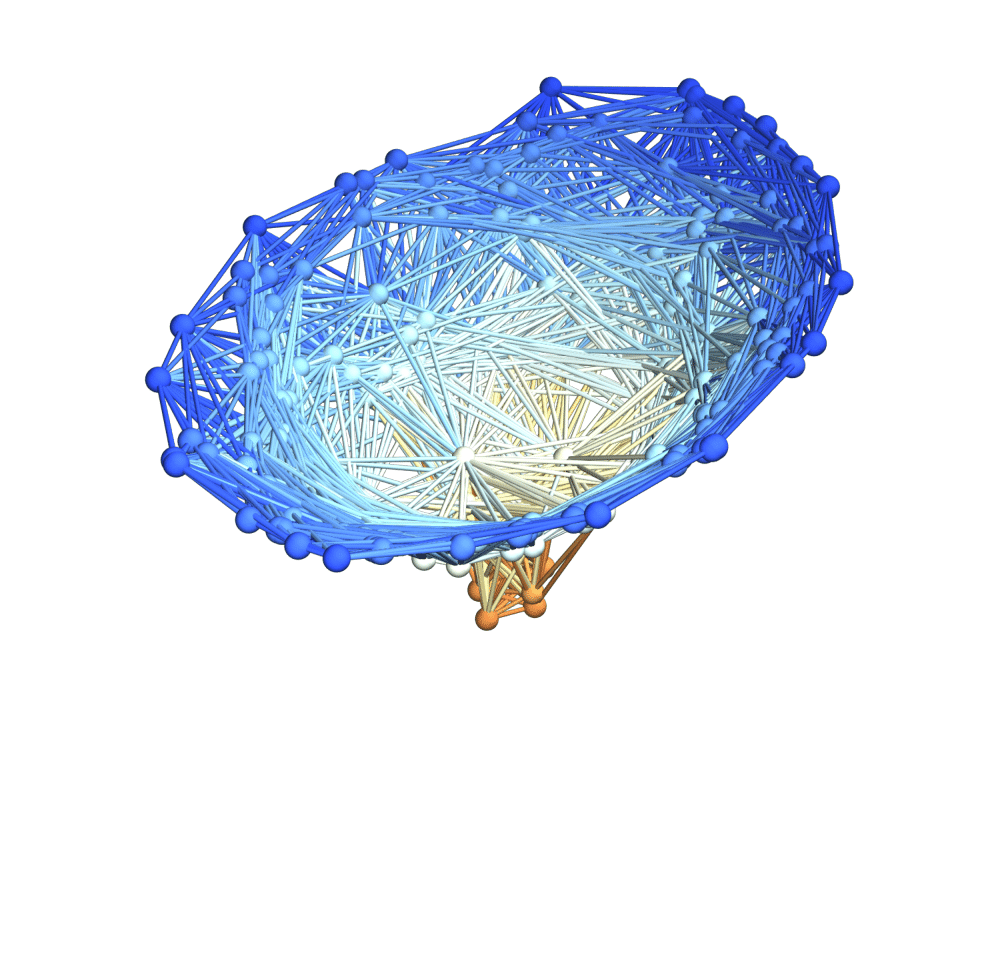}
\includegraphics[width=0.325\textwidth]{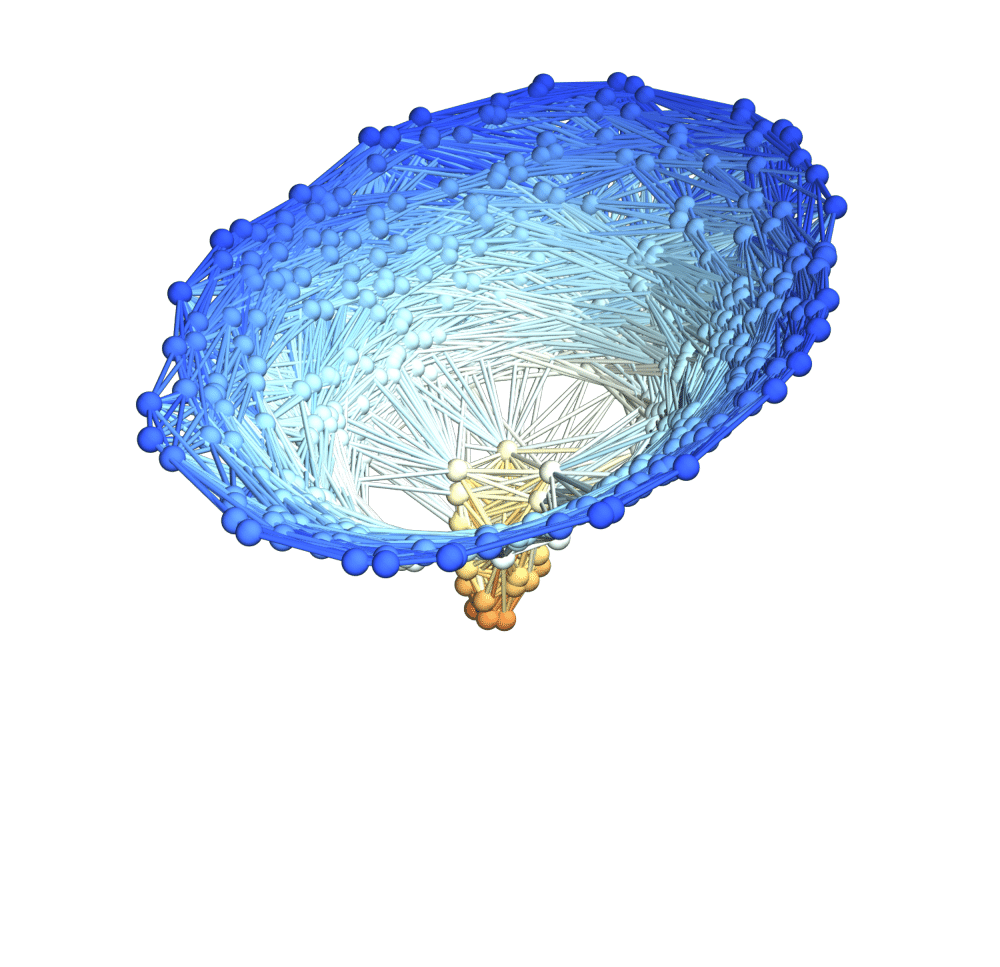}
\caption{Spatial hypergraphs corresponding to projections along the $z$-axis of the final hypersurface configuration of the rapidly rotating Kerr black hole test with ${a = 0.6}$ at time ${t = 20 M}$, with resolutions of 100, 200 and 400 vertices, respectively. The vertices have been assigned spatial coordinates according to the profile of the Boyer-Lindquist conformal factor ${\psi}$ through the spatial slice perpendicular to the $z$-axis, and the hypergraphs have been adapted and colored using the local curvature in ${\psi}$.}
\label{fig:Figure21}
\end{figure}

\begin{figure}[ht]
\centering
\includegraphics[width=0.325\textwidth]{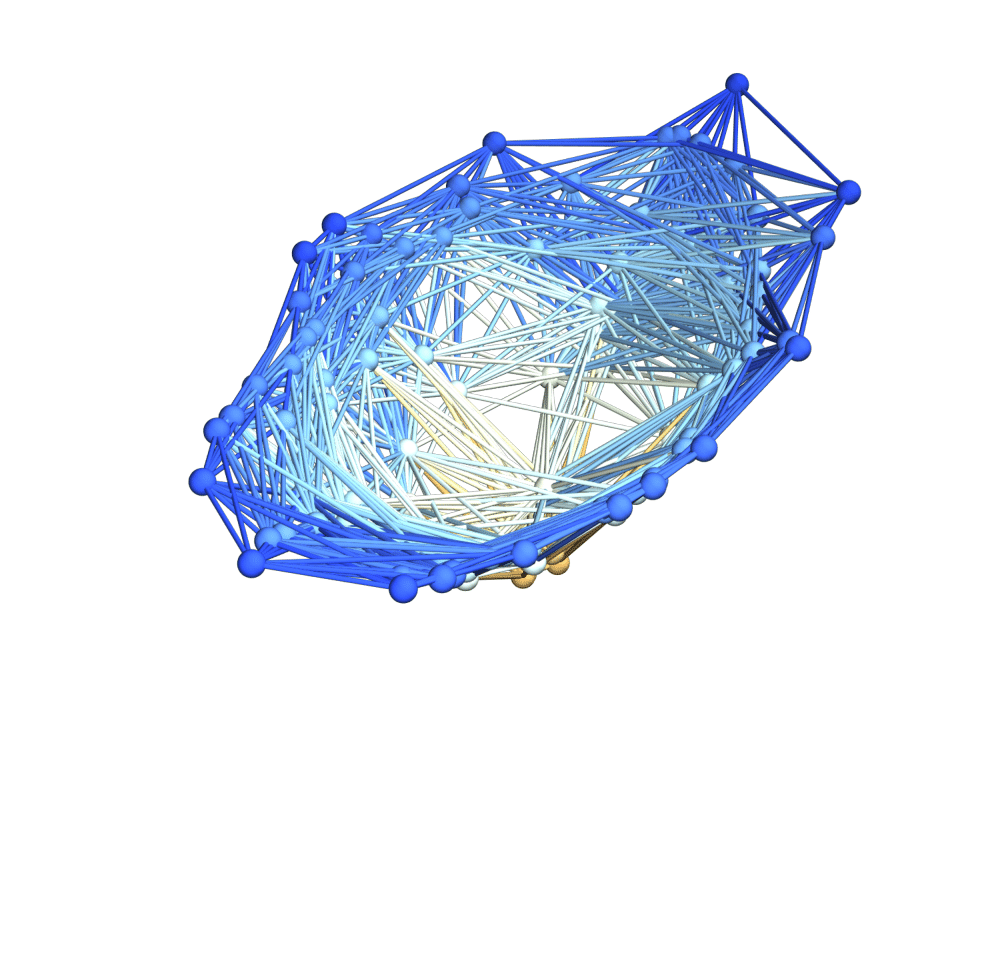}
\includegraphics[width=0.325\textwidth]{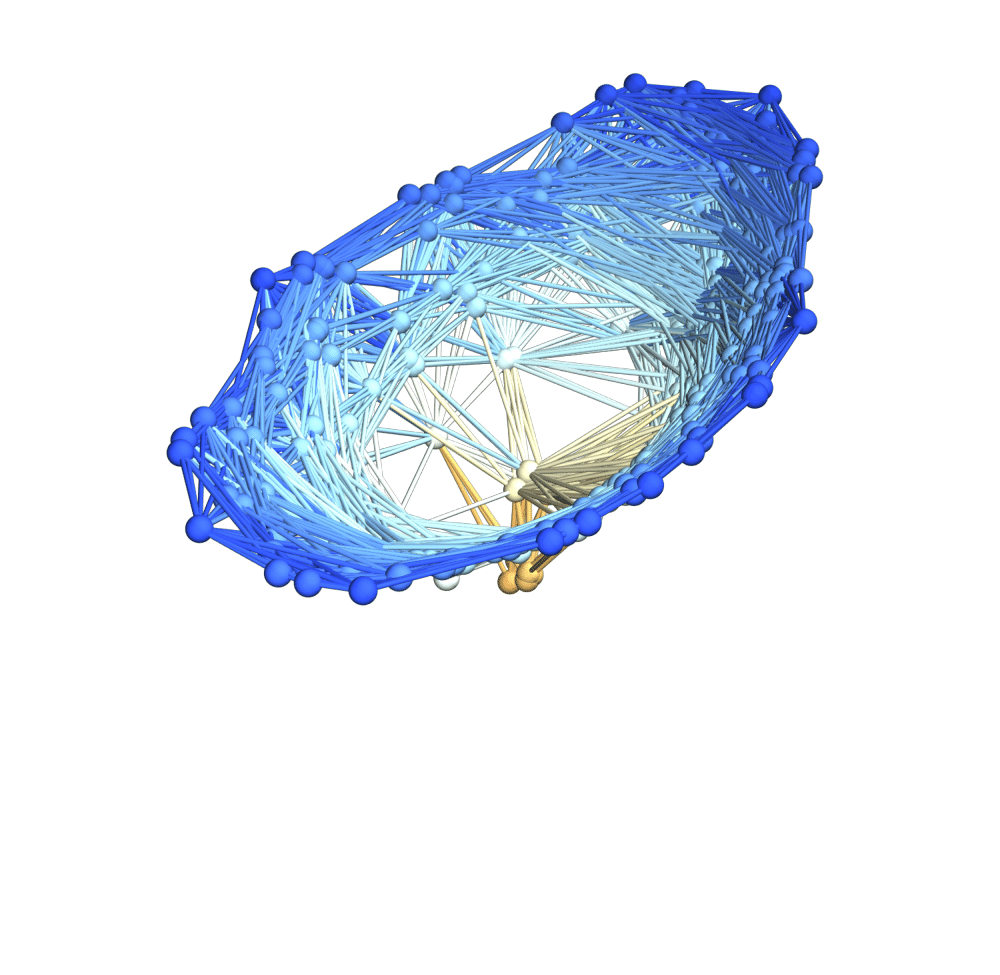}
\includegraphics[width=0.325\textwidth]{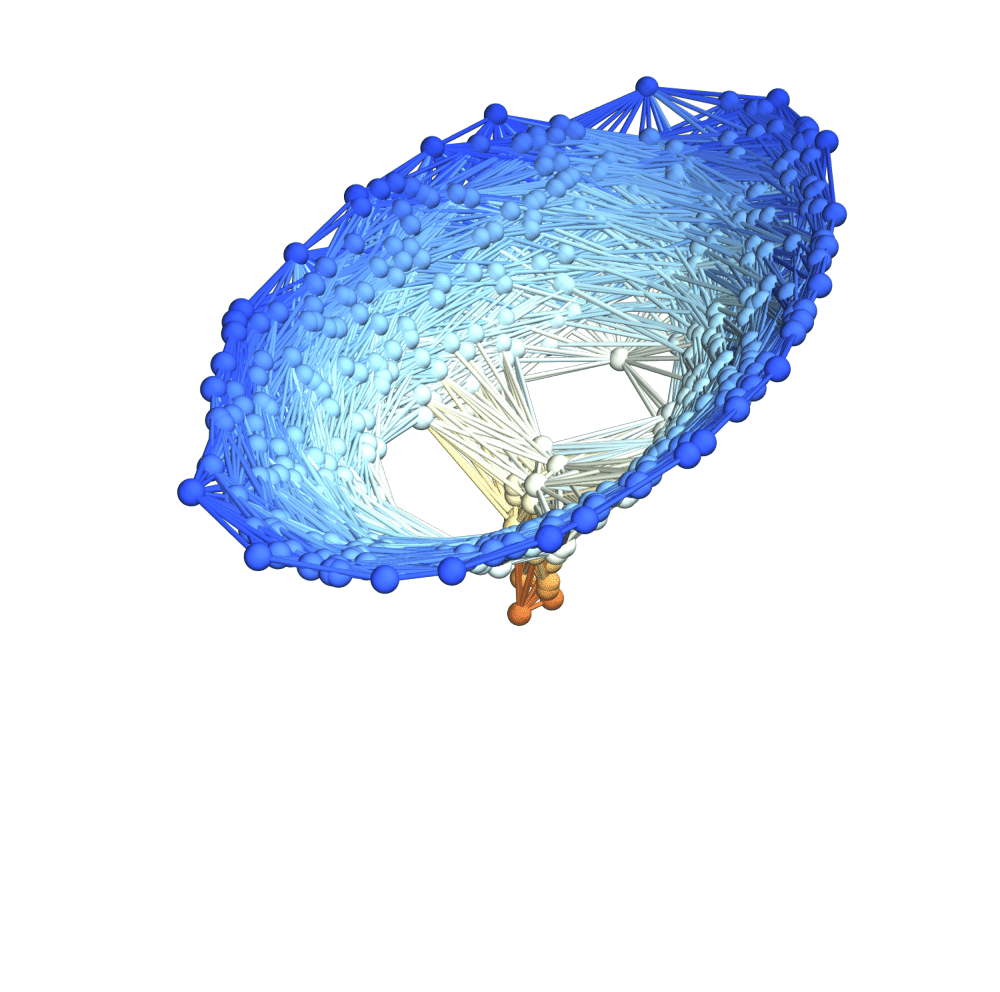}
\caption{Spatial hypergraphs corresponding to projections along the $z$-axis of the final hypersurface configuration of the rapidly rotating Kerr black hole test with ${a = 0.9}$ at time ${t = 20 M}$, with resolutions of 100, 200 and 400 vertices, respectively. The vertices have been assigned spatial coordinates according to the profile of the Boyer-Lindquist conformal factor ${\psi}$ through the spatial slice perpendicular to the $z$-axis, and the hypergraphs have been adapted and colored using the local curvature in ${\psi}$.}
\label{fig:Figure22}
\end{figure}

We choose a quasi-isotropic gauge for the initial data, as proposed by Brandt and Seidel\cite{brandt2}\cite{brandt3}\cite{brandt4}, in which a new radial coordinate ${\eta}$ is introduced to the Boyer-Lindquist coordinate system\cite{dain} using:

\begin{equation}
r = r_{+} \cosh^2 \left( \frac{\eta}{2} \right) - r_{-} \sinh^2 \left( \frac{\eta}{2} \right), \qquad \text{ with } \qquad r_{\pm} = M \pm \sqrt{M^2 - a^2},
\end{equation}
such that the spatial part of the Kerr metric can be written as:

\begin{equation}
dl^2 = \Psi_{0}^{4} \left[ e^{-2 q_0} \left( d \eta^2 + d \theta^2 \right) + \sin^2 \left( \theta \right) d \phi^2 \right],
\end{equation}
where:

\begin{equation}
\Psi_{0}^{4} = \frac{g_{\phi \phi}}{\sin^2 \left( \theta \right)}, \qquad \text{ and } \qquad \Psi_{0}^{4} e^{-2 q_0} = g_{r r} \left( \frac{dr}{d \eta} \right)^2 = g_{\theta \theta}.
\end{equation}
In each of the test cases, the solution is evolved until a final time of ${t = 20 M}$; the final hypersurface configurations for ${a = 0.3}$, ${a = 0.6}$ and ${a = 0.9}$ are shown in Figures \ref{fig:Figure13}, \ref{fig:Figure14} and \ref{fig:Figure15}, respectively, with resolutions of 100, 200 and 400 vertices, and with the hypergraphs adapted and colored using the Boyer-Lindquist conformal factor ${\psi}$. Figure \ref{fig:Figure16} shows the extreme separation of the interior and exterior horizons for ${a = 0.95}$ and ${a = 0.99}$ with a resolution of 400 vertices, with the interior region becoming completely disconnected at ${a = 1}$. Figures \ref{fig:Figure17}, \ref{fig:Figure18} and \ref{fig:Figure19} show the discrete characteristic structure of the ${a =0.3}$, ${a = 0.6}$ and ${a = 0.9}$ solutions, respectively, after time ${t = 20 M}$ (using directed acyclic graphs to show discrete characteristic lines), and Figures \ref{fig:Figure20}, \ref{fig:Figure21} and \ref{fig:Figure22} show projections along the $z$-axis of the final hypersurface configurations for ${a = 0.3}$, ${a = 0.6}$ and ${a = 0.9}$, respectively, with vertices assigned spatial coordinates according to the profile of the Boyer-Lindquist conformal factor ${\psi}$. We confirm that the ADM mass and the three components of the ADM angular momentum of the black hole (computed by integrating over a surface surrounding the boundary of asymptotic flatness) remain approximately constant, and the linear momentum of the black hole remains approximately zero, as expected. The convergence rates for the Hamiltonian constraint after time ${t = 20 M}$, with respect to the ${L_1}$, ${L_2}$ and ${L_{\infty}}$ norms, illustrating approximately fourth-order convergence of the finite difference scheme, are shown in Table \ref{tab:Table2}.

\begin{table}[ht]
\centering
\begin{tabular}{|c|c|c|c|c|c|c|}
\hline
Vertices & ${\epsilon \left( L_1 \right)}$ & ${\epsilon \left( L_2 \right)}$ & ${\epsilon \left( L_{\infty} \right)}$ & ${\mathcal{O} \left( L_1 \right)}$ & ${\mathcal{O} \left( L_2 \right)}$ & ${\mathcal{O} \left( L_{\infty} \right)}$\\
\hline\hline
100 & ${4.81 \times 10^{-2}}$ & ${1.98 \times 10^{-2}}$ & ${3.79 \times 10^{-2}}$ & - & - & -\\
\hline
200 & ${4.85 \times 10^{-3}}$ & ${1.55 \times 10^{-3}}$ & ${2.64 \times 10^{-3}}$ & 3.31 & 3.67 & 3.84\\
\hline
400 & ${5.51 \times 10^{-4}}$ & ${5.83 \times 10^{-5}}$ & ${2.05 \times 10^{-4}}$ & 3.14 & 4.73 & 3.69\\
\hline
800 & ${4.28 \times 10^{-5}}$ & ${2.83 \times 10^{-6}}$ & ${1.87 \times 10^{-5}}$ & 3.69 & 4.36 & 3.45\\
\hline
1600 & ${3.96 \times 10^{-6}}$ & ${9.05 \times 10^{-8}}$ & ${1.10 \times 10^{-6}}$ & 3.44 & 4.97 & 4.09\\
\hline
\end{tabular}
\caption{Convergence rates for the rapidly rotating Kerr black hole test with respect to the ${L_1}$, ${L_2}$ and ${L_{\infty}}$ norms for the Hamiltonian constraint $H$ after time ${t = 20 M}$, showing approximately fourth-order convergence.}
\label{tab:Table2}
\end{table}

\clearpage

\subsection{Maximally Extended Schwarzschild Black Hole (Einstein-Rosen Bridge)}

Next, we evolve an Einstein-Rosen bridge solution constructed via maximal analytic extension of the Schwarzschild metric. We begin by introducing the Kruskal-Szekeres coordinates ${\left( T, X, \theta, \phi \right)}$\cite{szekeres}\cite{kruskal}, in which the timelike coordinate $t$ and radial coordinate $r$ in the Schwarzschild coordinate system are replaced by a new timelike coordinate $T$ and a new spacelike coordinate $X$, respectively:

\begin{equation}
T = \sqrt{\frac{r}{2M} - 1} \exp \left( \frac{r}{2M} \right) \sinh \left( \frac{t}{4M} \right), \qquad \text{ and } \qquad X = \sqrt{\frac{r}{2M} - 1} \exp \left( \frac{r}{2M} \right) \cosh \left( \frac{t}{4M} \right),
\end{equation}
within the region ${r > r_s = 2M}$ on the exterior of the event horizon, and:

\begin{equation}
T = \sqrt{1 - \frac{r}{2M}} \exp \left( \frac{r}{4M} \right) \cosh \left( \frac{t}{4M} \right), \qquad \text{ and } \qquad X = \sqrt{1 - \frac{r}{2M}} \exp \left( \frac{r}{4M} \right) \sinh \left( \frac{t}{4M} \right),
\end{equation}
within the region ${0 < r < r_s = 2M}$ on the interior of the event horizon. Across the entire spacetime (i.e. the union of the interior and exterior regions with the horizon itself), the Schwarzschild radial coordinate $r$ can be defined as the unique solution of the equation:

\begin{equation}
T^2 - X^2 = \left( 1- \frac{r}{2M} \right) \exp \left( \frac{r}{2M} \right), \qquad \text{ where } \qquad T^2 - X^2 < 1,
\end{equation}
which, written in terms of the principal branch of the Lambert W function ${W_0}$, is simply:

\begin{equation}
r = 2M \left( 1 + W_0 \left( \frac{X^2 - T^2}{e} \right) \right).
\end{equation}
In a similar manner, the Schwarzschild timelike coordinate $t$ can be seen to be:

\begin{equation}
t = 4M \mathrm{arctanh} \left( \frac{T}{X} \right),
\end{equation}
within the region ${T^2 - X^2 < 0}$, ${X > 0}$ on the exterior of the event horizon, and:

\begin{equation}
t = 4M \mathrm{arctanh} \left( \frac{X}{T} \right),
\end{equation}
within the region ${0 < T^2 - X^2 < 1}$, ${T > 0}$ on the interior of the event horizon. Thus, we can write the full metric of a static Schwarzschild black hole in Kruskal-Szekeres coordinates as:

\begin{equation}
g = ds^2 = \frac{32 M^3}{r} \exp \left( - \frac{r}{2M} \right) \left( -dT^2 + dX^2 \right) + r^2 d \Omega^2,
\end{equation}
where ${d \Omega^2}$ denotes the usual induced Riemannian spatial metric on the 2-sphere ${S^2 \subset E^3}$:

\begin{equation}
d \Omega^2 = d \theta^2 + \sin^2 \left( \theta \right) d \phi^2.
\end{equation}

We note that, although the transformation between Schwarzschild coordinates and Kruskal-Szekeres coordinates described above is strictly only defined for ${r > r_s = 2M}$ and ${-\infty < t < \infty}$, since $r$ is an analytic function of $T$ and $X$, it follows that the transformation can be analytically continued up to the singularity at ${T^2 - X^2 = 1}$, such that:

\begin{equation}
-\infty < X < \infty, \qquad \text{ and } \qquad -\infty < T^2 - X^2 < \infty.
\end{equation}
Hence, the metric presented above is a valid solution to the Einstein field equations across the entirety of the (extended) spacetime, so long as the solution remains analytic across the whole domain. Note further that there exist two distinct singularities at ${r = 0}$: one for positive values of $T$ and one for negative, with the latter representing a time-reversed black hole (\textit{white hole}) solution. Thus, the entire spacetime is naturally divided up into four distinct regions described by the Kruskal-Szekeres coordinates, namely an exterior black hole region with:

\begin{equation}
-X < T < X, \qquad \text{ and } \qquad 2M = r_s < r,
\end{equation}
an interior black hole region with:

\begin{equation}
\left\lvert X \right\rvert < T < \sqrt{1 + X^2}, \qquad \text{ and } \qquad 0 < r < 2M = r_s,
\end{equation}
a \textit{parallel} exterior black hole region with:

\begin{equation}
X < T < -X, \qquad \text{ and } \qquad 2 M = r_s < r,
\end{equation}
and an interior \textit{white hole} region with:

\begin{equation}
- \sqrt{1 + X^2} < T < - \left\lvert X \right\rvert, \qquad \text{ and } \qquad 0 < r = r_s < 2M,
\end{equation}
with the transformation between Schwarzschild and Kruskal-Szekeres coordinates described above applying within the interior and exterior black hole regions, and with an analogous coordinate transformation applying within the parallel exterior black hole and interior white hole regions. The Schwarzschild time coordinate $t$ varies between positive and negative infinity (with infinities appearing at the two event horizons) within each region, with:

\begin{equation}
\tanh \left( \frac{t}{4M} \right) = \frac{T}{X},
\end{equation}
in the exterior and parallel exterior black hole regions, and:

\begin{equation}
\tanh \left( \frac{t}{4M} \right) = \frac{X}{T},
\end{equation}
in the interior black hole and interior white hole regions, as expected.

\begin{figure}[ht]
\centering
\includegraphics[width=0.325\textwidth]{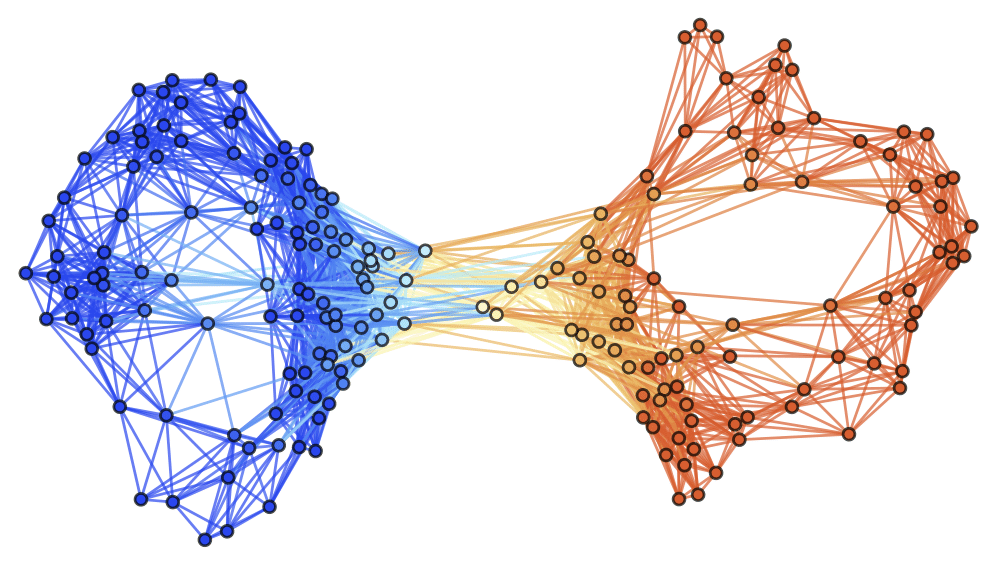}
\includegraphics[width=0.325\textwidth]{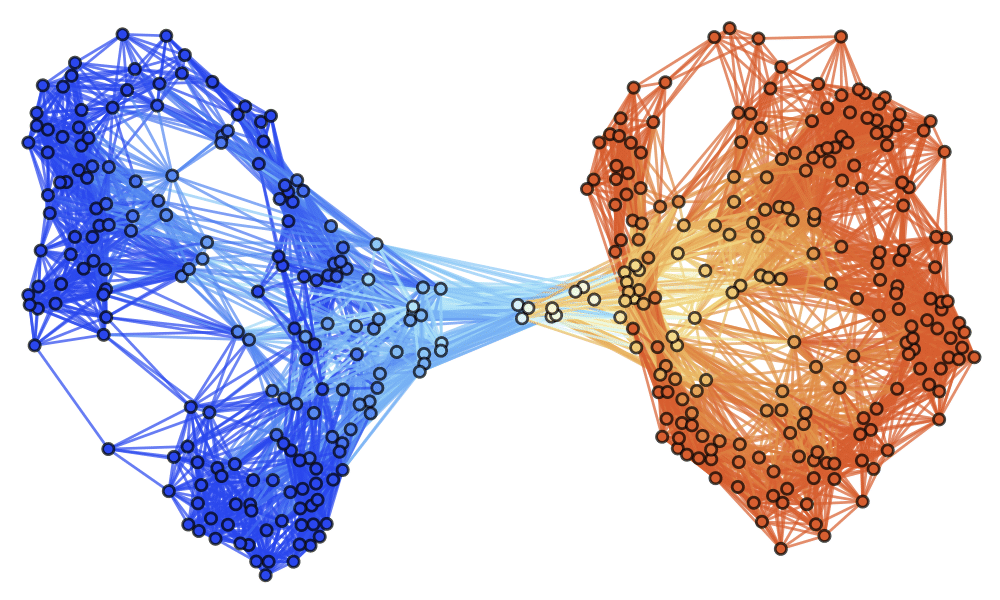}
\includegraphics[width=0.325\textwidth]{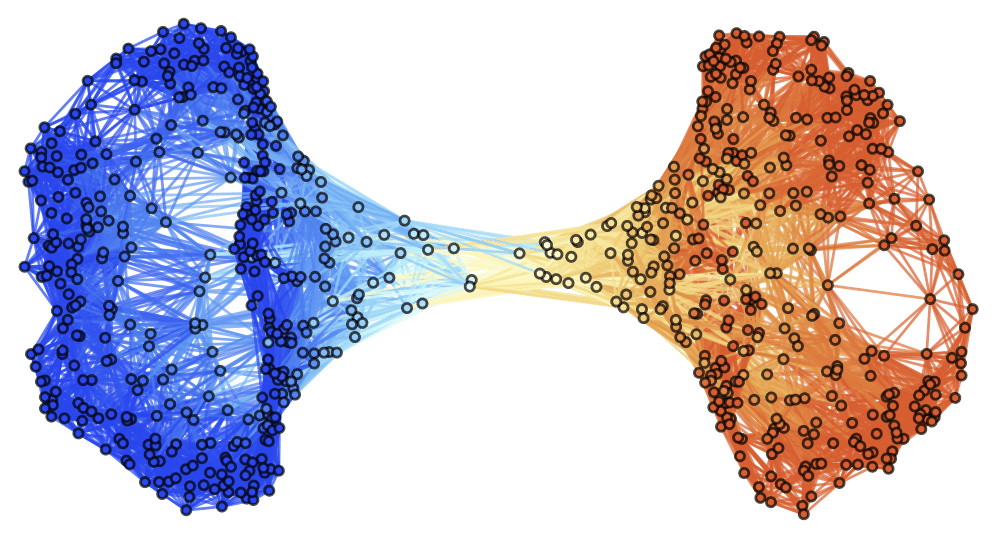}
\caption{Spatial hypergraphs corresponding to the initial hypersurface configuration of the maximally extended Schwarzschild black hole (Einstein-Rosen bridge) test at time ${t = 0 M}$, with resolutions of 200, 400 and 800 vertices, respectively. The hypergraphs have been adapted and colored using the local curvature in the Schwarzschild conformal factor ${\psi}$.}
\label{fig:Figure23}
\end{figure}

\begin{figure}[ht]
\centering
\includegraphics[width=0.325\textwidth]{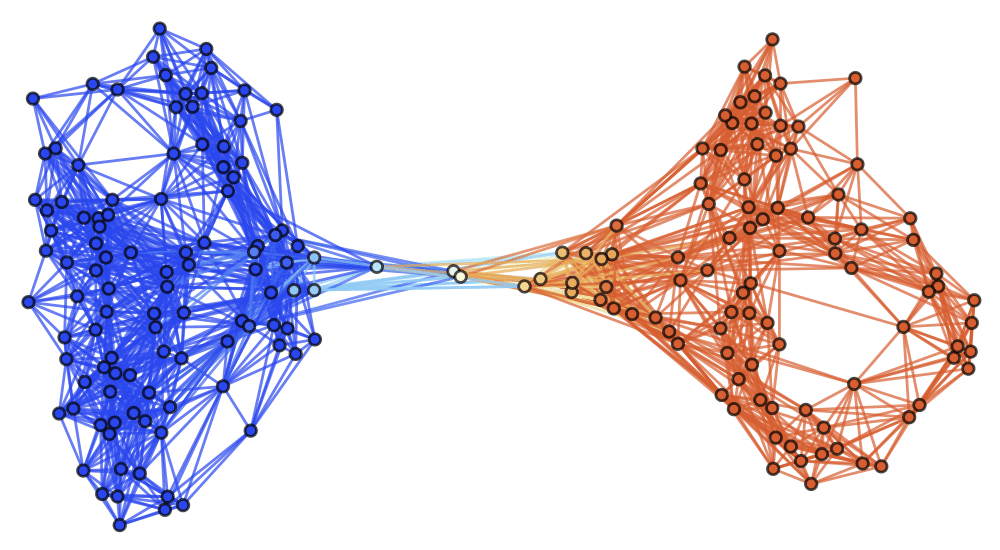}
\includegraphics[width=0.325\textwidth]{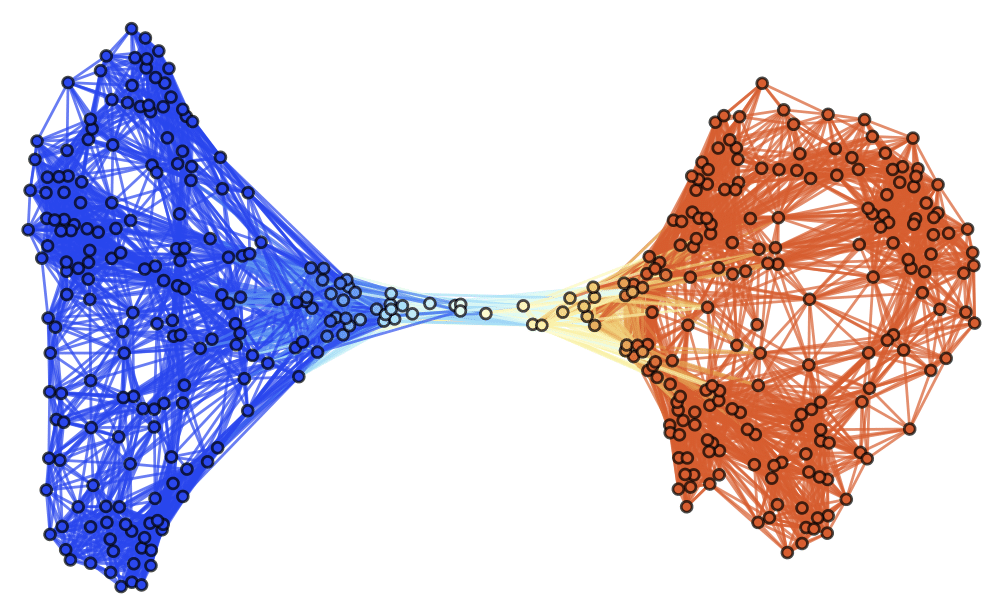}
\includegraphics[width=0.325\textwidth]{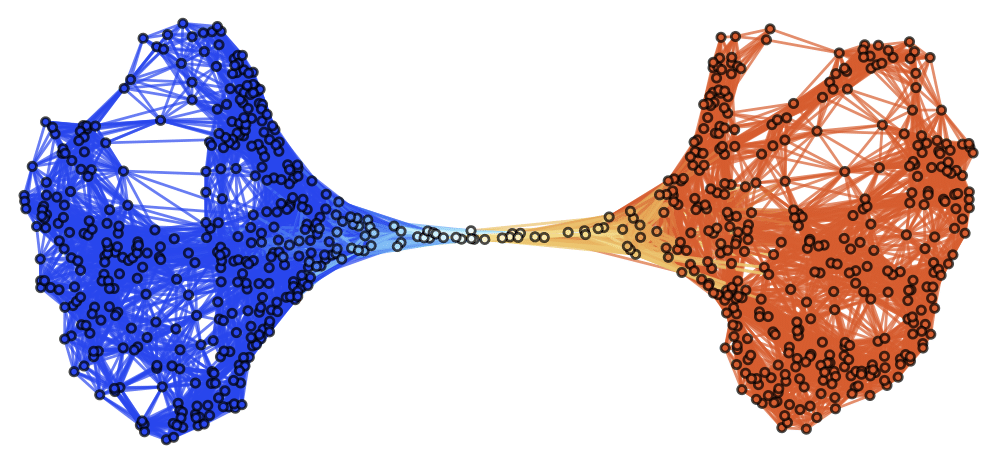}
\caption{Spatial hypergraphs corresponding to the intermediate hypersurface configuration of the maximally extended Schwarzschild black hole (Einstein-Rosen bridge) test at time ${t = 1.5 M}$, with resolutions of 200, 400 and 800 vertices, respectively. The hypergraphs have been adapted and colored using the local curvature in the Schwarzschild conformal factor ${\psi}$.}
\label{fig:Figure24}
\end{figure}

\begin{figure}[ht]
\centering
\includegraphics[width=0.325\textwidth]{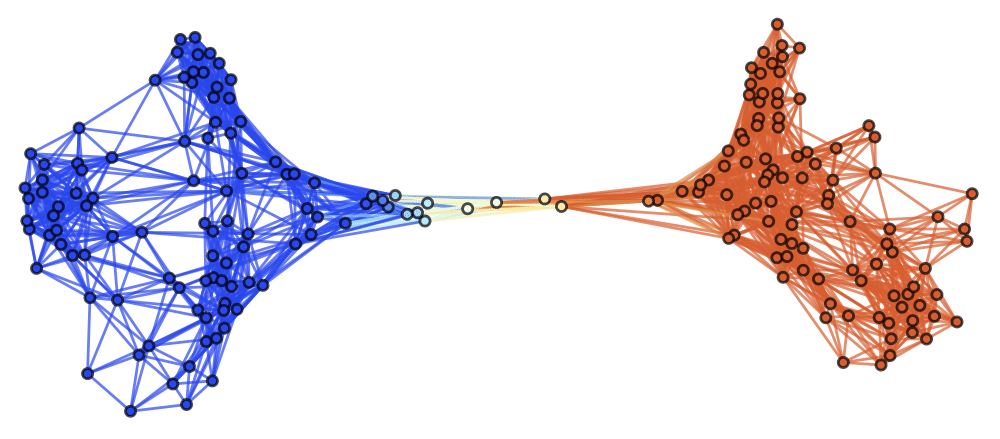}
\includegraphics[width=0.325\textwidth]{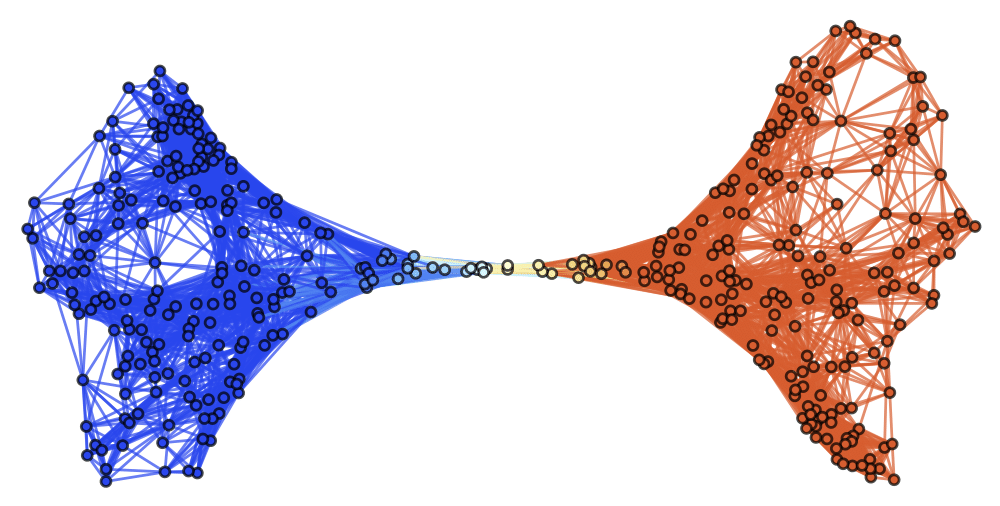}
\includegraphics[width=0.325\textwidth]{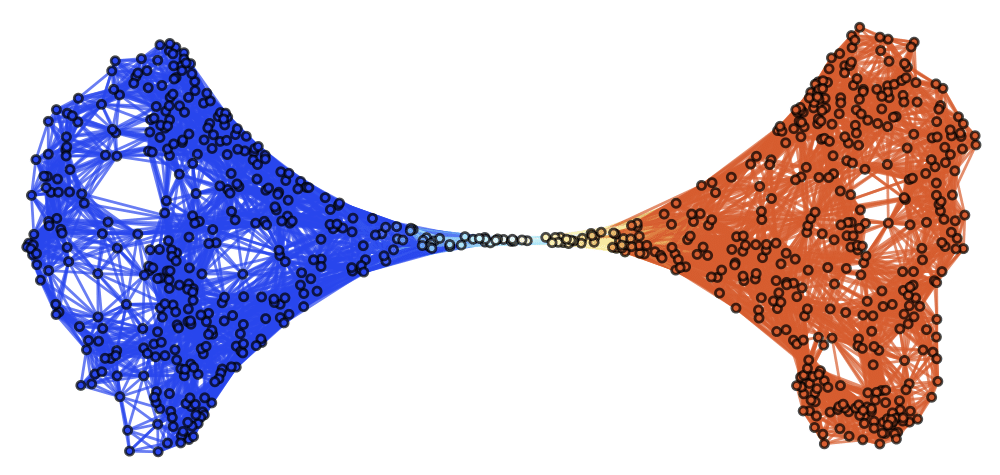}
\caption{Spatial hypergraphs corresponding to the final hypersurface configuration of the maximally extended Schwarzschild black hole (Einstein-Rosen bridge) test at time ${t = 3 M}$, with resolutions of 200, 400 and 800 vertices, respectively. The hypergraphs have been adapted and colored using the local curvature in the Schwarzschild conformal factor ${\psi}$.}
\label{fig:Figure25}
\end{figure}

\begin{figure}[ht]
\centering
\includegraphics[width=0.325\textwidth]{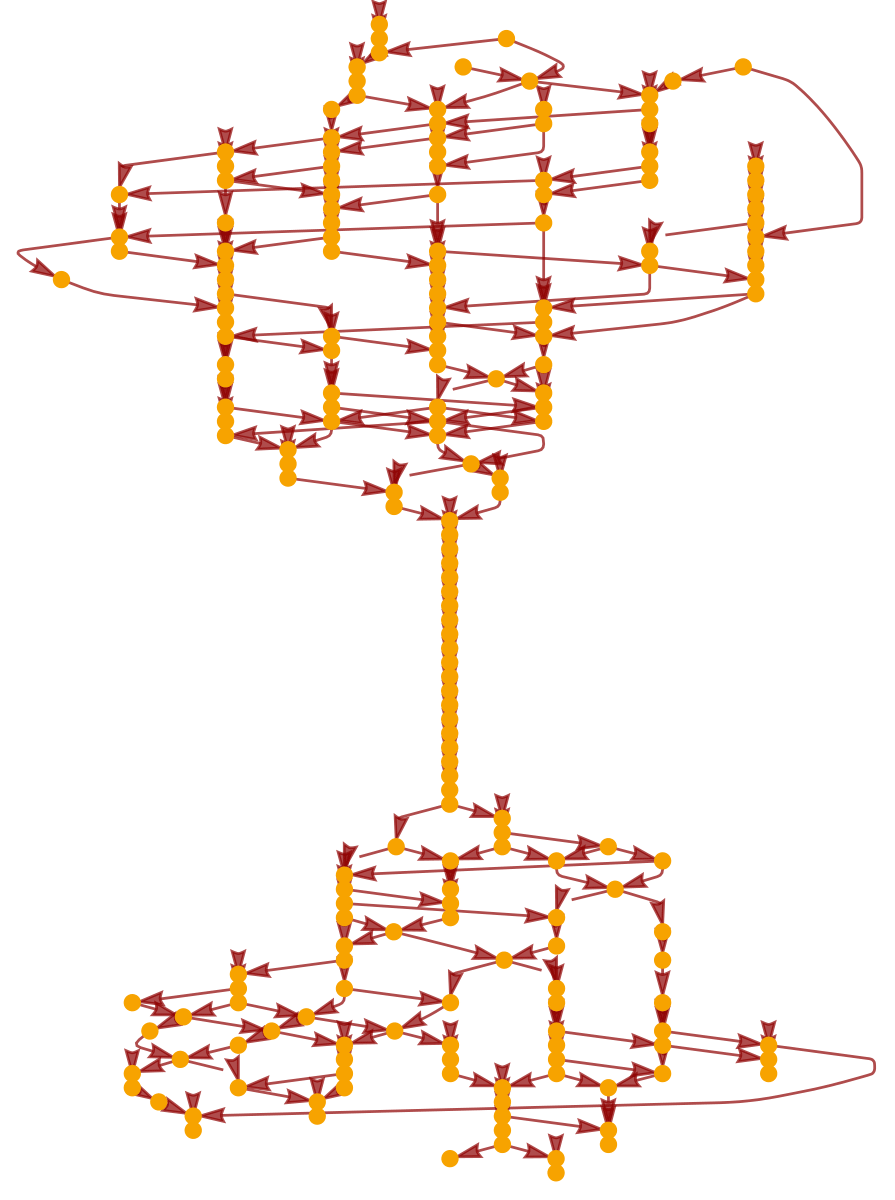}
\includegraphics[width=0.325\textwidth]{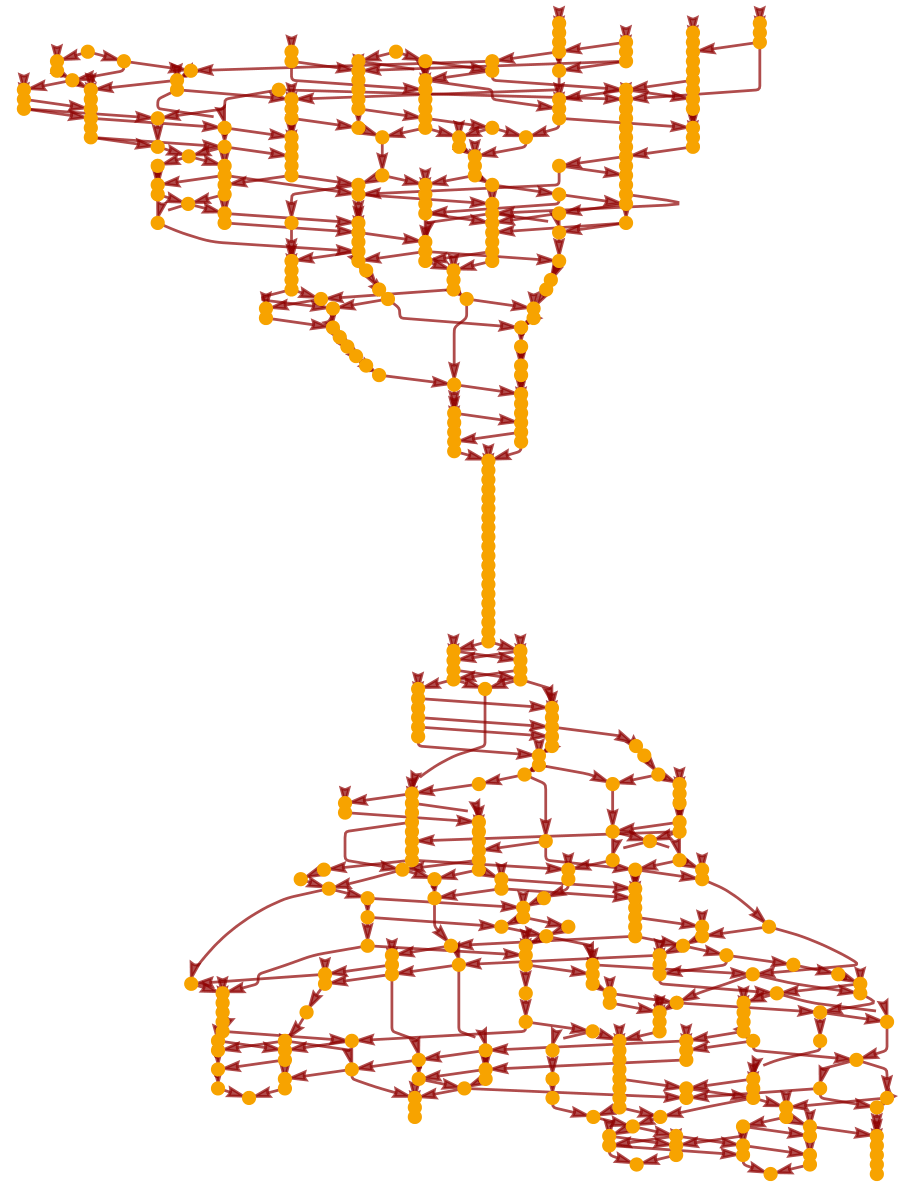}
\includegraphics[width=0.325\textwidth]{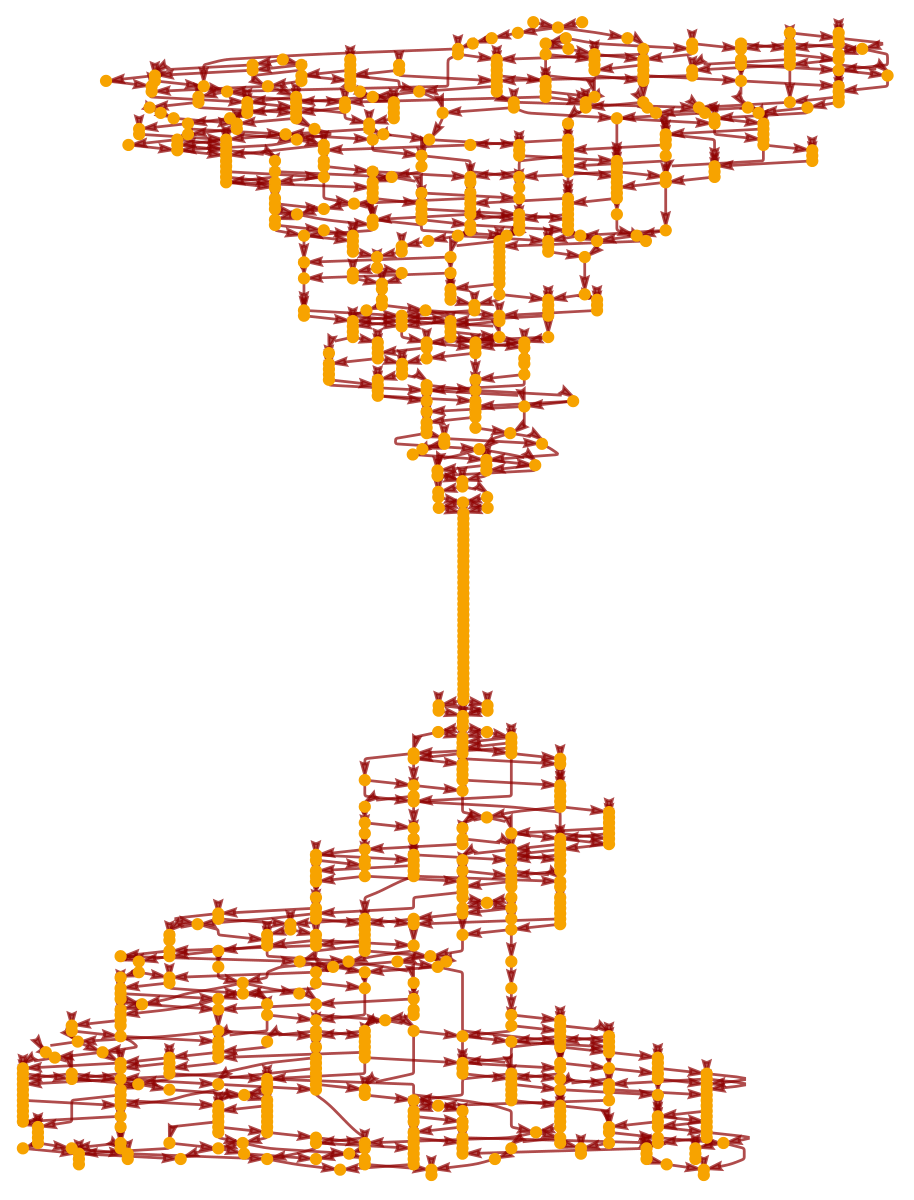}
\caption{Causal graphs corresponding to the discrete characteristic structure of the maximally extended Schwarzschild black hole (Einstein-Rosen bridge) test at time ${t = 3 M}$, with resolutions of 200, 400 and 800 hypergraph vertices, respectively.}
\label{fig:Figure26}
\end{figure}

\begin{figure}[ht]
\centering
\includegraphics[width=0.325\textwidth]{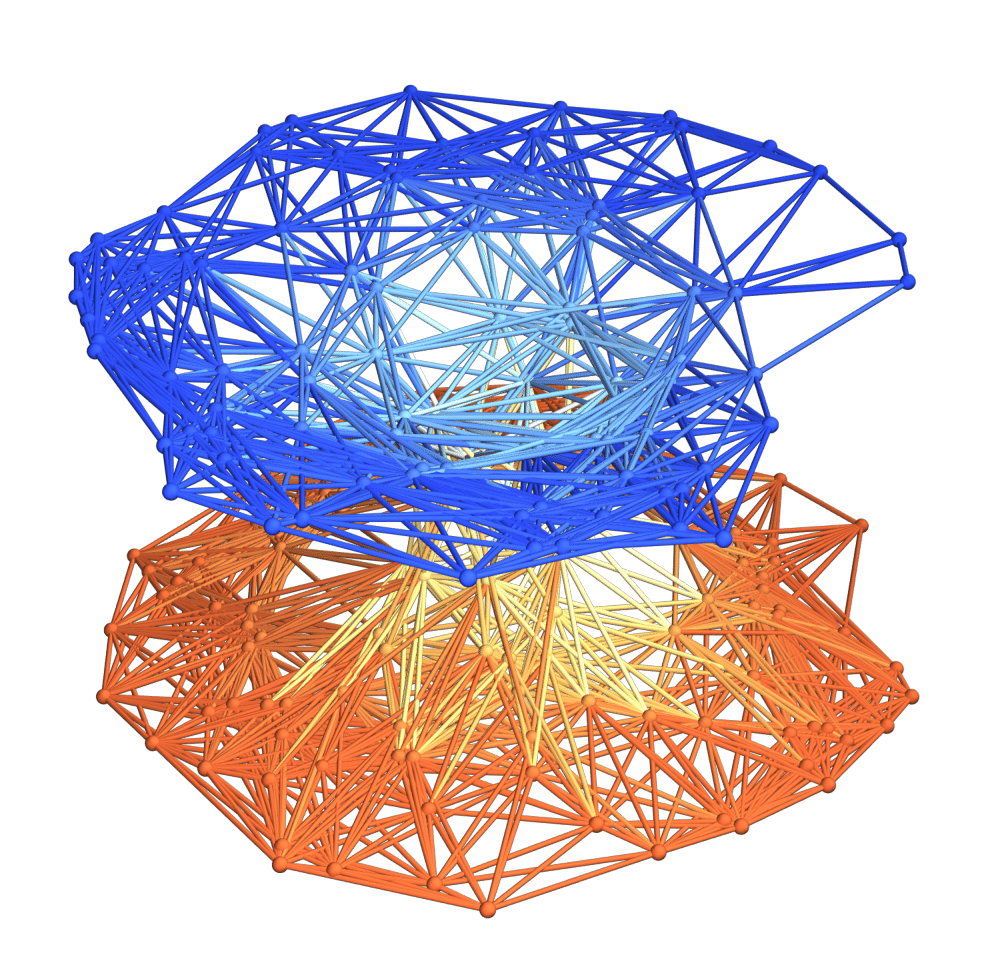}
\includegraphics[width=0.325\textwidth]{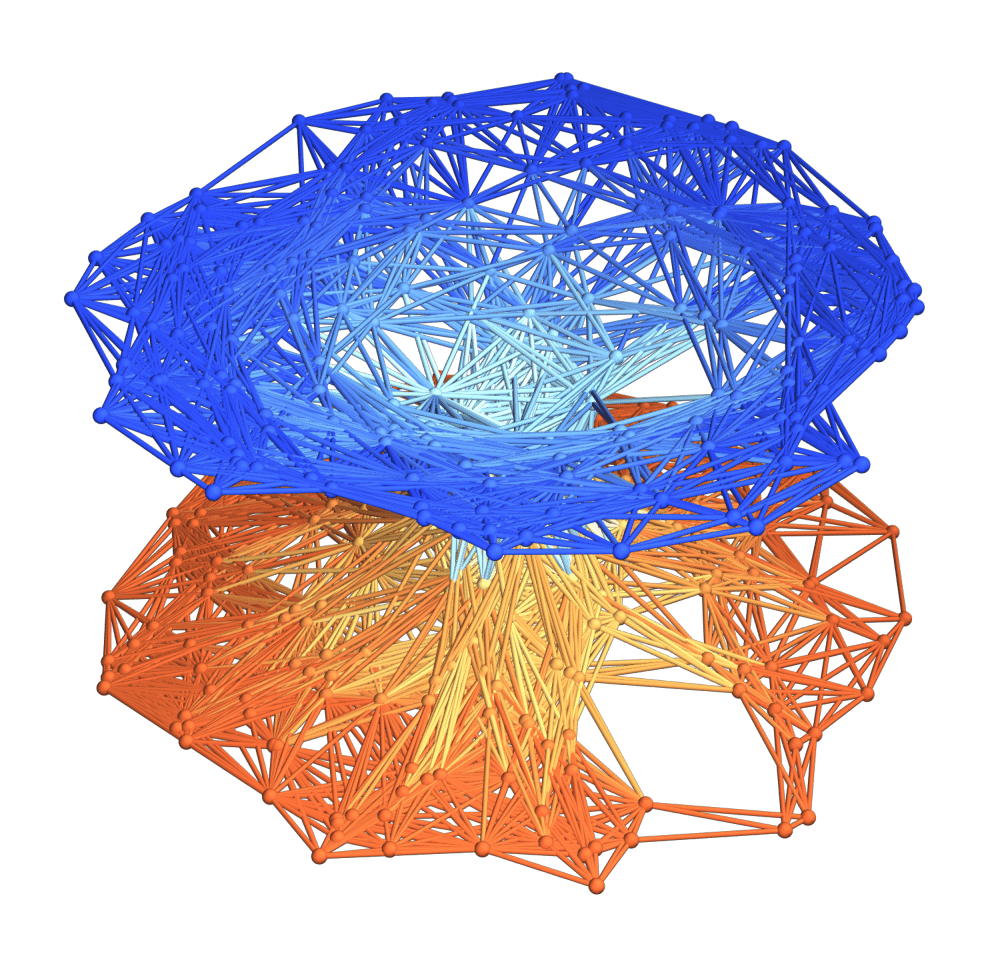}
\includegraphics[width=0.325\textwidth]{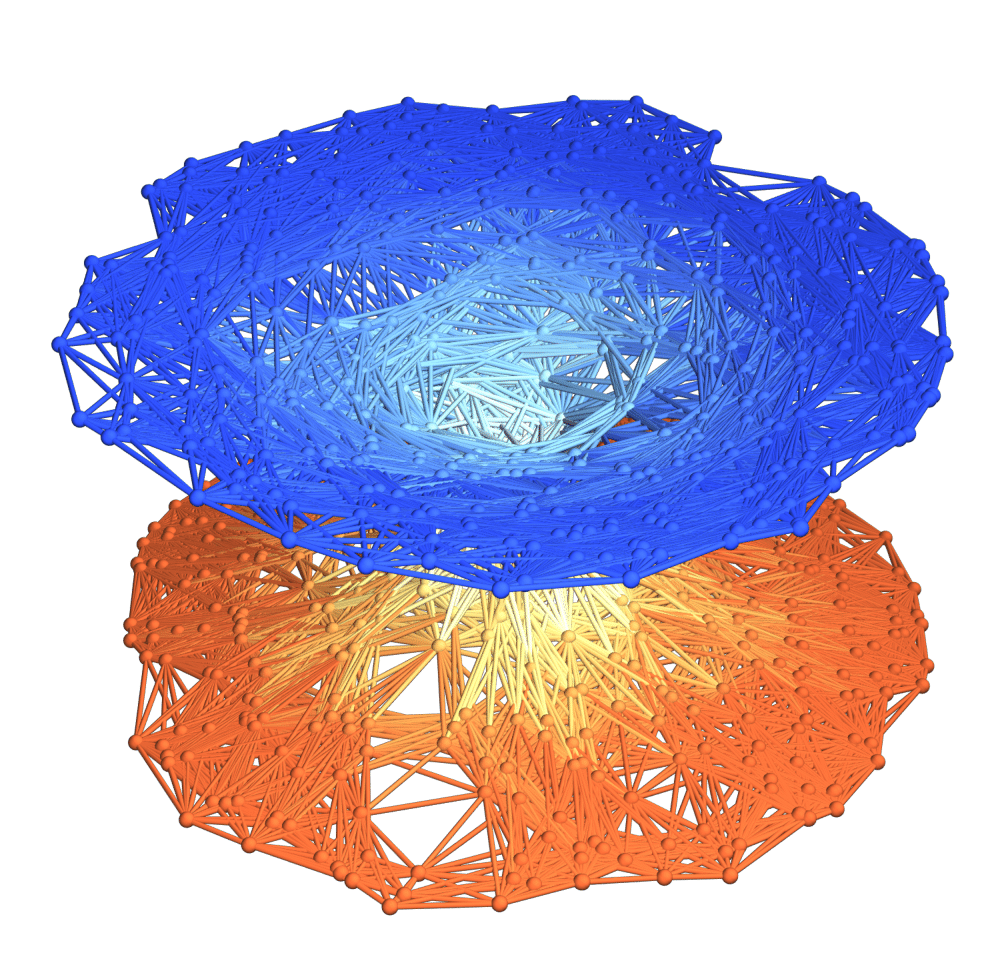}
\caption{Spatial hypergraphs corresponding to projections along the $z$-axis of the initial hypersurface configuration of the maximally extended Schwarzschild black hole (Einstein-Rosen bridge) test at time ${t = 0 M}$, with resolutions of 200, 400 and 800 vertices, respectively. The vertices have been assigned spatial coordinates according to the profile of the Schwarzschild conformal factor ${\psi}$ through a spatial slice perpendicular to the $z$-axis, and the hypergraphs have been adapted and colored using the local curvature in ${\psi}$.}
\label{fig:Figure27}
\end{figure}

\begin{figure}[ht]
\centering
\includegraphics[width=0.325\textwidth]{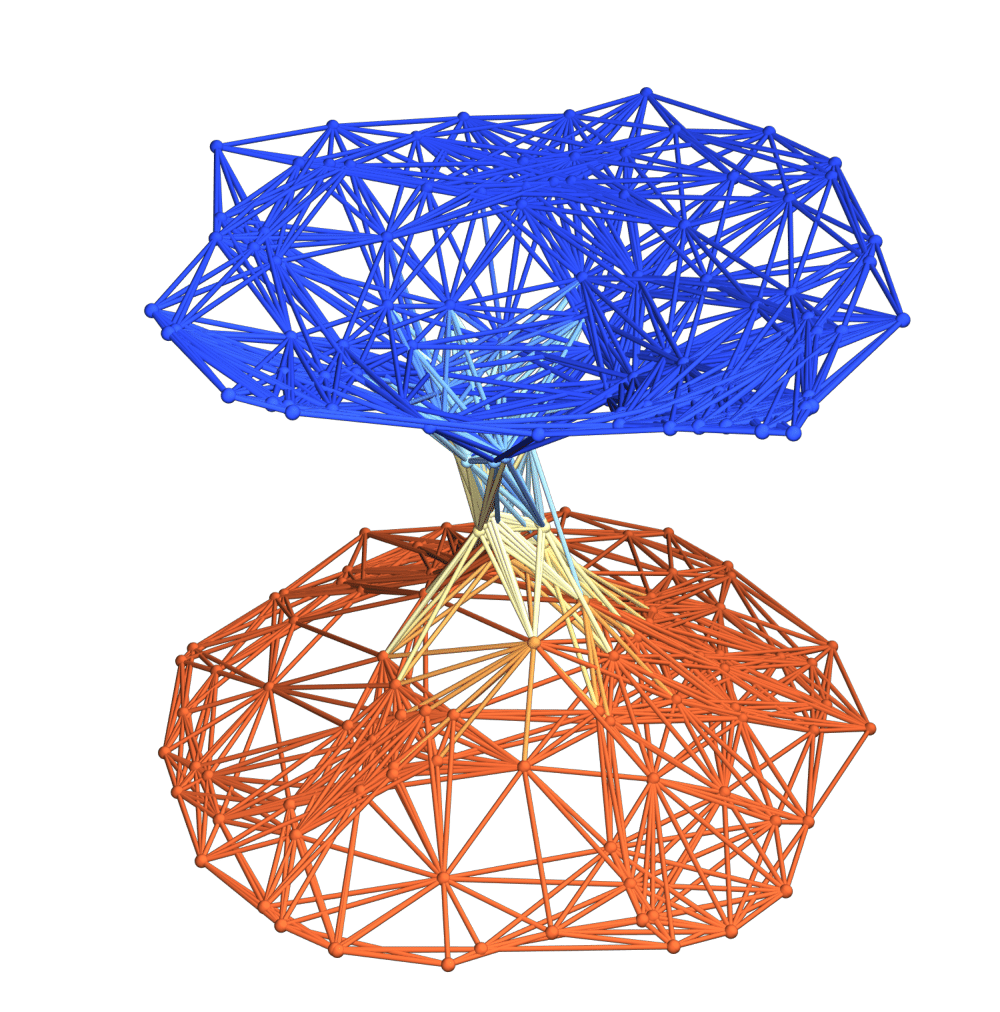}
\includegraphics[width=0.325\textwidth]{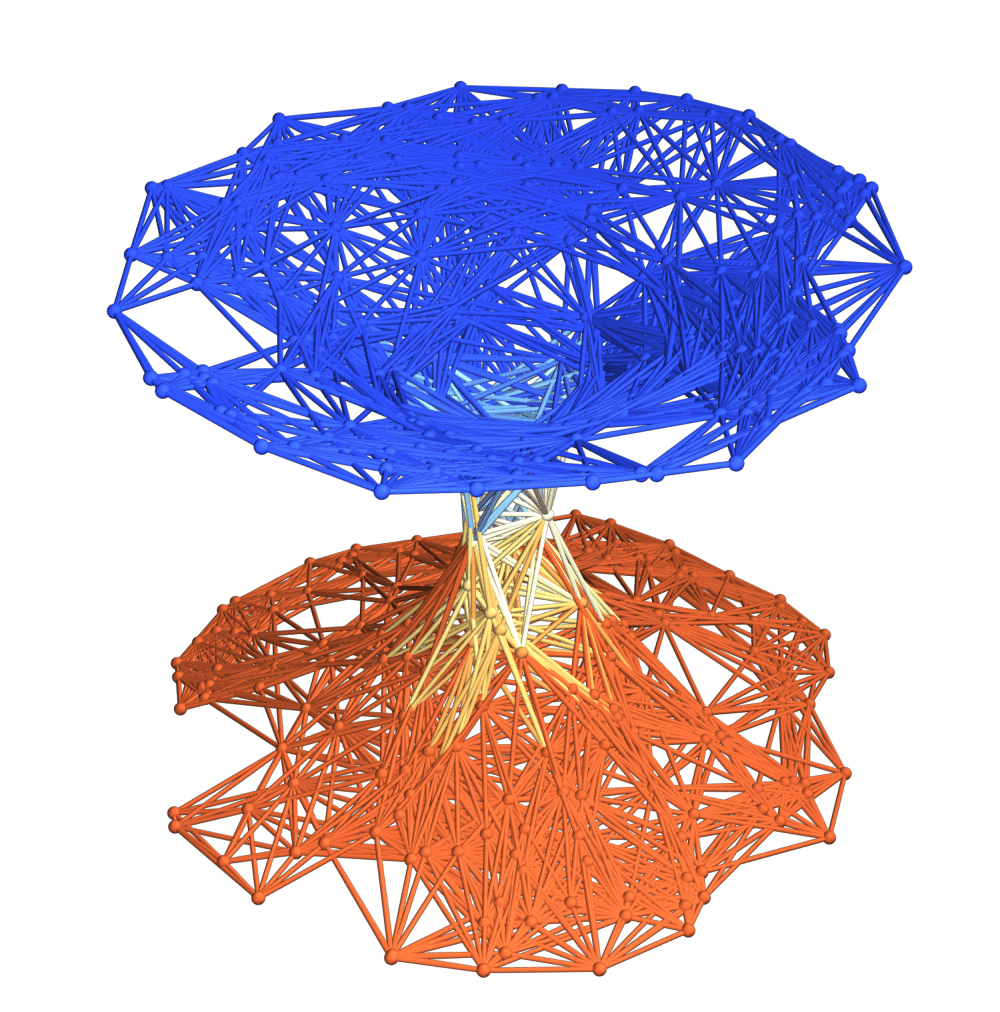}
\includegraphics[width=0.325\textwidth]{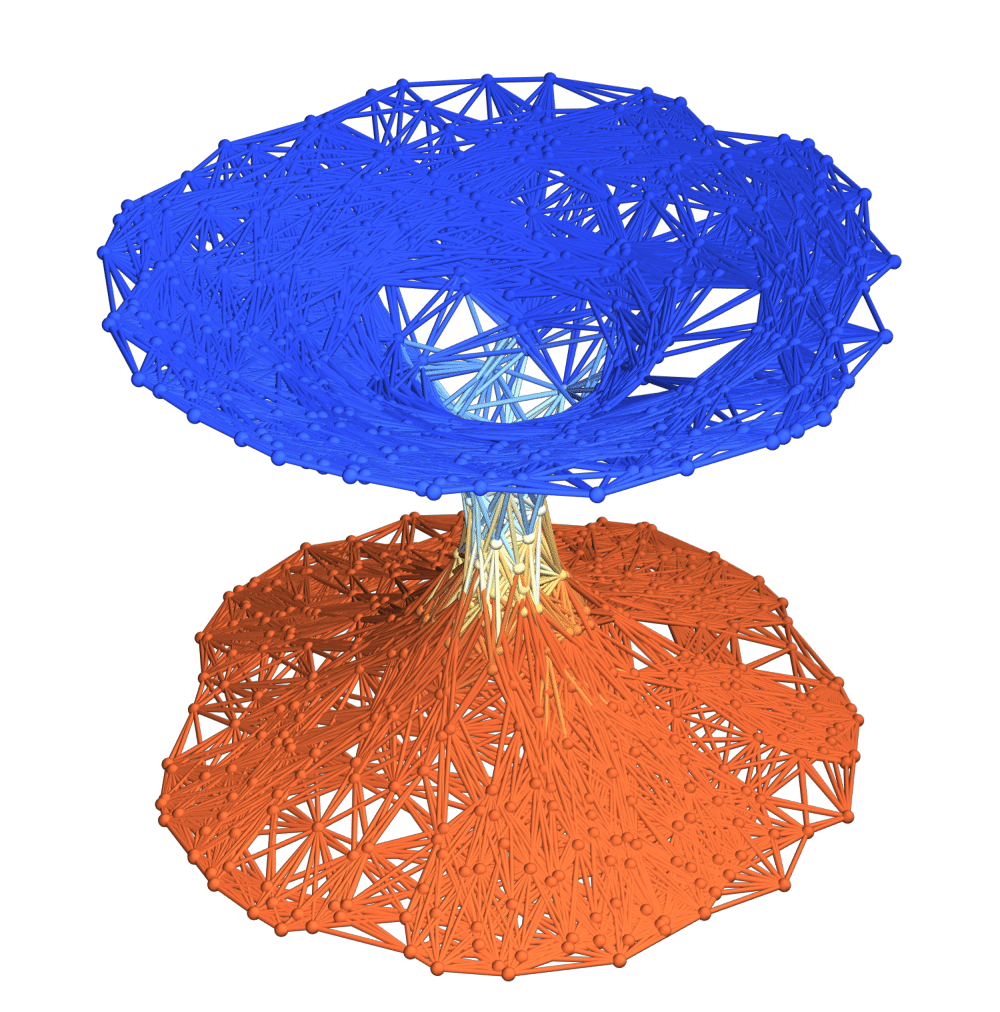}
\caption{Spatial hypergraphs corresponding to projections along the $z$-axis of the intermediate hypersurface configuration of the maximally extended Schwarzschild black hole (Einstein-Rosen bridge) test at time ${t = 1.5 M}$, with resolutions of 200, 400 and 800 vertices, respectively. The vertices have been assigned spatial coordinates according to the profile of the Schwarzschild conformal factor ${\psi}$ through a spatial slice perpendicular to the $z$-axis, and the hypergraphs have been adapted and colored using the local curvature in ${\psi}$.}
\label{fig:Figure28}
\end{figure}

\begin{figure}[ht]
\centering
\includegraphics[width=0.325\textwidth]{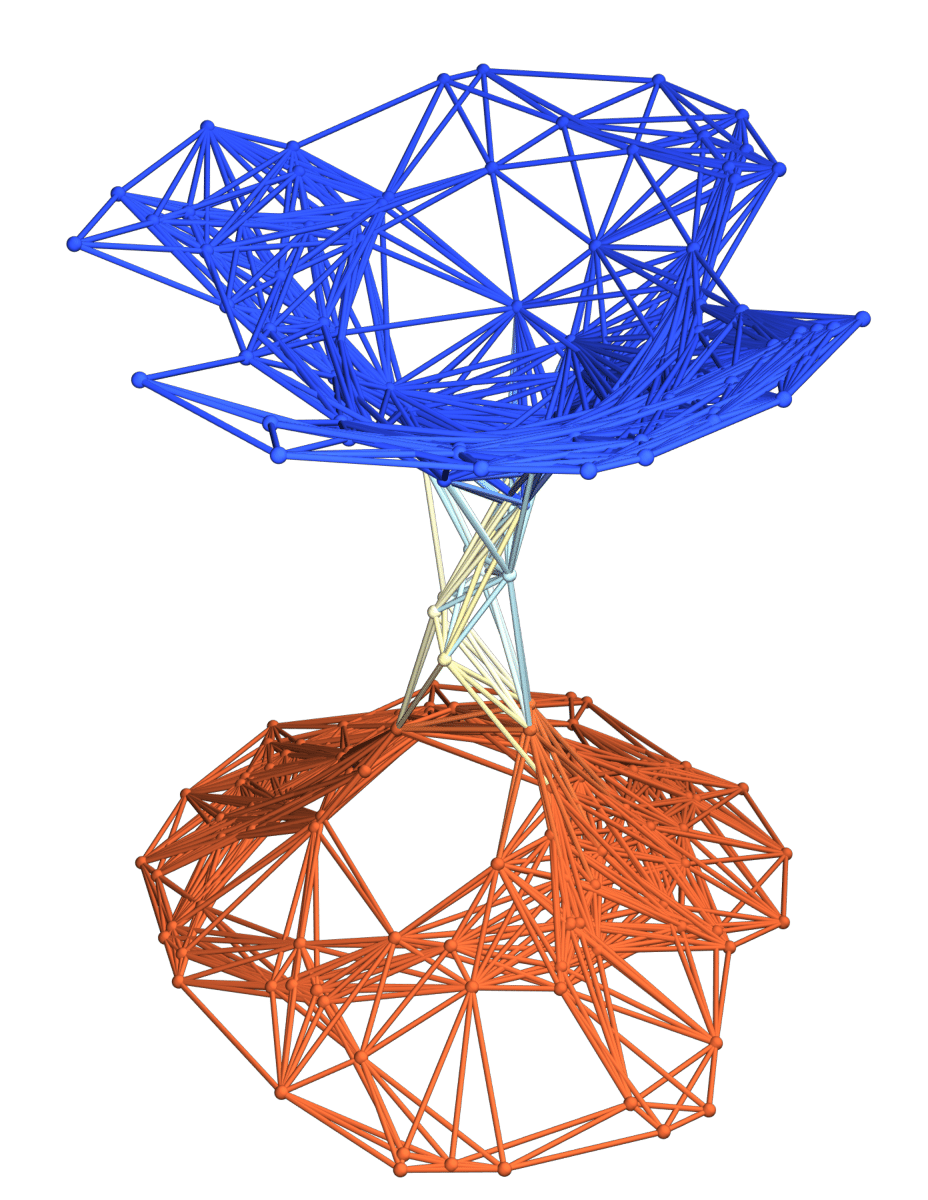}
\includegraphics[width=0.325\textwidth]{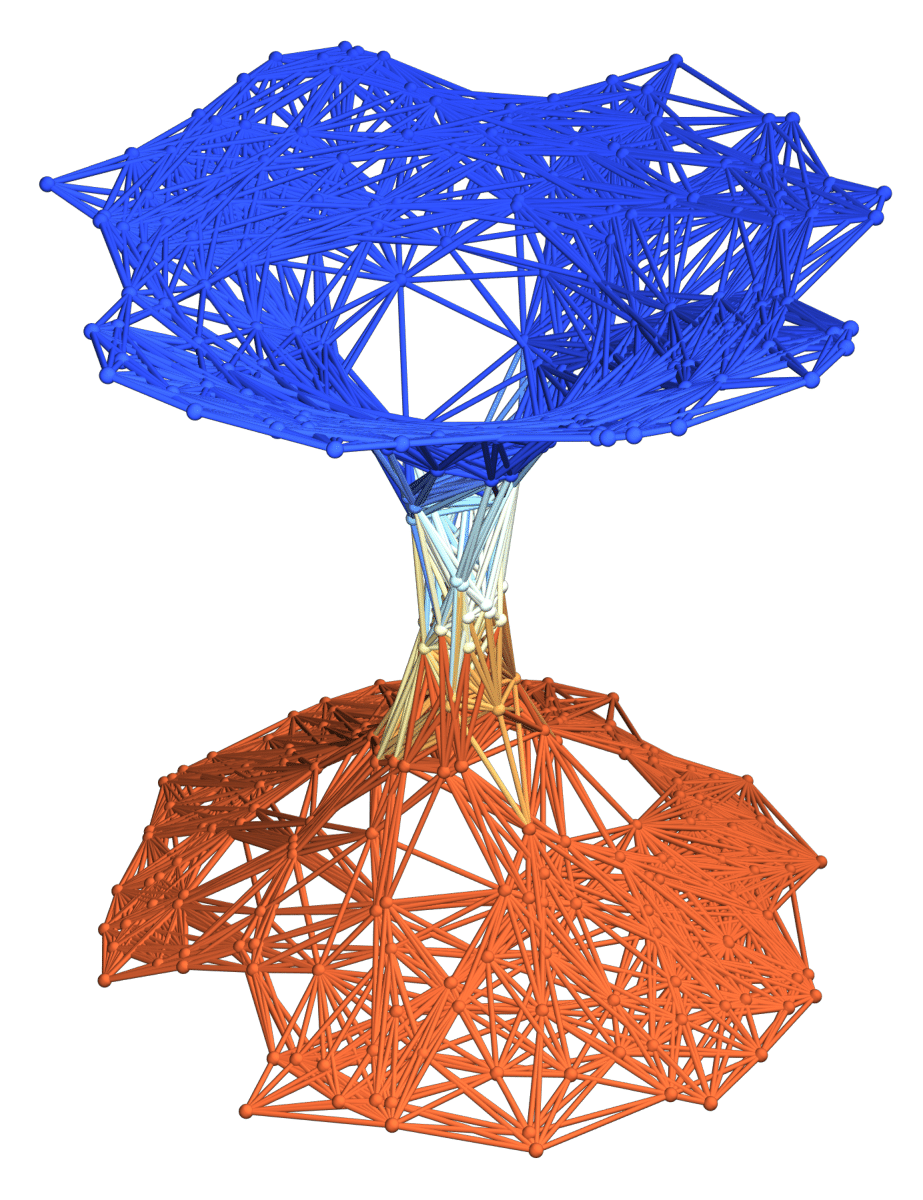}
\includegraphics[width=0.325\textwidth]{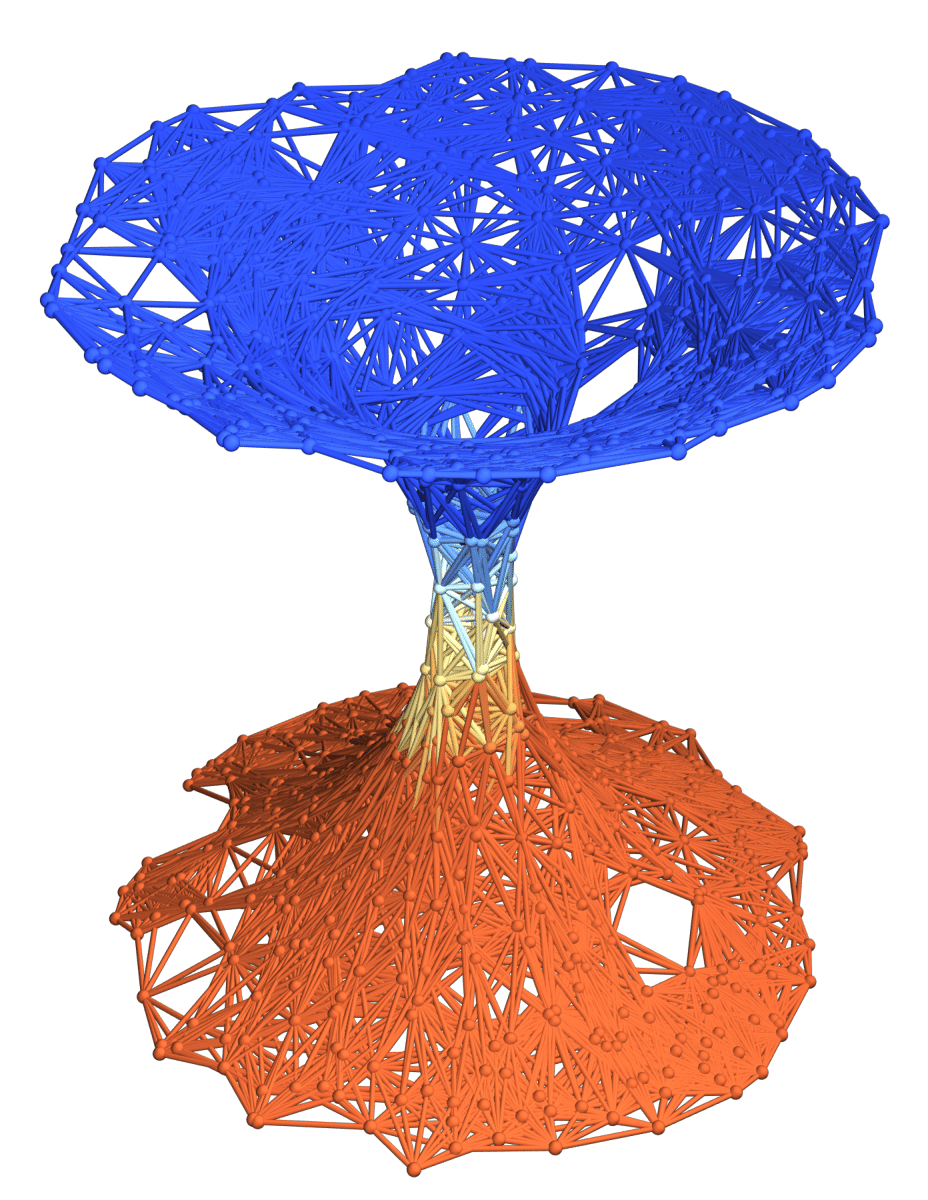}
\caption{Spatial hypergraphs corresponding to projections along the $z$-axis of the final hypersurface configuration of the maximally extended Schwarzschild black hole (Einstein-Rosen bridge) test at time ${t = 3 M}$, with resolutions of 200, 400 and 800 vertices, respectively. The vertices have been assigned spatial coordinates according to the profile of the Schwarzschild conformal factor ${\psi}$ through a spatial slice perpendicular to the $z$-axis, and the hypergraphs have been adapted and colored using the local curvature in ${\psi}$.}
\label{fig:Figure29}
\end{figure}

Observe that, in Schwarzschild coordinates, the original static Schwarzschild solution:

\begin{equation}
g = ds^2 = - \left( 1 - \frac{2M}{r} \right) dt^2 + \left( 1 - \frac{2M}{r} \right)^{-1} dr^2 + r^2 \left(d \theta^2 + \sin^2 \left( \theta \right) d \phi^2 \right),
\end{equation}
can be written in the form of an Einstein-Rosen bridge (\textit{wormhole}) metric over this maximally extended spacetime by replacing the Schwarzschild radial coordinate $r$ with ${u^2 = r - 2M}$, to obtain:

\begin{equation}
g = ds^2 = - \frac{u^2}{u^2 + 2M} dt^2 + 4 \left( u^2 + 2M \right) du^2 + \left( u^2 + 2M \right)^2 \left( d \theta^2 + \sin^2 \left( \theta \right) d \phi^2 \right),
\end{equation}
thus describing a spacetime with two sheets, namely ${u > 0}$ and ${u < 0}$, joined by a hyperplane with ${r = 2M}$ and ${u = 0}$, along which the metric $g$ vanishes. For this reason, we are able to use the same simple expression for the conformal factor ${\psi}$:

\begin{equation}
\psi = \left( 1 + \frac{M}{2r} \right)^{-2},
\end{equation}
as introduced in the static Schwarzschild black hole test above, and the four non-zero components of the Riemann curvature tensor ${R_{i j k l}}$ remain (up to symmetry and sign changes) identical. Using the Schwarzschild conformal factor as the scalar field in the refinement algorithm ${\phi = \psi}$, we place the outer boundary of the computational domain at ${60 M}$ as before, and enforce the usual Sommerfeld boundary conditions:

\begin{equation}
\frac{\partial f \left( x_i, t \right)}{\partial t} = - \frac{v x_i}{r} \frac{\partial f \left( x_i, t \right)}{\partial x_i} - v \frac{f \left( x_i, t \right) - f_0 \left( x_i, t \right)}{r},
\end{equation}
for arbitrary scalar field $f$, radial distance from the center of the domain ${r = \sqrt{x_{1}^{2} + x_{2}^{2} + x_{3}^{2}}}$, Minkowski space boundary values ${f_0}$ and radiative velocity ${v = 1}$. We evolve the solution until a final time of ${t = 3 M}$, with an intermediate check at time ${t = 1.5 M}$ (during which time the neck of the bridge stretches rapidly, as predicted by Fuller and Wheeler\cite{fuller}); the initial, intermediate and final hypersurface configurations are shown in Figures \ref{fig:Figure23}, \ref{fig:Figure24} and \ref{fig:Figure25}, respectively, with resolutions of 200, 400 and 800 vertices, and with the hypergraphs adapted and colored using the Schwarzschild conformal factor ${\psi}$. Figure \ref{fig:Figure26} shows the discrete characteristic structure of the solutions after time ${t = 3 M}$ (using directed acyclic causal graphs to show discrete characteristic lines), and Figures \ref{fig:Figure27}, \ref{fig:Figure28} and \ref{fig:Figure29} show projections along the $z$-axis of the initial, intermediate and final hypersurface configurations, respectively, with vertices assigned spatial coordinates according to the profile of the Schwarzschild conformal factor ${\psi}$. We confirm that the ADM mass of the black hole (computed by integrating over a pair of surfaces surrounding the two boundaries of asymptotic flatness) remains approximately constant, and that the linear and angular momenta both remain approximately zero, as expected. The convergence rates for the Hamiltonian constraint after time ${t = 3 M}$, with respect to the ${L_1}$, ${L_2}$ and ${L_{\infty}}$ norms, illustrating approximately fourth-order convergence of the finite difference scheme, are shown in Table \ref{tab:Table3}.

\begin{table}[ht]
\center
\begin{tabular}{|c|c|c|c|c|c|c|}
\hline
Vertices & ${\epsilon \left( L_1 \right)}$ & ${\epsilon \left( L_2 \right)}$ & ${\epsilon \left( L_{\infty} \right)}$ & ${\mathcal{O} \left( L_1 \right)}$ & ${\mathcal{O} \left( L_2 \right)}$ & ${\mathcal{O} \left( L_{\infty} \right)}$\\
\hline\hline
100 & ${4.91 \times 10^{-2}}$ & ${1.49 \times 10^{-2}}$ & ${6.83 \times 10^{-2}}$ & - & - & -\\
\hline
200 & ${6.06 \times 10^{-3}}$ & ${1.55 \times 10^{-3}}$ & ${6.92 \times 10^{-3}}$ & 3.02 & 3.27 & 3.30\\
\hline
400 & ${3.58 \times 10^{-4}}$ & ${6.53 \times 10^{-5}}$ & ${2.39 \times 10^{-4}}$ & 4.08 & 4.56 & 4.86\\
\hline
800 & ${2.33 \times 10^{-5}}$ & ${5.82 \times 10^{-6}}$ & ${1.22 \times 10^{-5}}$ & 3.94 & 3.49 & 4.29\\
\hline
1600 & ${1.05 \times 10^{-6}}$ & ${2.59 \times 10^{-7}}$ & ${9.34 \times 10^{-7}}$ & 4.47 & 4.49 & 3.71\\
\hline
\end{tabular}
\caption{Convergence rates for the maximally extended Schwarzschild black hole (Einstein-Rosen bridge) test with respect to the ${L_1}$, ${L_2}$ and ${L_{\infty}}$ norms for the Hamiltonian constraint $H$ after time ${t = 3 M}$, showing approximately fourth-order convergence.}
\label{tab:Table3}
\end{table}

\clearpage

\subsection{Head-On Collision of Two Schwarzschild Black Holes (Convergence Test)}

Penultimately, we stress-test the robustness and convergence properties of our simulation code by evolving a Brill-Lindquist binary black hole solution corresponding to a head-on collision of two equal mass static Schwarzschild black holes. Following Abrahams and Price\cite{abrahams}\cite{abrahams2}\cite{anninos2}, the Brill-Lindquist initial data describes a conformally flat three-dimensional spatial geometry of the form:

\begin{equation}
g = ds^2 = \Phi^4 d s_{flat}^{2},
\end{equation}
where ${d s_{flat}^{2}}$ designates the three-dimensional line element for flat space, and ${\Phi}$ is a solution to Laplace's equation:

\begin{equation}
\nabla^2 \Phi = 0,
\end{equation}
in flat space. The Brill-Lindquist metric can then be written in the same general form as the (also conformally flat) Misner metric\cite{misner3}, namely:

\begin{equation}
ds^2 = \Phi^4 \left( R, \theta ; \mu_0 \right) \left( dR^2 + R^2 \left[ d \theta^2 + \sin^2 \left( \theta \right) d \phi^2 \right] \right),
\end{equation}
with respect to the spherical coordinates ${\left( R, \theta, \phi \right)}$ for the flat three-dimensional space described by ${d s_{flat}^{2}}$, where ${\mu_0}$ is a dimensionless quantity parametrizing the initial separation of the two ``throats'' of the black holes in the Misner geometry. For the Brill-Lindquist geometry in particular, our Laplacian solution ${\Phi_{BL} \left( R, \theta \right)}$ takes the form of a Newtonian gravitational potential for a pair of masses $m$ located at points ${z = \pm z_0}$ along the $z$-axis (up to a factor of two), i.e:

\begin{equation}
\Phi_{BL} \left( R, \theta \right) = 1 + \frac{1}{2} \left( \frac{m}{\sqrt{R^2 \sin^2 \left( \theta \right) + \left( R \cos \left( \theta \right) - z_0 \right)^2}} + \frac{m}{\sqrt{R^2 \sin^2 \left( \theta \right) + \left( R \cos \left( \theta \right) + z_0 \right)^2}} \right).
\end{equation}
Introducing the combined mass term ${M = 2m}$, these square root terms in the denominators can be expanded out as a power series in ${\left( \frac{z_0}{R} \right)}$ for all radii ${R > z_0}$, to obtain:

\begin{equation}
d s_{BL}^{2} = \left[ 1 + \frac{M}{2R} \sum_{\ell = 0, 2, \dots} \left( \frac{z_0}{R} \right)^{\ell} P_{\ell} \left( \cos \left( \theta \right) \right) \right]^4 \left( d R^2 + R^2 \left[ d \theta^2 + \sin^2 \left( \theta \right) d \phi^2 \right] \right),
\end{equation}
where ${P_{\ell}}$ denote the standard Legendre polynomials, defined, as previously, in terms of the generating function\cite{arfken}:

\begin{equation}
\sum_{\ell = 0}^{\infty} P_{\ell} \left( x \right) t^n = \frac{1}{\sqrt{1 - 2 xt + t^2}}.
\end{equation}

If we now make a transformation from the isotropic radial coordinate $R$ to the Schwarzschild radial coordinate $r$, i.e:

\begin{equation}
R = \frac{\left( \sqrt{r} + \sqrt{r - 2M} \right)^2}{4},
\end{equation}
then the line element for the three-dimensional Brill-Lindquist geometry becomes:

\begin{equation}
d s_{BL}^{2} = \left[ 1 + \frac{\frac{M}{2R}}{1 + \frac{M}{2R}} \sum_{\ell = 2, 4, \dots} \left( \frac{z_0}{M} \right)^{\ell} \left( \frac{M}{R} \right)^{\ell} P_{\ell} \left( \cos \left( \theta \right) \right) \right]^4 \left( \frac{dr^2}{1 - \frac{2M}{r}} + r^2 \left[ d \theta^2 + \sin^2 \left( \theta \right) d \phi^2 \right] \right),
\end{equation}
which reduces to the three-dimensional Schwarzshild geometry if the sum over ${\ell}$ is suppressed (since this sum effectively encodes the deviation of the geometry from being perfectly spherical). Therefore, we can utilize perturbative techniques to describe the non-spherical geometry by treating the parameter ${\left( \frac{z_0}{M} \right) }$ as a perturbative expansion parameter, yielding (up to leading order in ${\left( \frac{z_0}{M} \right)}$ for each multipole index ${\ell}$) the following approximation to the initial Brill-Lindquist spatial geometry:

\begin{equation}
d s_{BL}^{2} \approx \left[ 1 + \frac{\frac{2M}{R}}{1 + \frac{M}{2R}} \sum_{\ell = 2, 4, \dots} \left( \frac{z_0}{M} \right)^{\ell} \left( \frac{M}{R} \right)^{\ell} P_{\ell} \left( \cos \left( \theta \right) \right) \right] \left( \frac{dr^2}{1 - \frac{2M}{r}} + r^2 \left[ d \theta^2 + \sin^2 \left( \theta \right) d \phi^2 \right] \right).
\end{equation}
The Misner initial spatial geometry is extremely similar, except that the coefficients in the multipole expansion are given by ${\kappa_{\ell} \left( \mu_0 \right)}$, with the ${\kappa_{\ell}}$ functions defined by:

\begin{equation}
\kappa_{\ell} \left( \mu_0 \right) = \frac{1}{\left( 4 \Sigma_1 \right)^{\ell + 1}} \sum_{n = 1}^{\infty} \frac{\left( \coth \left( n \mu_0 \right) \right)^{\ell}}{\sinh \left( n \mu_0 \right)},
\end{equation}
and with ${\Sigma_1}$ defined in terms of the total ADM mass of the system:

\begin{equation}
M = 4a \sum_{n = 1}^{\infty} \frac{1}{\sinh \left( n \mu_0 \right)} = 4a \Sigma_1,
\end{equation}
for arbitrary real parameter $a$. In other words, if we apply the simple substitution:

\begin{equation}
\left( \frac{z_0}{M} \right)^{\ell} \to 4 \kappa_{\ell} \left( \mu_0 \right),
\end{equation}
then we recover exactly the three-dimensional spatial line element of the Misner geometry.

We can then extract gravitational wave information from the resulting solution via the Moncrief formalism\cite{moncrief}\cite{moncrief2} (i.e. by applying metric perturbation theory to a three-dimensional Schwarzschild spatial background), or more generally via explicit computation of the complex Weyl scalar ${\Psi_4}$\cite{loffler}. In the former (Moncrief) case, we wish to perform a perturbative expansion of the spatial metric tensor onto a certain tensor basis, namely the basis of Regge-Wheeler harmonics\cite{regge}, with associated functions ${\left\lbrace c_{1}^{\times \ell m}, c_{2}^{\times \ell m}, h_{1}^{+ \ell m}, H_{2}^{+ \ell m}, K^{+ \ell m}, G^{+ \ell m} \right\rbrace}$. The derivation of the Regge-Wheeler formalism for perturbations of odd parity begins by applying a spatial metric perturbation of order ${\left( \ell, m \right)}$ of the form:

\begin{equation}
h_{i j} = h_1 \left( r, t \right) \left( \hat{e}_1 \right)_{i j} + h_2 \left( r, t \right) \left( \hat{e}_2 \right)_{i j},
\end{equation}
where ${\hat{e}_1}$ and ${\hat{e}_2}$ can be expressed in matrix form as:

\begin{equation}
\hat{e}_1 = \begin{bmatrix}
0 & - \frac{1}{\sin \left( \theta \right)} \frac{\partial Y_{\ell m} \left( \theta, \phi \right)}{\partial \phi} & \sin \left( \theta \right) \frac{\partial Y_{\ell m} \left( \theta, \phi \right)}{\partial \theta}\\
\mathrm{sym} & 0 & 0\\
\mathrm{sym} & 0 & 0
\end{bmatrix},
\end{equation}
and:

\begin{equation}
\hat{e}_2 = \begin{bmatrix}
0 & 0 & 0\\
0 & \frac{1}{\sin \left( \theta \right)} \left[ \frac{\partial^2}{\partial \theta \partial \phi} - \cot \left( \theta \right) \frac{\partial}{\partial \phi} \right] Y_{\ell m} \left( \theta, \phi \right) & \mathrm{sym}\\
0 & \frac{1}{2} \left[ \frac{1}{\sin \left( \theta \right)} \frac{\partial^2}{\partial \phi^2} + \cos \left( \theta \right) \frac{\partial}{\partial \theta} - \sin \left( \theta \right) \frac{\partial^2}{\partial \theta^2} \right] Y_{\ell m} \left( \theta, \phi \right) & - \left[ \sin \left( \theta \right) \frac{\partial^2}{\partial \theta \partial \phi} - \cos \left( \theta \right) \frac{\partial}{\partial \phi} \right] Y_{\ell m} \left( \theta, \phi \right)
\end{bmatrix},
\end{equation}
respectively. In the above, ${\mathrm{sym}}$ designates tensor symmetrization (i.e. ${h_{i j} = h_{j i}}$), and ${Y_{\ell m}}$ denote the standard Laplace spherical harmonic functions\cite{courant} of order ${\left( \ell, m \right)}$, namely:

\begin{equation}
Y_{\ell m} \left( \theta, \phi \right) = \sqrt{\frac{\left( 2 \ell + 1 \right)}{4 \pi} \frac{\left( \ell - m \right)!}{\left( \ell + m \right)!}} P_{\ell m} \left( \cos \left( \theta \right) \right) e^{i m \phi},
\end{equation}
where ${P_{\ell m}}$ denote the associated Legendre polynomials, including the Condon-Shortley phase\cite{condon}, which we can conveniently write in terms of derivatives of ordinary Legendre polynomials as:

\begin{equation}
P_{\ell m} \left( x \right) = \left( -1 \right)^m \left( 1 - x^2 \right)^{\frac{m}{2}} \frac{d^m}{d x^m} \left( P_{\ell} \left( x \right) \right),
\end{equation}
with the standard convention that:

\begin{equation}
P_{\ell \left( -m \right)} \left( x \right) = \left( -1 \right)^{m} \frac{\left( \ell - m \right)!}{\left( \ell + m \right)!} P_{\ell m} \left( x \right),
\end{equation}
for all ${m \geq 0}$ (i.e. the natural normalization condition implied by Rodrigues' formula). The perturbation ${\alpha^{\prime}}$ to the lapse function ${\alpha}$ then vanishes and the perturbation ${\beta_{i}^{\prime}}$ to the shift vector ${\beta_i}$ becomes:

\begin{equation}
\left[ \beta_{i}^{\prime} \right] = h_{0} \left( r, t \right) \left[ 0, - \frac{1}{\sin \left( \theta \right)} \frac{\partial Y_{\ell m} \left( \theta, \phi \right)}{\partial \phi}, \sin \left( \theta \right) \frac{\partial Y_{\ell m} \left( \theta, \phi \right)}{\partial \theta} \right].
\end{equation}

However, the case that primarily concerns us for our present purposes is the Regge-Wheeler formalism for perturbations of even parity, in which the spatial metric perturbation of order ${\left( \ell, m \right)}$ now has the form:

\begin{equation}
h_{i j} = h_{1}^{+ \ell m} \left( r, t \right) \left( \hat{f} \right)_{i j} + \frac{H_{2}^{+ \ell m} \left( r, t \right)}{\left[ 1 - \left( \frac{2M}{r} \right) \right]} \left( \hat{f}_2 \right)_{i j} + r^2 K^{+ \ell m} \left( r, t \right) \left( \hat{f}_3 \right)_{i j} + r^2 G^{+ \ell m} \left( r, t \right) \left( \hat{f}_4 \right)_{i j},
\end{equation}
with the perturbation ${\alpha^{\prime}}$ to the lapse function ${\alpha}$ becoming:

\begin{equation}
\alpha^{\prime} = - \frac{1}{2} \sqrt{1 - \frac{2M}{r}} H_0 \left( r, t \right) Y_{\ell m} \left( \theta, \phi \right),
\end{equation}
and the perturbation ${\beta_{i}^{\prime}}$ to the shift vector ${\beta_i}$ being:

\begin{equation}
\left[ \beta_{i}^{\prime} \right] = \left[ H_1 \left( r, t \right) Y_{\ell m} \left( \theta, \phi \right), h_0 \left( r, t \right) \frac{\partial Y_{\ell m} \left( \theta, \phi \right)}{\partial \theta}, h_0 \left( r, t \right) \frac{\partial Y_{\ell m} \left( \theta, \phi \right)}{\partial \phi} \right].
\end{equation}
In the above, ${\hat{f}_1}$, ${\hat{f}_2}$, ${\hat{f}_3}$ and ${\hat{f}_4}$ can be expressed in matrix form as:

\begin{equation}
\hat{f}_1 = \begin{bmatrix}
0 & \frac{\partial Y_{\ell m} \left( \theta, \phi \right)}{\partial \theta} & \frac{\partial Y_{\ell m} \left( \theta, \phi \right)}{\partial \phi}\\
\mathrm{sym} & 0 & 0\\
\mathrm{sym} & 0 & 0
\end{bmatrix}, \qquad \hat{f}_2 = \begin{bmatrix}
Y_{\ell m} \left( \theta, \phi \right) & 0 & 0\\
0 & 0 & 0\\
0 & 0 & 0
\end{bmatrix},
\end{equation}

\begin{equation}
\hat{f}_3 = \begin{bmatrix}
0 & 0 & 0\\
0 & Y_{\ell m} \left( \theta, \phi \right) & 0\\
0 & 0 & \sin^2 \left( \theta \right) Y_{\ell m} \left( \theta, \phi \right)
\end{bmatrix},
\end{equation}
and:

\begin{equation}
\hat{f}_4 = \begin{bmatrix}
0 & 0 & 0\\
0 & \frac{\partial^2 Y_{\ell m} \left( \theta, \phi \right)}{\partial \theta^2} & \mathrm{sym}\\
0 & \left[ \frac{\partial^2}{\partial \theta \partial \phi} - \cot \left( \theta \right) \frac{\partial}{\partial \phi} \right] Y_{\ell m} \left( \theta, \phi \right) & \left[ \frac{\partial^2}{\partial \phi^2} + \sin \left( \theta \right) \cos \left( \theta \right) \frac{\partial}{\partial \theta} \right] Y_{\ell m} \left( \theta, \phi \right)
\end{bmatrix},
\end{equation}
in the usual way. Arbitrary gauge transformations of the spatial metric perturbation ${h_{i j}}$ can then be written as:

\begin{equation}
\delta h_{i j} = C_{i ; j} + C_{j ; i} = C_{i \vert j} + C_{j \vert i},
\end{equation}
where the semicolons and vertical bars correspond to the ordinary notation for covariant differentiation of tensor fields in the Ricci calculus (being placed directly in front of an appended covariant index). Here, ${C_i}$ denotes an arbitrary vector harmonic of even parity, i.e:

\begin{equation}
\left[ C_i \right] = \left[ c_{2}^{\times \ell m} Y_{\ell m} \left( \theta, \phi \right), c_{1}^{\times \ell m} \frac{\partial Y_{\ell m}}{\partial \theta}, c_{1}^{\times \ell m} \frac{\partial Y_{\ell m}}{\partial \phi} \right],
\end{equation}
thus completing the description of the required tensor basis of six functions in Regge-Wheeler harmonics, namely ${\left\lbrace c_{1}^{\times \ell m}, c_{2}^{\times \ell m}, h_{1}^{+ \ell m}, H_{2}^{+ \ell m}, K^{+ \ell m}, G^{+ \ell m} \right\rbrace}$.

Given this set of basis functions, we are now able to define and calculate a new pair of gauge invariant polarization quantities ${Q_{\ell m}^{\times}}$ and ${Q_{\ell m}^{+}}$, with the explicit forms:

\begin{equation}
Q_{\ell m}^{\times} = \sqrt{\frac{2 \left( \ell + 2 \right)!}{\left( \ell - 2 \right)!}} \left[ c_{1}^{\times m} + \frac{1}{2} \left( \partial_r - \frac{2}{r} \right) c_{2}^{\times \ell m} \right] \frac{S}{r},
\end{equation}
and:

\begin{multline}
Q_{\ell m}^{+} = \frac{1}{\Lambda} \sqrt{\frac{2 \left( \ell - 1 \right) \left( \ell + 2 \right)}{\ell \left( \ell + 1 \right)}} \left( \ell \left( \ell + 1 \right) S \left( r^2 \partial_r G^{+ \ell m} - 2 h_{1}^{+ \ell m} \right) \right.\\
\left. + 2 r S \left( H_{2}^{+ \ell m} - r \partial_r K^{+ \ell m} \right) + \Lambda r K^{+ \ell m} \right),
\end{multline}
respectively, where we have introduced the intermediate quantities:

\begin{equation}
S = 1 - \frac{2M}{r}, \qquad \text{ and } \qquad \Lambda = \left( \ell - 1 \right) \left( \ell + 2 \right) + \frac{6M}{r}.
\end{equation}
Beyond simply being gauge invariant, the significance of the polarization quantities ${Q_{\ell m}^{\times}}$ and ${Q_{\ell m}^{+}}$ is that they can be shown to satisfy the following pair of wave equations:

\begin{equation}
\left( \partial_{t}^{2} - \partial_{r^*}^{2} \right) Q_{\ell m}^{\times} = -S \left[ \frac{\ell \left( \ell + 1 \right)}{r^2} - \frac{6M}{r^3} \right] Q_{\ell m}^{\times},
\end{equation}
and:

\begin{equation}
\left( \partial_{t}^{2} - \partial_{r^*}^{2} \right) Q_{\ell m}^{+} = -S \left[ \frac{1}{\Lambda^2} \left( \frac{72 M^3}{r^5} - \frac{12 M \left( \ell - 1 \right) \left( \ell + 2 \right)}{r^3} \left( 1 - \frac{3M}{r} \right) \right) + \frac{\ell \left( \ell^2 - 1 \right) \left( \ell + 2 \right)}{r^2 \Lambda} \right] Q_{\ell m}^{+},
\end{equation}
where we have introduced the new radial coordinate ${r^*}$:

\begin{equation}
r^* = r + 2M \log \left( \frac{r}{2M} - 1 \right),
\end{equation}
and therefore their solutions can be interpreted as being first-order gauge invariant gravitational waveforms. However, since the overall four-dimensional spacetime metric is treated here as being a perturbation of the spherically symmetric Schwarzschild metric, the interpretation of these solutions as gravitational waveforms is only valid when the background spacetime is still approximately spherically symmetric.

For the case of more general spacetimes in which this approximate symmetry assumption is not valid (such as the early-time case of the spacetime simulated here), we can opt instead to compute the complex Weyl scalar ${\Psi_4}$ explicitly. Within the Newman-Penrose formalism\cite{newman}, the Weyl scalar ${\Psi_4}$ is computed as:

\begin{equation}
\Psi_4 = C_{\alpha \beta \gamma \delta} n^{\alpha} \bar{m}^{\beta} n^{\gamma} \bar{m}^{\delta},
\end{equation}
with ${C_{i j k l}}$ being the Weyl curvature tensor, and ${n^i}$ and ${\bar{m}^i}$ being elements of a complex null tetrad ${\left\lbrace l^i, n^i, m^i, \bar{m}^i \right\rbrace}$, obeying the following normalization conditions:

\begin{equation}
l_i l^i = n_i n^i = m_i m^i = \bar{m}_i \bar{m}^i = 0, \qquad l_i m^i = l_i \bar{m}^i = n_i m^i = n_i \bar{m}^i = 0,
\end{equation}
and:

\begin{equation}
l_i n^i = l^i n_i = -1, \qquad m_i \bar{m}^i = m^i \bar{m}_i = 1,
\end{equation}
as well as the following compatibility conditions with respect to the spacetime metric tensor ${g_{i j}}$:

\begin{equation}
g_{i j} = -l_i n_j - n_i l_j + m_i \bar{m}_j + \bar{m}_i m_j, \qquad g^{i j} = -l^i n^j - n^i l^j + m^i \bar{m}^j + \bar{m}^i m^j.
\end{equation}
Physically, the ${\Psi_4}$ Weyl scalar has the interpretation of an outgoing gravitational (transverse) radiation term\cite{szekeres2}, with corresponding incoming gravitational (transverse) radiation term given by the ${\Psi_0}$ scalar:

\begin{equation}
\Psi_0 = C_{\alpha \beta \gamma \delta} l^{\alpha} m^{\beta} l^{\gamma} m^{\delta},
\end{equation}
and with the incoming and outgoing longitudinal radiation terms given by the ${\Psi_1}$ and ${\Psi_3}$ scalars:

\begin{equation}
\Psi_1 = C_{\alpha \beta \gamma \delta} l^{\alpha} n^{\beta} l^{\gamma} m^{\delta}, \qquad \Psi_3 = C_{\alpha \beta \gamma \delta} l^{\alpha} n^{\beta} \bar{m}^{\gamma} n^{\delta},
\end{equation}
respectively, and the gravitational monopole of the radiation source (in the form of a gravitational Coulomb term) given by the remaining ${\Psi_2}$ scalar:

\begin{equation}
\Psi_2 = C_{\alpha \beta \gamma \delta} l^{\alpha} m^{\beta} \bar{m}^{\gamma} n^{\delta}.
\end{equation}

Thus, we wish to construct a complex null tetrad which is exact (in the sense that it remains orthonormal within the numerical approximation to the spacetime), and which can be shown to reduce to the spatial Kinnersley triad (as commonly used in the perturbation analysis of the general Kerr black hole produced in the final stages of the merger) in the linear regime. For this purpose, we follow the techniques proposed by Baker, Campanelli and Lousto\cite{baker2}. The background Kinnersley tetrad vectors ${\left\lbrace l_{Kin}^{\mu}, n_{Kin}^{\mu}, m_{Kin}^{\mu} \right\rbrace}$ in Boyer-Lindquist coordinates are given by\cite{teukolsky}\cite{press}:

\begin{equation}
l_{Kin}^{\mu} = \frac{1}{\Delta} \left[ \left( r^2 + a^2 \right), \Delta, 0, a \right], \qquad n_{Kin}^{\mu} = \frac{1}{2 \Sigma} \left[ \left( r^2 + a^2 \right), - \Delta, 0, a \right],
\end{equation}
and:

\begin{equation}
m_{Kin}^{\mu} = \frac{1}{\sqrt{2} \left( r + i a \cos \left( \theta \right) \right)} \left[ i a \sin \left( \theta \right), 0, 1, \frac{i}{\sin \left( \theta \right)} \right],
\end{equation}
respectively, where $a$, ${\Sigma}$ and ${\Delta}$ designate the standard rotation-dependent parameters of the Kerr metric in Boyer-Lindquist coordinates:

\begin{equation}
a = \frac{J}{M}, \qquad \Sigma = r^2 + a^2 \cos^2 \left( \theta \right), \qquad \Delta = r^2 - r_s r + a^2.
\end{equation}
The significance of using the spatial Kinnersley tetrad for the purposes of perturbation analysis is that the vectors ${l_{Kin}^{\mu}}$ and ${n_{Kin}^{\mu}}$ are both guaranteed to lie along the background \textit{principal null directions} of the Weyl curvature tensor ${C_{i j k l}}$, i.e. they both lie in the direction of the eigenbivectors ${X^{i j}}$ associated with null vectors in the spacetime, as defined by the eigenvalue equation:

\begin{equation}
\frac{1}{2} C^{i j}_{m n} X^{m n} = \lambda X^{i j},
\end{equation}
for eigenvalue ${\lambda}$, thus allowing us to derive perturbation equations that are effectively decoupled along these two directions. If we now perform an ADM ${3 + 1}$ decomposition of the Kerr metric tensor, with ${\alpha_{\left( 0 \right)}}$ and ${\beta_{\left( 0 \right)}^{i}}$ denoting the zeroth-order lapse function and shift vector of the Kerr solution, respectively, i.e:

\begin{equation}
\alpha_{\left( 0 \right)} = \sqrt{\frac{\Delta \Sigma}{\Omega}}, \qquad \text{ and } \qquad \beta_{\left( 0 \right)}^{i} = \left[ 0, 0, - \frac{2 a M r}{\Omega} \right],
\end{equation}
where:

\begin{equation}
\Omega = \left( r^2 + a^2 \right) \Sigma + 2 M r a^2 \sin^2 \left( \theta \right),
\end{equation}
then the elements of our spatial tetrad in terms of this ${3 + 1}$ basis become:

\begin{equation}
l^{\mu} = \left[ \alpha_{\left( 0 \right)} l_{Kin}^{0}, l_{Kin}^{i} + \beta_{\left( 0 \right)}^{i} l_{Kin}^{0} \right], \qquad n^{\mu} = \left[ \alpha_{\left( 0 \right)} n_{Kin}^{0}, n_{Kin}^{i} + \beta_{\left( 0 \right)}^{i} n_{Kin}^{0} \right],
\end{equation}
and:

\begin{equation}
m^{\mu} = \left[ \alpha_{\left( 0 \right)} m_{Kin}^{0}, m_{Kin}^{i} + \beta_{\left( 0 \right)}^{i} m_{Kin}^{0} \right].
\end{equation}

In order to compute this exact complex null spatial tetrad numerically, we begin by assuming that the tetrad admits a ${3 + 1}$ decomposition of the form:

\begin{equation}
\tilde{l}^{\mu} = \frac{1}{\sqrt{2}} \left( u^{\mu} + r^{\mu} \right), \qquad \tilde{n}^{\mu} = \frac{1}{\sqrt{2}} \left( u^{\mu} - r^{\mu} \right), \qquad \tilde{m}^{\mu} = \frac{1}{\sqrt{2}} \left( \theta^{\mu} + i \phi^{\mu} \right),
\end{equation}
assuming orthonormal vectors ${r^{\mu}}$, ${\theta^{\mu}}$ and ${\phi^{\mu}}$ pointing along each of the coordinate axes:

\begin{equation}
r^{\mu} = \left[ 0, v_{2}^{i} \right], \qquad \theta^{\mu} = \left[ 0, v_{3}^{i} \right], \qquad \phi^{\mu} = \left[ 0, v_{1}^{i} \right],
\end{equation}
and a timelike unit vector (normalized and orthogonal to the spacelike hypersurface) ${u^{\mu}}$. This tetrad decomposition can be constructed by starting from a collection of real vectors that are aligned with the radial direction $r$ and the angular direction ${\phi}$, i.e. (in Cartesian coordinates):

\begin{equation}
v_{1}^{i} = \left[ -y, x, 0 \right], \qquad v_{2}^{i} = \left[ x, y, z \right], \qquad v_{3}^{i} = \sqrt{\mathrm{det} \left( g \right)} g^{i l} \epsilon_{l j k} v_{a}^{j} v_{2}^{k},
\end{equation}
and then applying the Gram-Schmidt process so as to guarantee orthonormality of the triad, i.e:

\begin{equation}
v_{1}^{i} \to \frac{v_{1}^{i}}{\sqrt{\omega_{1 1}}}, \qquad v_{2}^{i} \to \frac{\left( v_{2}^{i} - v_{1}^{i} \omega_{1 2} \right)}{\sqrt{\omega_{2 2}}}, \qquad v_{3}^{i} \to \frac{\left( v_{3}^{i} - v_{1}^{i} \omega_{1 3} - v_{2}^{i} \omega_{2 3} \right)}{\sqrt{\omega_{3 3}}},
\end{equation}
at each step in the computation, where we have introduced the new tensor quantity ${\omega_{i j}}$:

\begin{equation}
\omega_{i j} = \left( v_{i}^{\mu} v_{j}^{\nu} g_{\mu \nu} \right).
\end{equation}
Note that it is essential that the orthonormalization algorithm begins by applying the Gram-Schmidt process to the azimuthal direction vector ${v_{1}^{i}}$, since this is the only direction vector which is not affected by the frame dragging effect caused by the discrepancy between the non-zero Boyer-Lindquist shift vector ${\beta_{i}^{Kerr}}$ and our (approximately) zero numerical shift vector ${\beta^i}$, creating an off-diagonal distortion in the metric tensor due to coordinates being dragged in the direction of the angular coordinate ${\phi}$. However, we can counteract this effect by introducing an offset term into the definition of the angular coordinate ${\phi}$ in Cartesian coordinates:

\begin{equation}
\phi = \arctan \left( \frac{y}{x} \right),
\end{equation}
so as to restore the desired diagonal form of the metric tensor, namely by subtracting off the contributions from the ${g_{r \phi}}$ component of the metric tensor which is most significantly affected by the frame dragging effect:

\begin{equation}
\varphi = \arctan \left( \frac{y}{x} \right) + \int \left( \frac{\hat{g}_{r \phi}}{\hat{g}_{\phi \phi}} \right) dr,
\end{equation}
where ${\hat{g}_{i j}}$ denotes the metric tensor following numerical evolution. More specifically, the numerically evolved metric tensor ${\hat{g}_{i j}}$ with its shift magnitude equal to zero can be converted into a metric tensor ${g_{i j}}$ with shift magnitude equal to ${\varphi}$ using the transformation:

\begin{equation}
\phi \to \varphi = \phi + \phi_{offset} \left( t, r, \theta \right),
\end{equation}
where ${g_{\varphi \varphi} = \hat{g}_{\phi \phi}}$. From here, the diagonal nature of the three-dimensional Kerr spatial metric implies that:

\begin{equation}
g_{r \varphi} = 0 = \hat{g}_{r \phi} - \partial_r \phi_{offset} \hat{g}_{\phi \phi}, \qquad \implies \partial_r \phi_{offset} = \frac{\hat{g}_{r \phi}}{\hat{g}_{\phi \phi}},
\end{equation}
hence the form of the integral above.

Applying the Gram-Schmidt process described above for a general Kerr metric in Boyer-Lindquist coordinates, one consequently obtains the following orthogonal tetrad elements:

\begin{equation}
u^{\mu} = \sqrt{\frac{\Omega}{\Delta \Sigma}} \left[ 1, 0, 0, \frac{2 a M r}{\Omega} \right], \qquad r^{\mu} = \left[ 0, v_{2}^{a} \right] = \left[ 0, 
\sqrt{\frac{\Delta}{\Sigma}}, 0, 0 \right],
\end{equation}
and:

\begin{equation}
\theta^{\mu} = \left[ 0, v_{3}^{a} \right] = \left[ 0, 0, \frac{1}{\sqrt{\Sigma}}, 0 \right], \qquad \phi^{\mu} = \left[ 0, v_{1}^{a} \right] = \left[ 0, 0, 0, \frac{1}{\sin \left( \theta \right)} \sqrt{\frac{\Sigma}{\Omega}} \right],
\end{equation}
along with the appropriate normalization conditions:

\begin{equation}
- u_{\mu} u^{\mu} = r_{\mu} r^{\mu} = \theta_{\mu} \theta^{\mu} = \phi_{\mu} \phi^{\mu} = 1,
\end{equation}
such that the inverse metric tensor ${g^{\mu \nu}}$ can be written in the following rather straightforward way:

\begin{equation}
g^{\mu \nu} = 2 \left( m^{\left( \mu \right.} \bar{m}^{\left. \nu \right)} - l^{\left( \mu \right.} \bar{n}^{\left. \nu \right)} \right).
\end{equation}
We now wish to find a combination of null rotations, parametrized by a scalar variable $A$, as well as a boost rotation parametrized by scalar variables ${F_{A}}$ and ${F_{B}}$ (more precisely, the null rotations correspond to symmetries of Type I and Type II, with the boost rotation corresponding to a symmetry of Type III, with respect to the Petrov-Pirani-Penrose classification of algebraic symmetries of the Weyl curvature tensor in Lorentzian geometry\cite{petrov}\cite{pirani}\cite{penrose}), which will transform an orthonormal tetrad of the given form in the ${3 + 1}$ basis:

\begin{equation}
\tilde{l}^{\mu} = \frac{1}{\sqrt{2}} \left( u^{\mu} + r^{\mu} \right), \qquad \tilde{n}^{\mu} = \frac{1}{\sqrt{2}} \left( u^{\mu} - r^{\mu} \right), \qquad \tilde{m}^{\mu} = \frac{1}{\sqrt{2}} \left( \theta^{\mu} + i \phi^{\mu} \right),
\end{equation}
into a tetrad of the Kinnersley form (in Boyer-Lindquist coordinates) in the ${3 + 1}$ basis:

\begin{equation}
l^{\mu} = \left[ \alpha_{\left( 0 \right)} l_{Kin}^{0}, l_{Kin}^I + \beta_{\left( 0 \right)}^{i} l_{Kin}^{0} \right], \qquad n^{\mu} = \left[ \alpha_{\left( 0 \right)} n_{Kin}^{0}, n_{Kin}^{i} + \beta_{\left( 0 \right)}^{i} n_{Kin}^{0} \right],
\end{equation}
and:

\begin{equation}
m^{\mu} = \left[ \alpha_{\left(0 \right)}, m_{Kin}^{0}, m_{Kin}^{i} + \beta_{\left( 0 \right)}^{i} m_{Kin}^{0} \right],
\end{equation}
so as to ensure compatibility with the form of the tetrad assumed within perturbation analysis of a general Kerr black hole. It can easily be shown that the following coordinate transformation:

\begin{equation}
l^{\mu} = \frac{F_A}{2} \left[ \left( \sqrt{A^2 + 1} + 1 \right) \tilde{l}^{\mu} + \left( \sqrt{A^2 + 1} - 1 \right) \tilde{n}^{\mu} - i A \left( \tilde{m}^{\mu} - \tilde{\bar{m}}^{\mu} \right) \right],
\end{equation}

\begin{equation}
n^{\mu} = \frac{F_{A}^{-1}}{2} \left[ \left( \sqrt{A^2 + 1} - 1 \right) \tilde{l}^{\mu} + \left( \sqrt{A^2 + 1} + 1 \right) \tilde{n}^{\mu} - i A \left( \tilde{m}^{\mu} - \tilde{\bar{m}}^{\mu} \right) \right],
\end{equation}
and:

\begin{equation}
m^{\mu} = \frac{F_B}{2} \left[ \left( \sqrt{A^2 + 1} + 1 \right) \tilde{m}^{\mu} - \left( \sqrt{A^2 + 1} - 1 \right) \tilde{\bar{m}}^{\mu} + i A \left( \tilde{l}^{\mu} + \tilde{n}^{\mu} \right) \right],
\end{equation}
suffices for this purpose, with the scalar parameters $A$, ${F_A}$ and ${F_B}$ defined in terms of the Boyer-Lindquist coordinate parameters by:

\begin{equation}
A = a \sin \left( \theta \right) \sqrt{\frac{\Delta}{\Omega}}, \qquad F_A = \sqrt{\frac{2 \Sigma}{\Delta}}, \qquad F_B = \frac{\sqrt{\Sigma}}{\left( r + i a \cos \left( \theta \right) \right)},
\end{equation}
respectively.

With the exact complex null tetrad thus computed, we can now evaluate the Weyl scalar ${\Psi_4}$ by projecting the three-dimensional spatial Riemann curvature tensor ${R_{i j k l}^{\left( 3 \right)}}$ onto this tetrad basis, to obtain:

\begin{multline}
\Psi_4 = R_{i j k l}^{\left( 3 \right)} n^i \bar{m}^j n^k \bar{m}^l + 2 R_{0 j k l}^{\left( 3 \right)} \left( n^0 \bar{m}^j n^k \bar{m}^l - \bar{m}^0 n^j n^k \bar{m}^l \right)\\
+ R_{0 j 0 l}^{\left( 3 \right)} \left( n^0 \bar{m}^j n^0 \bar{m}^l + \bar{m}^0 n^j \bar{m}^0 n^l - 2 n^0 \bar{m}^j \bar{m}^0 n^l \right).
\end{multline}
Due to the properties of the computed tetrad, the ${\Psi_4}$ Weyl scalar can now be related directly to the strain $h$ associated with gravitational radiation in the outgoing radiative zone, since:

\begin{equation}
h = h_{+} - i h_{\times} = - \int_{- \infty}^{t} \int_{- \infty}^{t^{\prime}} \Psi_4 dt^{\prime \prime} dt^{\prime},
\end{equation}
where ${h_{+}}$ and ${h_{\times}}$ denote the \textit{plus polarization} and \textit{cross polarization} of the gravitational radiation, respectively. However, although the outlined procedure will successfully compute the Weyl scalar ${\Psi_4}$ as a complex function over the spatial hypergraph, for the purpose of gravitational wave extraction it is necessary to interpolate the scalar field ${\Psi_4 \left( t, r, \theta, \phi \right)}$ onto a coordinate sphere of a given radius, and to calculate the coefficients ${C^{\ell m}}$ of the projection:

\begin{equation}
C^{\ell m} \left( t, r \right) = \int {}_s Y_{\ell m}^{*} \left( \theta, \phi \right) \Psi_4 \left( t, r, \theta, \phi \right) r^2 d \Omega,
\end{equation}
onto the spin-weighted spherical harmonic functions ${{}_s Y_{\ell m} \left( \theta, \phi \right)}$, conventionally defined in terms of the ordinary spherical harmonic functions ${Y_{\ell m} \left(\theta, \phi \right)}$ by the relation\cite{dray}:

\begin{equation}
{}_s Y_{\ell m} \left( \theta, \phi \right) = \begin{cases}
\sqrt{\frac{\left( l - s \right)!}{\left( l + s \right)!}} \tilde{\partial}^s Y_{\ell m} \left( \theta, \phi \right), \qquad &\text{ if  } 0 \leq s \leq \ell,\\
\sqrt{\frac{\left( l + s \right)!}{\left( l - s \right)!}} \left( -1 \right)^s \bar{\tilde{\partial}}^{-s} Y_{\ell m} \left( \theta, \phi \right), \qquad &\text{ if } -l \leq s \leq 0,\\
0, \qquad &\text{ if } l < \left\lvert s \right\rvert,
\end{cases}
\end{equation}
with the covariant derivative operators ${\tilde{\partial}}$ and ${\bar{\tilde{\partial}}}$ on the sphere defined by:

\begin{equation}
\tilde{\partial} \eta \left( \theta, \phi \right) = - \left( \sin \left( \theta \right) \right)^s \left\lbrace \frac{\partial}{\partial \theta} + \frac{i}{\sin \left( \theta \right)} \frac{\partial}{\partial \phi} \right\rbrace \left[ \left( \sin \left( \theta \right) \right)^{-s} \eta \left( \theta, \phi \right) \right],
\end{equation}
and:

\begin{equation}
\bar{\tilde{\partial}} \eta \left( \theta, \phi \right) = - \left( \sin \left( \theta \right) \right)^{-s} \left\lbrace \frac{\partial}{\partial \theta} - \frac{i}{\sin \left( \theta \right)} \frac{\partial}{\partial \phi} \right\rbrace \left[ \left( \sin \left( \theta \right) \right)^s \eta \left( \theta, \phi \right) \right],
\end{equation}
respectively, for arbitrary functions ${\eta \left( \theta, \phi \right)}$.

\begin{figure}[ht]
\centering
\includegraphics[width=0.325\textwidth]{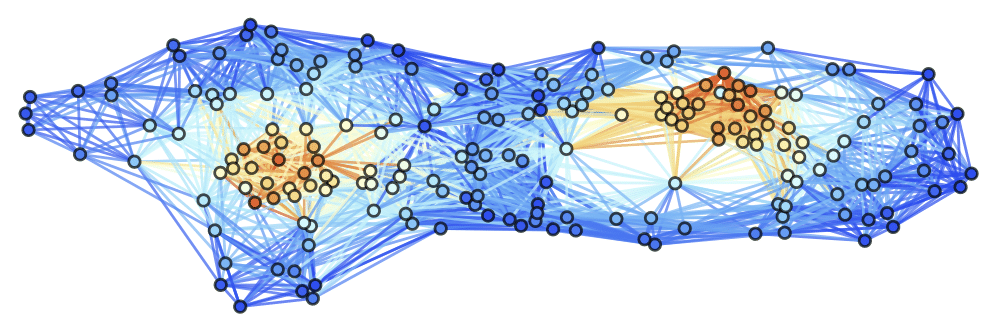}
\includegraphics[width=0.325\textwidth]{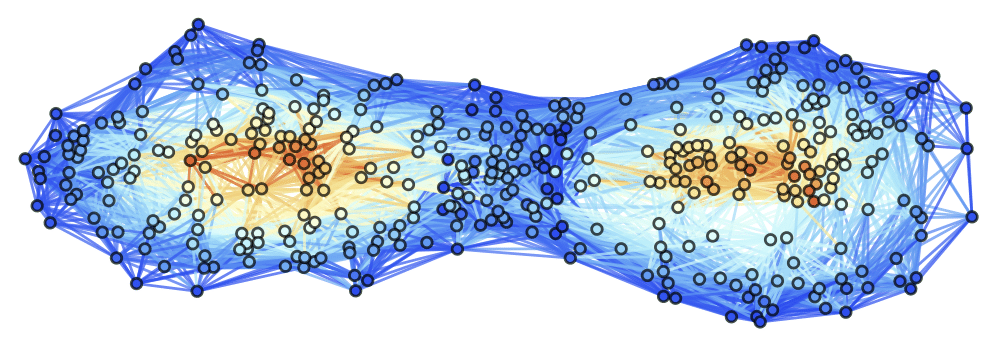}
\includegraphics[width=0.325\textwidth]{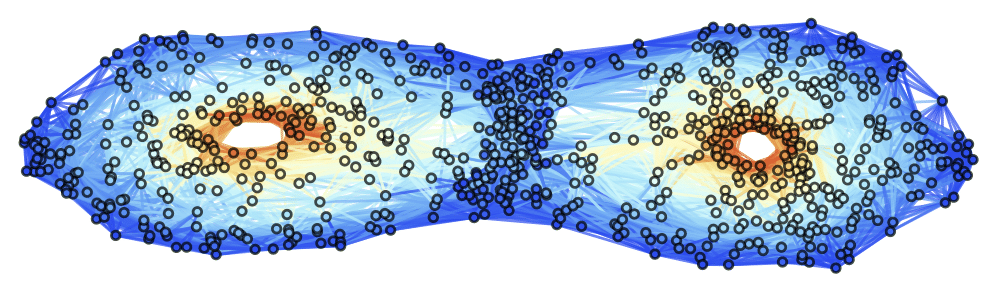}
\caption{Spatial hypergraphs corresponding to the initial hypersurface configuration of the head-on collision of Schwarzschild black holes (convergence) test at time ${t = 0 M}$, with resolutions of 200, 400 and 800 vertices, respectively. The hypergraphs have been adapted and colored using the local curvature in the Schwarzschild conformal factor ${\psi}$.}
\label{fig:Figure30}
\end{figure}

\begin{figure}[ht]
\centering
\includegraphics[width=0.325\textwidth]{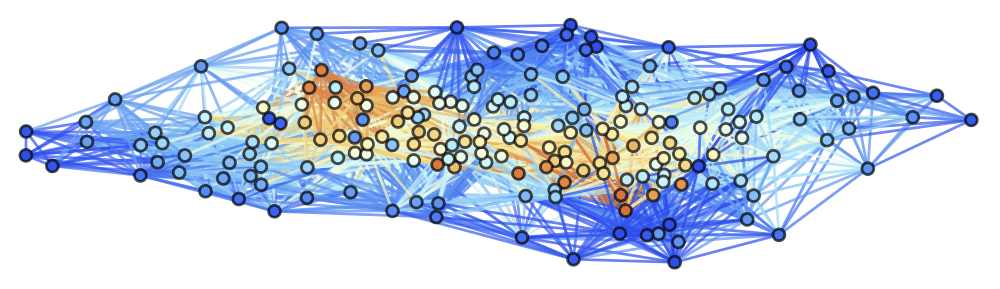}
\includegraphics[width=0.325\textwidth]{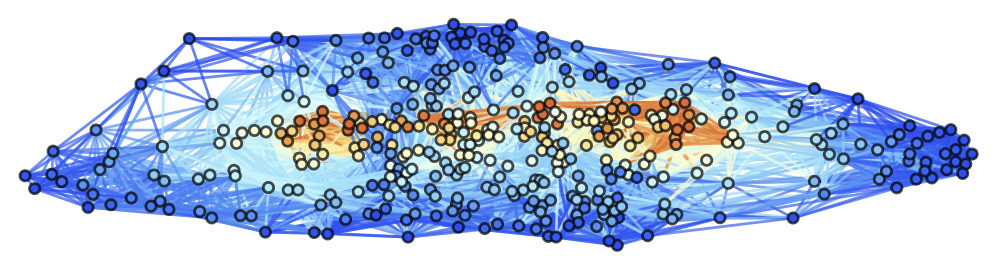}
\includegraphics[width=0.325\textwidth]{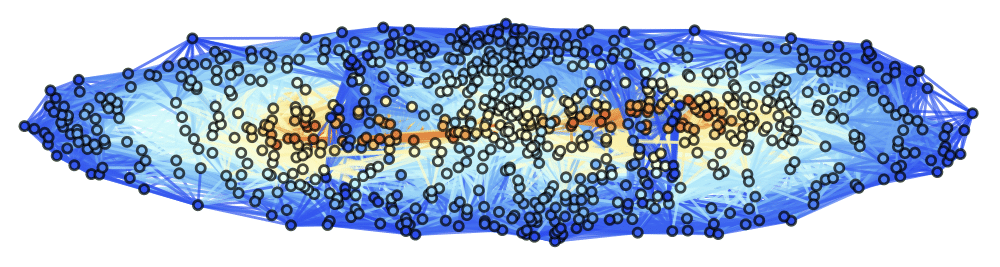}
\caption{Spatial hypergraphs corresponding to the intermediate hypersurface configuration of the head-on collision of Schwarzschild black holes (convergence) test at time ${t = 6 M}$, with resolutions of 200, 400 and 800 vertices, respectively. The hypergraphs have been adapted and colored using the local curvature in the Schwarzschild conformal factor ${\psi}$.}
\label{fig:Figure31}
\end{figure}

\begin{figure}[ht]
\centering
\includegraphics[width=0.325\textwidth]{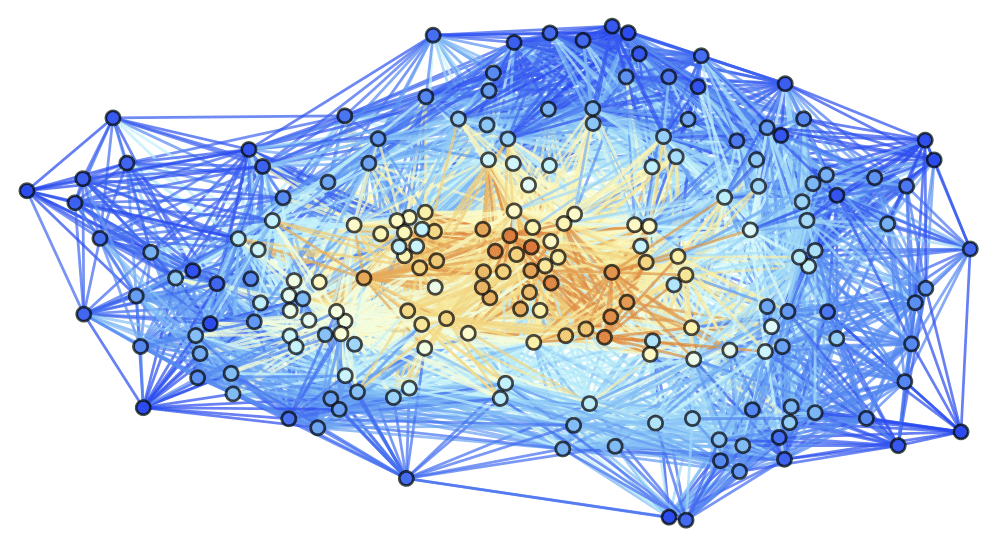}
\includegraphics[width=0.325\textwidth]{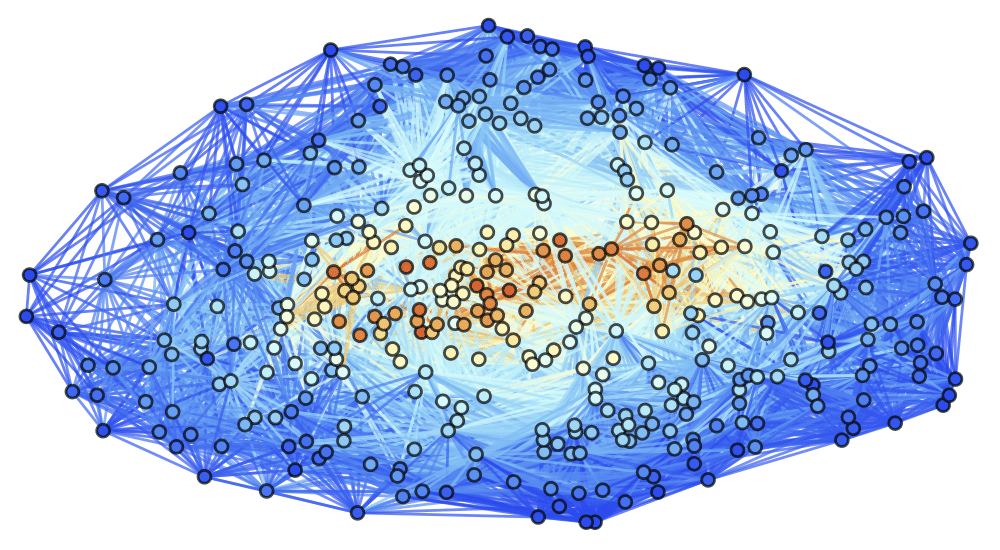}
\includegraphics[width=0.325\textwidth]{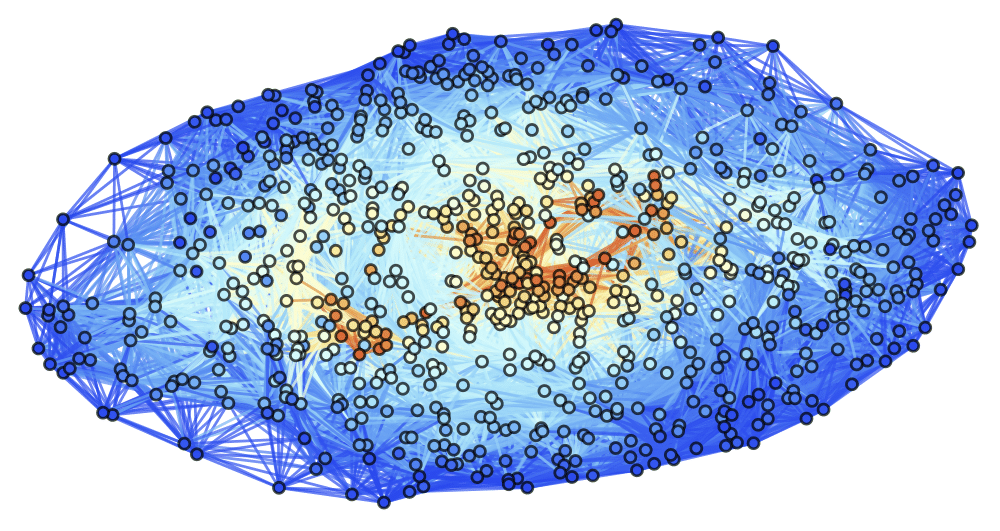}
\caption{Spatial hypergraphs corresponding to the final hypersurface configuration of the head-on collision of Schwarzschild black holes (convergence) test at time ${t = 12 M}$, with resolutions of 200, 400 and 800 vertices, respectively. The hypergraphs have been adapted and colored using the local curvature in the Schwarzschild conformal factor ${\psi}$.}
\label{fig:Figure32}
\end{figure}

\begin{figure}[ht]
\centering
\includegraphics[width=0.325\textwidth]{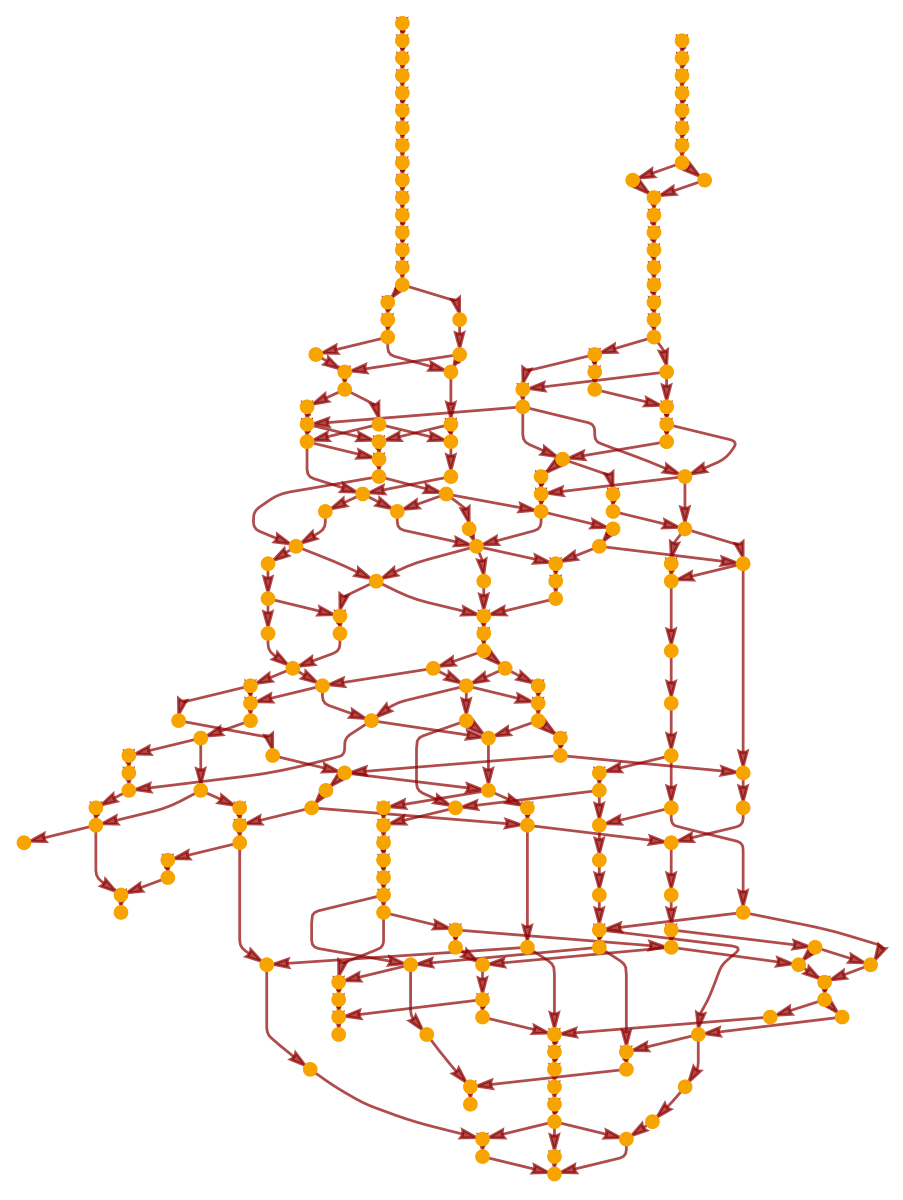}
\includegraphics[width=0.325\textwidth]{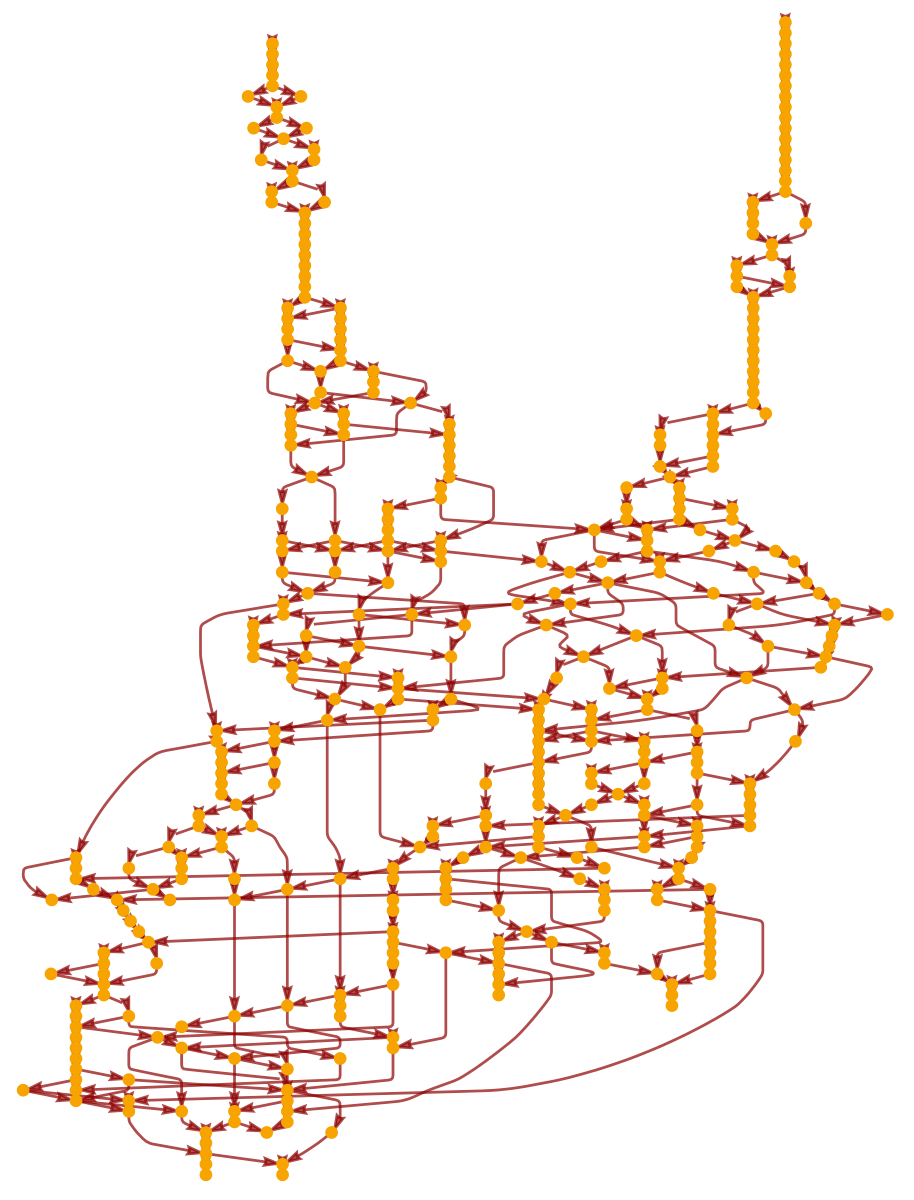}
\includegraphics[width=0.325\textwidth]{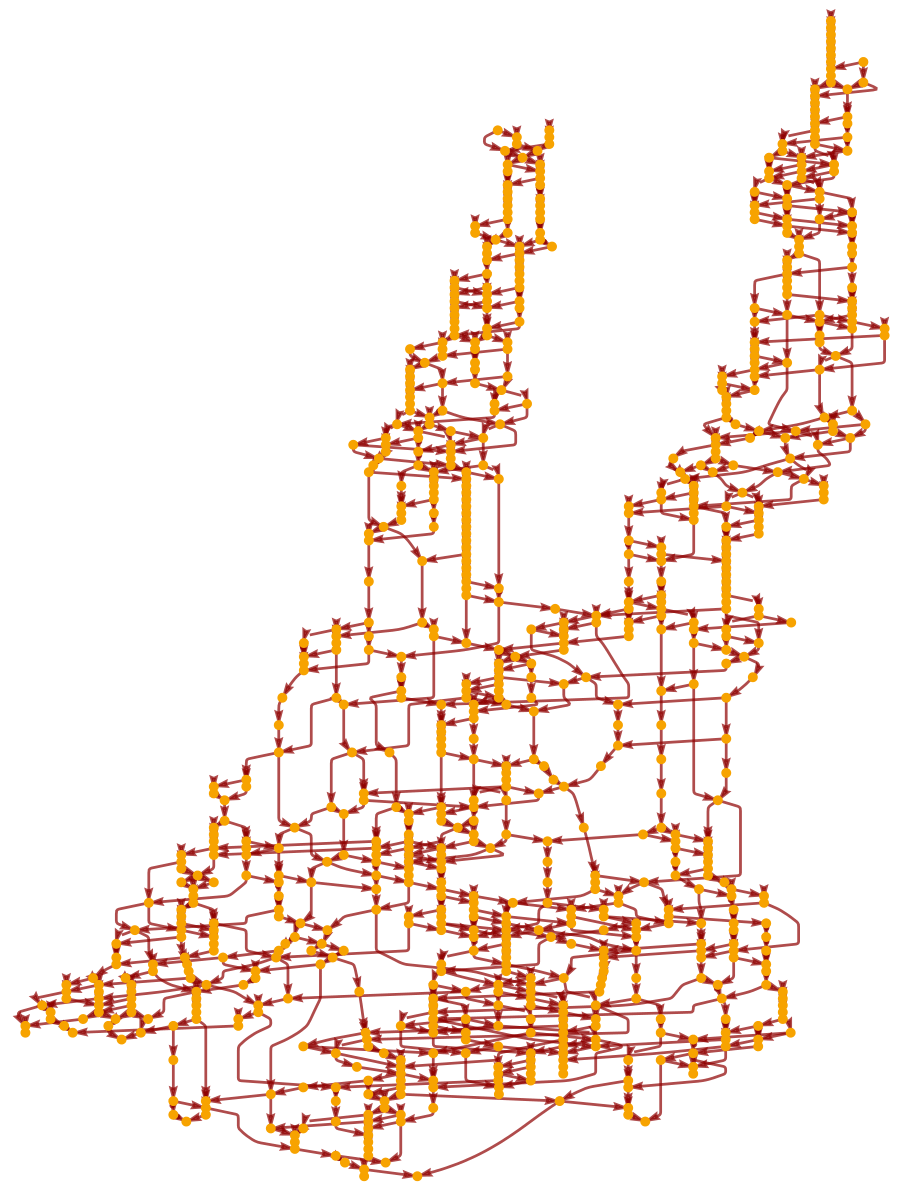}
\caption{Causal graphs corresponding to the discrete characteristic structure of the head-on collision of Schwarzschild black holes (convergence) test at time ${t = 12 M}$, with resolutions of 200, 400 and 800 hypergraph vertices, respectively.}
\label{fig:Figure33}
\end{figure}

\begin{figure}[ht]
\centering
\includegraphics[width=0.325\textwidth]{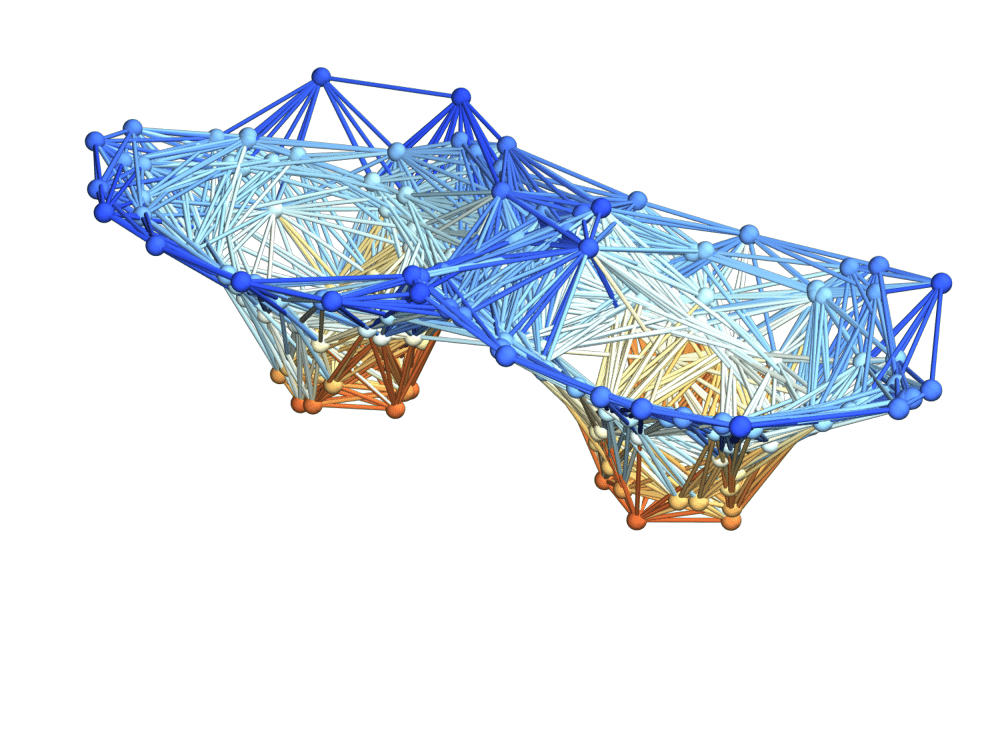}
\includegraphics[width=0.325\textwidth]{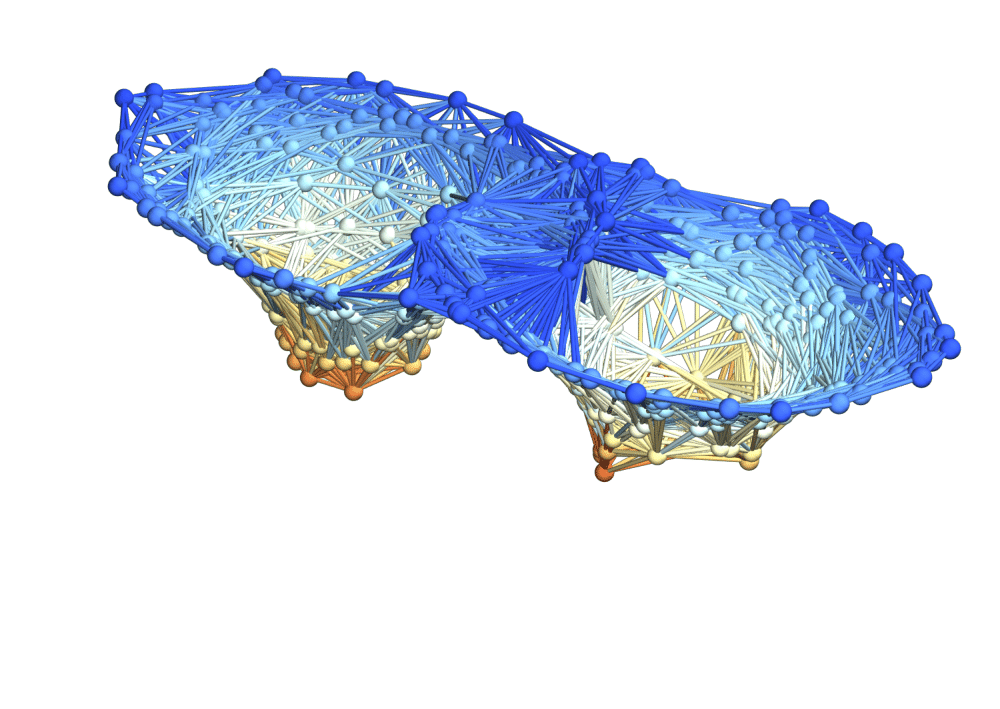}
\includegraphics[width=0.325\textwidth]{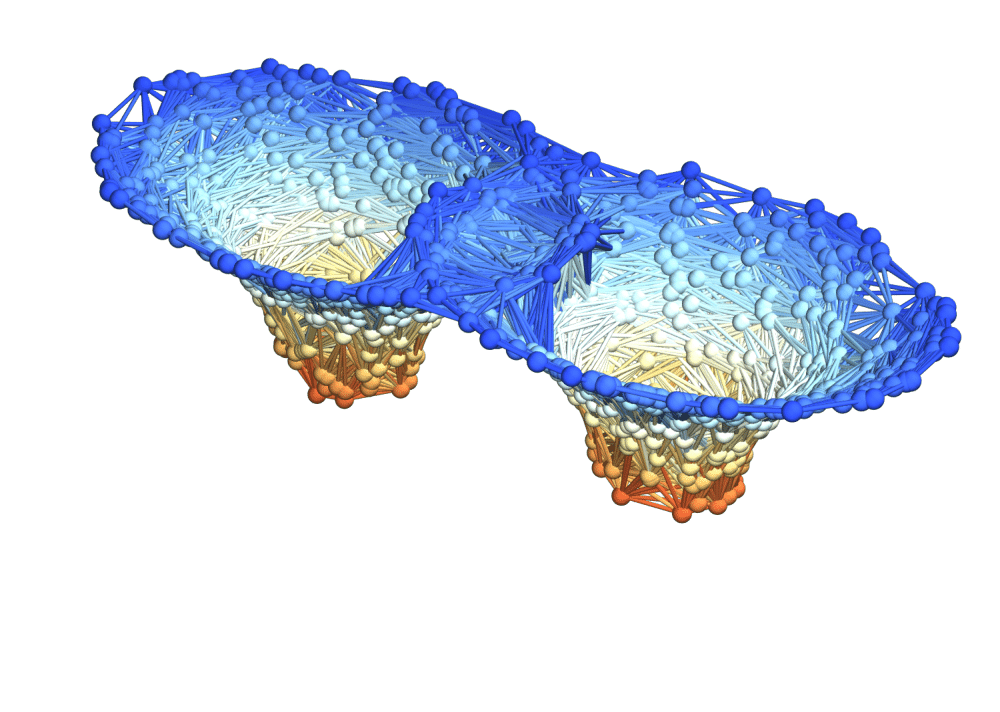}
\caption{Spatial hypergraphs corresponding to projections along the $z$-axis of the initial hypersurface configuration of the head-on collision of Schwarzschild black holes (convergence) test at time ${t = 0 M}$, with resolutions of 200, 400 and 800 vertices, respectively. The vertices have been assigned spatial coordinates according to the profile of the Schwarzschild conformal factor ${\psi}$ through a spatial slice perpendicular to the $z$-axis, and the hypergraphs have been adapted and colored using the local curvature in ${\psi}$.}
\label{fig:Figure34}
\end{figure}

\begin{figure}[ht]
\centering
\includegraphics[width=0.325\textwidth]{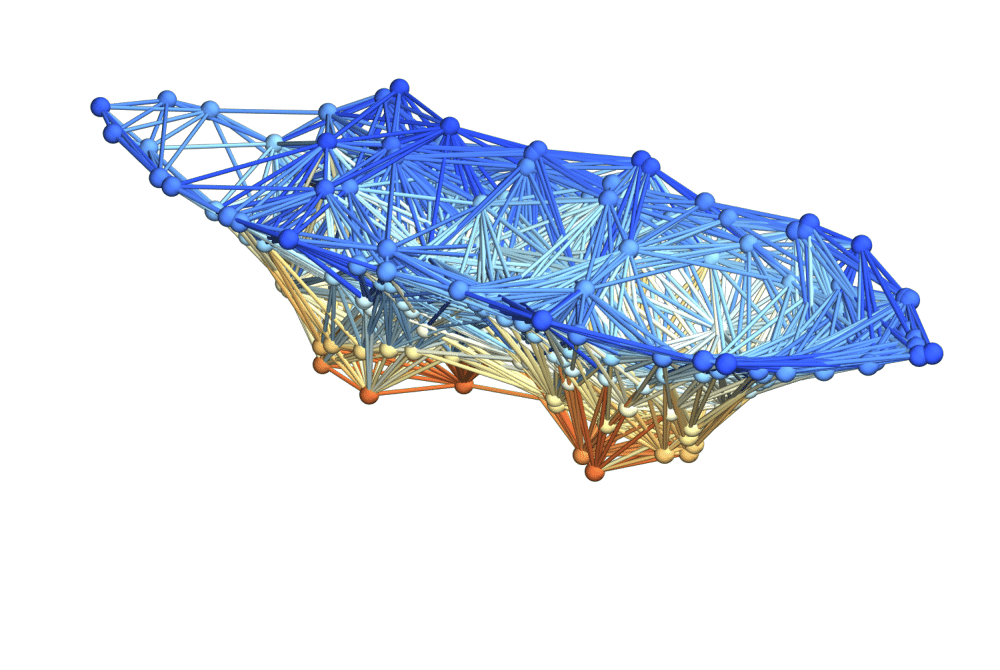}
\includegraphics[width=0.325\textwidth]{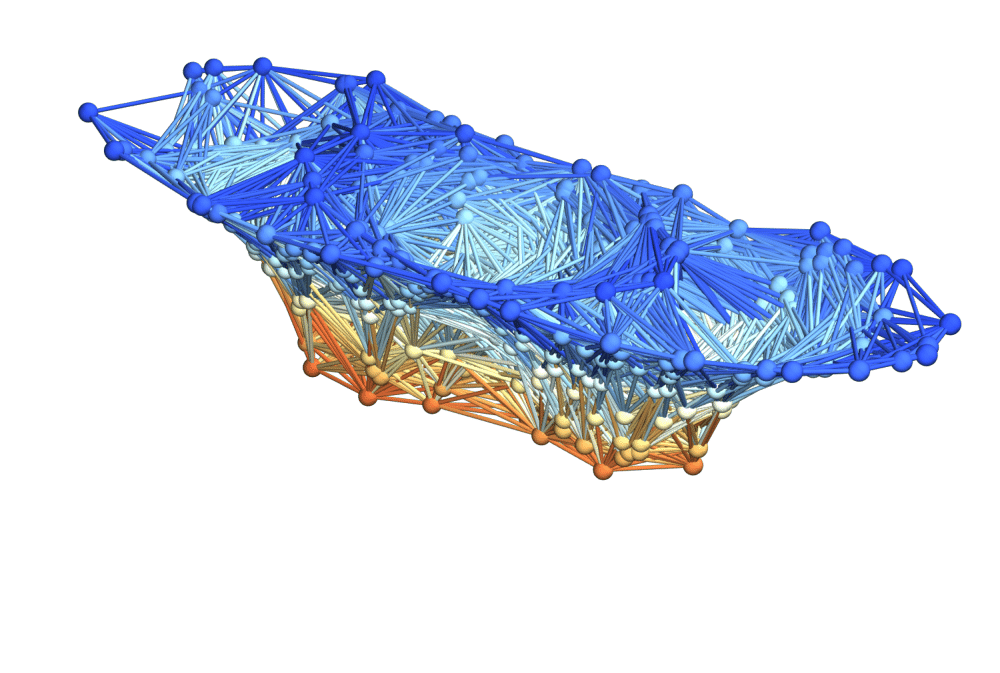}
\includegraphics[width=0.325\textwidth]{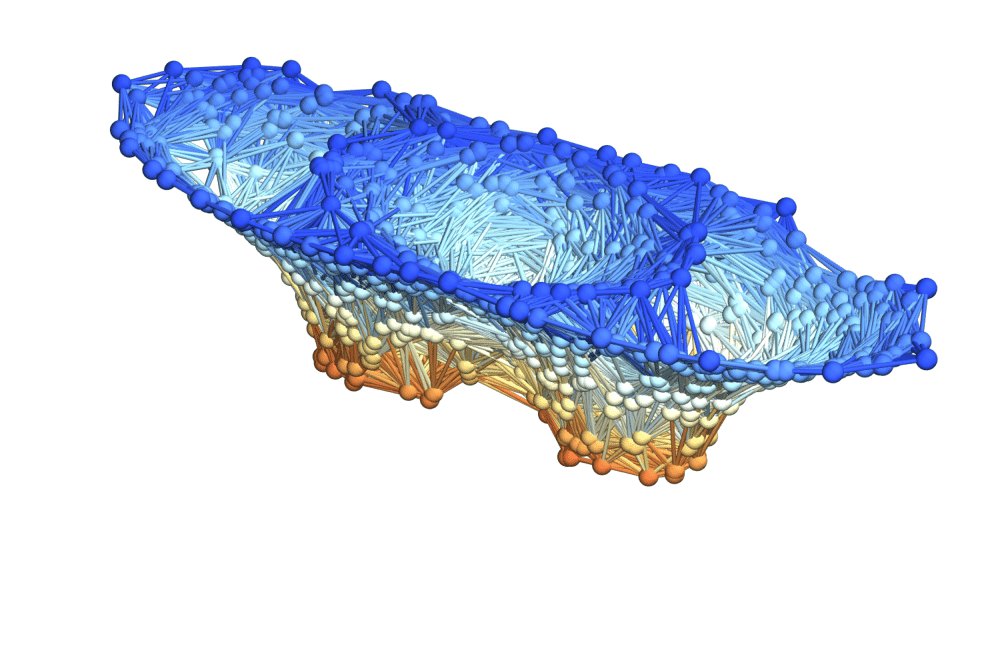}
\caption{Spatial hypergraphs corresponding to projections along the $z$-axis of the intermediate hypersurface configuration of the head-on collision of Schwarzschild black holes (convergence) at time ${t = 6 M}$, with resolutions of 200, 400 and 800 vertices, respectively. The vertices have been assigned spatial coordinates according to the profile of the Schwarzschild conformal factor ${\psi}$ through a spatial slice perpendicular to the $z$-axis, and the hypergraphs have been adapted and colored using the local curvature in ${\psi}$.}
\label{fig:Figure35}
\end{figure}

\begin{figure}[ht]
\centering
\includegraphics[width=0.325\textwidth]{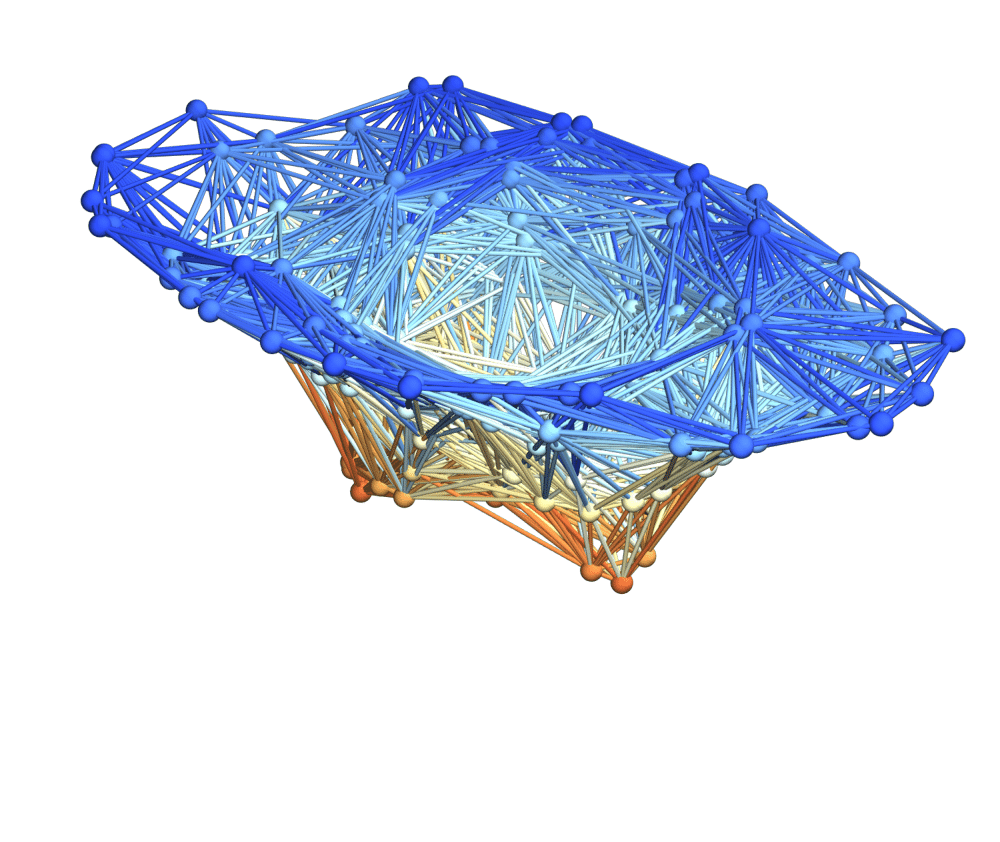}
\includegraphics[width=0.325\textwidth]{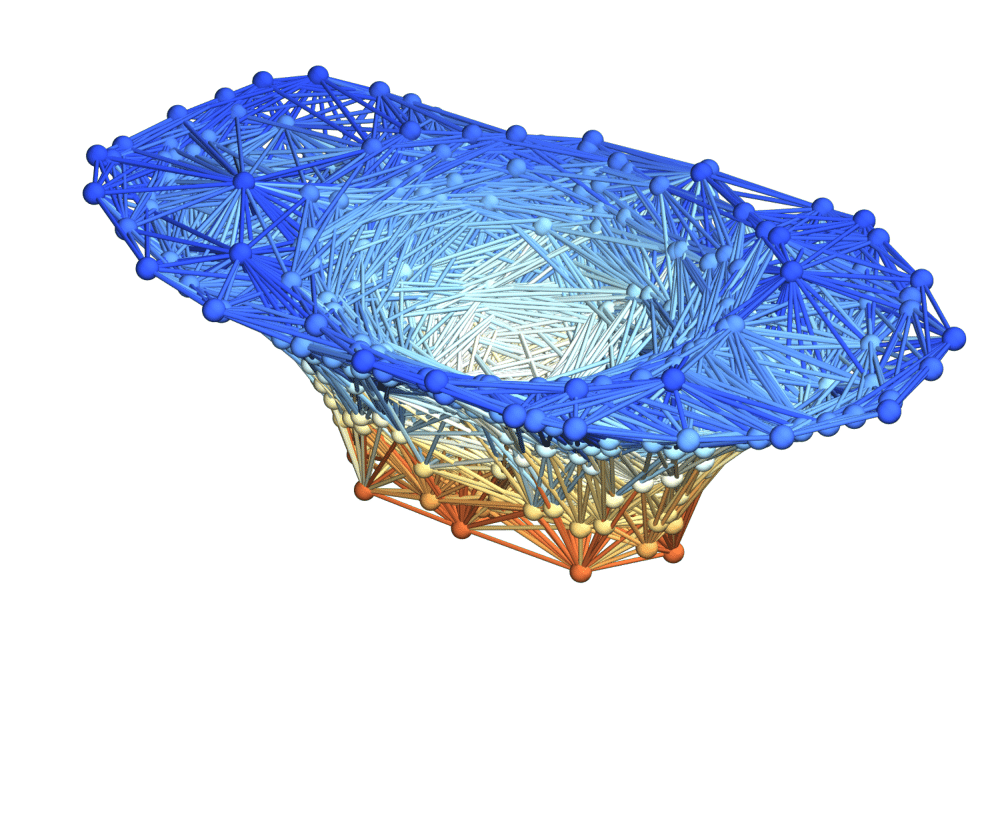}
\includegraphics[width=0.325\textwidth]{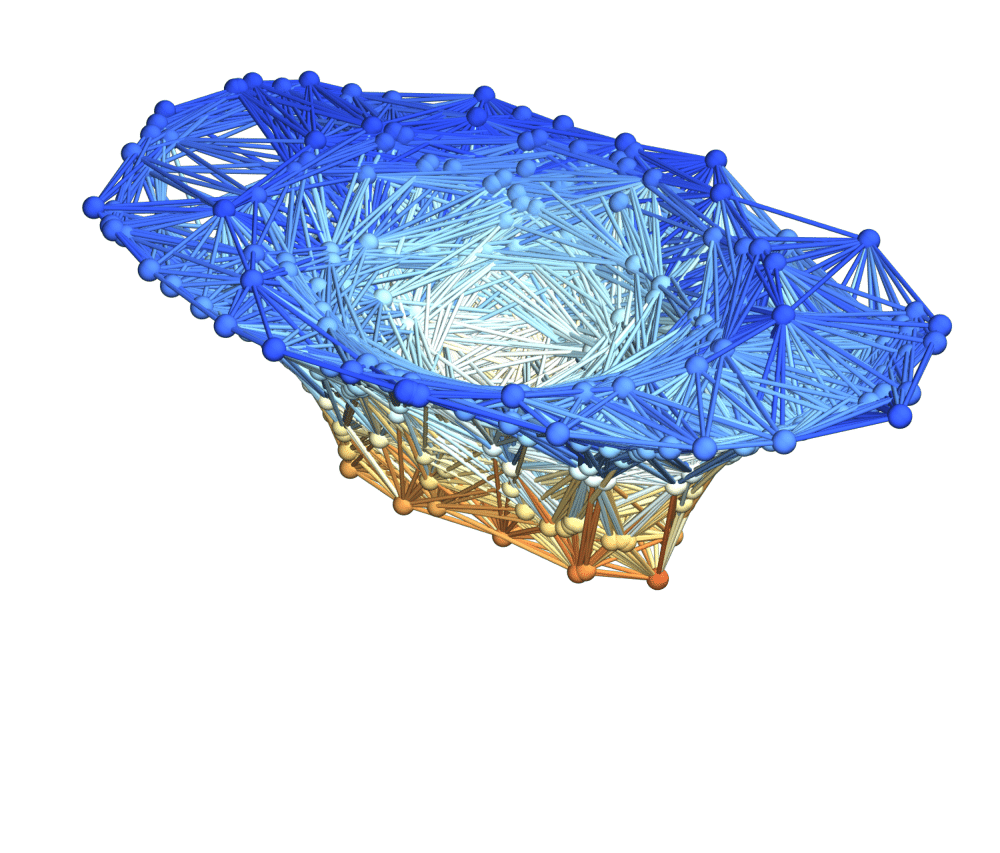}
\caption{Spatial hypergraphs corresponding to projections along the $z$-axis of the final hypersurface configuration of the head-on collision of Schwarzschild black holes (convergence) test at time ${t = 12 M}$, with resolutions of 200, 400 and 800 vertices, respectively. The vertices have been assigned spatial coordinates according to the profile of the Schwarzschild conformal factor ${\psi}$ through a spatial slice perpendicular to the $z$-axis, and the hypergraphs have been adapted and colored using the local curvature in ${\psi}$.}
\label{fig:Figure36}
\end{figure}

Having set up the initial data in Brill-Lindquist binary black hole form, for two static Schwarzschild black holes, each of mass ${0.5 M}$, with an initial separation of ${1 M}$, and with the center of mass of the binary placed at the center of the computational domain, we impose periodic boundary conditions for the sake of algorithmic simplicity (thus limiting the stable duration of the simulation due to the presence of non-negligible boundary effects). We use the standard expression for the Schwarzschild conformal factor ${\psi}$:

\begin{equation}
\psi = \left( 1 + \frac{M}{2r} \right)^{-2},
\end{equation}
as the scalar field in the refinement algorithm ${\phi = \psi}$ as usual, selecting a domain of size ${\left( 20 M \right)^3}$. We evolve the solution in the first instance until a final time of ${t = 12 M}$, with an intermediate check at time ${t = 6 M}$, during which time the two Schwarzschild black holes merge and the gravitational ringdown phase of the merger begins; the initial, intermediate and final hypersurface configurations are shown in Figures \ref{fig:Figure30}, \ref{fig:Figure31} and \ref{fig:Figure32}, respectively, with resolutions of 200, 400 and 800 vertices, and with the hypergraphs adapted and colored using the Schwarzschild conformal factor ${\psi}$. Figure \ref{fig:Figure33} shows the discrete characteristic structure of the solutions after time ${t = 120 M}$ (using directed acyclic causal graphs to show discrete characteristic lines), and Figures \ref{fig:Figure34}, \ref{fig:Figure35} and \ref{fig:Figure36} show projections along the $z$-axis of the initial, intermediate and final hypersurface configurations, respectively, with vertices assigned spatial coordinates according to the profile of the Schwarzschild conformal factor ${\psi}$.

\begin{figure}[ht]
\centering
\includegraphics[width=0.325\textwidth]{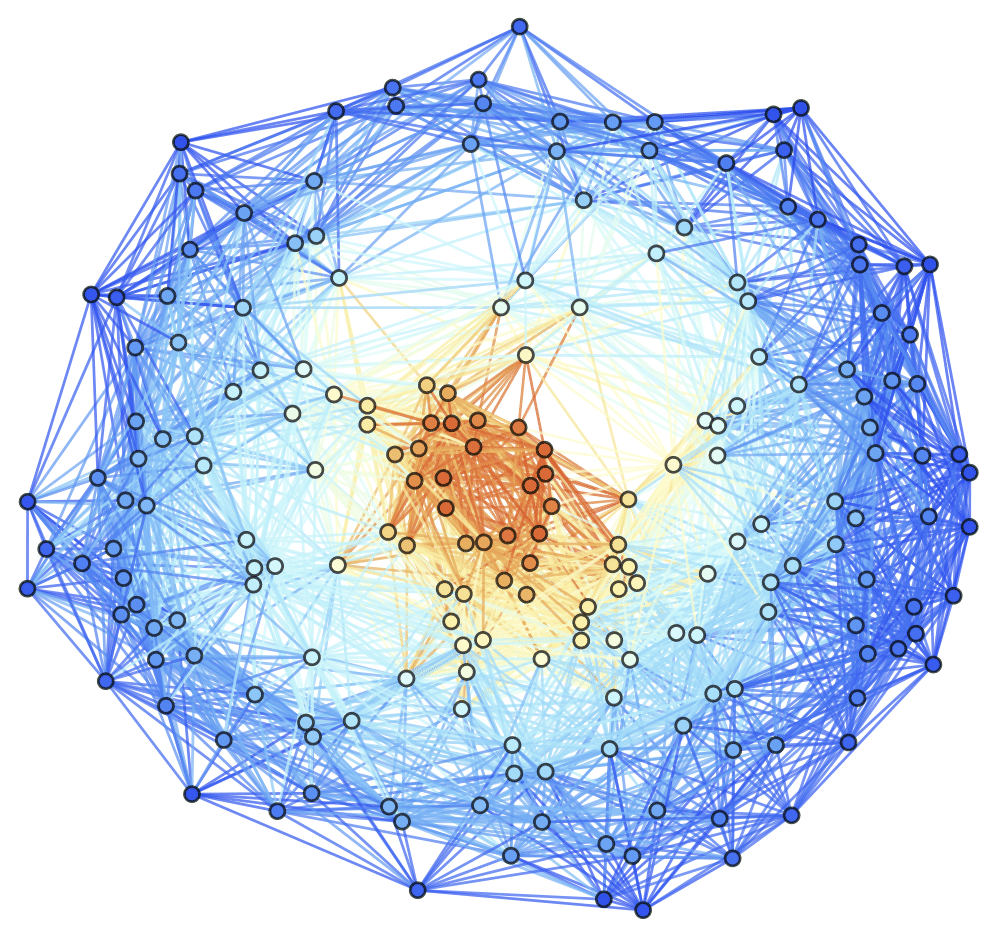}
\includegraphics[width=0.325\textwidth]{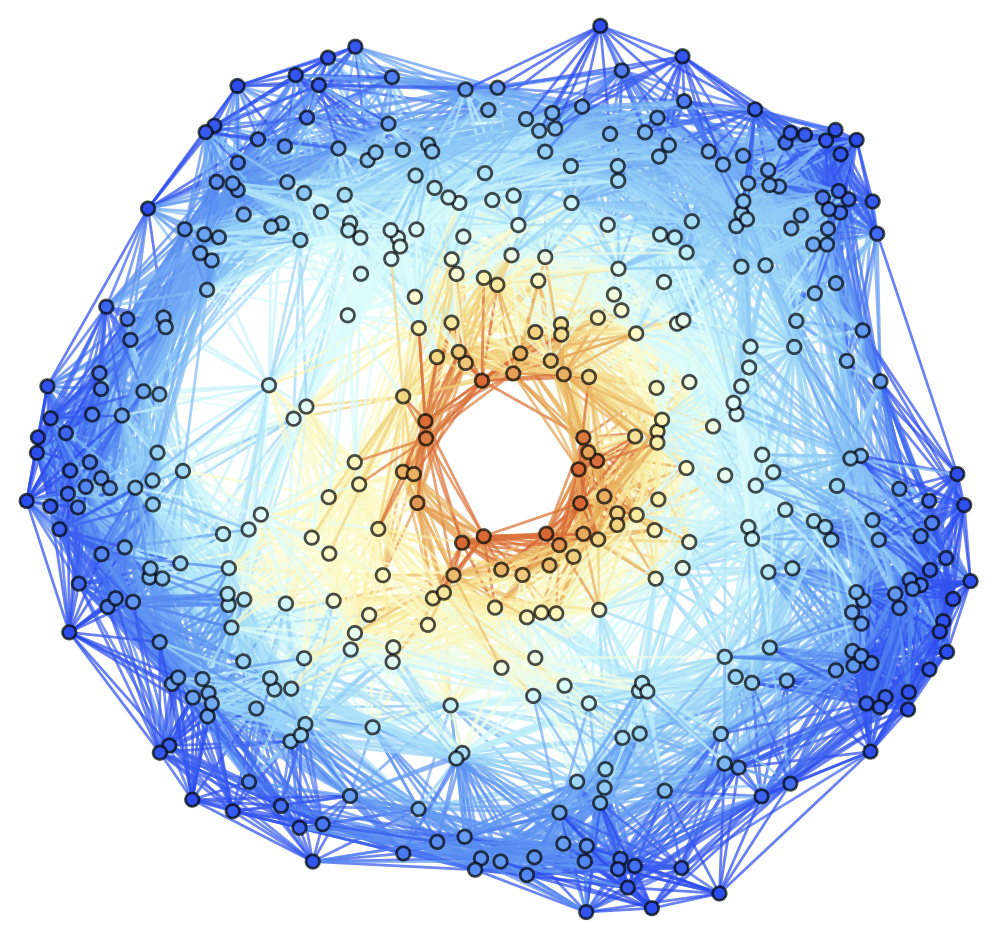}
\includegraphics[width=0.325\textwidth]{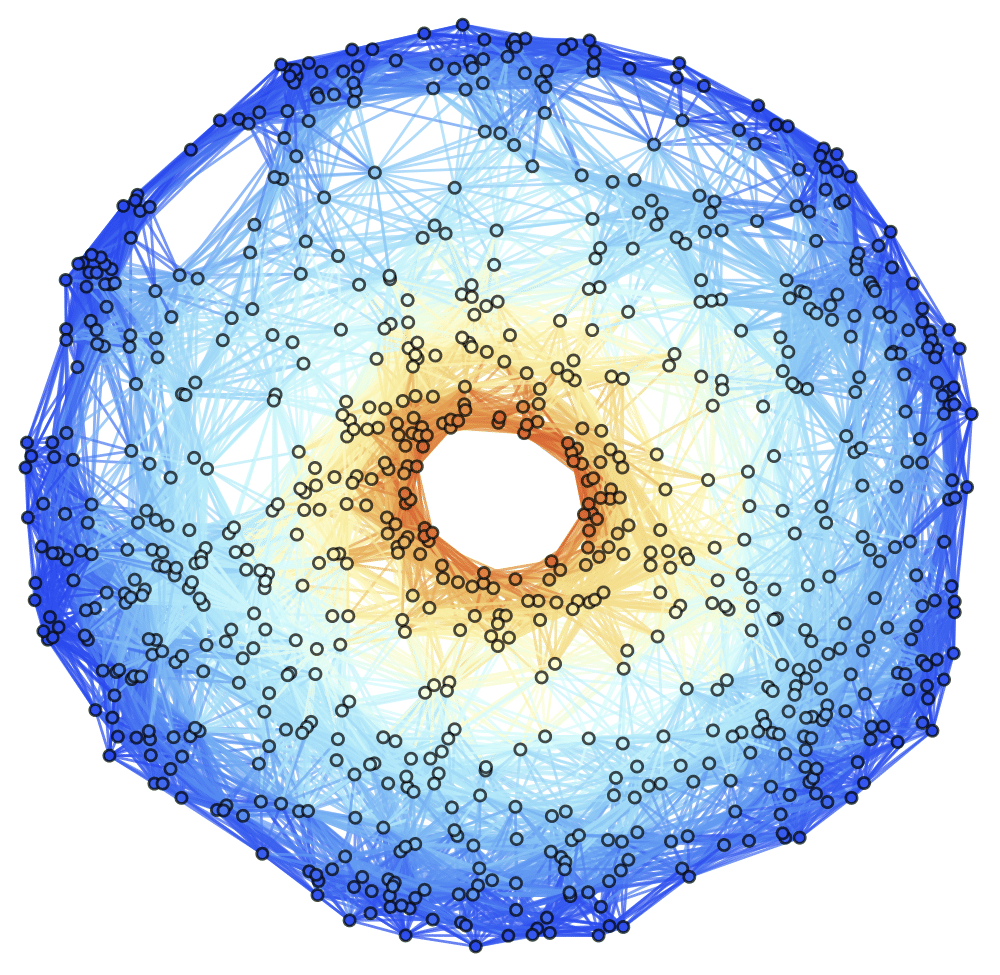}
\caption{Spatial hypergraphs corresponding to the post-ringdown hypersurface configuration of the head-on collision of Schwarzschild black holes (convergence) test at time ${t = 24 M}$, with resolutions of 200, 400 and 800 vertices, respectively. The hypergraphs have been adapted and colored using the local curvature in the Schwarzschild conformal factor ${\psi}$.}
\label{fig:Figure37}
\end{figure}

\begin{figure}[ht]
\centering
\includegraphics[width=0.325\textwidth]{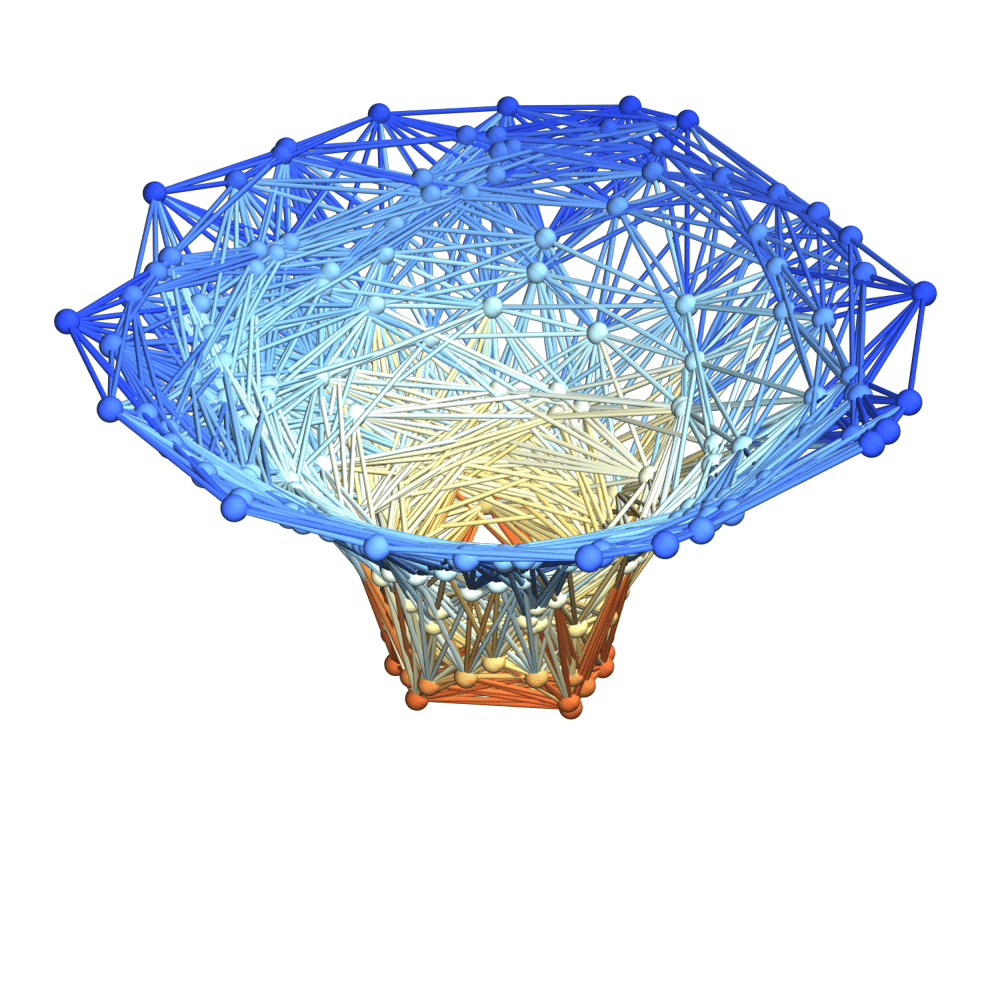}
\includegraphics[width=0.325\textwidth]{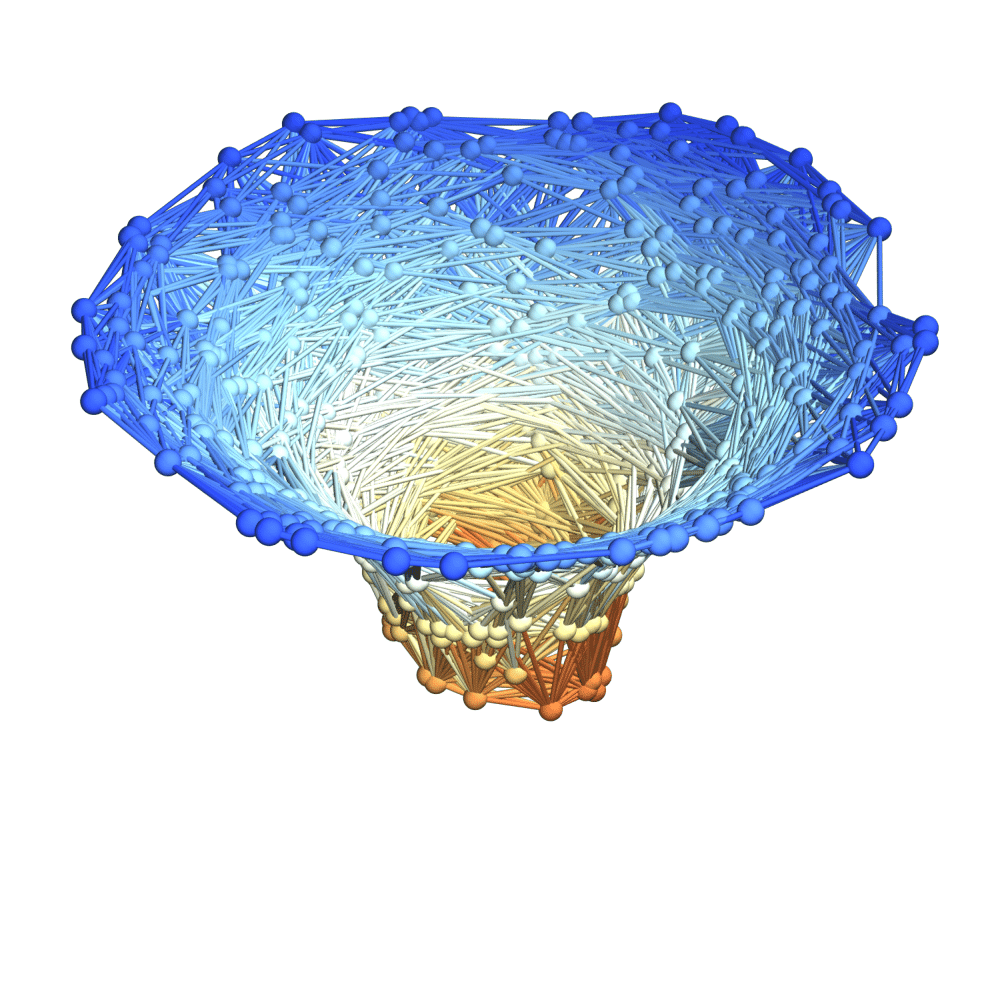}
\includegraphics[width=0.325\textwidth]{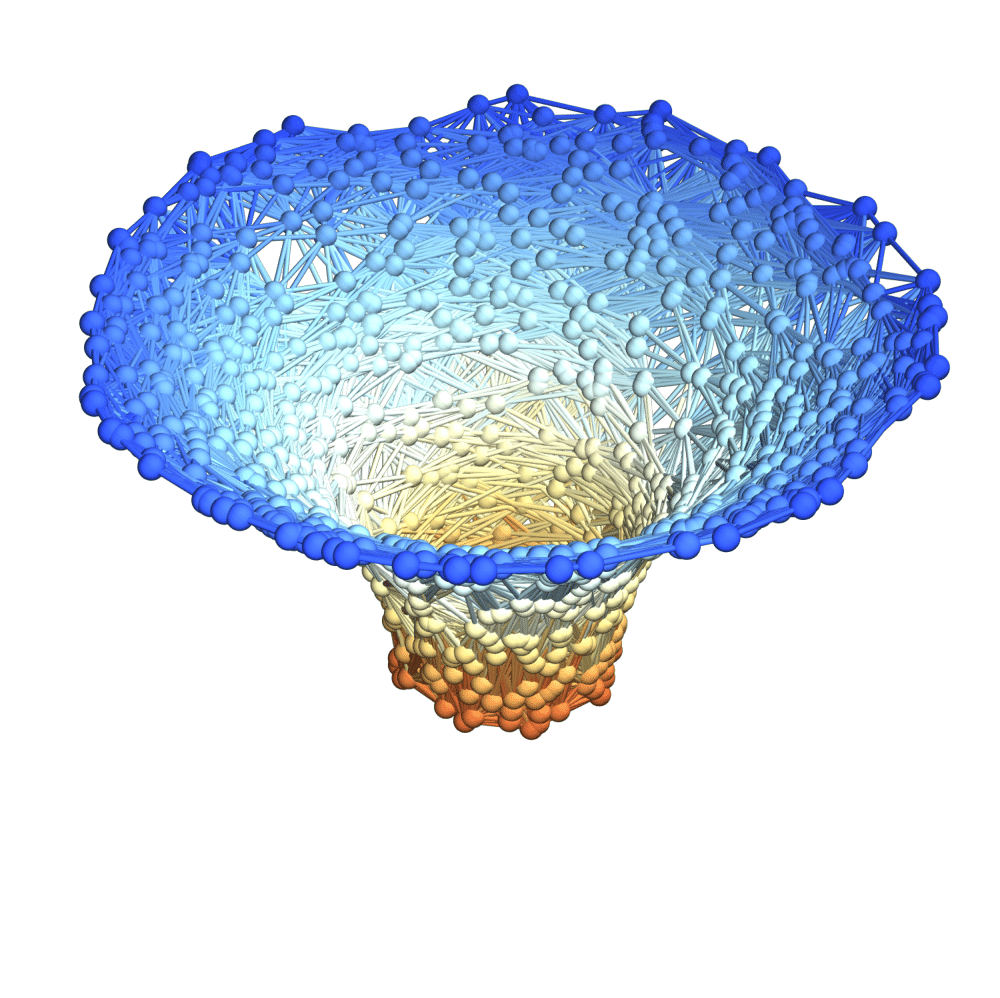}
\caption{Spatial hypergraphs corresponding to projections along the $z$-axis of the post-ringdown configuration of the head-on collision of Schwarzschild black holes (convergence) test at time ${t = 24 M}$, with resolutions of 200, 400 and 800 vertices, respectively. The vertices have been assigned spatial coordinates according to the profile of the Schwarzschild conformal factor ${\psi}$ through a spatial slice perpendicular to the $z$-axis, and the hypergraphs have been adapted and colored using the local curvature in ${\psi}$.}
\label{fig:Figure38}
\end{figure}

\begin{figure}[ht]
\centering
\includegraphics[width=0.895\textwidth]{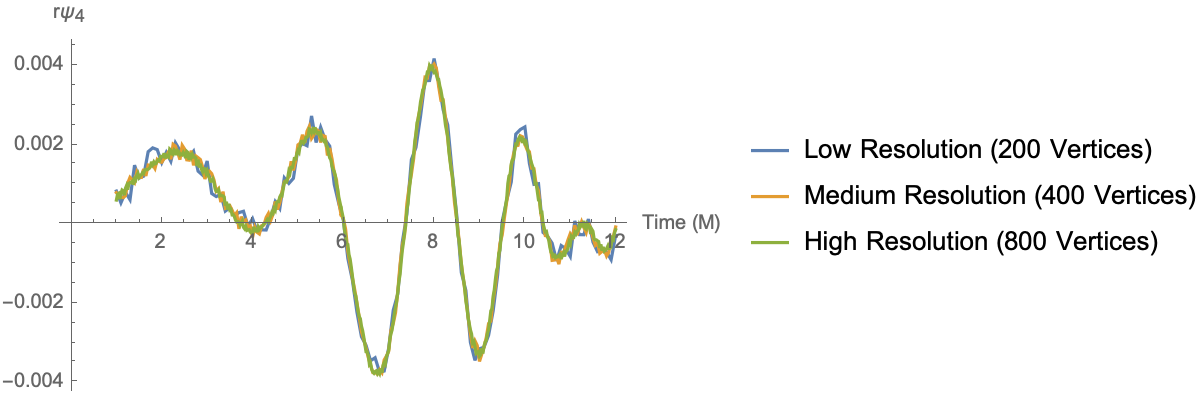}
\caption{Convergence test of the head-on collision of Schwarzschild black holes test after time ${t = 12 M}$, showing the real part of the ${\ell = 2}$, ${m = 0}$ mode of the radial Weyl scalar ${r \Psi_4}$ extrapolated on a sphere of radius ${R = 6 M}$, with resolutions of 200, 400 and 800 vertices, respectively.}
\label{fig:Figure39}
\end{figure}

The post-ringdown hypersurface configuration at time ${t = 24 M}$ is also shown in Figure \ref{fig:Figure37}, along with the projections along the $z$-axis of this configuration in Figure \ref{fig:Figure38}, both with resolutions of 200, 400 and 800 vertices. In order to demonstrate convergence and the successful extraction of gravitational wave data, we show the real part of the ${\ell = 2}$, ${m = 0}$ mode of the radial Weyl scalar ${r \Psi_4}$ at time ${t = 12 M}$, extrapolated on a sphere of radius ${R = 6 M}$ using standard fourth-order interpolation techniques across both the polar and azimuthal angular directions in Figure \ref{fig:Figure39}, with resolutions of 200, 400 and 800 vertices. We confirm that the ADM mass of the final black hole configuration (computed by integrating over a single surface surrounding the boundary of asymptotic flatness) is approximately equal to the sum of the ADM masses of the initial two black holes (computed by integrating over a pair of surfaces surrounding the two boundaries of asymptotic flatness), and that the linear and angular momenta converge to be approximately zero in the post-ringdown phase, as expected. The convergence rates for the Hamiltonian constraint after time ${t = 24 M}$, with respect to the ${L_1}$, ${L_2}$ and ${L_{\infty}}$ norms, illustrating approximately fourth-order convergence of the finite difference scheme, are shown in Table \ref{tab:Table4}.

\begin{table}
\centering
\begin{tabular}{|c|c|c|c|c|c|c|}
\hline
Vertices & ${\epsilon \left( L_1 \right)}$ & ${\epsilon \left( L_2 \right)}$ & ${\epsilon \left( L_{\infty} \right)}$ & ${\mathcal{O} \left( L_1 \right)}$ & ${\mathcal{O} \left( L_2 \right)}$ & ${\mathcal{O} \left( L_{\infty} \right)}$\\
\hline\hline
100 & ${2.74 \times 10^{-2}}$ & ${4.57 \times 10^{-2}}$ & ${6.07 \times 10^{-2}}$ & - & - & -\\
\hline
200 & ${1.30 \times 10^{-3}}$ & ${3.26 \times 10^{-3}}$ & ${5.98 \times 10^{-3}}$ & 4.40 & 3.81 & 3.34\\
\hline
400 & ${1.39 \times 10^{-4}}$ & ${3.57 \times 10^{-4}}$ & ${2.35 \times 10^{-4}}$ & 3.22 & 3.19 & 4.67\\
\hline
800 & ${1.22 \times 10^{-5}}$ & ${3.19 \times 10^{-5}}$ & ${1.86 \times 10^{-5}}$ & 3.51 & 3.48 & 3.66\\
\hline
1600 & ${8.36 \times 10^{-7}}$ & ${3.94 \times 10^{-6}}$ & ${1.73 \times 10^{-6}}$ & 3.87 & 3.02 & 3.43\\
\hline
\end{tabular}
\caption{Convergence rates for the head-on collision of Schwarzschild black holes test with respect to the ${L_1}$, ${L_2}$ and ${L_{\infty}}$ norms for the Hamiltonian constraint $H$ after time ${t = 24 M}$, showing approximately fourth-order convergence.}
\label{tab:Table4}
\end{table}

\clearpage

\subsection{Head-On Collision of Two Rapidly Rotating Kerr Black Holes (Angular Momentum Test)}

\begin{figure}[ht]
\centering
\includegraphics[width=0.325\textwidth]{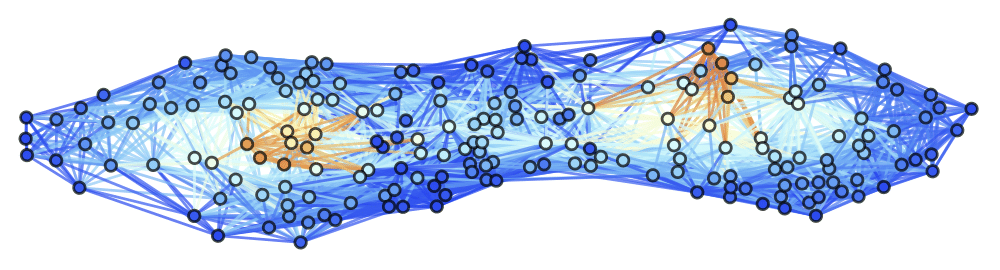}
\includegraphics[width=0.325\textwidth]{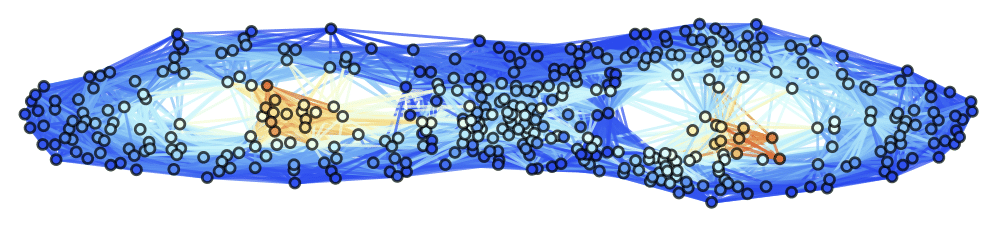}
\includegraphics[width=0.325\textwidth]{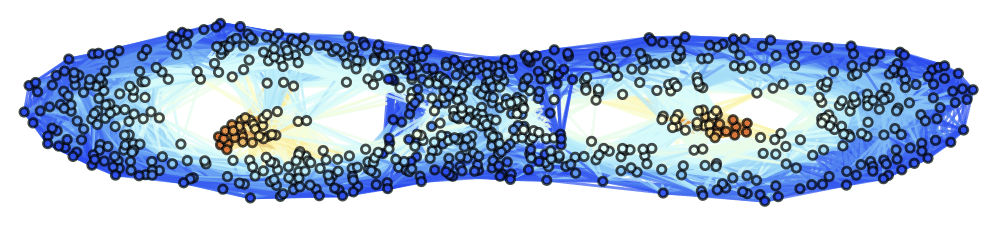}
\caption{Spatial hypergraphs corresponding to the initial hypersurface configuration of the head-on collision of rapidly rotating Kerr black holes (angular momentum) test with ${a = 0.6}$ at time ${t = 0 M}$, with resolutions of 200, 400 and 800 vertices, respectively. The hypergraphs have been adapted and colored using the local curvature in the Boyer-Lindquist conformal factor ${\psi}$.}
\label{fig:Figure40}
\end{figure}

\begin{figure}[ht]
\centering
\includegraphics[width=0.325\textwidth]{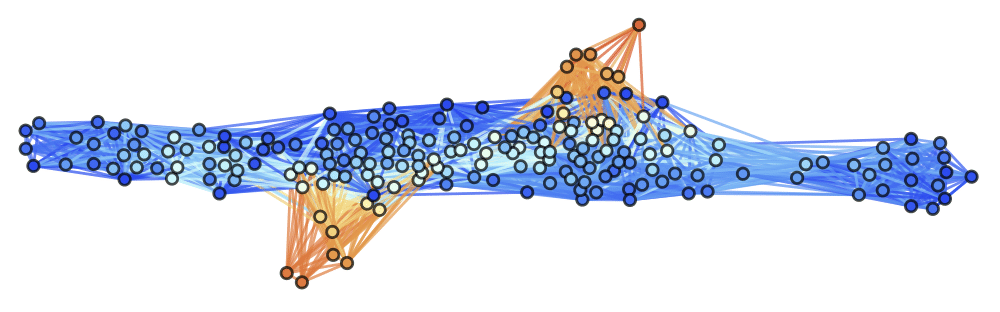}
\includegraphics[width=0.325\textwidth]{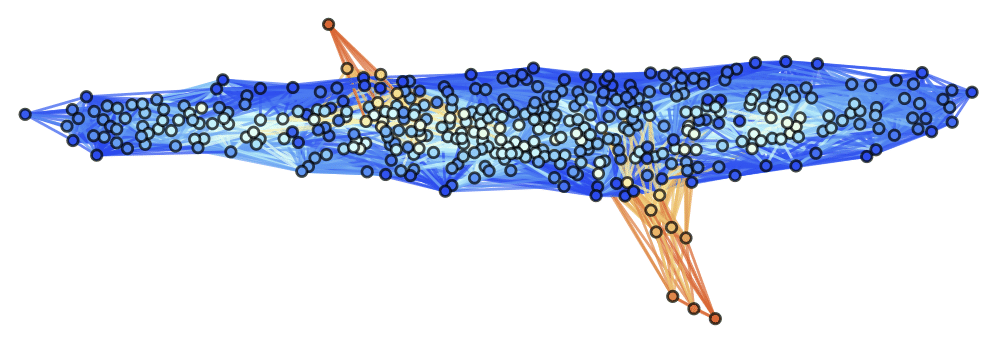}
\includegraphics[width=0.325\textwidth]{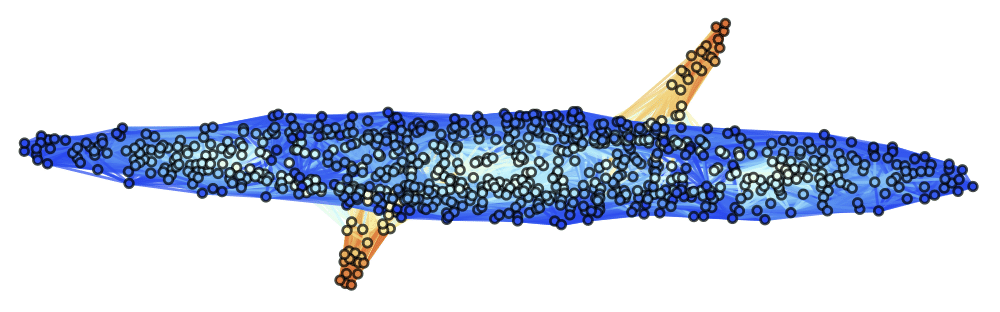}
\caption{Spatial hypergraphs corresponding to the intermediate hypersurface configuration of the head-on collision of rapidly rotating Kerr black holes (angular momentum) test with ${a = 0.6}$ at time ${t = 6 M}$, with resolutions of 200, 400 and 800 vertices, respectively. The hypergraphs have been adapted and colored using the local curvature in the Boyer-Lindquist conformal factor ${\psi}$.}
\label{fig:Figure41}
\end{figure}

\begin{figure}[ht]
\centering
\includegraphics[width=0.325\textwidth]{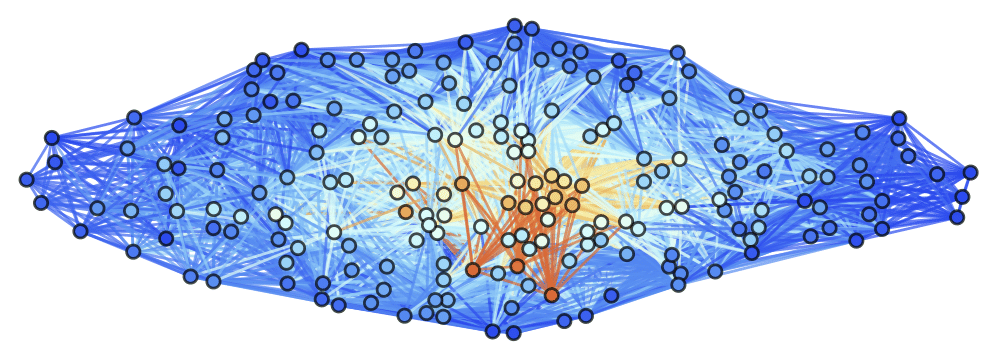}
\includegraphics[width=0.325\textwidth]{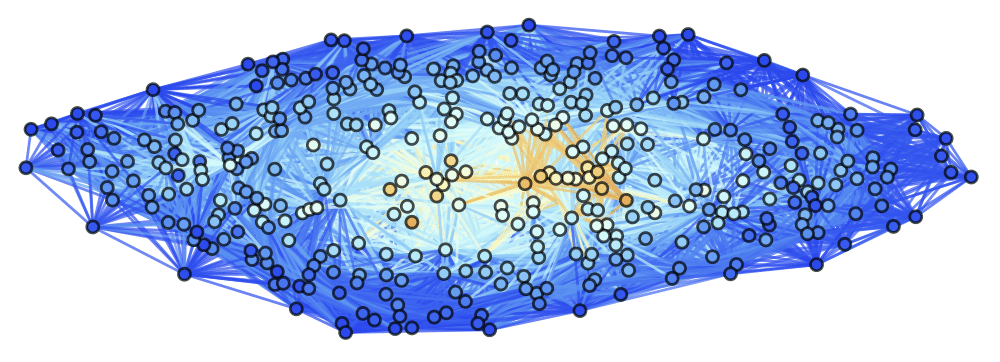}
\includegraphics[width=0.325\textwidth]{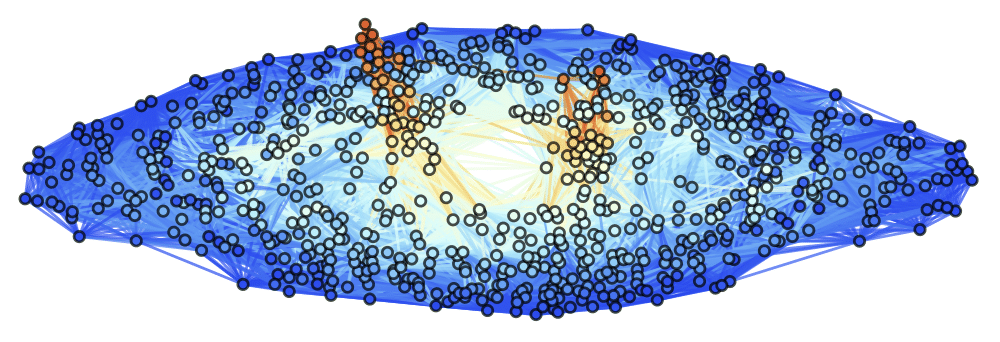}
\caption{Spatial hypergraphs corresponding to the final hypersurface configuration of the head-on collision of rapidly rotating Kerr black holes (angular momentum) test with ${a = 0.6}$ at time ${t = 12 M}$, with resolutions of 200, 400 and 800 vertices, respectively. The hypergraphs have been adapted and colored using the local curvature in the Boyer-Lindquist conformal factor ${\psi}$.}
\label{fig:Figure42}
\end{figure}

\begin{figure}[ht]
\centering
\includegraphics[width=0.325\textwidth]{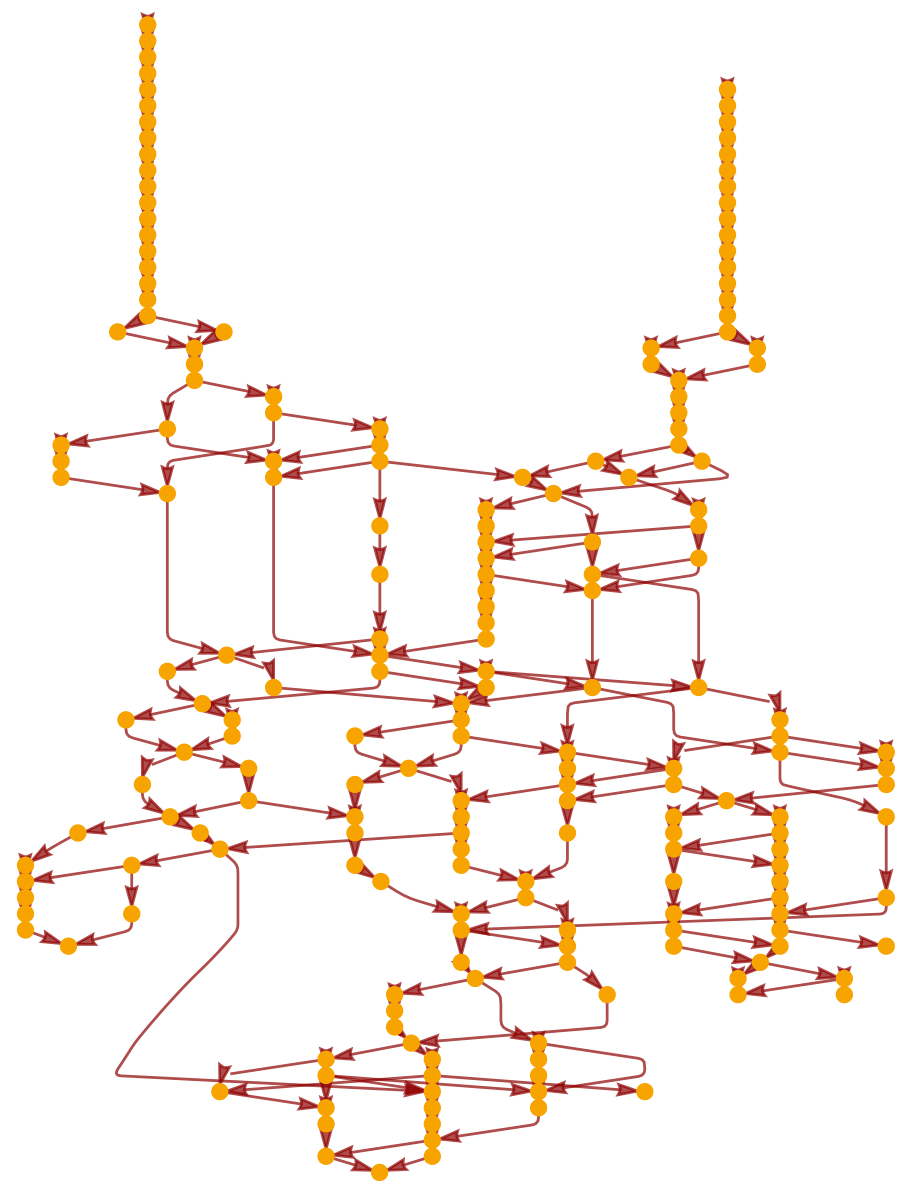}
\includegraphics[width=0.325\textwidth]{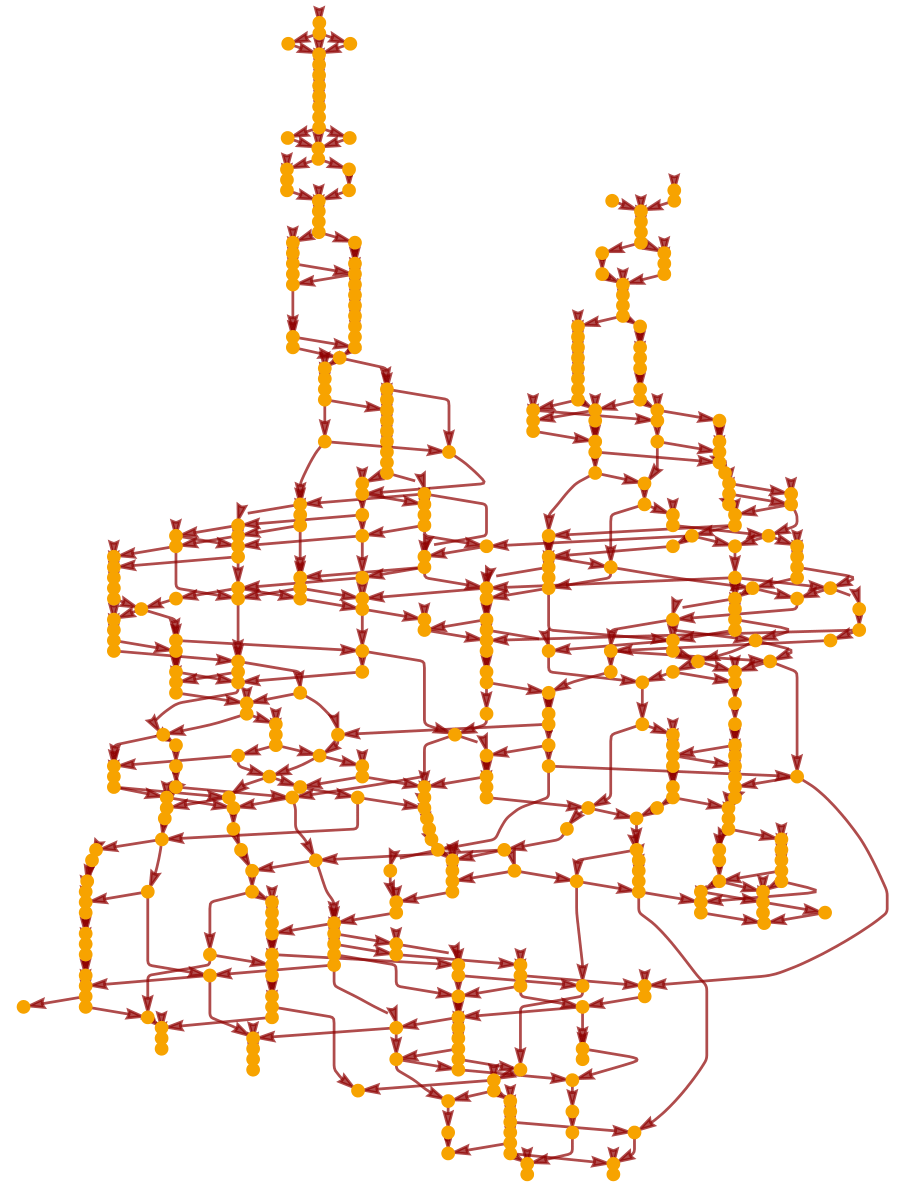}
\includegraphics[width=0.325\textwidth]{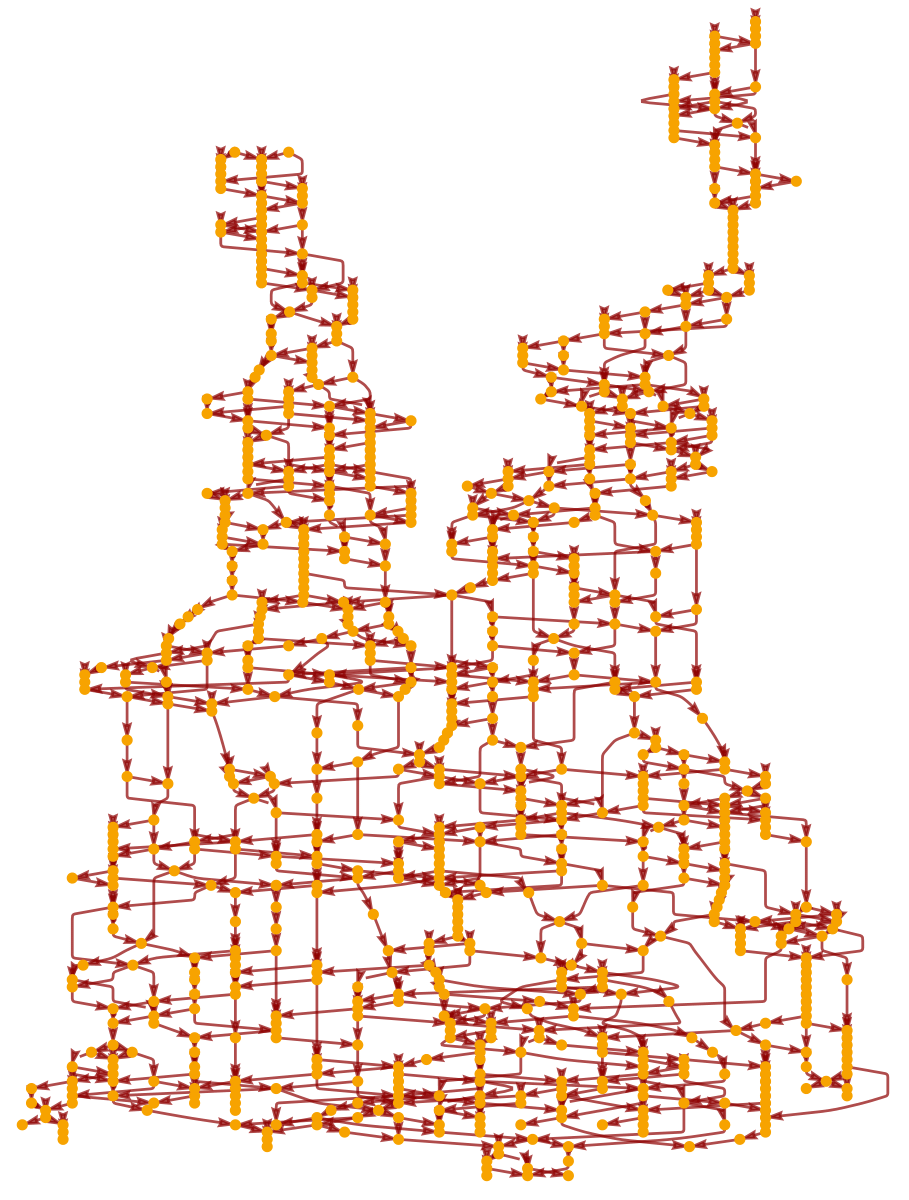}
\caption{Causal graphs corresponding to the discrete characteristic structure of the head-on collision of rapidly rotating Kerr black holes (angular momentum) test with ${a = 0.6}$ at time ${t = 12 M}$, with resolutions of 200, 400 and 800 hypergraph vertices, respectively.}
\label{fig:Figure43}
\end{figure}

\begin{figure}[ht]
\centering
\includegraphics[width=0.325\textwidth]{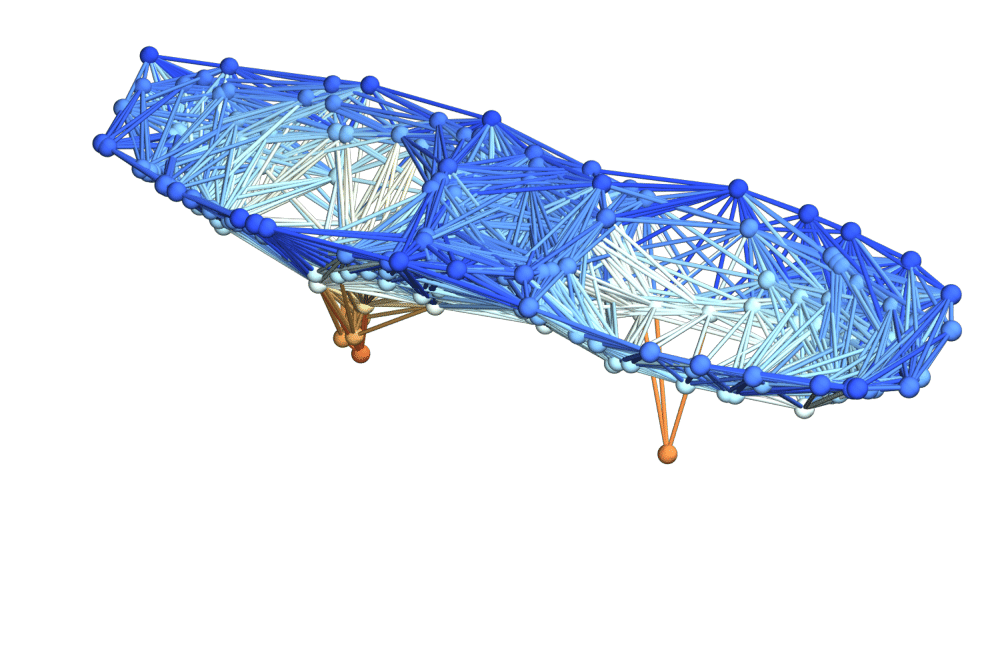}
\includegraphics[width=0.325\textwidth]{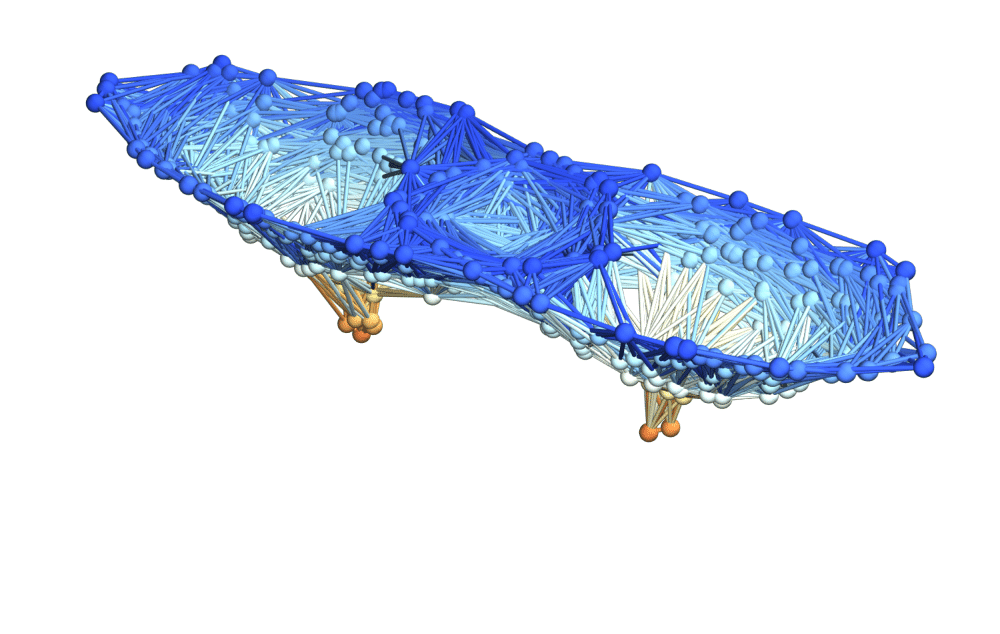}
\includegraphics[width=0.325\textwidth]{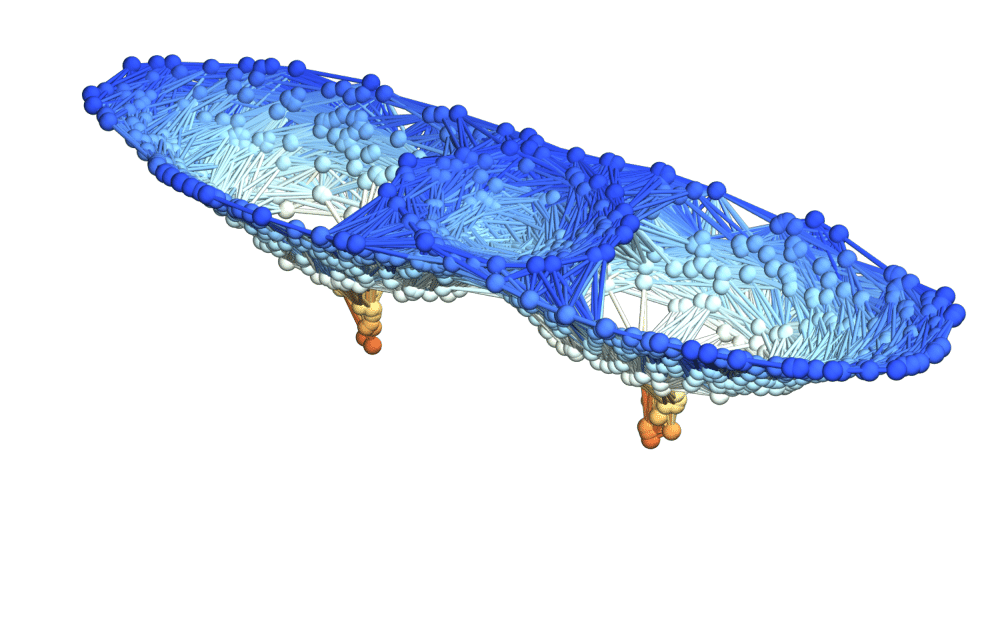}
\caption{Spatial hypergraphs corresponding to projections along the $z$-axis of the initial hypersurface configuration of the head-on collision of rapidly rotating Kerr black holes (angular momentum) test with ${a = 0.6}$ at time ${t = 0 M}$, with resolutions of 200, 400 and 800 vertices, respectively. The vertices have been assigned spatial coordinates according to the profile of the Boyer-Lindquist conformal factor ${\psi}$ through a spatial slice perpendicular to the $z$-axis, and the hypergraphs have been adapted and colored using the local curvature in ${\psi}$.}
\label{fig:Figure44}
\end{figure}

\begin{figure}[ht]
\centering
\includegraphics[width=0.325\textwidth]{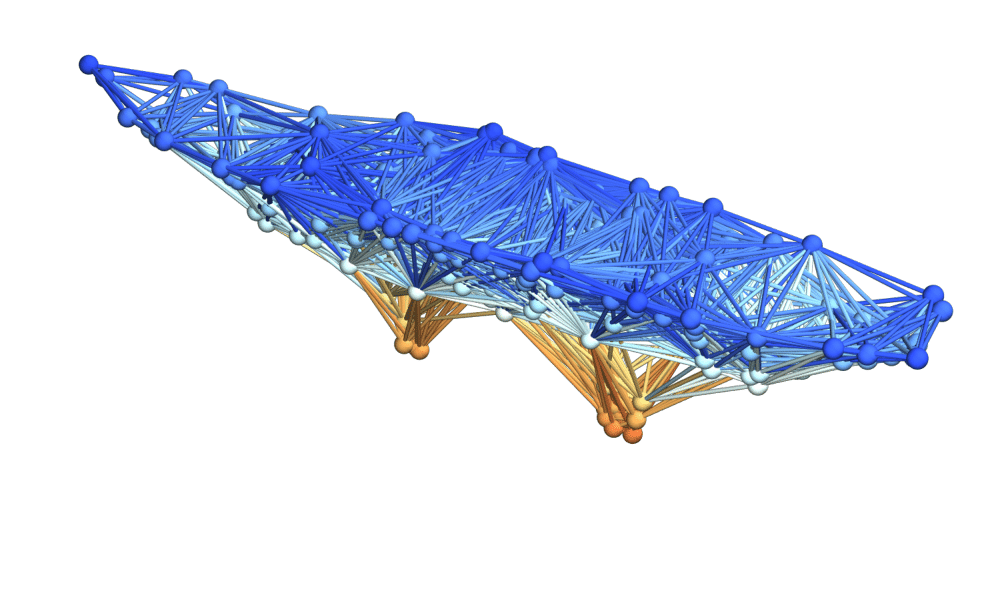}
\includegraphics[width=0.325\textwidth]{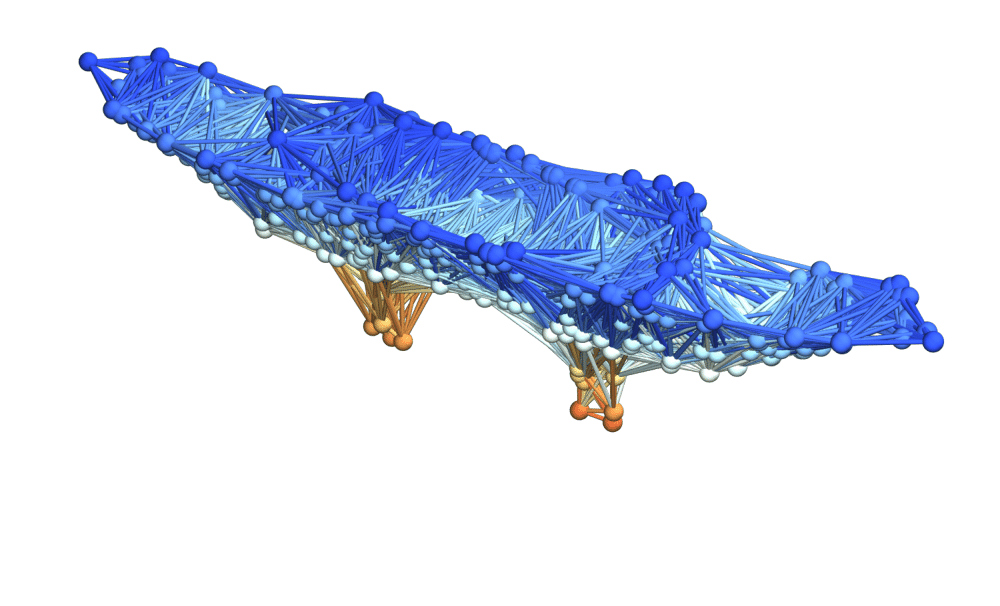}
\includegraphics[width=0.325\textwidth]{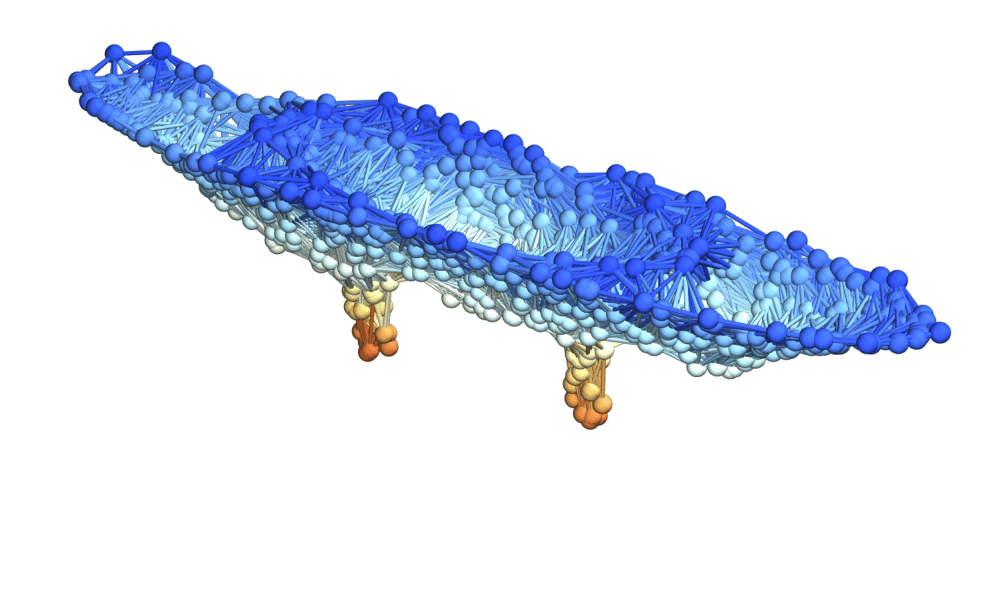}
\caption{Spatial hypergraphs corresponding to projections along the $z$-axis of the intermediate hypersurface configuration of the head-on collision of rapidly rotating Kerr black holes (angular momentum) test with ${a = 0.6}$ at time ${t = 6 M}$, with resolutions of 200, 400 and 800 vertices, respectively. The vertices have been assigned spatial coordinates according to the profile of the Boyer-Lindquist conformal factor ${\psi}$ through a spatial slice perpendicular to the $z$-axis, and the hypergraphs have been adapted and colored using the local curvature in ${\psi}$.}
\label{fig:Figure45}
\end{figure}

\begin{figure}[ht]
\centering
\includegraphics[width=0.325\textwidth]{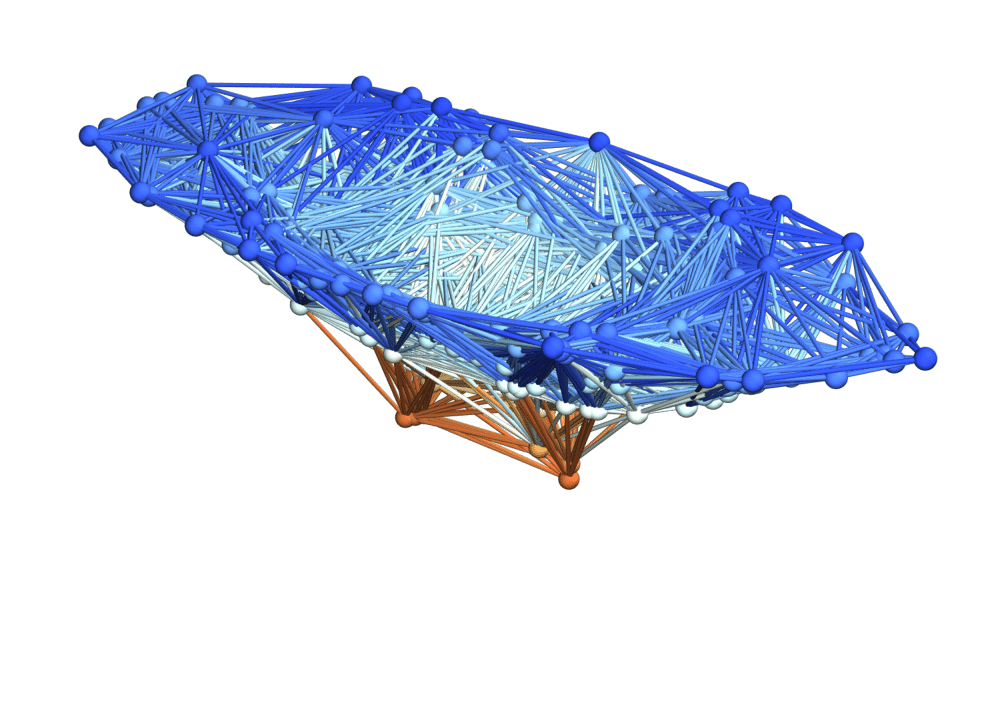}
\includegraphics[width=0.325\textwidth]{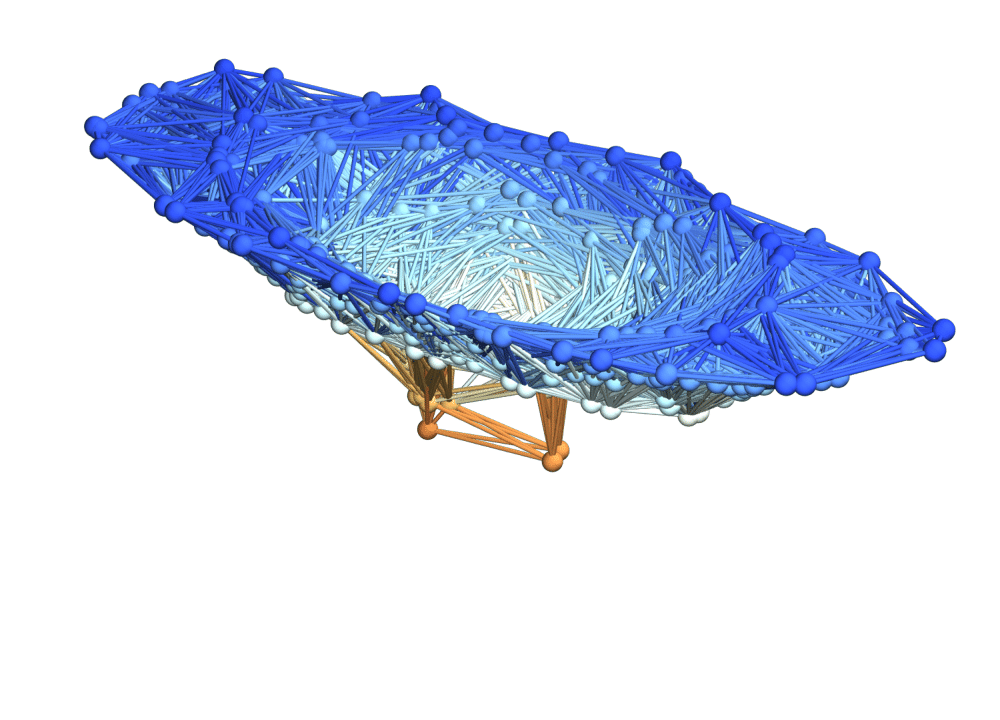}
\includegraphics[width=0.325\textwidth]{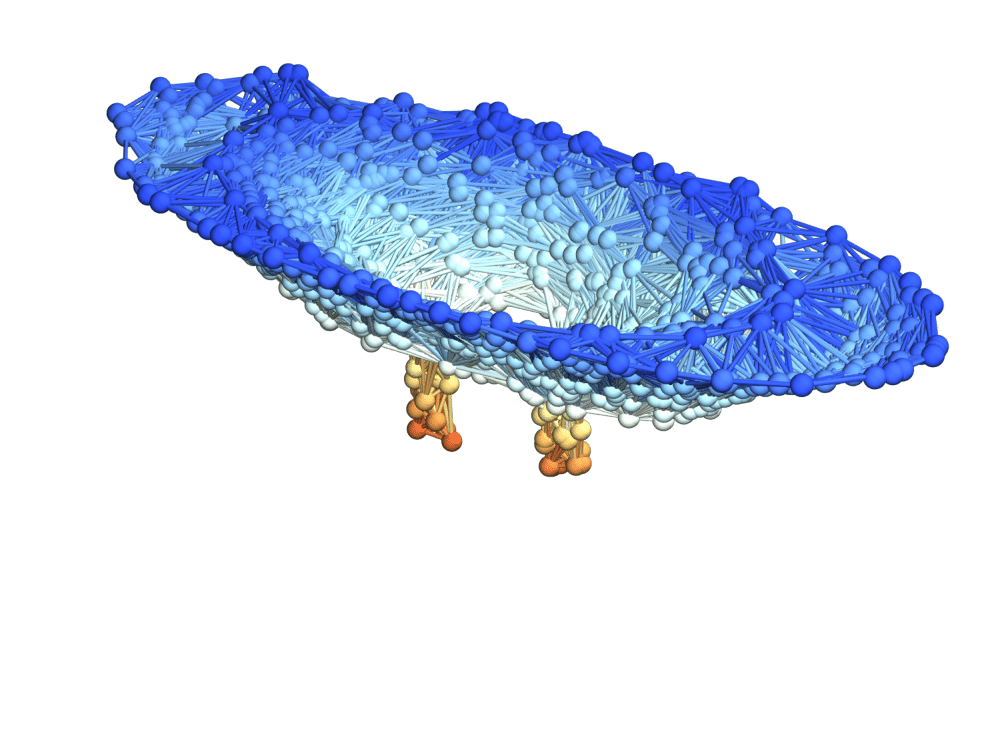}
\caption{Spatial hypergraphs corresponding to projections along the $z$-axis of the final hypersurface configuration of the head-on collision of rapidly rotating Kerr black holes (angular momentum) test with ${a = 0.6}$ at time ${t = 12 M}$, with resolutions of 200, 400 and 800 vertices, respectively. The vertices have been assigned spatial coordinates according to the profile of the Boyer-Lindquist conformal factor ${\psi}$ through a spatial slice perpendicular to the $z$-axis, and the hypergraphs have been adapted and colored using the local curvature in ${\psi}$.}
\label{fig:Figure46}
\end{figure}

Finally, we stress-test the robustness and convergence properties of our simulation code by evolving a more general Brill-Lindquist binary black hole solution corresponding to a head-on collision of two equal mass rapidly rotating Kerr black holes. Since the perturbation analyses presented for the previous test case apply to arbitrary Kerr solutions as well as to static Schwarzschild ones, the construction of the initial data and the extraction of the gravitational wave information via either the Moncrief formalism or the computation of the complex Weyl scalar ${\Psi_4}$ both remain unchanged from the previous example. As before, the two rapidly Kerr black holes will each have a mass of ${0.5 M}$, with an initial separation of ${1 M}$, with the center of the binary placed at the center of the computational domain, and with periodic boundary conditions imposed (along with all of the resultant limitations on stable simulation time). Each black hole is initialized with the angular momentum parameter ${a = \frac{J}{M} = 0.6}$. The simulation is run using quasi-isotropic gauge conditions for the initial data, with the new radial coordinate ${\eta}$, defined by:

\begin{equation}
r = r_{+} \cosh^2 \left( \frac{\eta}{2} \right) - r_{-} \sin^2 \left( \frac{\eta}{2} \right), \qquad \text{ with } \qquad r_{\pm} = M \pm \sqrt{M^2 - a^2},
\end{equation}
introduced into the Boyer-Lindquist coordinate system, as described in the setup for the previous Kerr black hole test. We also choose the standard Boyer-Lindquist conformal factor ${\psi}$:

\begin{equation}
\psi = \frac{\Sigma^{\frac{1}{4}}}{\sqrt{\sin \left( \lambda \right)}},
\end{equation}
where ${\lambda}$ is the parameter used in the definition of a new radial coordinate ${\tilde{r}}$:

\begin{equation}
\tilde{r} = \frac{\alpha^2 \cos^2 \left( \frac{\lambda}{2} \right) + \delta^2 \sin^2 \left( \frac{\lambda}{2} \right)}{\alpha \sin \left( \lambda \right)} + M,
\end{equation}
with:

\begin{equation}
\delta = \sqrt{M^2 - a^2}, \qquad 0 \leq \lambda \leq \pi, \qquad \alpha > 0,
\end{equation}
and where:

\begin{equation}
\Sigma = r^2 + a^2 \cos^2 \left( \theta \right),
\end{equation}
as usual, as the scalar field employed within the refinement algorithm ${\phi = \psi}$, with a domain size of ${\left( 20 M \right)^3}$. As with the head-on collision of static Schwarzschild black holes, we evolve the solution in the first instance until a final time of ${t = 12 M}$, with an intermediate check at time ${t = 6 M}$, during which time the two rapidly rotating Kerr black holes merge and the gravitational ringdown phase of the merger begins; the initial, intermediate and final hypersurface configurations are shown in Figures \ref{fig:Figure40}, \ref{fig:Figure41} and \ref{fig:Figure42}, respectively, with resolutions of 200, 400 and 800 vertices, and with the hypergraphs adapted and colored using the Boyer-Lindquist conformal factor ${\psi}$. Figure \ref{fig:Figure43} shows the discrete characteristic structure of the solutions after time ${t = 12 M}$ (using directed acyclic causal graphs to show discrete characteristic lines), and Figures \ref{fig:Figure44}, \ref{fig:Figure45} and \ref{fig:Figure46} show projections along the $z$-axis of the initial, intermediate and final hypersurface configurations, respectively, with vertices assigned spatial coordinates according to the profile of the Boyer-Lindquist conformal factor ${\psi}$.

\begin{figure}[ht]
\centering
\includegraphics[width=0.325\textwidth]{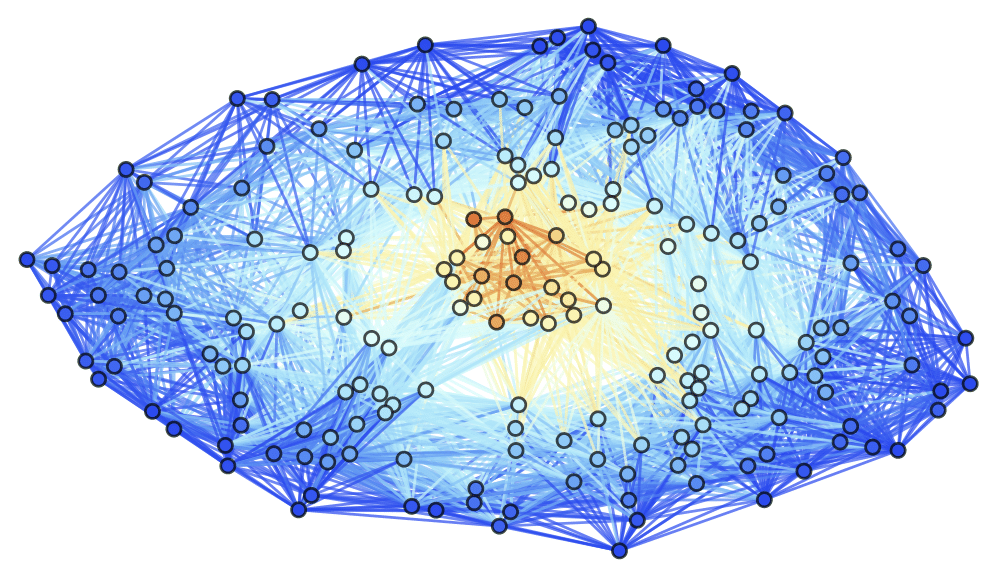}
\includegraphics[width=0.325\textwidth]{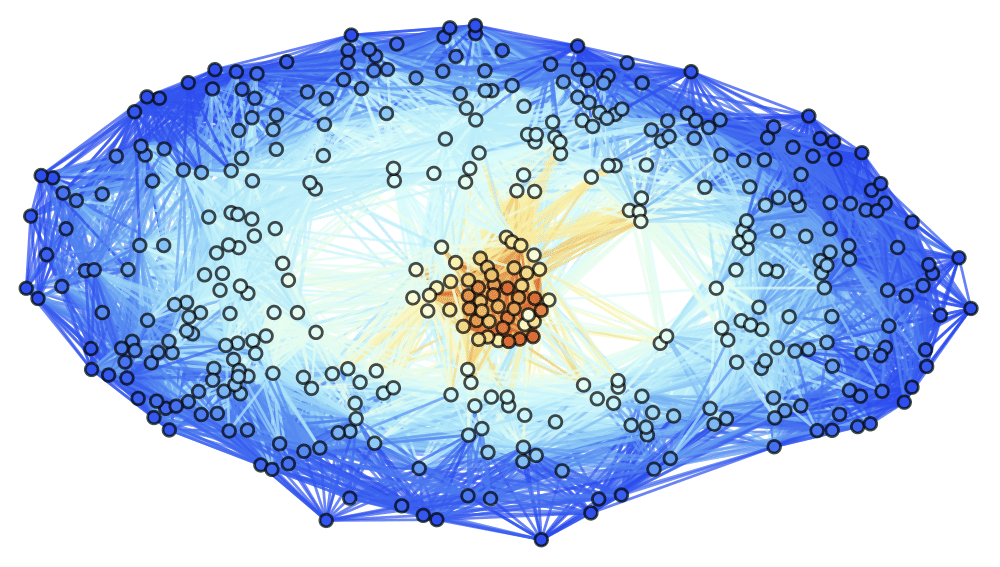}
\includegraphics[width=0.325\textwidth]{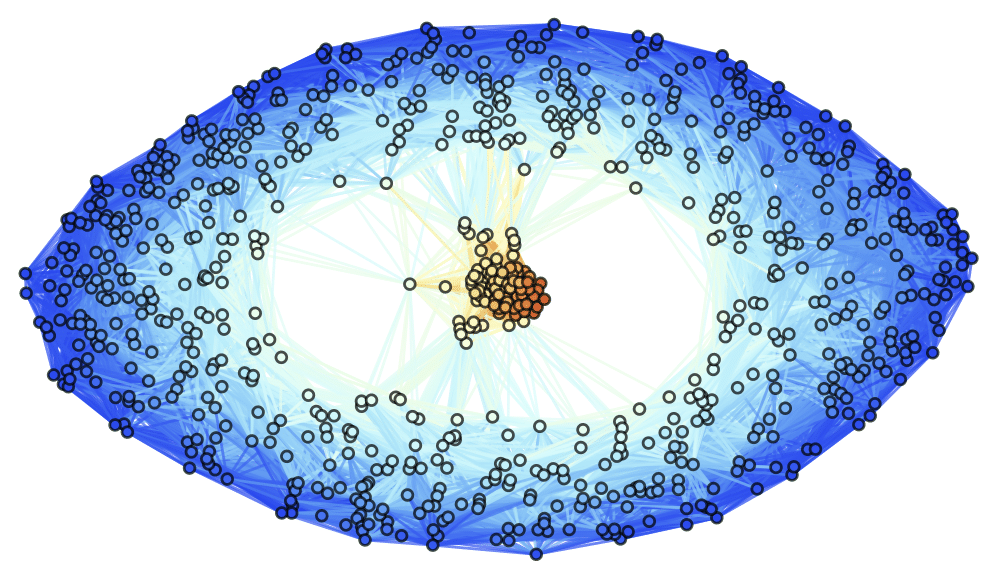}
\caption{Spatial hypergraphs corresponding to the post-ringdown hypersurface configuration of the head-collision of rapidly rotating Kerr black holes (angular momentum) test with ${a = 0.6}$ at time ${t = 24 M}$, with resolutions of 200, 400 and 800 vertices, respectively. The hypergraphs have been adapted and colored using the local curvature in the Boyer-Lindquist conformal factor ${\psi}$.}
\label{fig:Figure47}
\end{figure}

\begin{figure}[ht]
\centering
\includegraphics[width=0.325\textwidth]{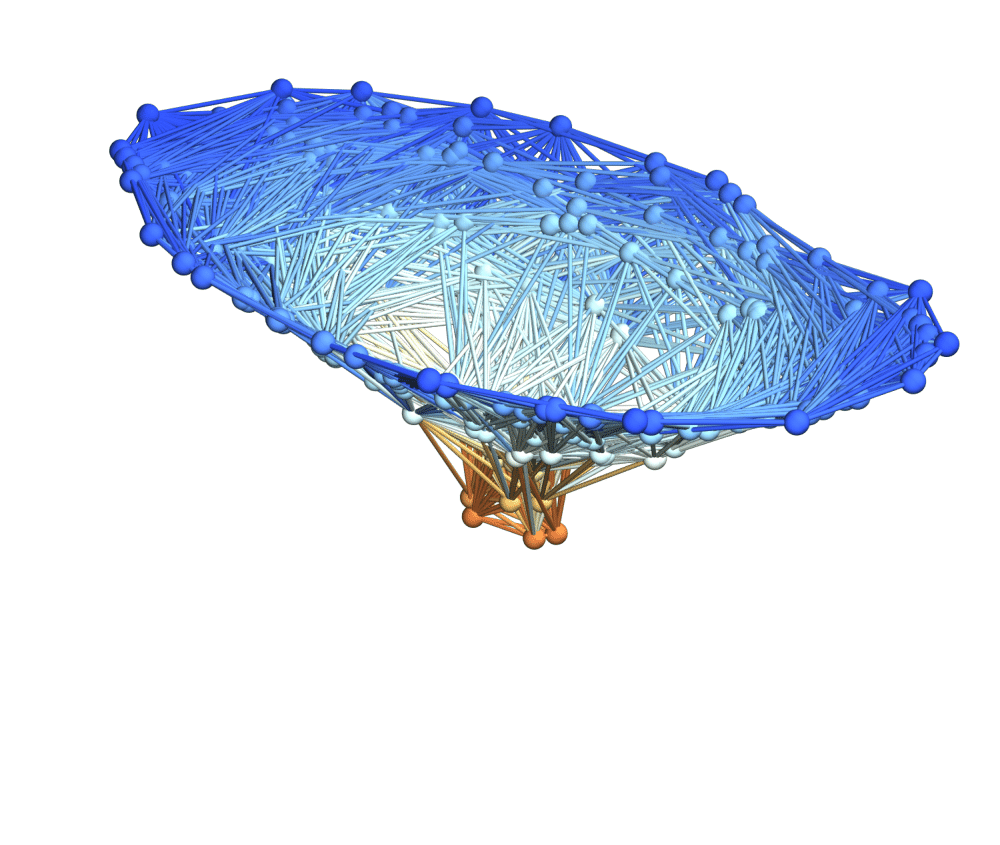}
\includegraphics[width=0.325\textwidth]{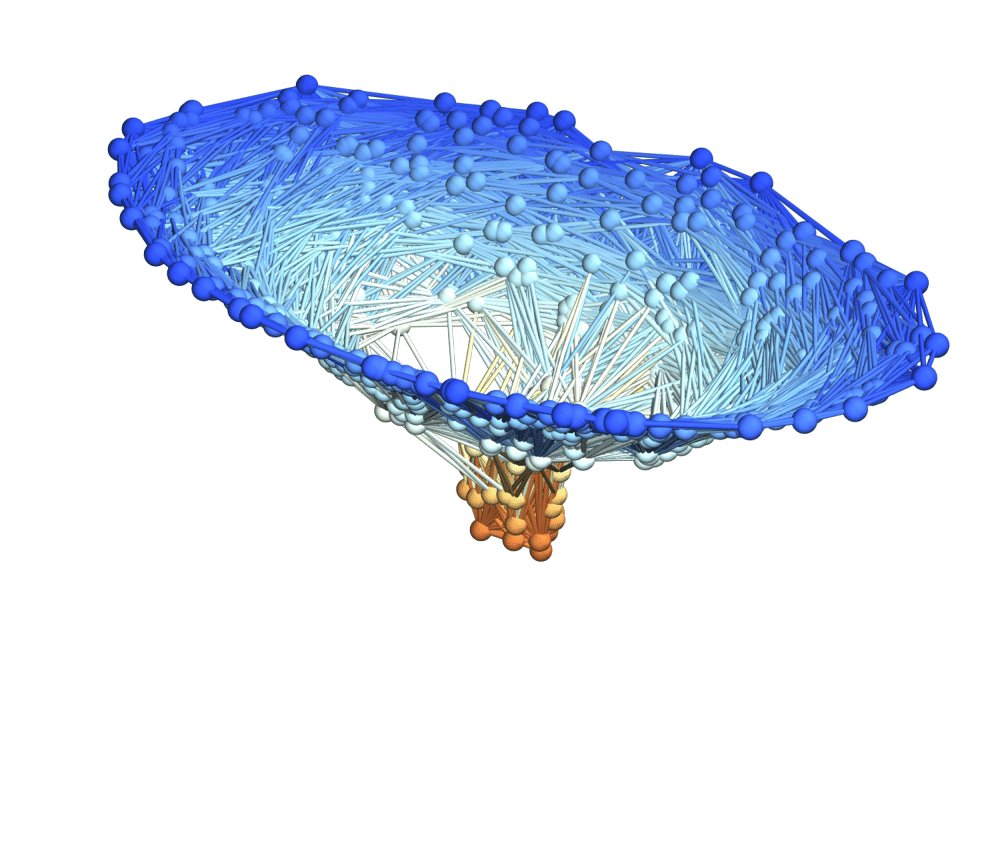}
\includegraphics[width=0.325\textwidth]{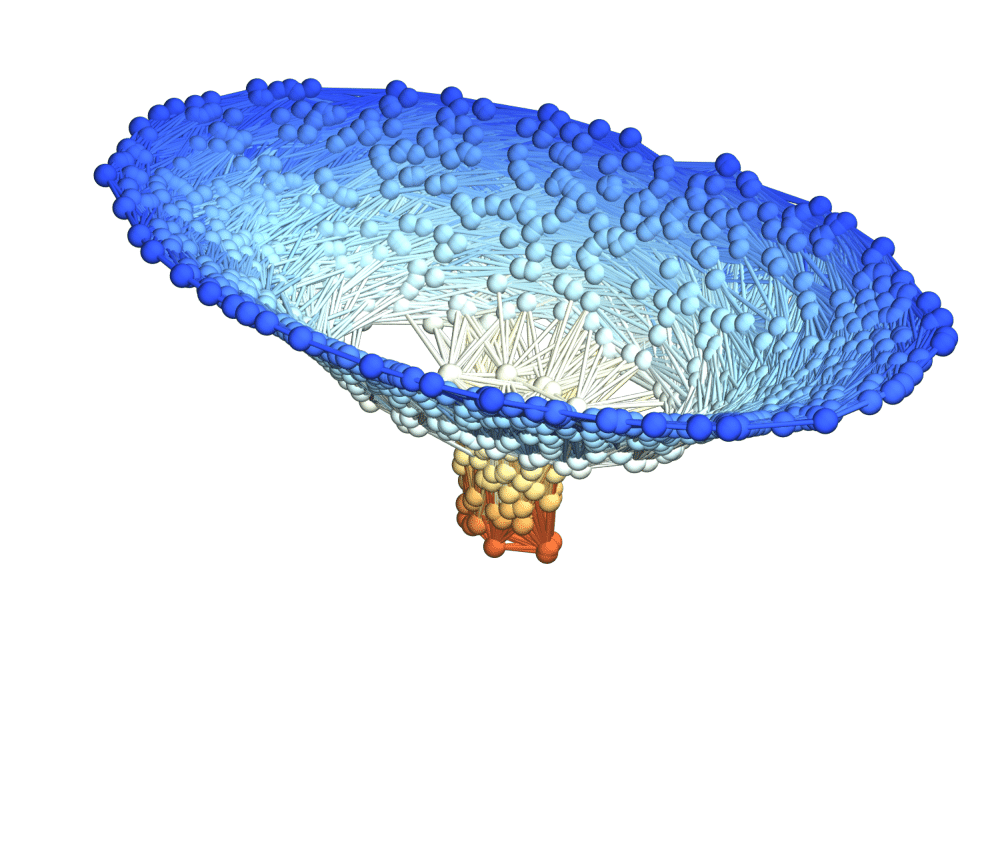}
\caption{Spatial hypergraphs corresponding to projections along the $z$-axis of the post-ringdown configuration of the head-on collision of rapidly rotating Kerr black holes (angular momentum) test with ${a = 0.6}$ at time ${t = 24 M}$, with resolutions of 200, 400 and 800 vertices, respectively. The vertices have been assigned spatial coordinates according to the profile of the Boyer-Lindquist conformal factor ${\psi}$ through a spatial slice perpendicular to the $z$-axis, and the hypergraphs have been adapted and colored using the local curvature in ${\psi}$.}
\label{fig:Figure48}
\end{figure}

\begin{figure}[ht]
\centering
\includegraphics[width=0.895\textwidth]{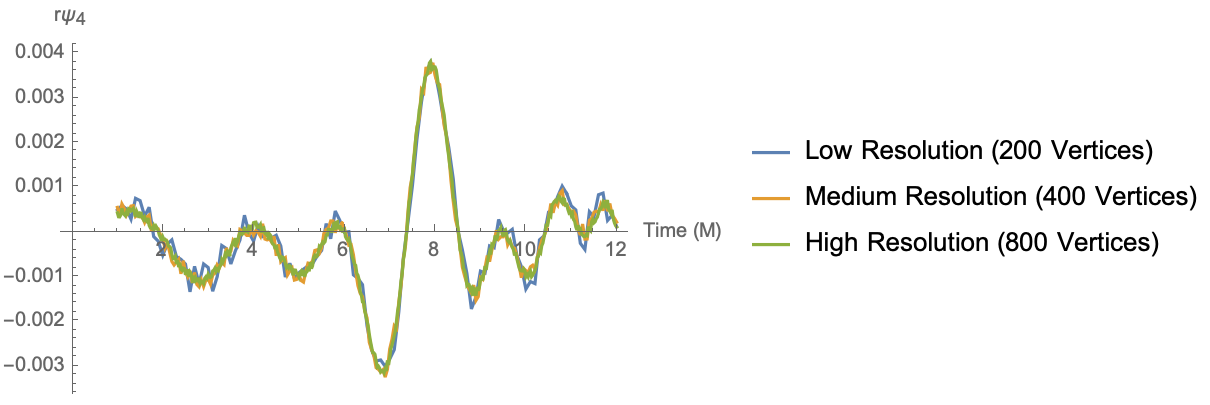}
\caption{Convergence test of the head-on collision of rapidly rotating Kerr black holes (angular momentum) test after time ${t = 12 M}$, showing the real part of the ${\ell = 2}$, ${m = 0}$ mode of the radial Weyl scalar ${r \Psi_4}$ extrapolated on a sphere of radius ${R = 60 M}$, with resolutions of 200, 400 and 800 vertices, respectively.}
\label{fig:Figure49}
\end{figure}

As with the static Schwarzschild black hole collision, the post-ringdown hypersurface configuration at time ${t = 24 M}$ is shown in Figure \ref{fig:Figure47}, along with the projections along the $z$-axis of this configuration in Figure \ref{fig:Figure48}, both with resolutions of 200, 400 and 800 vertices. In order to demonstrate convergence and the successful extraction of gravitational wave data as before, we show the real part of the ${\ell = 2}$, ${m = 0}$ mode of the radial Weyl scalar ${r \Psi_4}$ at time ${t = 12 M}$, extrapolated on a sphere of radius ${6 M}$ using standard fourth-order interpolation techniques across both the polar and azimuthal angular directions in Figure \ref{fig:Figure49}, with resolutions of 200, 400 and 800 vertices. We confirm that the ADM mass of the final Kerr black hole configuration (computed by integrating over a single surface surrounding the boundary of asymptotic flatness) is approximately equal, as previously, to the sum of ADM masses of the initial two Kerr black holes (computed by integrating over a pair of surfaces surrounding the two boundaries of asymptotic flatness), that the ADM linear momentum converges to be approximately zero in the post-ringdown phase, and that the angular momentum parameter ${a = \frac{J}{M}}$ of the final Kerr black hole converges to be approximately the same as the angular momentum parameters of the two initial Kerr black holes, as expected. The convergence rates for the Hamiltonian constraint after time ${t = 24 M}$, with respect to the ${L_1}$, ${L_2}$ and ${L_{\infty}}$ norms, illustrating approximately fourth-order convergence of the finite difference scheme, are shown in Table \ref{tab:Table5}.

\begin{table}
\centering
\begin{tabular}{|c|c|c|c|c|c|c|}
\hline
Vertices & ${\epsilon \left( L_1 \right)}$ & ${\epsilon \left( L_2 \right)}$ & ${\epsilon \left( L_{\infty} \right)}$ & ${\mathcal{O} \left( L_1 \right)}$ & ${\mathcal{O} \left( L_2 \right)}$ & ${\mathcal{O} \left( L_{\infty} \right)}$\\
\hline\hline
100 & ${8.72 \times 10^{-2}}$ & ${3.83 \times 10^{-2}}$ & ${5.53 \times 10^{-2}}$ & - & - & -\\
\hline
200 & ${5.96 \times 10^{-3}}$ & ${3.43 \times 10^{-3}}$ & ${6.56 \times 10^{-3}}$ & 3.87 & 3.48 & 3.07\\
\hline
400 & ${4.56 \times 10^{-4}}$ & ${2.71 \times 10^{-4}}$ & ${4.79 \times 10^{-4}}$ & 3.71 & 3.61 & 3.78\\
\hline
800 & ${2.80 \times 10^{-5}}$ & ${2.27 \times 10^{-5}}$ & ${5.29 \times 10^{-5}}$ & 4.03 & 3.62 & 3.18\\
\hline
1600 & ${9.86 \times 10^{-7}}$ & ${9.53 \times 10^{-7}}$ & ${4.46 \times 10^{-6}}$ & 4.83 & 4.57 & 3.57\\
\hline
\end{tabular}
\caption{Convergence rates for the head-on collision of rapidly rotating Kerr black holes (angular momentum) test with respect to the ${L_1}$, ${L_2}$ and ${L_{\infty}}$ norms for the Hamiltonian constraint $H$ after time ${t = 24 M}$, showing approximately fourth-order convergence.}
\label{tab:Table5}
\end{table}

\clearpage

\section{Comparison to Pure Wolfram Model Evolution}
\label{sec:Section5}

\begin{figure}[ht]
\centering
\includegraphics[width=0.295\textwidth]{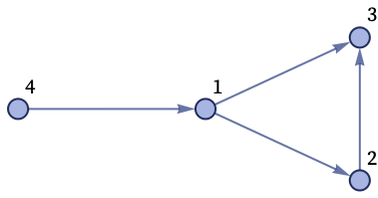}\hspace{0.1\textwidth}
\includegraphics[width=0.295\textwidth]{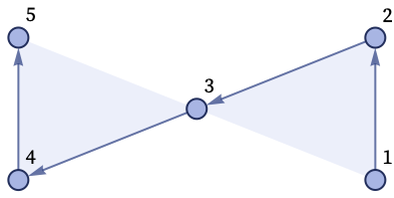}
\caption{Two spatial hypergraphs, represented internally as finite collections of (potentially ordered) relations between abstract elements, namely ${\left\lbrace \left\lbrace 1, 2 \right\rbrace, \left\lbrace 1, 3 \right\rbrace, \left\lbrace 2, 3 \right\rbrace, \left\lbrace 4, 1 \right\rbrace \right\rbrace}$ and ${\left\lbrace \left\lbrace 1, 2, 3 \right\rbrace, \left\lbrace 3, 4, 5 \right\rbrace \right\rbrace}$, respectively. Example taken from \cite{wolfram2}.}
\label{fig:Figure50}
\end{figure}

\begin{figure}[ht]
\centering
\includegraphics[width=0.495\textwidth]{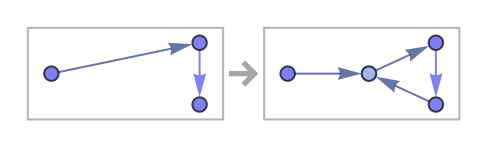}
\caption{A hypergraph transformation rule, represented internally as a set substitution system ${\left\lbrace \left\lbrace x, y \right\rbrace, \left\lbrace y, z \right\rbrace \right\rbrace \to \left\lbrace \left\lbrace w, y \right\rbrace, \left\lbrace y, z \right\rbrace, \left\lbrace z, w \right\rbrace, \left\lbrace x, w \right\rbrace \right\rbrace}$. Example taken from \cite{wolfram2}.}
\label{fig:Figure51}
\end{figure}

\begin{figure}[ht]
\centering
\includegraphics[width=0.695\textwidth]{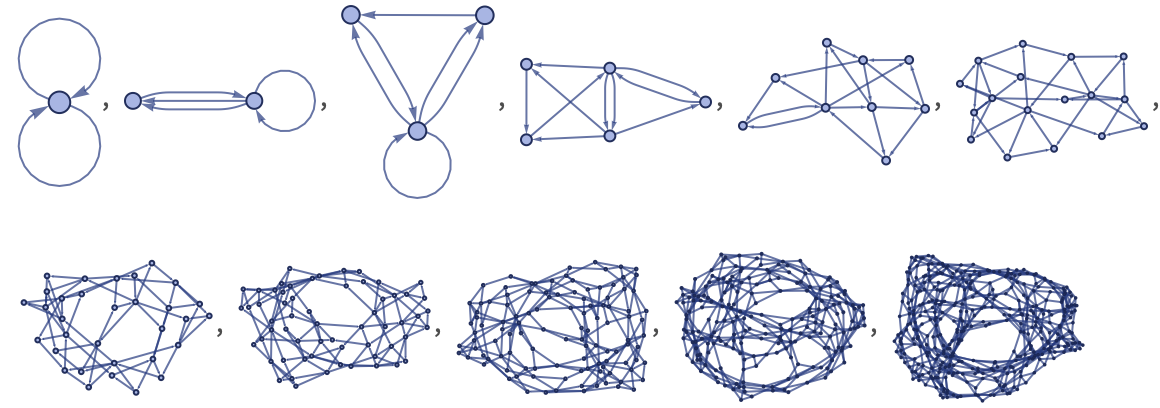}
\caption{The hypergraphs obtained during the first 10 steps of evolution of the set substitution system ${\left\lbrace \left\lbrace x, y \right\rbrace, \left\lbrace y, z \right\rbrace \right\rbrace \to \left\lbrace \left\lbrace w, y \right\rbrace, \left\lbrace y, z \right\rbrace, \left\lbrace z, w \right\rbrace, \left\lbrace x, w \right\rbrace \right\rbrace}$, assuming a simple initial condition consisting of a single vertex with two self-loops. Example taken from \cite{wolfram2}.}
\label{fig:Figure52}
\end{figure}

\begin{figure}[ht]
\centering
\includegraphics[width=0.495\textwidth]{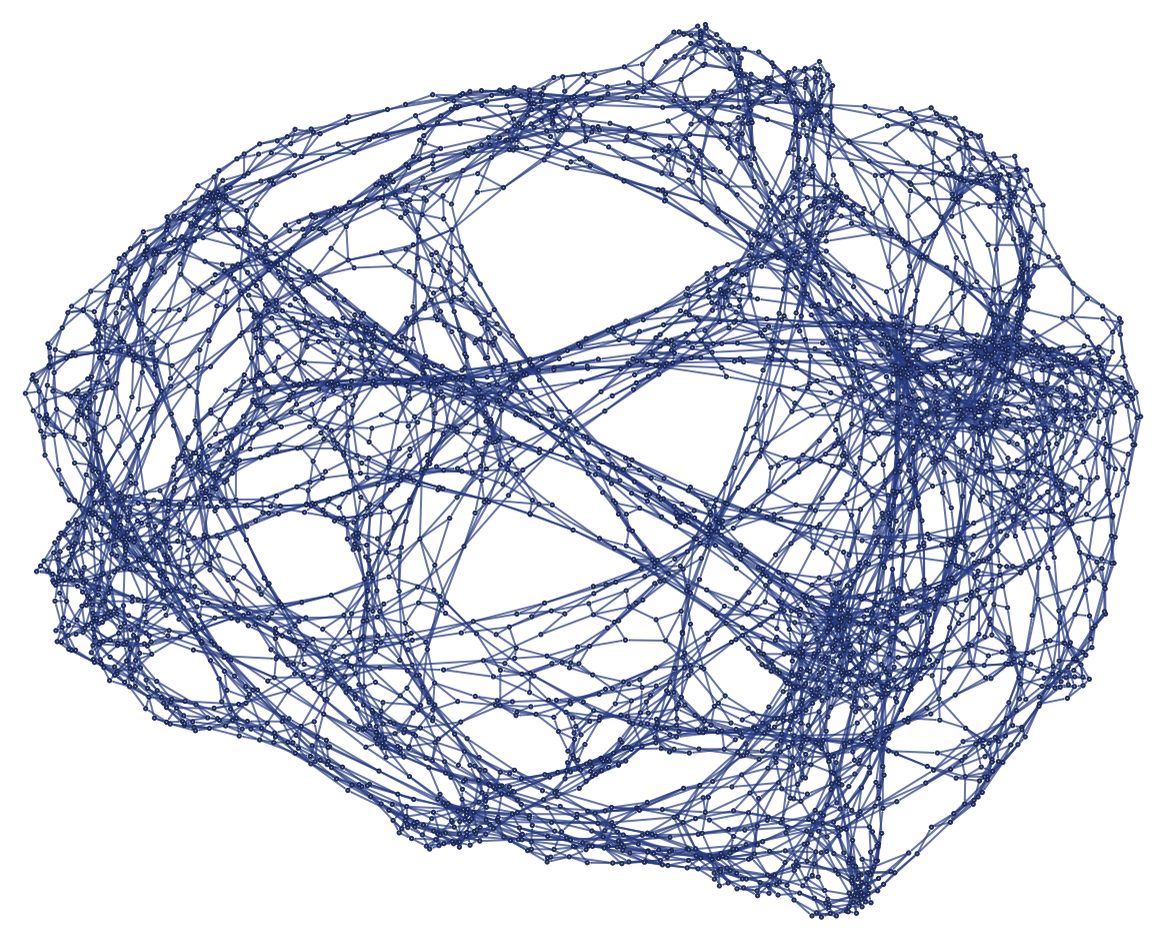}
\caption{The hypergraph obtained after 14 steps of evolution of the set substitution system ${\left\lbrace \left\lbrace x, y \right\rbrace, \left\lbrace y, z \right\rbrace \right\rbrace \to \left\lbrace \left\lbrace w, y \right\rbrace, \left\lbrace y, z \right\rbrace, \left\lbrace z, w \right\rbrace, \left\lbrace x, w \right\rbrace \right\rbrace}$, assuming a simple initial condition consisting of a single vertex with two self-loops. Example taken from \cite{wolfram2}.}
\label{fig:Figure53}
\end{figure}

In this final section, we proceed to compare the numerical results presented in the section above with results obtained using pure Wolfram model evolution (i.e. pure hypergraph substitution dynamics)\cite{gorard}\cite{gorard2}. We begin with a discrete approximation to a spacelike hypersurface, namely a \textit{spatial hypergraph} ${H = \left( V, E \right)}$, which is both finite and undirected:

\begin{equation}
E \subset \mathcal{P} \left( V \right) \setminus \left\lbrace \emptyset \right\rbrace,
\end{equation}
where ${\mathcal{P} \left( V \right)}$ denotes the power set of $V$. Such hypergraphs are represented internally as finite collections of (potentially ordered) relations of abstract elements, as illustrated in Figure \ref{fig:Figure50}. An \textit{update rule} $R$ on such a hypergraph ${H = \left( V, E \right)}$ is then an abstract rewrite rule of the general form ${H_1 \to H_2}$ (indicating that a subhypergraph matching the pattern ${H_1}$ is to be replaced by a distinct hypergraph matching the pattern ${H_2}$), which can be formulated equivalently as a \textit{set substitution system}, as shown in Figure \ref{fig:Figure51}. Although there is no a priori order in which these rewrite rules should be applied (as discussed later), we can nevertheless choose to apply them to every possible matching and non-overlapping subhypergraph in accordance with some pre-defined evaluation strategy, and thus succeed in obtaining a deterministic evolution, as shown in Figures \ref{fig:Figure52} and \ref{fig:Figure53}.

\begin{figure}[ht]
\centering
\includegraphics[width=0.495\textwidth]{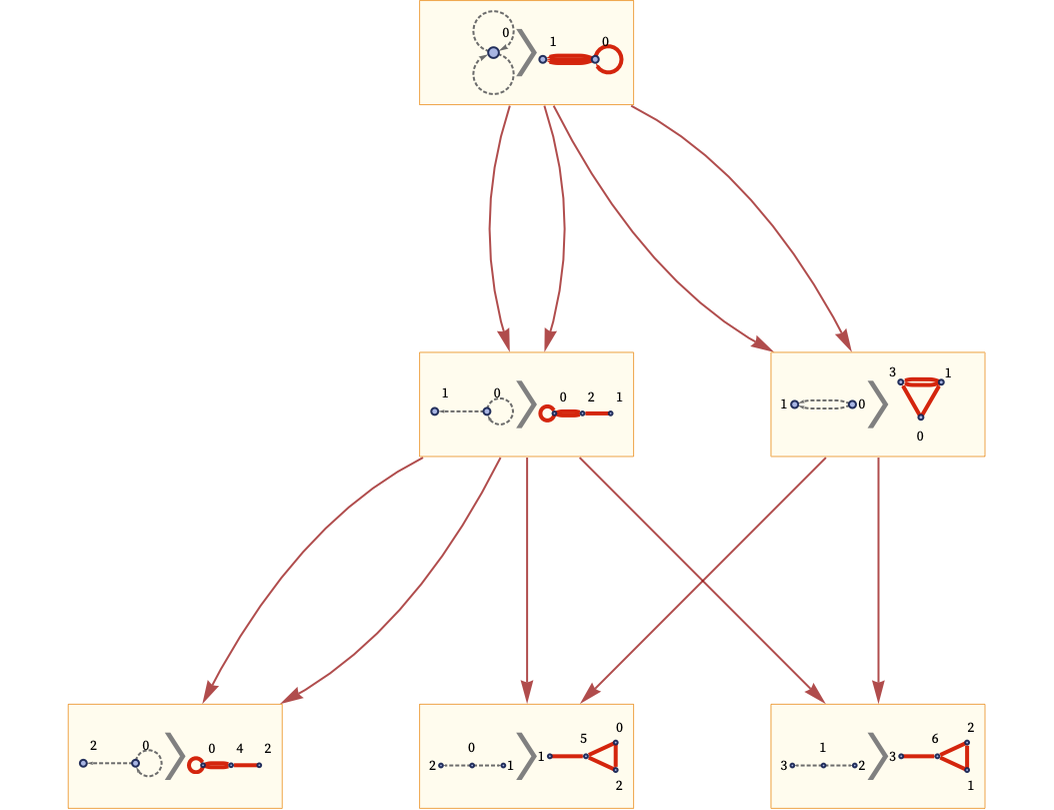}
\caption{The causal graph obtained after three steps of evolution of the set substitution system ${\left\lbrace \left\lbrace x, y \right\rbrace, \left\lbrace x, z \right\rbrace \right\rbrace \to \left\lbrace \left\lbrace x, y \right\rbrace, \left\lbrace x, w \right\rbrace, \left\lbrace y, w \right\rbrace, \left\lbrace z, w \right\rbrace \right\rbrace}$, assuming a simple initial condition consisting of a single vertex with two self-loops. Example taken from \cite{wolfram2}.}
\label{fig:Figure54}
\end{figure}

\begin{figure}[ht]
\centering
\includegraphics[width=0.495\textwidth]{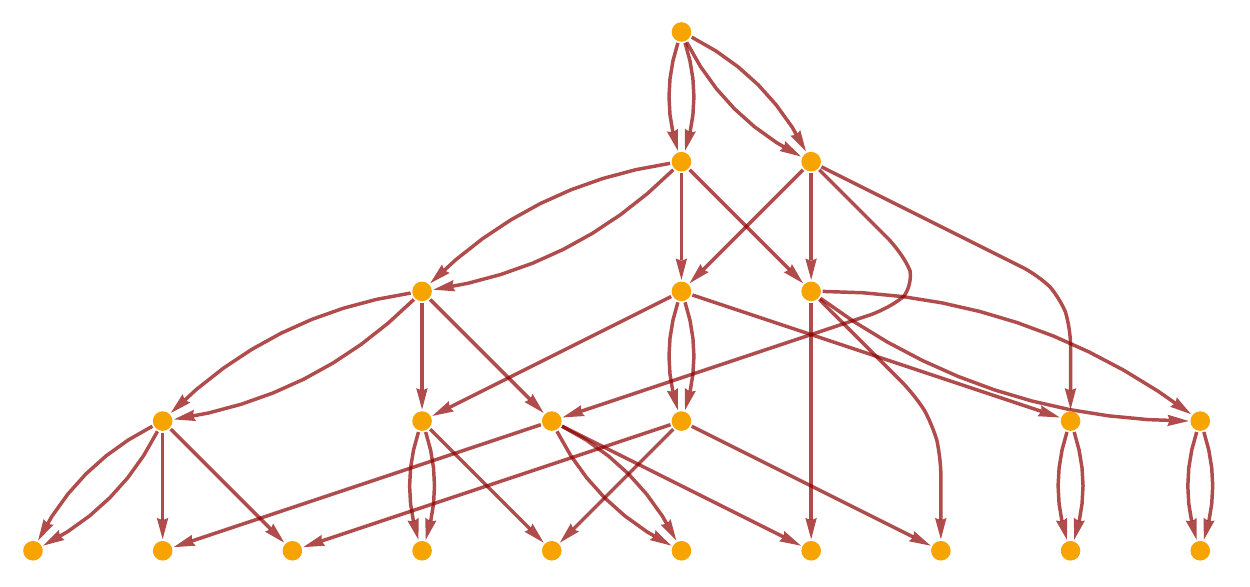}
\includegraphics[width=0.495\textwidth]{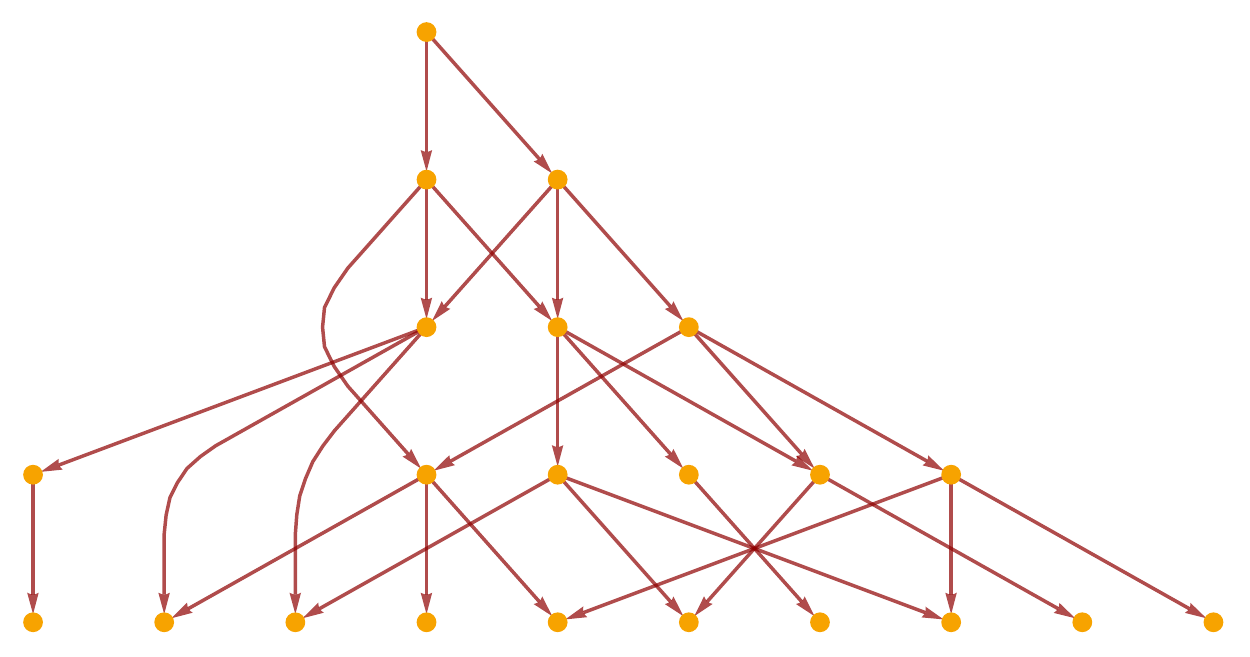}
\caption{The full causal graph obtained after five steps of evolution of the set substitution system ${\left\lbrace \left\lbrace x, y \right\rbrace, \left\lbrace x, z \right\rbrace \right\rbrace \to \left\lbrace \left\lbrace x, y \right\rbrace, \left\lbrace x, w \right\rbrace, \left\lbrace y, w \right\rbrace, \left\lbrace z, w \right\rbrace \right\rbrace}$ (left), along with its transitive reduction (right), assuming a simple initial condition consisting of a single vertex with two self-loops.}
\label{fig:Figure55}
\end{figure}

With a deterministic evolution obtained in this manner, we can now use the sequence of applications of rewrite rules (known henceforth as rewrite \textit{events}) to construct a directed, acyclic graph representing their causal relationships, known as the \textit{causal graph}. Each vertex in the causal graph ${G_{causal}}$ corresponds to a particular rewrite event, and a directed edge ${A \to B}$ exists if and only if:

\begin{equation}
\mathrm{In} \left( B \right) \cap \mathrm{Out} \left( A \right) \neq \emptyset,
\end{equation}
i.e. if and only if the hyperedges in the input for event $B$ overlap with the hyperedges in the output of event $A$. An example of such a causal graph, generated by the first three steps of evolution of a simple Wolfram model system, is shown in Figure \ref{fig:Figure54}; the transitive reduction of such a causal graph thus forms the Hasse diagram for a causal partial order relation, as shown in Figure \ref{fig:Figure55}. More precisely, similar to an ordinary causal set, the set of rewrite events ${\mathcal{C}}$ is now equipped with a binary relation, denoted ${\prec}$, satisfying a subset of the axioms of a partial order relation\cite{gorard3}, namely antisymmetry/acyclicity:

\begin{equation}
\nexists x, y \in \mathcal{C}, \qquad \text{ such that } \qquad x \prec y \text{ and } y \prec x,
\end{equation}
and transitivity:

\begin{equation}
\forall x, y, z \in \mathcal{C}, \qquad x \prec y \text{ and } y \prec z \implies x \prec z,
\end{equation}
as well as either \textit{local finiteness} (in the stronger case):

\begin{equation}
\forall x, y \in \mathcal{C}, \qquad \left\lvert \mathbf{I} \left[ x, y \right] \right\rvert < \infty,
\end{equation}
or \textit{local countability} (in the weaker case):

\begin{equation}
\forall x, y \in \mathcal{C}, \qquad \left\lvert \mathbf{I} \left[ x, y \right] \right\rvert \leq \aleph_0,
\end{equation}
so as to guarantee discreteness. In the above, ${\mathbf{I} \left[ x, y \right]}$ corresponds to the discrete analog of the \textit{Alexandrov topology} or \textit{interval topology} on continuous spacetime, namely:

\begin{equation}
\mathbf{I} \left[ x, y \right] = \mathrm{Fut} \left( x \right) \cap \mathrm{Past} \left( y \right),
\end{equation}
for (exclusive) discrete sets of future and past events ${\mathrm{Fut}}$ and ${\mathrm{Past}}$:

\begin{equation}
\mathrm{Fut} \left( x \right) = \left\lbrace w \in \mathcal{C} : x \prec w \text{ and } x \neq w \right\rbrace, \qquad \text{ and } \qquad \mathrm{Past} \left( x \right) = \left\lbrace w \in \mathcal{C} : w \prec x \text{ and } x \neq w \right\rbrace,
\end{equation}
in direct analogy to the Alexandrov interval ${\mathbf{A} \left[ x, y \right]}$ on a smooth Lorentzian manifold ${\left( \mathcal{M}, g \right)}$, namely:

\begin{equation}
\mathbf{A} \left[ x, y \right] = I^{+} \left( x \right) \cap I^{-} \left( y \right),
\end{equation}
where ${I^{+} \left( x \right)}$ and ${I^{-} \left( x \right)}$ denote the \textit{chronological future} and \textit{chronological past} of the event ${x \in \mathcal{M}}$:

\begin{equation}
I^{+} \left( x \right) = \left\lbrace y \in \mathcal{M} : x \ll y \right\rbrace, \qquad \text{ and } \qquad I^{-} \left( x \right) = \left\lbrace y \in \mathcal{M} : y \ll x \right\rbrace,
\end{equation}
and where ${\ll}$ denotes the usual chronological precedence relation on the smooth manifold ${\mathcal{M}}$.

In certain restricted cases, such a causal graph will thus form a discrete approximation to the conformal structure of a continuous Lorentzian manifold; specifically, we restrict ourselves to the cases in which the set of rewrite events ${\mathcal{C}}$ is a proper subset of an uncountable set of spacetime events ${\mathcal{C} \subset \mathcal{M}}$, where ${\left( \mathcal{M}, g \right)}$ is a continuous Lorentzian manifold. In such cases, there will exist a \textit{faithful embedding}\cite{bombelli} of the set ${\mathcal{C}}$ within ${\mathcal{M}}$, i.e. there will exist an injective embedding map ${\Phi}$:

\begin{equation}
\Phi : \mathcal{C} \to \left( \mathcal{M}, g \right), \qquad \text{ such that } \forall x, y \in \mathcal{C}, \qquad x \prec_{\mathcal{C}} y \Leftrightarrow \Phi \left( x \right) \prec_{\mathcal{M}} \Phi \left( y \right),
\end{equation}
where ${\prec_{\mathcal{C}}}$ and ${\prec_{\mathcal{M}}}$ denote the standard causal partial order relations on the sets ${\mathcal{C}}$ and ${\mathcal{M}}$, respectively. The embedding ${\Phi : \mathcal{C} \to \mathcal{M}}$ is then \textit{faithful} in the sense that the points in the image ${\Phi \left( \mathcal{C} \right)}$ are uniformly distributed (with respect to the continuous Lorentzian volume measure on ${\left( \mathcal{M}, g \right)}$), with density:

\begin{equation}
\rho_c = V_{c}^{-1},
\end{equation}
where ${V_c}$ has an interpretation as a characteristic discreteness scale of spacetime (i.e. a volume cut-off below which the continuum approximation of spacetime is no longer valid). From here, it has been conjectured (although not yet proved) that a single causal graph whose set of rewrite events is ${\mathcal{C}}$ can be faithfully embedded (assuming a fixed density ${\rho_c}$) into a pair of distinct Lorentzian manifolds, denoted ${\left( \mathcal{M}, g \right)}$ and ${\left( \mathcal{M}^{\prime}, g^{\prime} \right)}$, if and only if the manifolds ${\left( \mathcal{M}, g \right)}$ and ${\left( \mathcal{M}^{\prime}, g^{\prime} \right)}$ are themselves \textit{approximately isometric}\cite{bombelli2}, meaning intuitively that their metric differs only at scales smaller than that of the characteristic discreteness volume ${V_c}$ scale. We can formalize the notion of \textit{approximate isometry} as a map ${f : \mathcal{M} \to \mathcal{M}}$ (where ${\left( \mathcal{M}, g \right)}$ is a Lorentzian manifold) which distorts distances $d$ between pairs of points $x$ and $y$ by no more than some bounded ${\epsilon > 0}$:

\begin{equation}
\forall x, y \in \mathcal{M}, \qquad \left\lvert d \left( f \left( x \right), f \left( y \right) \right) - d \left( x, y \right) \right\rvert \leq \epsilon.
\end{equation}

\begin{figure}[ht]
\centering
\includegraphics[width=0.495\textwidth]{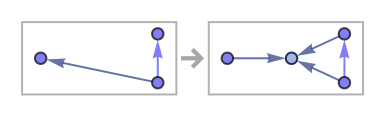}
\caption{A hypergraph transformation rule, represented internally as a set substitution system ${\left\lbrace \left\lbrace x, y \right\rbrace, \left\lbrace x, z \right\rbrace \right\rbrace \to \left\lbrace \left\lbrace x, y \right\rbrace, \left\lbrace x, w \right\rbrace, \left\lbrace y, w \right\rbrace, \left\lbrace z, w \right\rbrace \right\rbrace}$. Example taken from \cite{wolfram2}.}
\label{fig:Figure56}
\end{figure}

\begin{figure}[ht]
\centering
\includegraphics[width=0.695\textwidth]{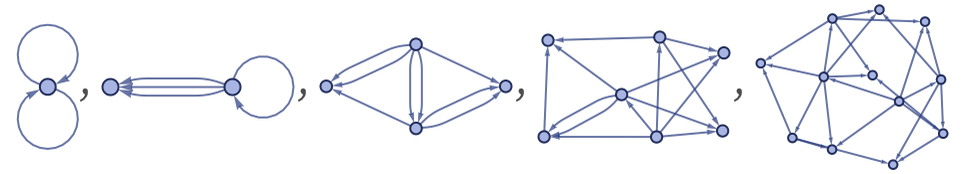}
\caption{The hypergraphs obtained during the first four steps of evolution of the set substitution system ${\left\lbrace \left\lbrace x, y \right\rbrace, \left\lbrace x, z \right\rbrace \right\rbrace \to \left\lbrace \left\lbrace x, y \right\rbrace, \left\lbrace x, w \right\rbrace, \left\lbrace y, w \right\rbrace, \left\lbrace z, w \right\rbrace \right\rbrace}$, assuming a simple initial condition consisting of a single vertex with two self-loops. Example taken from \cite{wolfram2}.}
\label{fig:Figure57}
\end{figure}

\begin{figure}[ht]
\centering
\includegraphics[width=0.595\textwidth]{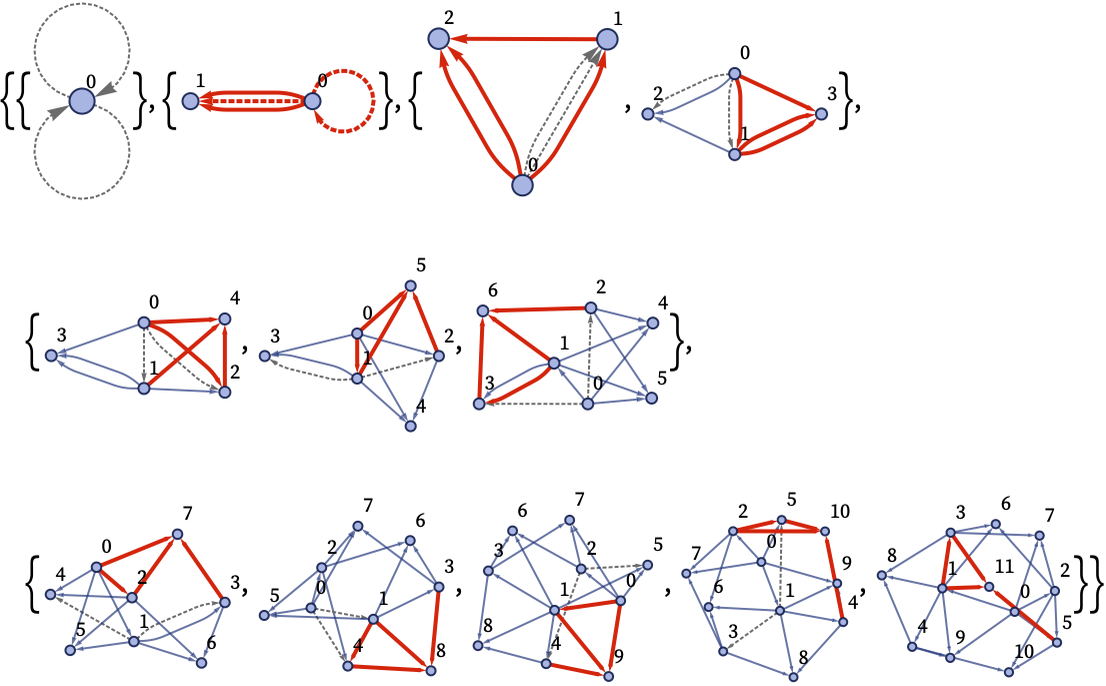}
\caption{The non-overlapping (and therefore spacelike-separated) subhypergraphs involved in the rewrite events applied during the first four steps of evolution of the set substitution system ${\left\lbrace \left\lbrace x, y \right\rbrace, \left\lbrace x, z \right\rbrace \right\rbrace \to \left\lbrace \left\lbrace x, y \right\rbrace, \left\lbrace x, w \right\rbrace, \left\lbrace y, w \right\rbrace, \left\lbrace z, w \right\rbrace \right\rbrace}$, assuming a simple initial condition consisting of a single vertex with two self-loops. Example taken from \cite{wolfram2}.}
\label{fig:Figure58}
\end{figure}

\begin{figure}[ht]
\centering
\includegraphics[width=0.695\textwidth]{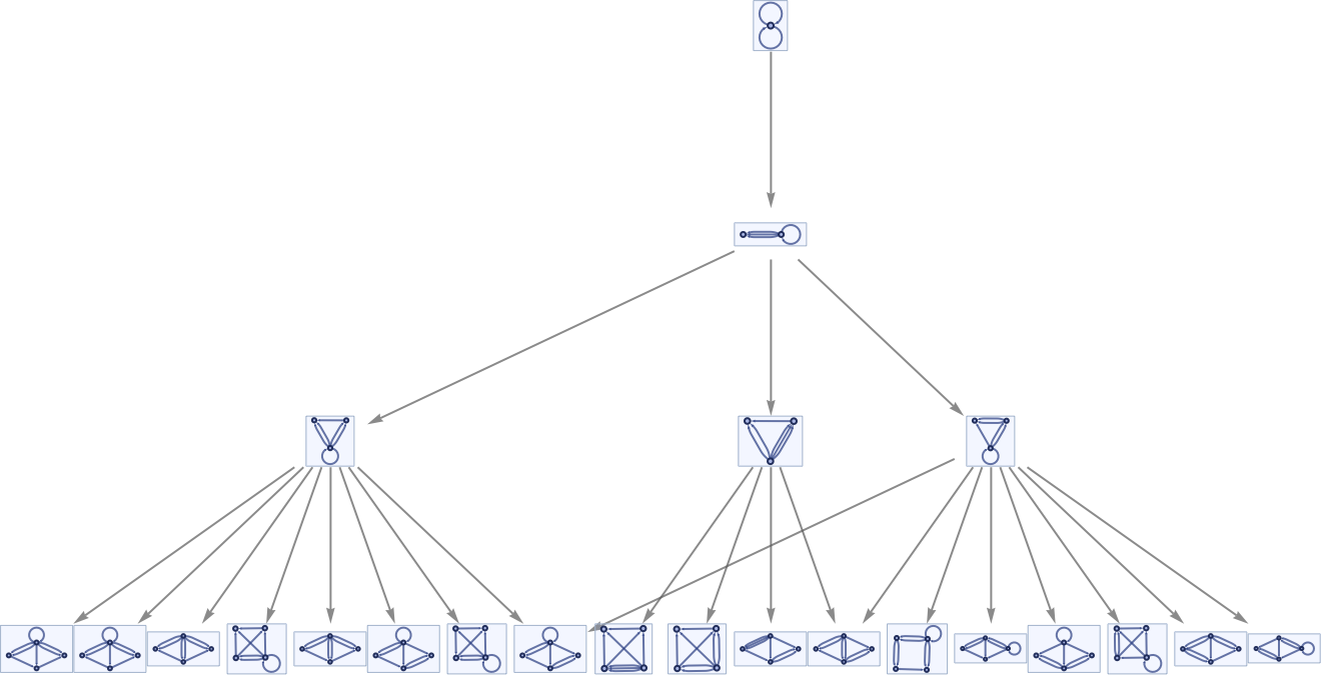}
\caption{The multiway evolution graph obtained via the non-deterministic evolution of the set substitution system ${\left\lbrace \left\lbrace x, y \right\rbrace, \left\lbrace x, z \right\rbrace \right\rbrace \to \left\lbrace \left\lbrace x, y \right\rbrace, \left\lbrace x, w \right\rbrace, \left\lbrace y, w \right\rbrace, \left\lbrace z, w \right\rbrace \right\rbrace}$, assuming a simple initial condition consisting of a single vertex with two self-loops. Example taken from \cite{wolfram2}.}
\label{fig:Figure59}
\end{figure}

\begin{figure}[ht]
\centering
\includegraphics[width=0.495\textwidth]{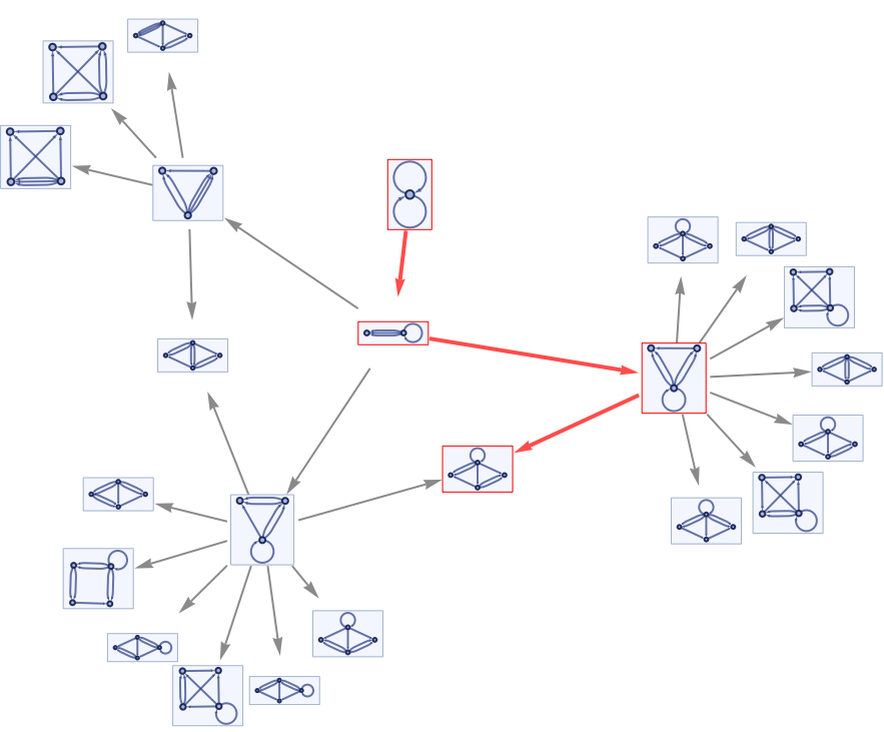}
\caption{The default evaluation order for the set substitution system ${\left\lbrace \left\lbrace x, y \right\rbrace, \left\lbrace x, z \right\rbrace \right\rbrace \to \left\lbrace \left\lbrace x, y \right\rbrace, \left\lbrace x, w \right\rbrace, \left\lbrace y, w \right\rbrace, \left\lbrace z, w \right\rbrace \right\rbrace}$, represented as a single path in the multiway evolution graph for the same system. Example taken from \cite{wolfram2}.}
\label{fig:Figure60}
\end{figure}

\begin{figure}[ht]
\centering
\includegraphics[width=0.345\textwidth]{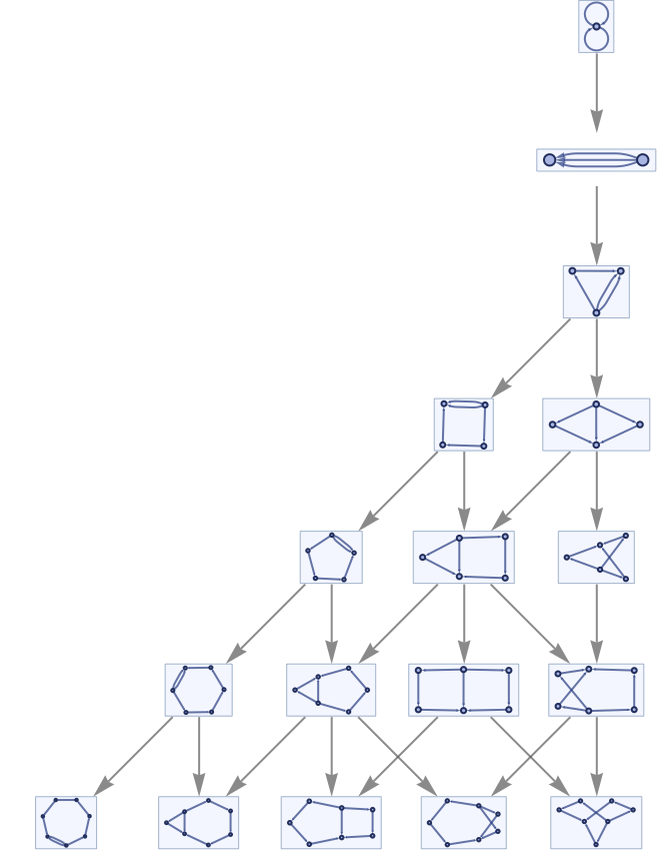}
\caption{The multiway evolution graph obtained via the non-deterministic evolution of a (globally) confluent set substitution system ${\left\lbrace \left\lbrace x, y \right\rbrace, \left\lbrace z, y \right\rbrace \right\rbrace \to \left\lbrace \left\lbrace x, w \right\rbrace, \left\lbrace y, w \right\rbrace, \left\lbrace z, w \right\rbrace \right\rbrace}$, with the property that every bifurcation converges after a single step. Example taken from \cite{wolfram2}.}
\label{fig:Figure61}
\end{figure}

\begin{figure}[ht]
\centering
\includegraphics[width=0.345\textwidth]{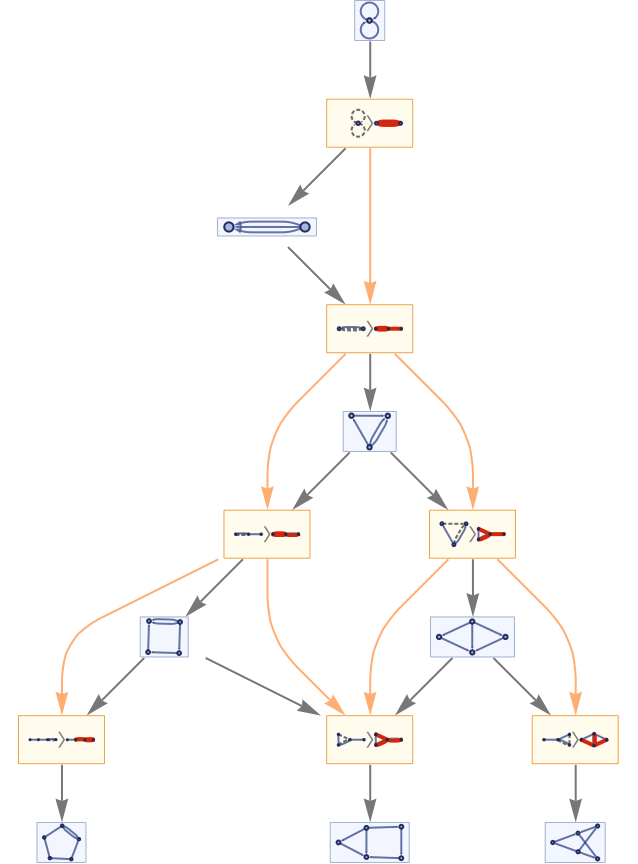}
\caption{The multiway evolution causal graph (showing evolution edges in gray and causal edges in orange) obtained via the non-determinsitic evolution of a (globally) confluent set substitution system ${\left\lbrace \left\lbrace x, y \right\rbrace, \left\lbrace z, y \right\rbrace \right\rbrace \to \left\lbrace \left\lbrace x, w \right\rbrace, \left\lbrace y, w \right\rbrace, \left\lbrace z, w \right\rbrace \right\rbrace}$. Example taken from \cite{wolfram2}.}
\label{fig:Figure62}
\end{figure}

Since each possible choice of evaluation order for the hypergraph corresponds to a different possible choice of maximally non-overlapping (i.e. spacelike-separated) subhypergraphs to which the rewrite events can be applied, as illustrated by the examples in Figures \ref{fig:Figure56}, \ref{fig:Figure57} and \ref{fig:Figure58}, we can describe the family of all possible such choices by means of a directed, acyclic graph known as a \textit{multiway evolution graph}. Each vertex in the multiway evolution graph ${G_{multiway}}$ is a particular hypergraph state, and the directed edge ${A \to B}$ exists if and only if there exists a rewrite event which transforms hypergraph $A$ to hypergraph $B$. The default choice of evaluation order, as shown above, therefore corresponds to a particular path in the multiway evolution graph, as demonstrated in Figures \ref{fig:Figure59} and \ref{fig:Figure60}, with vertices merged on the basis of hypergraph isomorphism, using a slight generalization of the algorithm presented in \cite{gorard5}. To avoid having to deal with the added complexities of multiway evolution, for the purposes of this article we shall consider only Wolfram model system that are (globally) \textit{confluent}\cite{dershowitz}\cite{huet}, meaning that, if ${\to}$ denotes the binary rewrite relation and $A$ denotes the set of multiway states, then:

\begin{equation}
\forall a, b, c \in A , \qquad \text{ such that } a \to^{*} b \text{ and } a \to^{*} c, \qquad \exists d \in A \text{ such that } b \to^{*} d \text{ and } c \to^{*} d,
\end{equation}
where ${\to^{*}}$ corresponds to the reflexive transitive closure of ${\to}$, i.e. the smallest binary relation containing ${\to}$ and also satisfying the axioms of reflexivity and transitivity:

\begin{equation}
\forall a, b, c \in A, \qquad a \to^{*} a, \qquad \text{ and } \qquad a \to^{*} b, b \to^{*} c \implies a \to^{*} c.
\end{equation}
This confluence condition is generally necessary (though not sufficient) to guarantee \textit{causal invariance} of the evolution, namely the condition that the causal graphs associated with all possible evaluation orders are isomorphic as directed, acyclic graphs. An example of a (globally) confluent set substitution system is shown in Figure \ref{fig:Figure61}, and its associated \textit{multiway evolution causal graph} (with rewrite events shown in yellow, state vertices shown in blue, evolution edges shown in gray and causal edges shown in orange), demonstrating the phenomenon of causal invariance, is shown in Figure \ref{fig:Figure62}.

\begin{figure}[ht]
\centering
\includegraphics[width=0.545\textwidth]{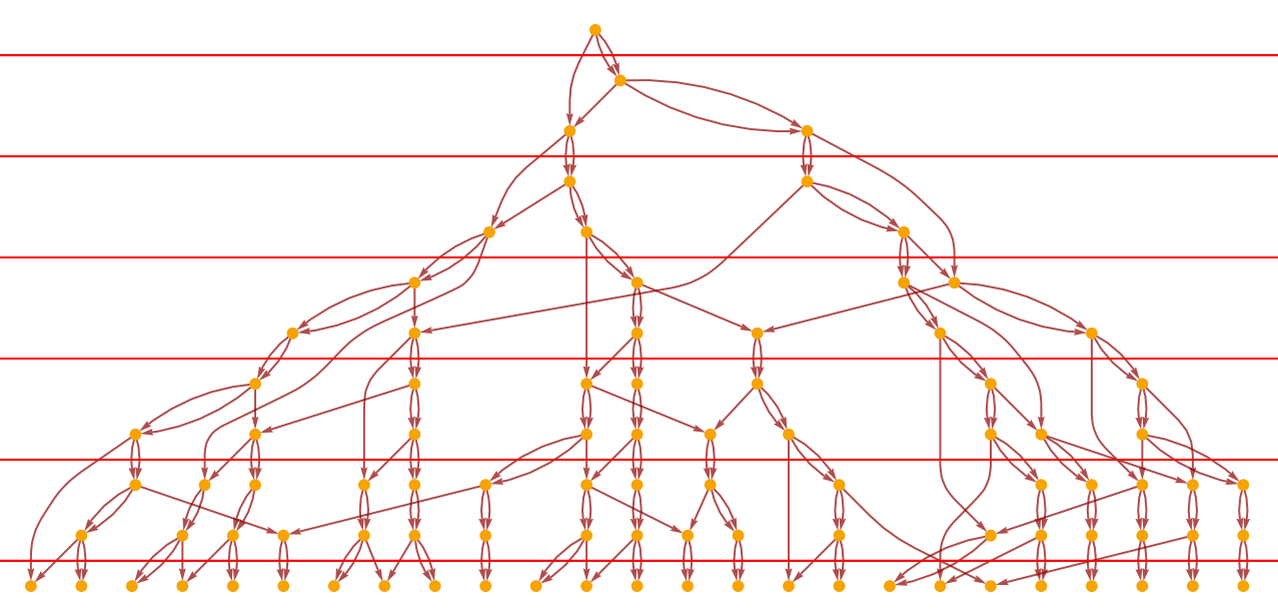}
\includegraphics[width=0.445\textwidth]{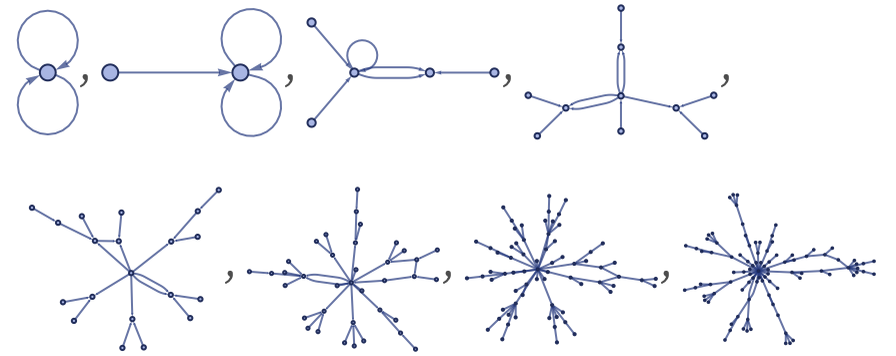}
\caption{The default gauge choice for the causal graph obtained via the evolution of the set substitution system ${\left\lbrace \left\lbrace x, y \right\rbrace, \left\lbrace z, y \right\rbrace \right\rbrace \to \left\lbrace \left\lbrace x, z \right\rbrace, \left\lbrace y, w \right\rbrace, \left\lbrace w, z \right\rbrace \right\rbrace}$, corresponding to the default choice of evaluation order. Example taken from \cite{wolfram2}.}
\label{fig:Figure63}
\end{figure}

\begin{figure}[ht]
\centering
\includegraphics[width=0.545\textwidth]{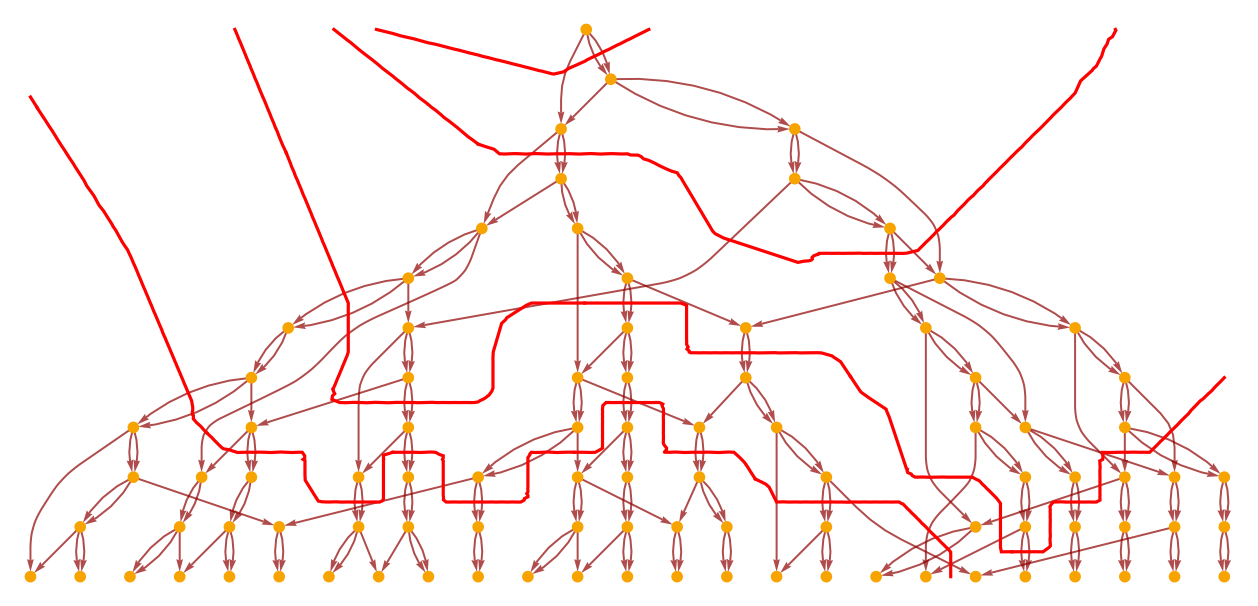}
\includegraphics[width=0.445\textwidth]{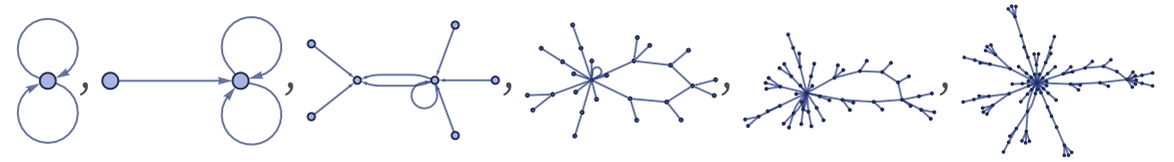}
\caption{An alternative possible gauge choice for the causal graph obtained via the evolution of the set substitution system ${\left\lbrace \left\lbrace x, y \right\rbrace, \left\lbrace z, y \right\rbrace \right\rbrace \to \left\lbrace \left\lbrace x, z \right\rbrace, \left\lbrace y, w \right\rbrace, \left\lbrace w, z \right\rbrace \right\rbrace}$, corresponding to a different possible choice of evaluation order. Example taken from \cite{wolfram2}.}
\label{fig:Figure64}
\end{figure}

The relevance of this causal invariance property for our present purposes lies in the fact that there exists a bijective correspondence between gauge choices in general relativity and evaluation orders on hypergraphs, since in order to specify a particular set of spacelike-separated rewrite events to be applied simultaneously at each evolution step, it suffices to define a non-degenerate \textit{universal time function}, denoted $t$, mapping rewrite events to integers\cite{gorard}\cite{gorard3}:

\begin{equation}
t : \mathcal{C} \to \mathbb{Z}, \qquad \text{ such that } \qquad \Delta t \neq 0 \text{ everywhere},
\end{equation}
with ${\mathcal{C}}$ denoting, as usual, the discrete set of rewrite events, and where the level sets of $t$ effectively \textit{foliate} the causal graph into a non-intersecting collection of spacelike hypersurfaces:

\begin{equation}
\forall t_1, t_2 \in \mathbb{Z}, \qquad \Sigma_{t_1} = \left\lbrace p \in \mathcal{C} : t \left( p \right) = t_1 \right\rbrace, \qquad \text{ and } \qquad \Sigma_{t_1} \cap \Sigma_{t_2} = \emptyset \Leftrightarrow t_1 \neq t_2.
\end{equation}
Each such level set corresponds to a particular hypergraph configuration, and forms the discrete analog of a Cauchy surface in Lorentzian geometry. More formally, if the discrete future/past Cauchy development of a set of rewrite events ${S \in \mathcal{C}}$, denoted ${D^{\pm} \left( S \right)}$, defines the set of all rewrite events for which every past/future-directed, inextendible path through the causal graph intersects at least one event in $S$, then the discrete Cauchy development of the set ${S \in \mathcal{C}}$, denoted ${D \left( S \right)}$, becomes the union of the discrete future and past Cauchy developments:

\begin{equation}
D \left( S \right) = D^{+} \left( S \right) \cup D^{-} \left( S \right).
\end{equation}
Then, a discrete Cauchy surface in our context is an achronal set of rewrite events ${S \in \mathcal{C}}$ whose discrete Cauchy development ${D \left( S \right)}$ is equal to ${\mathcal{C}}$, where achronality refers to the constraint that $S$ is disjoint from its own chronological future:

\begin{equation}
\nexists q, r \in S, \qquad \text{ such that } \qquad r \in I^{+} \left( q \right).
\end{equation}
In fact, since the standard hypergraph distance ${\Delta l}$ (i.e. the number of hyperedges that must be traversed when constructing a path from one vertex to another) can be computed between any pair of spacelike-separated rewrite events on a hypergraph, each spacelike hypersurface is equipped with a natural spatial metric tensor ${\gamma_{i j}}$:

\begin{equation}
\Delta l^2 = \gamma_{i j} \Delta x^i \Delta x^j,
\end{equation}
for some choice of assignments of spatial coordinates ${x^i \left( t \right)}$ to the vertices. Similarly, a directed graph distance ${\Delta \tau}$ in the causal graph can be computed between any pair of timelike-separated rewrite events, which, assuming that these events lie in spacelike hypersurfaces whose values of the universal time function correspond to $t$ and ${t + \Delta t}$, respectively, also induces a natural discrete gauge variable ${\alpha}$:

\begin{equation}
\Delta \tau = \alpha \Delta t.
\end{equation}
By rewriting the spatial coordinates on the hypergraph between the $t$ and ${t + \Delta t}$ hypersurfaces in accordance with the scheme:

\begin{equation}
x^i \left( t + \Delta t \right) = x^i \left( t \right) - \beta^i \left( t \right),
\end{equation}
for the family of discrete gauge variables ${\beta^i}$, we thus obtain a discrete form the spacetime line element for the ${\left( 3 + 1 \right)}$ ADM decomposition of the metric, namely:

\begin{equation}
\Delta s^2 = \left( - \alpha^2 + \beta^i \beta_i \right) \Delta t^2 + 2 \beta_i \Delta x^i \Delta t + \gamma_{i j} \Delta x^i \Delta x^j.
\end{equation}
This elegant correspondence between gauge choices on causal graphs and evaluation orders on hypergraphs is illustrated explicitly by the examples shown in Figures \ref{fig:Figure63} and \ref{fig:Figure64}.

From here, one is able to probe certain features of the limiting Riemannian geometry of hypergraphs (as well as the limiting Lorentzian geometry of causal graphs), such as limiting dimension and discrete curvature, by investigating the growth rates of small geodesic balls, tubes and cones. Consider the analogy with a $d$-dimensional Riemannian manifold ${\left( \mathcal{M}, g \right)}$, in which we can expand out the infinitesimal metric volume element ${d \mu_g}$ at a point ${p \in \mathcal{M}}$ as the square root of the metric tensor determinant\cite{jost}\cite{gray}:

\begin{equation}
d \mu_g \left( p \right) = \sqrt{\mathrm{det} \left( g \left( p \right) \right)},
\end{equation}
such that, assuming that the underlying manifold ${\left( \mathcal{M}, g \right)}$ is analytic, we can write the volume element at a nearby point ${p + \delta x}$ as a power series in ${\delta x}$:

\begin{equation}
\sqrt{\mathrm{det} \left( g \left( p + \delta x \right) \right)} = \sqrt{\mathrm{det} \left( g \left( p \right) \right)} \left( 1 - \frac{1}{6} \sum_{i = 1}^{d} R_{i j} \left( p \right) \delta x^i \delta x^j + O \left( \delta x^3 \right) + \dots \right).
\end{equation}
Here, ${R_{i j}}$ denotes the usual Riemannian Ricci curvature tensor, and ${\delta x^i}$, ${\delta x^j}$ correspond to a pair of covariant vectors representing the orthogonal components of the perturbation ${\delta x}$. From here, we can integrate over a ball ${B_{\epsilon} \left( p \right)}$ centered at point $p$, and with an infinitesimal radius ${\epsilon}$, in order to obtain the usual volume formula for a small geodesic ball:

\begin{equation}
\mathrm{Vol} \left( B_{\epsilon} \left( p \right) \right) = \int_{B_{\epsilon} \left( p \right)} \sqrt{\mathrm{det} \left( g \left( p + \delta x \right) \right)} d^d \left( \delta x \right),
\end{equation}
thus yielding, up to second-order in ${\epsilon}$:

\begin{equation}
\int_{B_{\epsilon} \left( p \right)} \sqrt{\mathrm{det} \left( g \left( p + \delta x \right) \right)} d^d \left( \delta x \right) = \frac{\pi^{\frac{d}{2}}}{\Gamma \left( \frac{d}{2} + 1 \right)} \epsilon^d \left( 1 - \frac{\epsilon^2}{6 \left( d + 2 \right)} \sum_{i = 1}^{d} R_{i}^{i} + O \left( \epsilon^4 \right) \right),
\end{equation}
in which $R$ denotes the usual Riemannian Ricci scalar, which we can write in trace form as:

\begin{equation}
\sum_{i = 1}^{d} R_{i}^{i} = R.
\end{equation}
Therefore, the growth rate of a small geodesic ball in a hypergraph allows us effectively to probe the value of its limiting Ricci scalar\cite{wolfram2}\cite{gorard}. Likewise, if we consider instead integrating over a tube ${T_{\epsilon, \delta x} \left( p \right)}$, starting at point $p$ and extending along a geodesic oriented in direction ${\delta x}$ over an infinitesimal distance ${\delta}$, and with infinitesimal radius ${\epsilon}$, then we obtain the corresponding volume formula for a small geodesic tube:

\begin{equation}
\mathrm{Vol} \left( T_{\epsilon, \delta x} \left( p \right) \right) = \frac{\pi^{\frac{d - 1}{2}}}{\Gamma \left( \frac{d + 1}{2} \right)} \epsilon^{d - 1} \delta x \left( 1 - \left( \frac{d - 1}{d + 1} \right) \left( R - \sum_{i, j = 1}^{d} R_{i j} \hat{\delta x}^{i} \hat{\delta x}^{j} \right) \epsilon^2 + O \left( \epsilon^3 + \epsilon^2 \delta x \right) + \dots \right),
\end{equation}
where now ${\hat{\delta x}^i}$, ${\hat{\delta x}^j}$ denote components of the unit vectors along the geodesic. As such, the growth rate of a small geodesic tube in a hypergraph allows us effectively to probe higher-order contractions of its limiting Riemann curvature tensor\cite{gorard3}.

\begin{figure}[ht]
\centering
\includegraphics[width=0.895\textwidth]{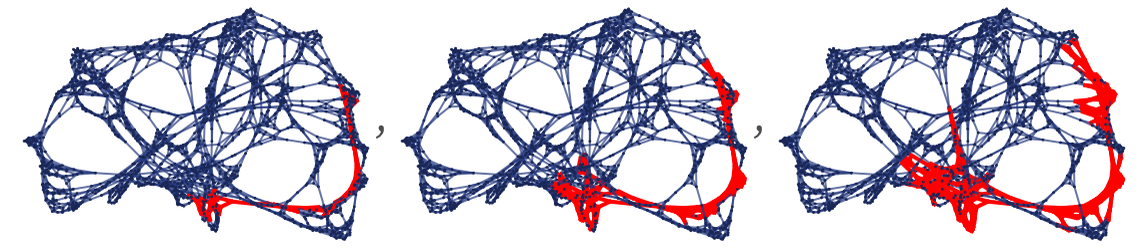}
\caption{Infinitesimal geodesic ``tubes'' of varying radii (1, 3 and 5, respectively) embedded in the hypergraph obtained by the evolution of the set substitution system ${\left\lbrace \left\lbrace x, y \right\rbrace, \left\lbrace x, z \right\rbrace \right\rbrace \to \left\lbrace \left\lbrace x, z \right\rbrace, \left\lbrace x, w \right\rbrace, \left\lbrace y, w \right\rbrace, \left\lbrace z, w \right\rbrace \right\rbrace}$. Example taken from \cite{wolfram2}.}
\label{fig:Figure65}
\end{figure}

To make this more explicit, we can exploit the usual geometric intuition for the Riemannian Ricci curvature scalar $R$, namely as a measure of the discrepancy between the volume of an infinitesimal geodesic ball in a manifold ${\left( \mathcal{M}, g \right)}$ of dimension $d$ and the volume of a ball of equivalent radius in flat (Euclidean) space ${\mathbb{R}^d}$:

\begin{equation}
\frac{\mathrm{Vol} \left( B_{\epsilon} \left( p \right) \subset \mathcal{M} \right)}{\mathrm{Vol} \left( B_{\epsilon} \left( 0 \right) \subset \mathbb{R}^d \right)} = 1 - \frac{R}{6 \left( d + 2 \right)} \epsilon^2 + O \left( \epsilon^4 \right),
\end{equation}
in the limit as the radius goes to zero, i.e. ${\epsilon \to 0}$, where we use ${B_{\epsilon} \left( p \right) \subset \mathcal{M}}$ to indicate a ball of radius ${\epsilon}$ in the Riemannian manifold ${\left( \mathcal{M}, g \right)}$, and ${B_{\epsilon} \left( 0 \right) \subset \mathbb{R}^d}$ to indicate a ball of radius ${\epsilon}$ in flat (Euclidean) space ${\mathbb{R}^d}$. We can equivalently reformulate this definition as a measure of the discrepancy between the distance ${\delta}$ between the centers $p$ and $q$ of two geodesic balls ${B_{\epsilon} \left( p \right) \subset \mathcal{M}}$ and ${B_{\epsilon} \left( q \right) \subset \mathcal{M}}$ (obtained before and after parallel transport, respectively), and the average distance $W$ between points on the surface of the ball ${B_{\epsilon} \left( p \right)}$ and corresponding points on the surface of the ball ${B_{\epsilon} \left( q \right)}$ after parallel transport:

\begin{equation}
W \left( B_{\epsilon} \left( p \right), B_{\epsilon} \left( q \right) \right) = \delta \left( 1 - \frac{\epsilon^2}{2 \left( d + 2 \right)} R + O \left( \epsilon^3 + \epsilon^2 \delta \right) \right),
\end{equation}
in the limit as both the radii and the parallel transport distance go to zero, i.e. ${\epsilon, \delta \to 0}$, where we have used the shorthand ${\delta = d \left( p, q \right)}$ as a shorthand to designate the distance between the two centers $p$ and $q$. From here, we can build on the prior work of Forman\cite{forman}, Ollivier\cite{ollivier}\cite{ollivier2}\cite{ollivier3}, and Eidi and Jost\cite{eidi}, amongst several others, in order to extend these purely geometrical definitions on Riemannian manifolds ${\left( \mathcal{M}, g \right)}$ to the more general case of arbitrary metric-measure spaces ${\left( X, d \right)}$ (which include, as an important special case, directed hypergraphs). In this generalization, we replace the notion of a Riemannian metric volume element ${\mu_g}$ by a probability measure ${m_x}$, and we generalize the concept of average distance between corresponding points on two balls (obtained before and after parallel transport) to the 1-Wasserstein (transportation) distance between the corresponding probability measures. Specifically, if ${\left( X, d \right)}$ designates a Polish metric space equipped with a Borel ${\sigma}$-algebra, then a family of probability of measures $m$:

\begin{equation}
m = \left\lbrace m_x : x \in X \right\rbrace,
\end{equation}
such that each of the ${m_x}$ has a finite first moment, and such that the map ${m \to m_x}$ is measurable, can be interpreted as a \textit{random walk} on the space $X$. The \textit{1-Wasserstein distance}, denoted ${W_1 \left( m_x, m_y \right)}$, between the probability measures ${m_x}$ and ${m_y}$ defined on the metric space $X$, is then the optimal transport distance between these measures:

\begin{equation}
W_1 \left( m_x, m_y \right) = \inf_{\epsilon \in \Pi \left( m_x, m_y \right)} \left[ \int_{\left( x, y \right) \in X \times Y} d \left( x, y \right) d \epsilon \left( x, y \right) \right],
\end{equation}
in which ${\Pi \left( m_x, m_y \right)}$ designates the set of measures defined on the metric product space ${X \times Y}$ that project onto ${m_x}$ and ${m_y}$ (or, equivalently, the coupling between random walks projecting to ${m_x}$ and ${m_y}$). More informally, if ${\Pi \left( m_x, m_y \right)}$ signifies the set of possible transportations of the measure ${m_x}$ over to point $y$, then the 1-Wasserstein distance quantifies the minimum cost of ``disassembling'' the measure at point $x$, transporting it to point $y$, and reassembling it at point $y$.

For a metric space ${\left( X, d \right)}$ equipped with a random walk $m$, we can therefore define the \textit{Ollivier-Ricci curvature scalar} ${\kappa}$ in direction ${\left( x, y \right)}$ by analogy with the Riemannian case, namely as the discrepancy between the 1-Wasserstein distance between measures ${m_x}$ and ${m_y}$ and the regular metric distance between points $x$ and $y$:

\begin{equation}
\kappa \left( x, y \right) = 1 - \frac{W_1 \left( m_x, m_y \right)}{d \left( x, y \right)},
\end{equation}
assuming that points ${x, y \in X}$ are distinct. In the case where ${\left( X, d \right)}$ reduces to a Riemannian manifold ${\left( \mathcal{M}, g \right)}$, then the probability measure $m$ reduces to the usual Riemannian volume measure ${\mathrm{Vol}}$, and so the Ollivier-Ricci curvature scalar ${\kappa \left(p, q \right)}$ reduces to the usual Riemannian Ricci curvature scalar ${R \left( p, q \right)}$ (up to a multiplicative constant), hence providing a formal justification for our prior geometrical intuition. For the purpose of analyzing Wolfram model systems, however, we are interested in the case in which ${\left( X, d \right)}$ is a discrete metric-measure space, and therefore in which the discrete 1-Wasserstein transportation distance ${W_1 \left( m_x, m_y \right)}$ between discrete probability measures ${m_x}$ and ${m_y}$ defined on the discrete metric space $X$ reduces to the multi-marginal (discrete) optimal transportation distance between these measures:

\begin{equation}
W_1 \left( m_x, m_y \right) = \inf_{\mu_{x, y} \in \Pi \left( m_x, m_y \right)} \left[ \sum_{\left( x^{\prime}, y^{\prime} \right) \in X \times X} d \left( x^{\prime}, y^{\prime} \right) \mu_{x, y} \left( x^{\prime}, y^{\prime} \right) \right],
\end{equation}
where ${\Pi \left( m_x, m_y \right)}$ here designates the set of discrete probability measures ${\mu_{x, y}}$ that satisfy the coupling conditions:

\begin{equation}
\sum_{y^{\prime} \in X} \mu_{x, y} \left( x^{\prime}, y^{\prime} \right) = m_x \left( x^{\prime} \right), \qquad \text{ and } \qquad \sum_{x^{\prime} \in X} \mu_{x, y} \left( x^{\prime}, y^{\prime} \right) = m_y \left( y^{\prime} \right).
\end{equation}
For a directed hypergraph ${H = \left( V, E \right)}$, we shall assume for the sake of simplicity that each hyperedge ${e \in E}$ corresponds to a directional relation between two independent sets of vertices $A$ and $B$, which we refer to as the \textit{tail set} and the \textit{head set}, respectively. In this case, the discrete 1-Wasserstein transportation distance ${W_1 \left( \mu_{A^{in}}, \mu_{B^{out}} \right)}$ between discrete probability measures ${\mu_{A^{in}}}$ and ${\mu_{B^{out}}}$ defined on the directed hypergraph ${H = \left( V, E \right)}$ reduces again to the multi-marginal (discrete) optimal transportation distance between these measures:

\begin{equation}
W_1 \left( \mu_{A^{in}}, \mu_{B^{out}} \right) = \min_{u \in A^{in} \left( u \to A \right), v \in B^{out} \left( B \to v \right)} \left[ \sum_{u \to A} \sum_{B \to v} d \left( u, v \right) \epsilon \left( u, v \right) \right],
\end{equation}
where now we are minimizing over the set of all couplings ${\epsilon}$ between the measures ${\mu_{A^{in}}}$ and ${\mu_{B^{out}}}$ that satisfy the coupling constraints:

\begin{equation}
\sum_{u \to A} \epsilon \left( u, v \right) = \sum_{j = 1}^{m} \mu_{y_j} \left( v \right), \qquad \text{ and } \qquad \sum_{B \to v} \epsilon \left( u, v \right) = \sum_{i = 1}^{n} \mu_{x_i} \left( u \right).
\end{equation}
In the above, ${d \left( u, v \right)}$ refers to the standard hypergraph distance metric (i.e. the minimum number of directed hyperedges that one must traverse in order to travel from vertex ${u \in A^{in} \left( u \to A \right)}$ to vertex ${v \in B^{out} \left( B \to v \right)}$), and the coupling ${\epsilon \left( u, v \right)}$ may be informally interpreted as the ``mass'' being moved from vertex $u$ to vertex $v$. By choosing coupling constants ${\epsilon \left( u, v \right) = 1}$ for all such vertices $u$ and $v$, such that each directed hyperedge corresponds to a single unit of spatial distance, we thus force the induced metric tensor to be torsion-free, and therefore we implicitly enforce the Levi-Civita connection.

\begin{figure}[ht]
\centering
\includegraphics[width=0.395\textwidth]{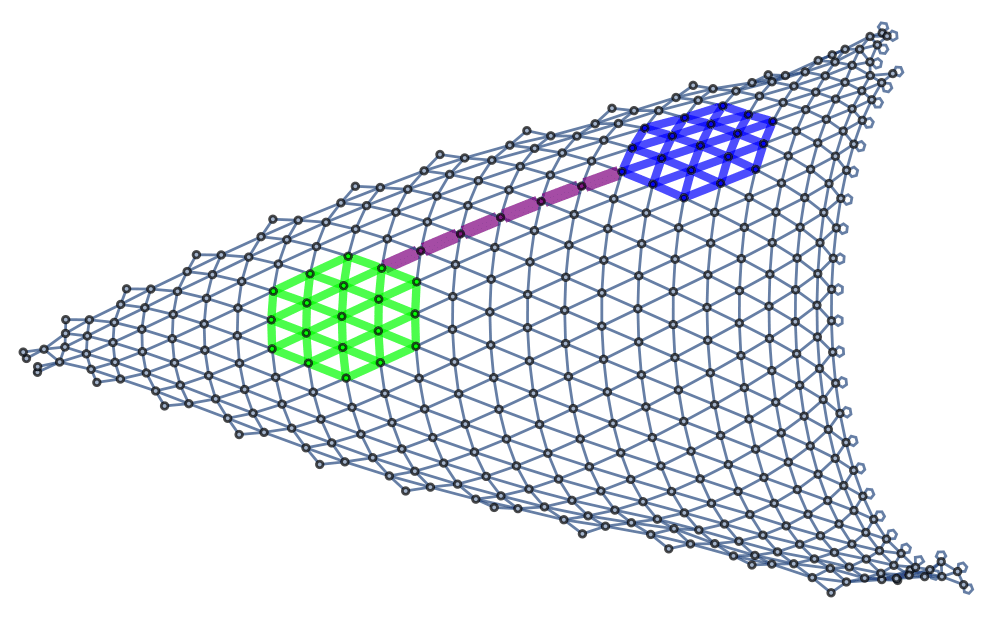}\hspace{0.1\textwidth}
\includegraphics[width=0.395\textwidth]{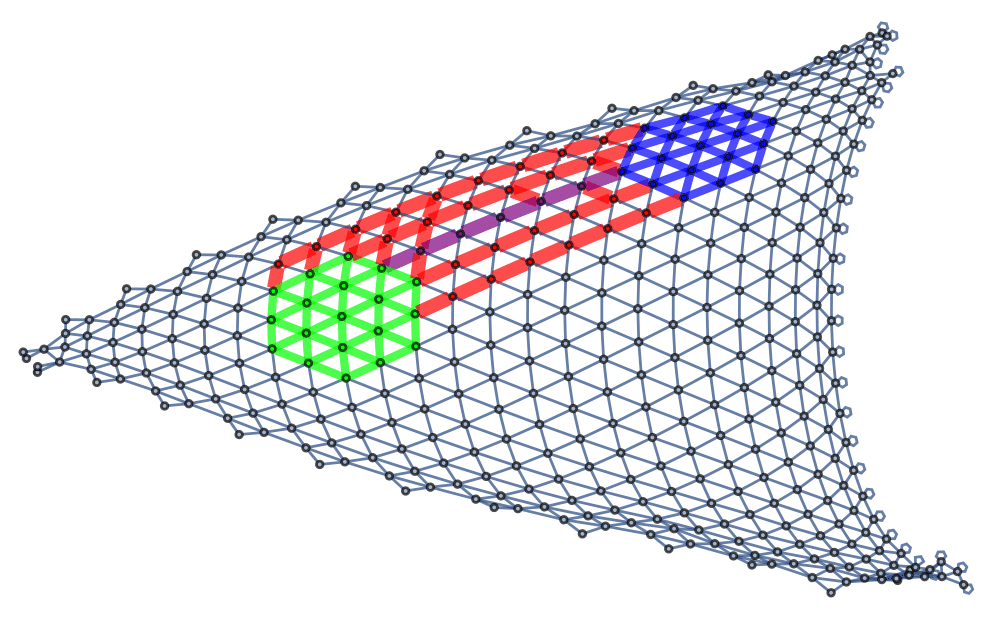}
\caption{On the left, a purple path showing the distance between the centers of two nearby finite geodesic balls embedded in an asymptotically-flat spatial hypergraph with the limiting structure of a two-dimensional Riemannian manifold, as obtained by the evolution of the set substitution system ${\left\lbrace \left\lbrace x, y, y \right\rbrace, \left\lbrace z, x, w \right\rbrace \right\rbrace \to \left\lbrace \left\lbrace y, v, y \right\rbrace, \left\lbrace y, z, v \right\rbrace, \left\lbrace w, v, v \right\rbrace \right\rbrace}$. On the right, a collection of red paths between corresponding points on the surfaces of the two balls after parallel transport. The discrete Ollivier-Ricci curvature scalar ${\kappa}$ is equal to zero along the purple path, since there is no net divergence or convergence of the red geodesics.}
\label{fig:Figure66}
\end{figure}

\begin{figure}[ht]
\centering
\includegraphics[width=0.395\textwidth]{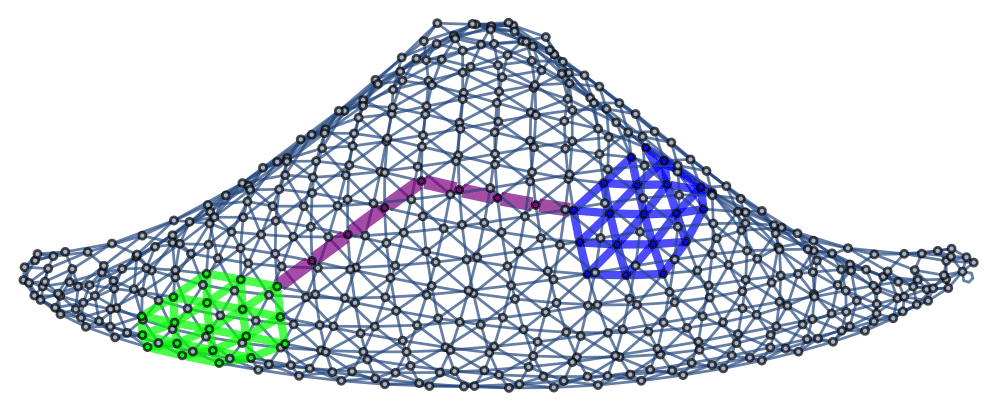}\hspace{0.1\textwidth}
\includegraphics[width=0.395\textwidth]{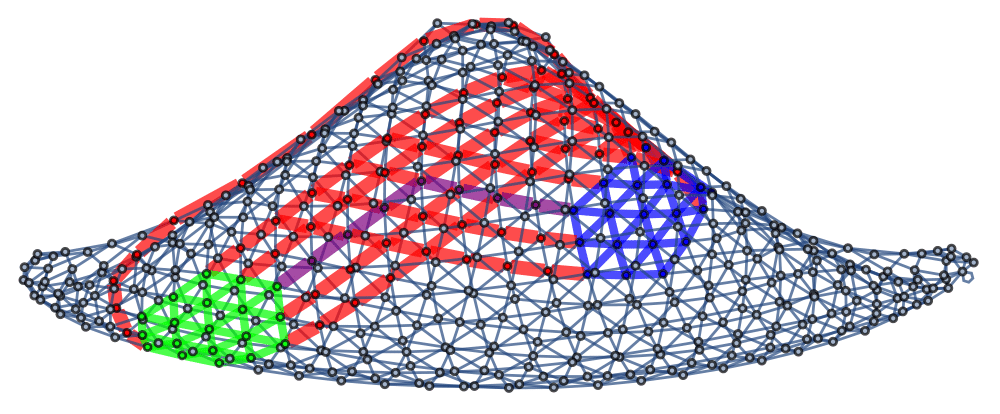}
\caption{On the left, a purple path showing the distance between the centers of two nearby finite geodesic balls embedded in a spatial hypergraph with the limiting structure of a two-dimensional Riemannian manifold with globally positive curvature, as obtained by the evolution of the set substitution system ${\left\lbrace \left\lbrace x, x, y \right\rbrace, \left\lbrace x, z, w \right\rbrace \right\rbrace \to \left\lbrace \left\lbrace w, w, v \right\rbrace, \left\lbrace v, w, y \right\rbrace, \left\lbrace z, y, v \right\rbrace \right\rbrace}$. On the right, a collection of red paths between corresponding points on the surfaces of the two balls after parallel transport. The discrete Ollivier-Ricci curvature scalar ${\kappa}$ is positive along the purple path, since there is a net divergence of the red geodesics.}
\label{fig:Figure67}
\end{figure}

\begin{figure}[ht]
\centering
\includegraphics[width=0.395\textwidth]{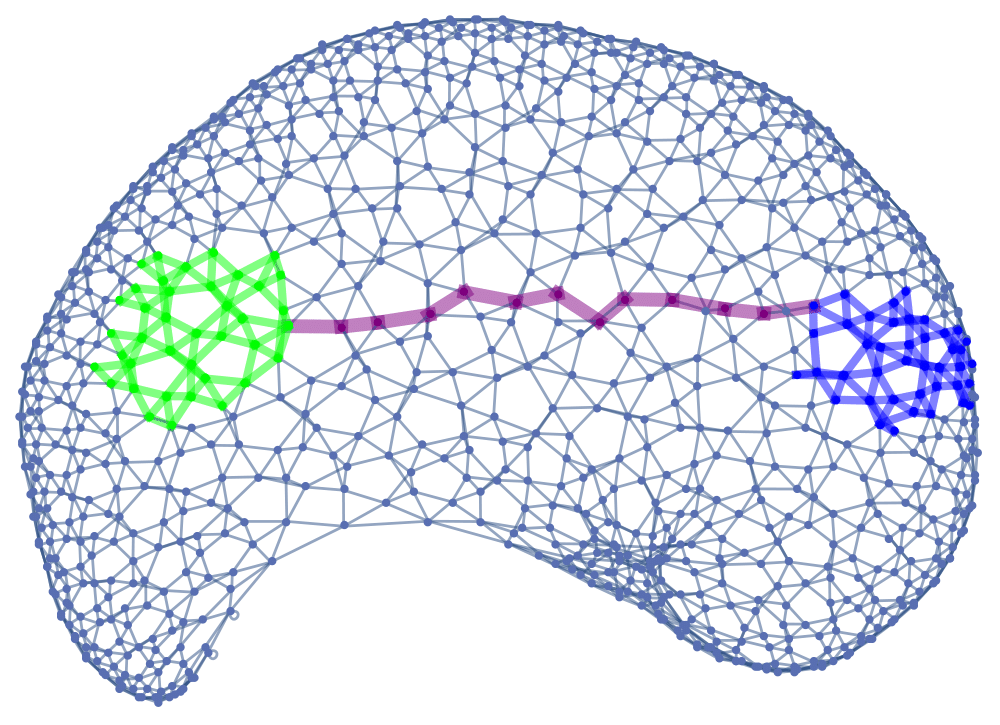}\hspace{0.1\textwidth}
\includegraphics[width=0.395\textwidth]{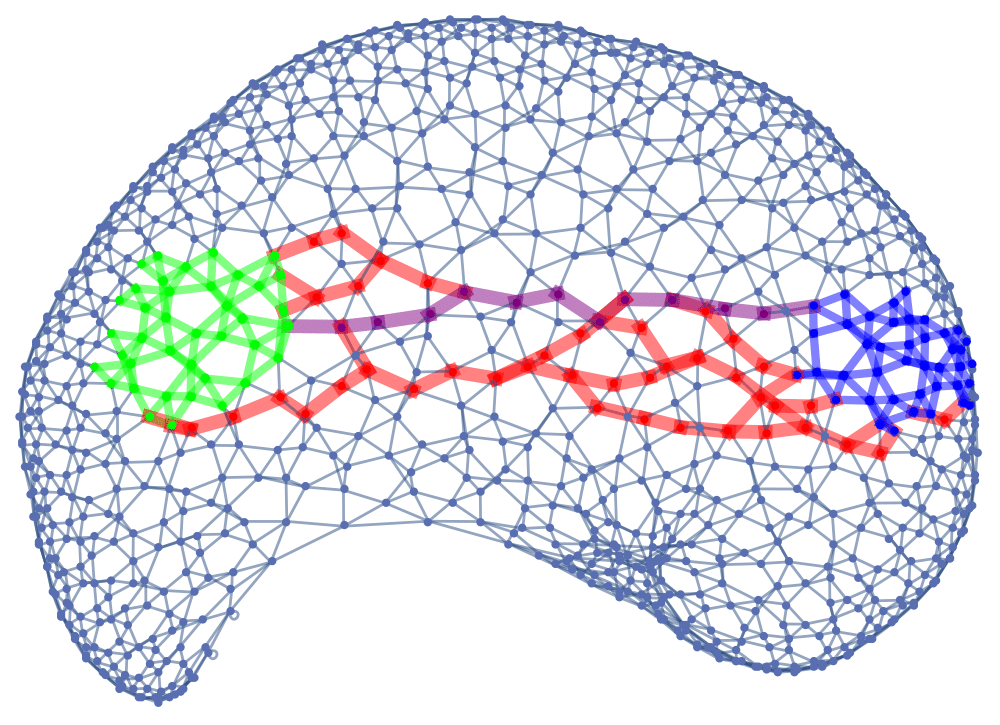}
\caption{On the left, a purple path showing the distance between the centers of two nearby finite geodesic balls embedded in a spatial hypergraph with the limiting structure of a two-dimensional Riemannian manifold with globally negative curvature, as obtained by the evolution of the set substitution system ${\left\lbrace \left\lbrace x, y, x \right\rbrace, \left\lbrace x, z, u \right\rbrace \right\rbrace \to \left\lbrace \left\lbrace u, v, u \right\rbrace, \left\lbrace v, u, z \right\rbrace, \left\lbrace x, y, v \right\rbrace \right\rbrace}$. On the right, a collection of red paths between corresponding points on the surfaces of the two balls after parallel transport. The discrete Ollivier-Ricci curvature scalar ${\kappa}$ is negative along the purple path, since there is a net convergence of the red geodesics.}
\label{fig:Figure68}
\end{figure}

This allows us to compute the Ollivier-Ricci curvature scalar ${\kappa}$ for a single directed hyperedge ${e \in E}$ of the form:

\begin{equation}
A = \left\lbrace x_1, \dots, x_n \right\rbrace \to^{e} B = \left\lbrace y_1, \dots, y_m \right\rbrace,
\end{equation}
where ${n, m \leq \left\lvert V \right\rvert}$, namely

\begin{equation}
\kappa \left( e \right) = 1 - W_1 \left( \mu_{A^{in}}, \mu_{B^{out}} \right),
\end{equation}
subject to the usual assumption that the probability measures ${\mu_{A^{in}}}$ and ${\mu_{B^{out}}}$ satisfy the coupling constraints:

\begin{equation}
\mu_{A^{in}} = \sum_{i = 1}^{n} \mu_{x_i}, \qquad \text{ and } \qquad \mu_{B^{out}} = \sum_{j = 1}^{m} \mu_{y_j}.
\end{equation}
By adopting the notational convention that, for any given vertex in the tail set ${x_i \in A}$, ${d_{x_{i}^{in}}}$ denotes the number of hyperedges that include ${x_i}$ as an element of their head set, and for any given vertex in the head set ${y_j \in B}$, ${d_{y_{j}^{out}}}$ denotes the number of hyperedges that include ${y_j}$ as an element of their tail set, then we can write the probability measures ${\mu_{x_i} \left( z \right)}$ and ${\mu_{y_j} \left( z \right)}$ more explicitly as:

\begin{equation}
\forall 1 \leq i \leq n, z \in V, \qquad \mu_{x_i} \left( z \right) = \begin{cases}
0, \qquad &\text{ if } z = x_i \text{ and } d_{x_{i}^{in}} \neq 0,\\
\frac{1}{n}, \qquad &\text{ if } z = x_i \text{ and } d_{x_{i}^{in}} = 0,\\
\sum\limits_{e^{\prime} : z \to x_i} \frac{1}{n \times d_{x_{i}^{in}} \times \left\lvert \mathrm{tail} \left( e^{\prime} \right) \right\rvert}, \qquad &\text{ if } z \neq x_i \text{ and } \exists e^{\prime} : z \to x_i,\\
0, \qquad &\text{ if } z \neq x_i \text{ and } \nexists e^{\prime} : z \to x_i,
\end{cases}
\end{equation}
and:

\begin{equation}
\forall 1 \leq j \leq m, z \in V, \qquad \mu_{y_j} \left( z \right) = \begin{cases}
0, \qquad &\text{ if } z = y_j \text{ and } d_{y_{j}^{out}} \neq 0,\\
\frac{1}{m}, \qquad &\text{ if } z = y_j \text{ and } d_{y_{j}^{out}} = 0,\\
\sum\limits_{e^{\prime} : y_j \to z} \frac{1}{m \times d_{y_{j}^{out}} \times \left\lvert \mathrm{head} \left( e^{\prime} \right) \right\rvert}, \qquad &\text{ if } z \neq y_j \text{ and } \exists e^{\prime} : y_j \to z,\\
0, \qquad &\text{ if } z \neq y_j \text{ and } \nexists e^{\prime} : y_j \to z.
\end{cases}
\end{equation}
Illustrative examples of how the discrete Ollivier-Ricci curvature scalar ${\kappa}$ is computed in hypergraphs whose limiting structures correspond to asymptotically-flat, asymptotically positively curved and asymptotically negatively curved Riemannian manifolds are shown in Figures \ref{fig:Figure66}, \ref{fig:Figure67} and \ref{fig:Figure68}, respectively.

If we wish to extend this analysis to encompass higher-order contractions of the Riemann curvature tensor, and in particular if we wish to compute projections of the Ricci curvature tensor (in both the Riemannian hypergraph case, and the Lorentzian causal graph case), we can begin by choosing a \textit{geodesic normal coordinate} system surrounding point ${p \in \mathcal{M}}$ in some Riemannian or Lorentzian manifold ${\left( \mathcal{M}, g \right)}$. More specifically, if ${\left( \mathcal{M}, g \right)}$ is a ${C^{\infty}}$ manifold, then for any point ${x \in \mathcal{M}}$, we can introduce a linear map:

\begin{equation}
D : C^{\infty} \left( \mathcal{M} \right) \to \mathbb{R},
\end{equation}
that satisfies the Leibniz identity:

\begin{equation}
\forall f, g \in C^{\infty} \left( \mathcal{M} \right), \qquad D \left( f g \right) = D \left( f \right) \cdot g \left( x \right) + f \left( x \right) \cdot D \left( g \right),
\end{equation}
where ${\cdot}$ denotes the pointwise product and ${+}$ denotes the sum of functions, hence forming a real associative algebra; moreover, in the above, a function ${f : \mathcal{M} \to \mathbb{R}}$ is assumed to be an element of ${C^{\infty} \left( \mathcal{M} \right)}$ if and only if:

\begin{equation}
\forall \varphi : U \to \mathbb{R}^n, \qquad \text{ the map } f \circ \varphi^{-1} : \varphi \left[ U \right] \subseteq \mathbb{R}^n \to \mathbb{R} \text{ is infinitely differentiable},
\end{equation}
where ${\varphi : U \to \mathbb{R}^n}$ is a coordinate chart for an open subset ${U \subset \mathcal{M}}$. If we additionally define the operations of addition and scalar multiplication, using:

\begin{equation}
\left( D_1 + D_2 \right) \left( f \right) = D_1 \left( f \right) + D_2 \left( f \right),
\end{equation}
and:

\begin{equation}
\left( \lambda \cdot D \right) \left( f \right) = \lambda \cdot D \left( f \right),
\end{equation}
respectively, then the maps ${D : C^{\infty} \left( \mathcal{M} \right) \to \mathbb{R}}$ become \textit{derivations} in the algebraic sense, and thus form a vector space which we interpret as the \textit{tangent space} of the manifold ${\left( \mathcal{M}, g \right)}$ at point ${x \in \mathcal{M}}$, denoted ${T_x \mathcal{M}}$. Geodesic normal coordinates then form the natural coordinate system within a \textit{normal neighborhood}, i.e. within a neighborhood ${U \in \mathcal{M}}$ of a point ${p \in \mathcal{M}}$ whose coordinate chart is of the form:

\begin{equation}
\varphi = E^{-1} \circ \exp_{p}^{-1} : U \to \mathbb{R}^n,
\end{equation}
such that ${\exp_p}$ acts as a diffeomorphism between the neighborhood ${U \in \mathcal{M}}$ and a proper neighborhood ${V \subset T_p \mathcal{M}}$ of the tangent space. In the above, the exponential map ${\exp_p}$ is defined as the map from the tangent space of ${\left( \mathcal{M}, g \right)}$ at ${p \in \mathcal{M}}$ back to the manifold ${\left( \mathcal{M}, g \right)}$ itself:

\begin{equation}
\exp_p : T_p \mathcal{M} \supset V \to \mathcal{M},
\end{equation}
while $E$ is a generic isomorphism between flat (Euclidean) space ${\mathbb{R}^n}$ and the tangent space of ${\left( \mathcal{M}, g \right)}$ at ${p \in \mathcal{M}}$:

\begin{equation}
E : \mathbb{R}^n \to T_p \mathcal{M}.
\end{equation}

With the geodesic normal coordinates at ${p \in \mathcal{M}}$ thus defined, the metric tensor now becomes approximately Euclidean:

\begin{equation}
g_{i j} = \delta_{i j} + O \left( \left\lVert \mathbf{x} \right\rVert^2 \right),
\end{equation}
with the Euclidean metric tensor given by the Kronecker delta function ${\delta_{i j}}$, allowing us to perform a Taylor expansion of the metric tensor along a radial geodesic in the normal coordinate system:

\begin{equation}
g_{i j} = \delta_{i j} - \frac{1}{3} R_{i j k l} x^k x^l + O \left( \left\lVert \mathbf{x} \right\rVert^2 \right),
\end{equation}
i.e. the first-order correction term is proportional to a projection of the full Riemann curvature tensor ${R_{i j k l}}$. By Taylor expanding with respect to the tangent space of a chosen geodesic, denoted ${\boldsymbol\gamma_0}$, within this affine space of possible geodesics (which we can describe as a simple one-parameter family of geodesics ${\boldsymbol\gamma_{\tau}}$), we are effectively expanding in terms of Jacobi fields, namely:

\begin{equation}
\mathbf{J} \left( t \right) = \left. \left( \frac{\partial \boldsymbol\gamma_{\tau} \left( t \right)}{\partial \tau} \right) \right\rvert_{\tau = 0}.
\end{equation}
As before, we can now expand the square root of the determinant of the metric tensor to obtain a relationship between the metric volume element ${d \mu_g}$ of the (Riemannian or Lorentzian) manifold ${\left( \mathcal{M}, g \right)}$, and the metric volume element ${d \mu_{Euclidean}}$ of flat (Euclidean) space:

\begin{equation}
d \mu_g = \left[ 1 - \frac{1}{6} R_{j k} x^j x^k + O \left( \left\lVert \mathbf{x} \right\rVert^3 \right) \right] d \mu_{Euclidean},
\end{equation}
thus allowing us to interpret the projections of the Ricci curvature tensor as measuring the discrepancy between the volume of an infinitesimal geodesic cone ${C_t}$ of extent $t$ and a cone of equivalent extent in flat (Euclidean) space ${\mathbb{R}^n}$:

\begin{equation}
\mathrm{Vol} \left( C_t \left( p \right) \right) = a t^n \left[ 1 - \frac{1}{6} R_{i j} t^i t^j + O \left( \left\lVert \mathbf{t} \right\rVert^3 \right) \right],
\end{equation}
as desired. In order to be able to compute such a quantity in an arbitrary (and potentially discrete) metric-measure space, however, requires us first to generalize the definition of the Riemann curvature tensor to such a space. For this purpose, we consider the sectional curvature tensor $K$ of a Riemannian manifold ${\left( \mathcal{M}, g \right)}$, typically defined via:

\begin{equation}
K \left( \mathbf{u}, \mathbf{v} \right) = \frac{\left\langle R \left( \mathbf{u}, \mathbf{v} \right) \mathbf{v}, \mathbf{u} \right\rangle}{\left\langle \mathbf{u}, \mathbf{u} \right\rangle \left\langle \mathbf{v}, \mathbf{v} \right\rangle - \left\langle \mathbf{u}, \mathbf{v} \right\rangle^2},
\end{equation}
for linearly independent tangent vectors ${\mathbf{u}}$ and ${\mathbf{v}}$ at point ${p \in \mathcal{M}}$, and where ${R \left( \mathbf{u}, \mathbf{v} \right)}$ denotes the projection of the full Riemann curvature tensor in direction ${\left( \mathbf{u}, \mathbf{v} \right)}$. If we consider the case in which ${\mathbf{u}}$ and ${\mathbf{v}}$ are orthonormal, this definition simplifies to:

\begin{equation}
K \left( \mathbf{u}, \mathbf{v} \right) = \left\langle R \left( \mathbf{u}, \mathbf{v} \right) \mathbf{v}, \mathbf{u} \right),
\end{equation}
indicating that the components of the sectional curvature tensor ${K \left( \mathbf{u}, \mathbf{v} \right)}$ completely determine the corresponding components of the Riemann curvature tensor ${R \left( \mathbf{u}, \mathbf{v} \right)}$, as required.

\begin{figure}[ht]
\centering
\includegraphics[width=0.395\textwidth]{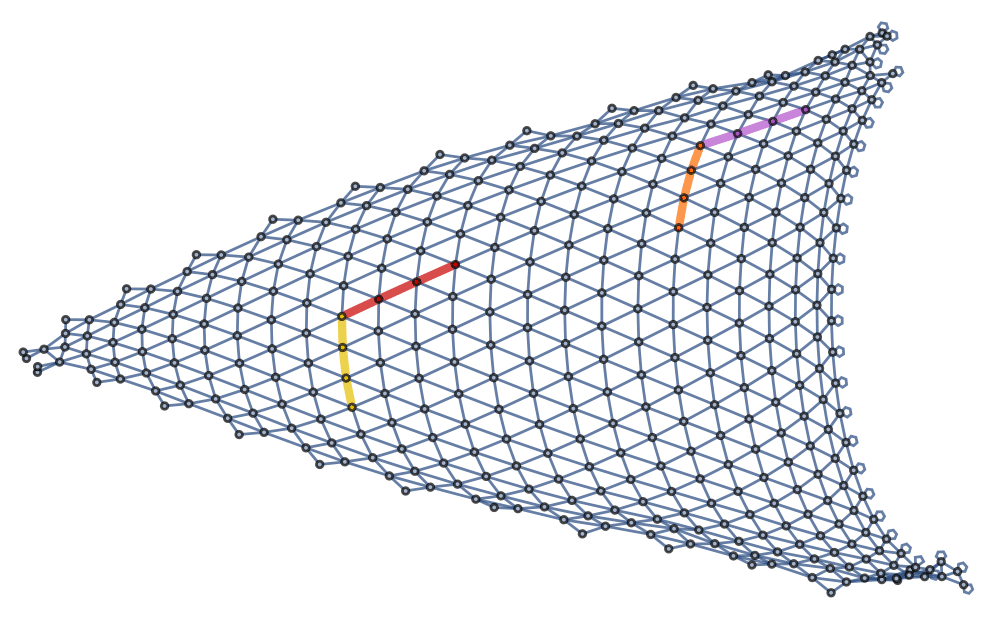}\hspace{0.1\textwidth}
\includegraphics[width=0.395\textwidth]{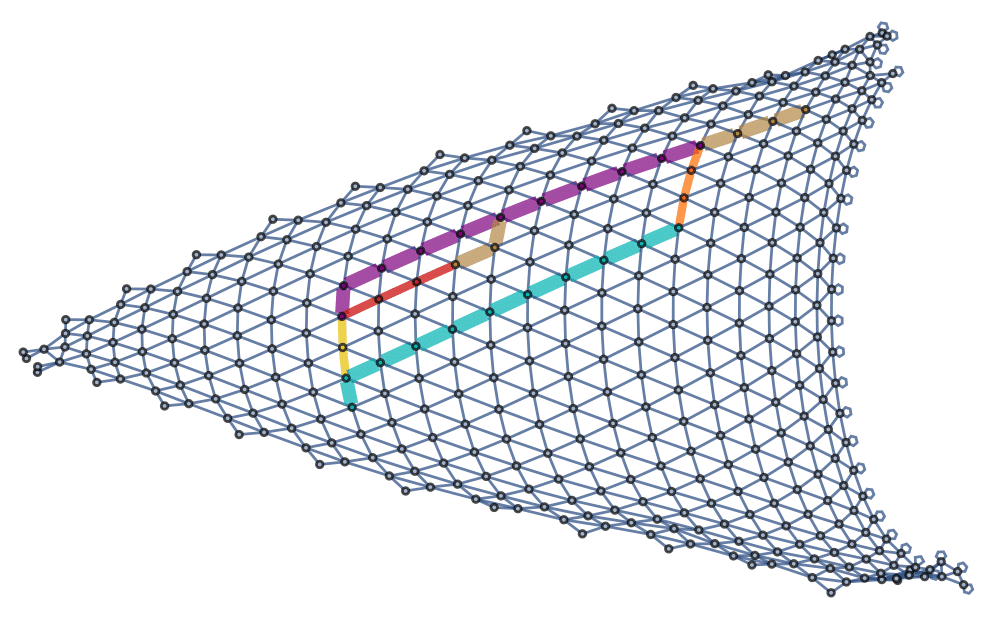}
\caption{On the left, a pair of nearby pairs of tangent vectors embedded in an asymptotically-flat spatial hypergraph with the limiting structure of a two-dimensional Riemannian manifold, as obtained by the evolution of the set substitution system ${\left\lbrace \left\lbrace x, y, y \right\rbrace, \left\lbrace z, x, w \right\rbrace \right\rbrace \to \left\lbrace \left\lbrace y, v, y \right\rbrace, \left\lbrace y, z, v \right\rbrace, \left\lbrace w, v, v \right\rbrace \right\rbrace}$. On the right, a purple path showing the distance between the origin points of the two pairs of tangent vectors, along with additional light blue and brown paths showing the distance between the corresponding endpoints of the two pairs of tangent vectors. The projection of the discrete sectional curvature tensor $K$ onto the yellow and red geodesics is equal to zero along the purple path, since there is no net divergence or convergence of the light blue and brown geodesics}.
\label{fig:Figure69}
\end{figure}

\begin{figure}[ht]
\centering
\includegraphics[width=0.395\textwidth]{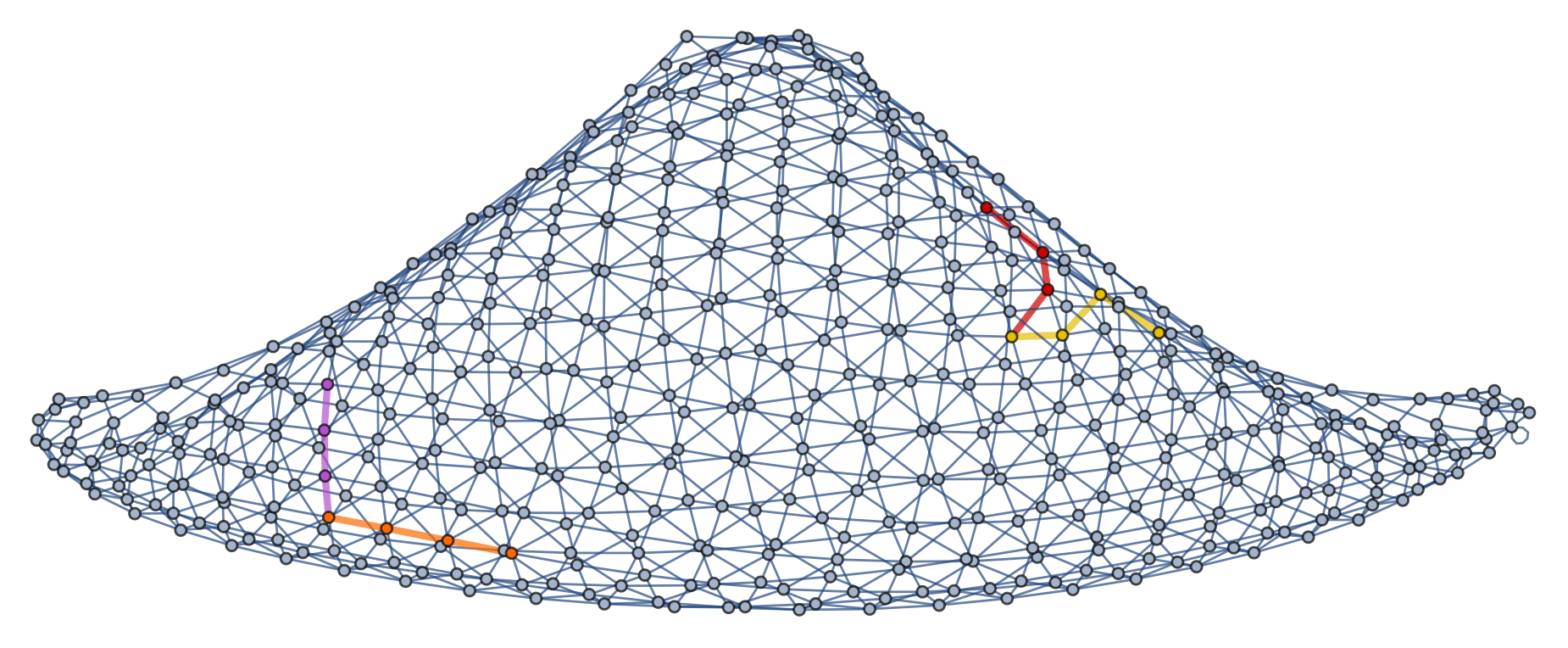}\hspace{0.1\textwidth}
\includegraphics[width=0.395\textwidth]{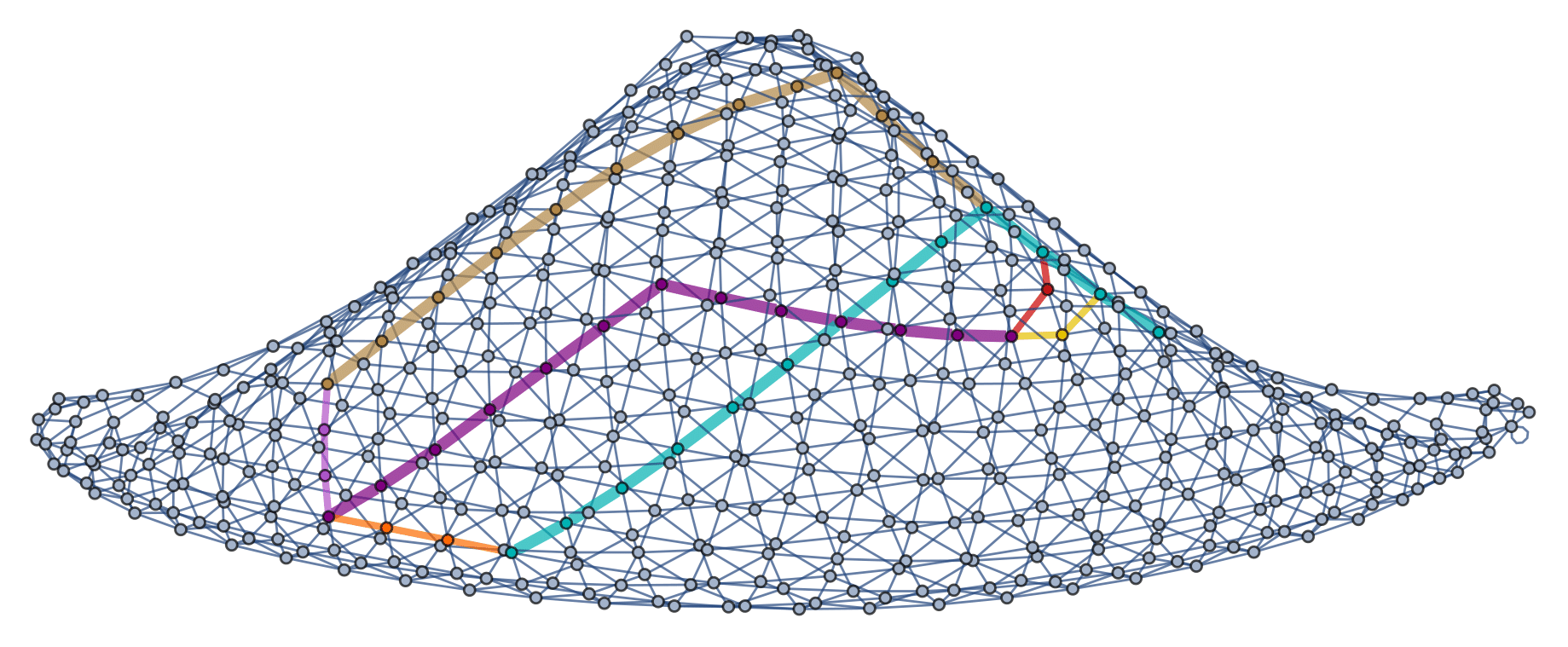}
\caption{On the left, a pair of nearby pairs of tangent vectors embedded in a spatial hypergraph with the limiting structure of a two-dimensional Riemannian manifold with globally positive curvature, as obtained by the evolution of the set substitution system ${\left\lbrace \left\lbrace x, x, y \right\rbrace, \left\lbrace x, z, w \right\rbrace \right\rbrace \to \left\lbrace \left\lbrace w, w, v \right\rbrace, \left\lbrace v, w, y \right\rbrace, \left\lbrace z, y, v \right\rbrace \right\rbrace}$. On the right, a purple path showing the distance between the origin points of the two pairs of tangent vectors, along with additional light blue and brown paths showing the distance between the endpoints of the two pairs of tangent vectors. The projection of the discrete sectional curvature tensor $K$ onto the yellow and red geodesics is positive along the purple path, since there is a net divergence of the light blue and brown geodesics.}
\label{fig:Figure70}
\end{figure}

\begin{figure}[ht]
\centering
\includegraphics[width=0.395\textwidth]{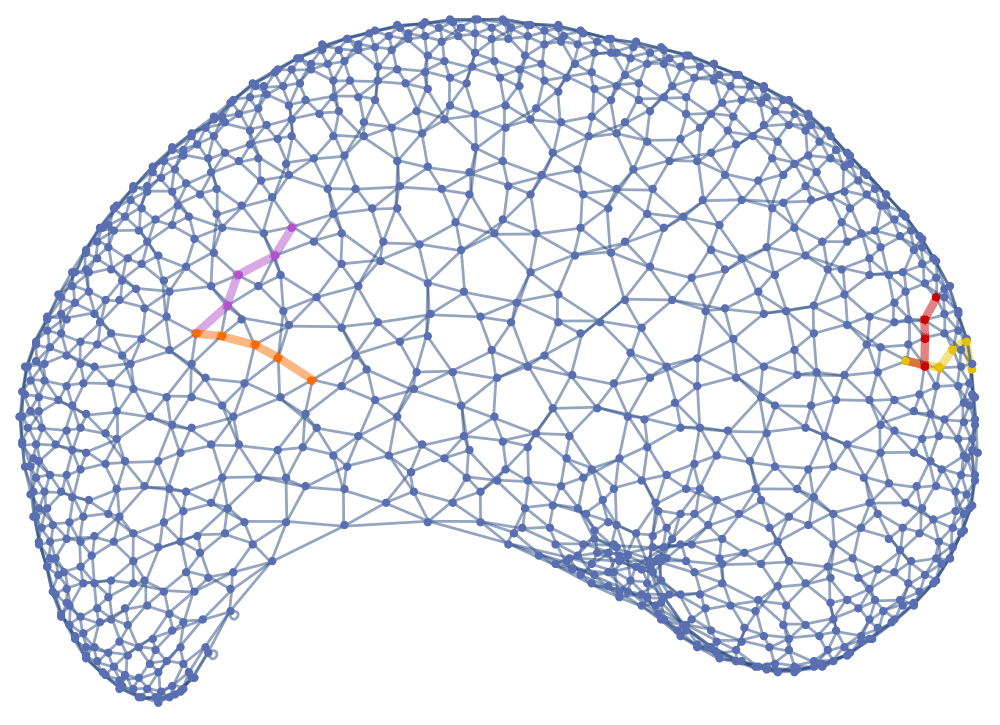}\hspace{0.1\textwidth}
\includegraphics[width=0.395\textwidth]{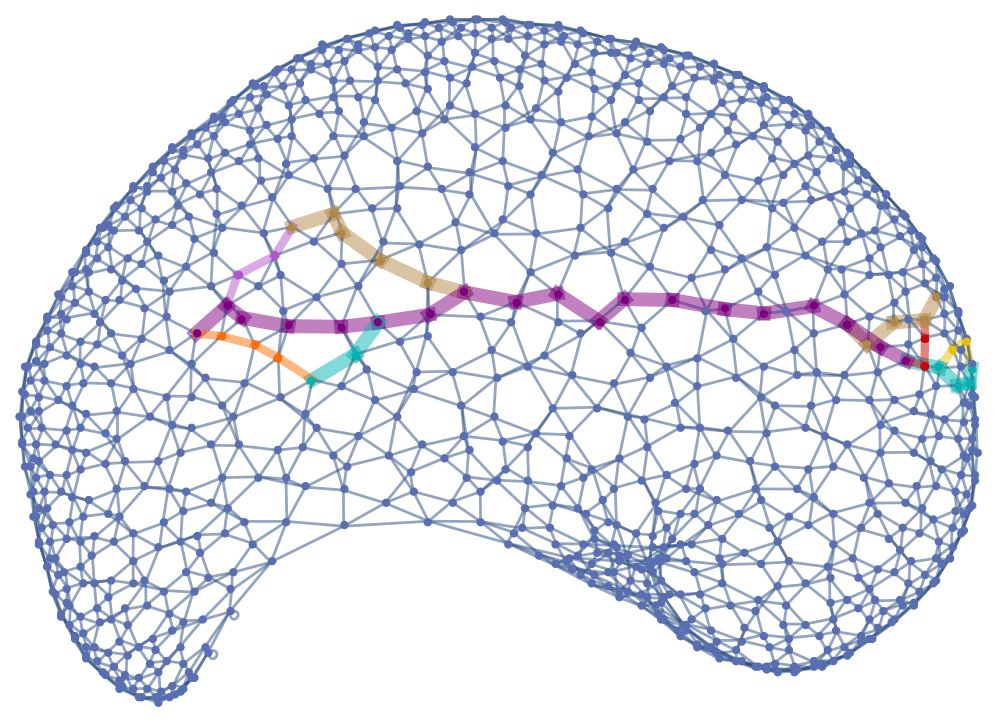}
\caption{On the left, a pair of nearby pairs of tangent vectors embedded in a spatial hypergraph with the limiting structure of a two-dimensional Riemannian manifold with globally negative curvature, as obtained by the evolution of the set substitution system ${\left\lbrace \left\lbrace x, y, x \right\rbrace, \left\lbrace x, z, u \right\rbrace \right\rbrace \to \left\lbrace \left\lbrace u, v, u \right\rbrace, \left\lbrace v, u, z \right\rbrace, \left\lbrace x, y, v \right\rbrace \right\rbrace}$. On the right, a purple path showing the distance between the origin points of the two pairs of tangent vectors, along with additional light blue and brown paths showing the distance between the endpoints of the two pairs of tangent vectors. The projection of the discrete sectional curvature tensor $K$ onto the yellow and red geodesics is negative along the purple path, since there is a net convergence of the light blue and brown geodesics.}
\label{fig:Figure71}
\end{figure}

\begin{figure}[ht]
\centering
\includegraphics[width=0.395\textwidth]{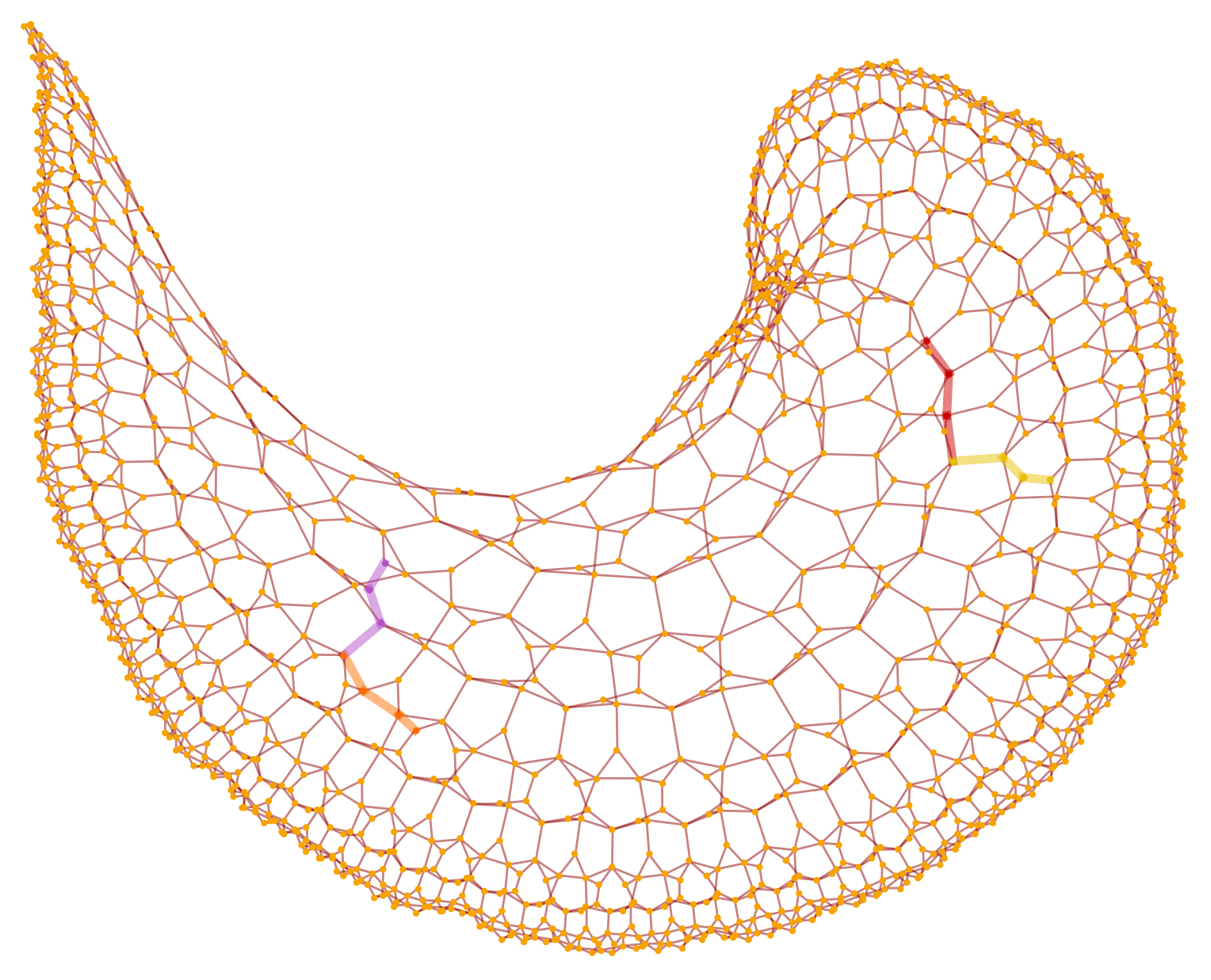}\hspace{0.1\textwidth}
\includegraphics[width=0.395\textwidth]{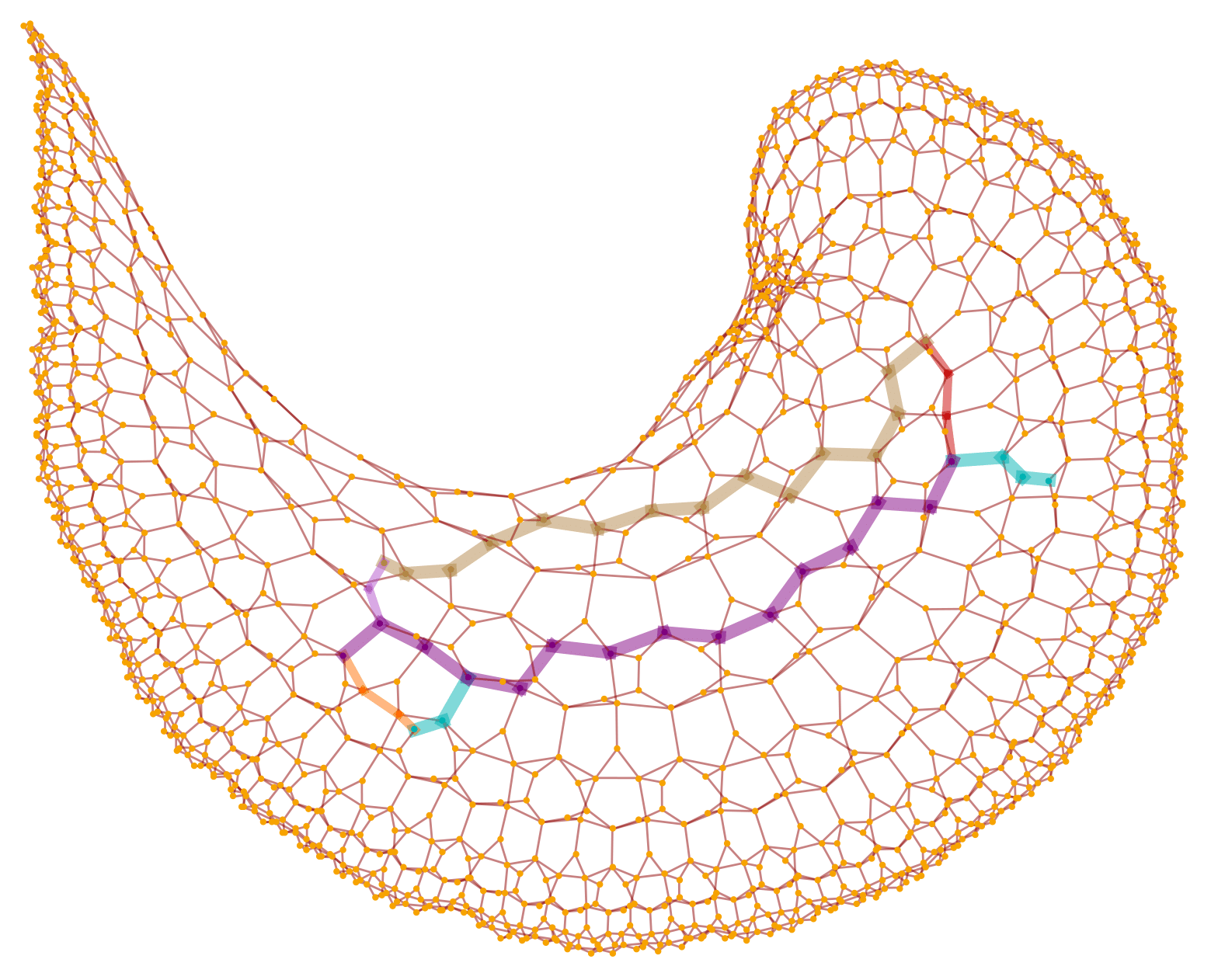}
\caption{On the left, a pair of nearby pairs of tangent vectors embedded in a causal graph with the limiting structure of a two-dimensional Lorentzian manifold with globally negative curvature, as obtained by the evolution of the set substitution system ${\left\lbrace \left\lbrace x, y, x \right\rbrace, \left\lbrace x, z, u \right\rbrace \right\rbrace \to \left\lbrace \left\lbrace u, v, u \right\rbrace, \left\lbrace v, u, z \right\rbrace, \left\lbrace x, y, v \right\rbrace \right\rbrace}$. On the right, a purple path showing the distance between the origin points of the two pairs of tangent vectors, along with additional light blue and brown paths showing the distance between the endpoints of the two pairs of tangent vectors. The projection of the discrete sectional curvature tensor $K$ onto the yellow and red geodesics is negative along the purple path, since there is a net convergence of the light blue and brown geodesics.}
\label{fig:Figure72}
\end{figure}

The geometrical intuition behind the sectional curvature tensor ${K \left( \mathbf{u}, \mathbf{v} \right)}$ is quite straightforward, namely that it quantifies the Gaussian curvature of the surface swept out by the family of geodesics emanating from point ${p \in \mathcal{M}}$, with initial directions defined by the tangent plane ${\sigma_p}$ (which is, in turn, defined by the pair of tangent vectors ${\mathbf{u}}$ and ${\mathbf{v}}$). In the case of a general metric-measure space ${\left( X, d \right)}$, such an intuition may easily be generalized by quantifying the discrepancy between the distance separating points $x$ and $y$ in $X$, and the average distance separating the points ${\exp \left( \epsilon \mathbf{v}_x \right)}$ and ${\exp_y \left( \epsilon \mathbf{v}_y \right)}$, where point $y$ is obtained by traveling a distance of ${\delta}$ away from point $x$ in the direction of the tangent vector ${\mathbf{u}}$, and where points ${\exp_x \left( \epsilon \mathbf{v}_x \right)}$ and ${\exp_y \left( \epsilon \mathbf{v}_y \right)}$ are obtained by traveling a distance of ${\epsilon}$ in the directions defined by the vectors ${\mathbf{v}_x}$ and ${\mathbf{v}_y}$, respectively:

\begin{equation}
d \left( \exp_x \left( \mathbf{v}_x \right), \exp_y \left( \epsilon \mathbf{v}_y \right) \right) = \delta \left( 1 - \frac{\epsilon^2}{2} K \left( \mathbf{u}, \mathbf{v} \right) + O \left( \epsilon^3 + \epsilon^2 \delta \right) \right),
\end{equation}
where this identity holds in the limit as both separation distances converge to zero, ${\epsilon, \delta \to 0}$. Here, the notation ${\exp_x \left( \mathbf{v} \right)}$ signifies the endpoint of a unit-speed geodesic ${\boldsymbol\gamma}$ whose origin point is ${x \in X}$ and whose initial direction is ${\mathbf{v}}$, and ${\delta = d \left( x, y \right)}$ refers to the standard (hyper)graph metric distance between points $x$ and $y$. As discussed in previous sections, we may consider two discrete geodesics ${\gamma_1}$ and ${\gamma_2}$ emanating from a common point $p$ to be \textit{tangent} if the shortest path from the endpoint of ${\gamma_1}$ to any point on ${\gamma_2}$ (or vice versa) terminates at $p$. Illustrative examples of how projections of the discrete sectional curvature tensor $K$ are computed in hypergraphs whose limiting structures correspond to asymptotically-flat, asymptotically positively curved and asymptotically negatively curved Riemannian manifolds are shown in Figures \ref{fig:Figure69}, \ref{fig:Figure70} and \ref{fig:Figure71}, respectively. The extension of this sectional curvature computation from the Riemannian hypergraph case to the Lorentzian causal graph case is shown in Figure \ref{fig:Figure72}, in which the sectional curvature tensor is computed for a directed causal graph whose limiting structure corresponds to an asymptotically negatively curved Lorentzian manifold. Moreover, as a useful consistency test, one can confirm that computing the average of the sectional curvature tensor ${K \left( \mathbf{u}, \mathbf{v} \right)}$, projected across all vectors ${\mathbf{v}}$ (equivalent to taking a trace between the corresponding indices), indeed yields the prior definition of the Ollivier-Ricci curvature scalar, via:

\begin{equation}
\delta \left( 1 - \frac{\epsilon^2}{2 \left( d + 2 \right)} R + O \left( \epsilon^3 + \epsilon^2 \delta \right) \right),
\end{equation}
as required, since the collections of tangent vectors of length ${\epsilon}$ emanating from points ${x, y \in X}$ form finite geodesic balls ${S_x}$ and ${S_y}$ of radius ${\epsilon}$, such that the average sectional curvature quantifies precisely the discrepancy between the distance between the points $x$ and $y$ and the average distance between the corresponding points on ${S_x}$ and ${S_y}$ (before and after parallel transport).

Since, in the above, we have set the couplings ${\epsilon}$ between the measures ${\mu_{A^{in}}}$ and ${\mu_{B^{out}}}$, namely:

\begin{equation}
\sum_{u \to A} \epsilon \left( u, v \right) = \sum_{j = 1}^{m} \mu_{y_j} \left( v \right), \qquad \text{ and } \qquad \sum_{B \to v} \epsilon \left( u, v \right) = \sum_{i = 1}^{n} \mu_{x_i} \left( u \right),
\end{equation}
to be equal to 1 everywhere, thus endowing the events in the causal graph with a constant, real scalar field ${\phi : \mathcal{C} \to \mathbb{R}}$ of (in principle) compact support. As discussed by Dowker and Glazer\cite{dowker}, we can write a discrete version of the d'Alembertian operator ${B^{\left( d \right)}}$ for this scalar field, over an asymptotically-flat ${\left( d - 1 \right) + 1}$-dimensional causal graph, using the following general ansatz:

\begin{equation}
B^{\left( d \right)} \phi \left( e \right) = \frac{1}{l^2} \left( \alpha_d \phi \left( e \right) + \beta_{d} \sum_{i = 1}^{n_d - 1} C_{i}^{\left( d \right)} \sum_{y \in L_i} \phi \left( e \right) \right),
\end{equation}
for undetermined dimension-dependent constants ${\alpha_d}$, ${\beta_d}$ and ${C_{i}^{\left( d \right)}}$ (for ${i = 0, \dots, n_d - 1}$, and with the initial coefficient ${C_{0}^{\left( d \right)}}$ fixed to be equal to 1), and assuming a characteristic length scale $l$. Here, ${L_k \left( e \right)}$ designates a causal \textit{layer}, i.e. a set of $k$-nearest neighbors in the causal past of an event ${e \in \mathcal{C}}$):

\begin{equation}
L_k \left( e \right) = \left\lbrace e^{\prime} \prec e : n \left[ e, e^{\prime} \right] = k \right\rbrace,
\end{equation}
${n_d}$ indicates the number of \textit{layers} which should be summed over, and ${n \left[ p, q \right]}$ quantifies the cardinality of the discrete spacetime order interval ${\mathbf{I} \left[ p, q \right]}$:

\begin{equation}
n \left[ p, q \right] = \left\lvert \mathbf{I} \left[ p, q \right] \right\rvert - 2, \qquad \text{ where } \qquad \mathbf{I} \left[ p, q \right] = \left\lbrace r \in \mathcal{C} : q \prec r \prec p \right\rbrace.
\end{equation}
By substituting ${d = 2}$ and ${d = 4}$, we recover the ${1 + 1}$-dimensional discrete d'Alembertian proposed by Sorkin and Henson\cite{sorkin}\cite{henson}:

\begin{equation}
B^{\left( 2 \right)} \phi \left( e \right) = \frac{1}{l^2} \left[ - 2 \phi \left( x \right) + 4 \left( \sum_{e^{\prime} \in L_0 \left( e \right)} \phi \left( e^{\prime} \right) - 2 \sum_{e^{\prime} \in L_1 \left( e \right)} \phi \left( e^{\prime} \right) + \sum_{e^{\prime} \in L_2 \left( e \right)} \phi \left( e^{\prime} \right) \right) \right],
\end{equation}
and the ${3 + 1}$-dimensional discrete d'Alembertian proposed by Benincasa and Dowker\cite{benincasa}:

\begin{multline}
B^{\left( 4 \right)} \phi \left( e \right) = \frac{1}{l^2} \left[ - \frac{4}{\sqrt{6}} \phi \left( e \right) + \frac{4}{\sqrt{6}} \left( \sum_{e^{\prime} \in L_0 \left( e \right)} \phi \left( e^{\prime} \right) - 9 \sum_{e^{\prime} \in L_1 \left( e \right)} \phi \left( e^{\prime} \right) \right. \right.\\
\left. \left. + 16 \sum_{e^{\prime} \in L_2 \left( e \right)} \phi \left( e^{\prime} \right) - 8 \sum_{e^{\prime} \in L_3 \left( e \right)} \phi \left( e^{\prime} \right) \right) \right],
\end{multline}
respectively.

The significance of this operator for our present purposes lies in the fact that it allows us to define a discrete analog of the Ricci curvature scalar over causal graphs that is provably compatible with the aforementioned Ollivier-Ricci definition. More specifically, if we introduce an intermediate length scale ${l_k \geq l}$ for the purpose of dampening certain microscopic fluctuations in the value of ${B^{\left( d \right)} \phi \left( x \right)}$, then we can construct a ``smoothed'' version of the discrete d'Alembertian, denoted ${B_{k}^{\left( d \right)}}$, whose expectation value ${\left\langle \hat{B}_{k}^{\left( d \right)} \phi \left( x \right) \right\rangle}$ is now given by the integral:

\begin{multline}
\left\langle \hat{B}_{k}^{\left( d \right)} \phi \left( x \right) \right\rangle = \alpha_d l_{k}^{-2} \phi \left( x \right)\\
+ \beta_d l_{k}^{- \left( d + 2 \right)} \int_{J^{-} \left( x \right)} \sqrt{-g \left( y \right)} \phi \left( y \right) \sum_{i = 0}^{n_d - 1} C_{i}^{\left( d \right)} \frac{\left( \mathrm{Vol} \left( \mathbf{I} \left[ y, x \right] \right) l_{k}^{-d} \right)^{i - 2}}{\Gamma \left( i - 1 \right)} \exp \left( - \mathrm{Vol} \left( \mathbf{I} \left[ y, x \right] \right) l_{k}^{-d} \right) d^d y.
\end{multline}
For instance, in the familiar ${d = 2}$ (${1 + 1}$-dimensional) and ${d = 4}$ (${3 + 1}$-dimensional) cases discussed above, one has:

\begin{equation}
\left\langle \hat{B}_{k}^{\left( 2 \right)} \phi \left( x \right) \right\rangle = \frac{2}{l_{k}^{2}} \left[ - \phi \left( x \right) + \frac{2}{l_{k}^{2}} \int_{y \in J^{-} \left( x \right)} \sqrt{-g} \exp \left( - \xi_2 \right) \left( 1 - 2 \xi_2 + \frac{1}{2} x_{2}^{2} \right) \phi \left( y \right) d^2 y \right],
\end{equation}
and:

\begin{equation}
\left\langle \hat{B}_{k}^{\left( 4 \right)} \phi \left( x \right) \right\rangle = \frac{4}{\sqrt{6} l_{k}^{2}} \left[ - \phi \left( x \right) + \frac{1}{l_{k}^{2}} \int_{y \in J^{-} \left( x \right)} \sqrt{-g} \exp \left( - \xi_4 \right) \left( 1 - 9 \xi_4 + 8 \xi_{4}^{2} - \frac{4}{3} \xi_{4}^{3} \right) \phi \left( y \right) d^4 y \right],
\end{equation}
respectively. In the continuum limit ${\rho_c \to \infty}$, the contributions to this integral from within the Riemann normal neighborhood ${\mathcal{W}_1}$ of event $x$ will be given, up to first-order, by the continuum d'Alembertian ${\Box \phi \left( x \right)}$, and up to second-order by the spacetime Ricci curvature scalar ${R \left( x \right)}$ at event $x$, i.e:

\begin{equation}
\lim_{\rho_c \to \infty} \left[ \frac{1}{\sqrt{\rho_c}} \left. \left\langle \hat{B}_{k}^{\left( d \right)} \phi \left( x \right) \right\rangle \right\rvert_{\mathcal{W}_1} \right] = \Box \phi \left( x \right) - \frac{1}{2} R \left( x \right) \phi \left( x \right),
\end{equation}
where the ${\Box \phi \left( x \right)}$ vanishes in our case, since the scalar field ${\phi}$ is constant across all events in the causal graph (or, at least, all events within the Riemann normal neighborhood of $x$). Thus, starting from the assumption that the coupling constants satisfy ${\epsilon \left( u, v \right) = 1}$ for all pairs of vertices $u$ and $v$, and replacing the assumption of weak ergodicity of the causal graph dynamics with the much stronger assumption of uniform Poisson distribution of the rewrite events, our construction of the discrete Einstein-Hilbert action for Wolfram model systems reduces to the general $d$-dimensional form of the Benincasa-Dowker action on causal sets:

\begin{equation}
S^{\left( 4 \right)} \left( \mathcal{C} \right) = \sum_{e \in \mathcal{C}} R \left( e \right),
\end{equation}
which, in the ${d = 4}$ (${3 + 1}$-dimensional) case, we can write explicitly in terms of the cardinalities of sets of $k$-nearest neighbors in the causal past of event ${e \in \mathcal{C}}$:

\begin{equation}
N_k \left( e \right) = \left\lvert L_k \left( e \right) \right\rvert = \left\lvert \left\lbrace e^{\prime} \prec e : n \left[ e, e^{\prime} \right] = k \right\rbrace \right\rvert,
\end{equation}
by using the explicit form of the ${d = 4}$ (${3 + 1}$-dimensional) discrete d'Alembertian ${B^{\left( 4 \right)}}$ as:

\begin{equation}
S^{\left( 4 \right)} \left( \mathcal{C} \right) = \frac{4}{\sqrt{6}} \left[ n - N_0 \left( \mathcal{C} \right) + 9 N_1 \left( \mathcal{C} \right) - 16 N_2 \left( \mathcal{C} \right) + 8 N_3 \left( \mathcal{C} \right) \right],
\end{equation}
where ${N_k \left( \mathcal{C} \right)}$ signifies the number of discrete spacetime order intervals in ${\mathcal{C}}$ that contain exactly $k$ elements, and $n$ denotes the cardinality of the set of events ${n = \left\lvert \mathcal{C} \right\rvert}$. Then, as proposed by Sorkin, Benincasa and Dowker, and demonstrated numerically by Cunningham\cite{cunningham}, so long as one assumes zero surface terms, and moreover so long as the contributions to the integral from further down the light cone (i.e. from the region outside the Riemann normal neighborhood ${\mathcal{W}_1}$) vanish in the limit, then the expectation value of the discrete action ${S^{\left( 4 \right)}}$ will converge to the continuum Einstein-Hilbert action ${S_{EH} \left( g \right)}$ on the manifold ${\left( \mathcal{M}, g \right)}$ in the continuum limit ${\rho_c \to \infty}$:

\begin{equation}
\lim_{\rho_c \to \infty} \left[ \hbar \frac{l^2}{l_{p}^{2}} \left\langle \hat{S}^{\left( 4 \right)} \left( \mathcal{C} \right) \right\rangle \right] = S_{EH} \left( g \right),
\end{equation}
where ${\hbar}$ denotes the reduced Planck constant, ${l_p}$ denotes the Planck length, and where the intermediate length scale ${l_k}$ has been intentionally chosen such that:

\begin{equation}
l_k \gg \left( l^2 L \right)^{\frac{1}{3}},
\end{equation}
for some length $L$ on the order of the Hubble scale, thus ensuring that the classical action is only approximately valid for spacetime regimes in which the Ricci curvature is approximately constant over distance scales on the order of ${\left( l^2 L \right)^{\frac{1}{3}}}$.

\begin{figure}[ht]
\centering
\includegraphics[width=0.395\textwidth]{NumericalRelativity79}\hspace{0.1\textwidth}
\includegraphics[width=0.395\textwidth]{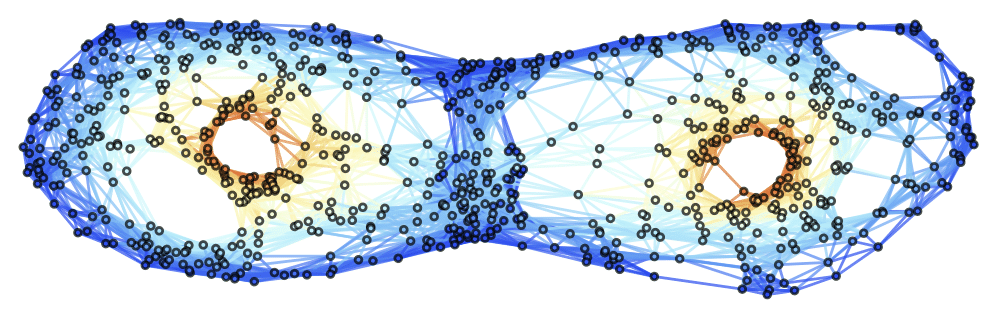}
\caption{Spatial hypergraphs corresponding to the initial hypersurface configuration of the head-on collision of Schwarzschild black holes (convergence) test at time ${t = 0M}$, produced via numerical solution of the Einstein field equations and by pure Wolfram model evolution (with set substitution rule ${\left\lbrace \left\lbrace x, y \right\rbrace, \left\lbrace y, z \right\rbrace, \left\lbrace z, w \right\rbrace, \left\lbrace w, v \right\rbrace \right\rbrace \to \left\lbrace \left\lbrace y, u \right\rbrace, \left\lbrace u, v \right\rbrace, \left\lbrace w, x \right\rbrace, \left\lbrace x, u \right\rbrace \right\rbrace}$), respectively, with a resolution of 800 vertices. The hypergraphs have been colored using the local curvature in the Schwarzschild conformal factor ${\psi}$.}
\label{fig:Figure73}
\end{figure}

\begin{figure}[ht]
\centering
\includegraphics[width=0.395\textwidth]{NumericalRelativity82}\hspace{0.1\textwidth}
\includegraphics[width=0.395\textwidth]{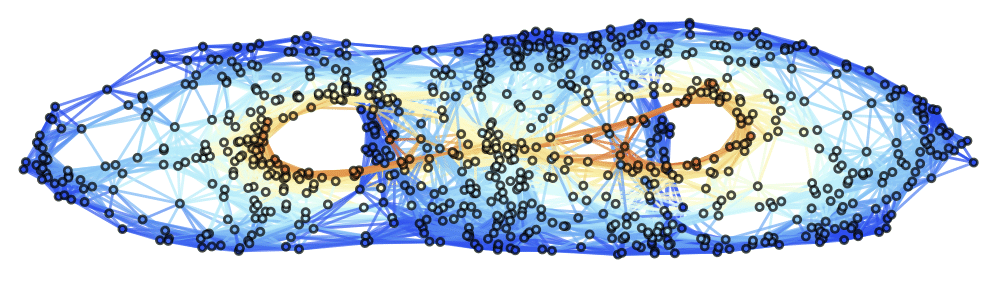}
\caption{Spatial hypergraphs corresponding to the intermediate hypersurface configuration of the head-on collision of Schwarzschild black holes (convergence) test at time ${t = 6 M}$, produced via numerical solution of the Einstein field equations and by pure Wolfram model evolution (with set substitution rule ${\left\lbrace \left\lbrace x, y \right\rbrace, \left\lbrace y, z \right\rbrace, \left\lbrace z, w \right\rbrace, \left\lbrace w, v \right\rbrace \right\rbrace \to \left\lbrace \left\lbrace y, u \right\rbrace, \left\lbrace u, v \right\rbrace, \left\lbrace w, x \right\rbrace, \left\lbrace x, u \right\rbrace \right\rbrace}$), respectively, with a resolution of 800 vertices. The hypergraphs have been colored using the local curvature in the Schwarzschild conformal factor ${\psi}$.}
\label{fig:Figure74}
\end{figure}

\begin{figure}[ht]
\centering
\includegraphics[width=0.395\textwidth]{NumericalRelativity85}\hspace{0.1\textwidth}
\includegraphics[width=0.395\textwidth]{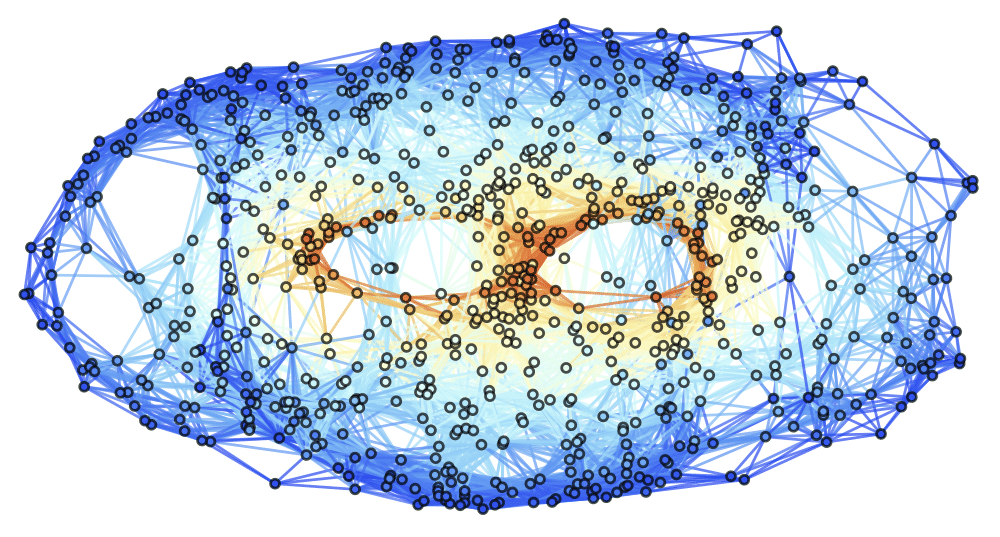}
\caption{Spatial hypergraphs corresponding to the final hypersurface configuration of the head-on collision of Schwarzschild black holes (convergence) test at time ${t = 12 M}$, produced via numerical solution of the Einstein field equations and by pure Wolfram model evolution (with set substitution rule ${\left\lbrace \left\lbrace x, y \right\rbrace, \left\lbrace y, z \right\rbrace, \left\lbrace z, w \right\rbrace, \left\lbrace w, v \right\rbrace \right\rbrace \to \left\lbrace \left\lbrace y, u \right\rbrace, \left\lbrace u, v \right\rbrace, \left\lbrace w, x \right\rbrace, \left\lbrace x, u \right\rbrace \right\rbrace}$), respectively, with a resolution of 800 vertices. The hypergraphs have been colored using the local curvature in the Schwarzschild conformal factor ${\psi}$.}
\label{fig:Figure75}
\end{figure}

\begin{figure}[ht]
\centering
\includegraphics[width=0.395\textwidth]{NumericalRelativity88}\hspace{0.1\textwidth}
\includegraphics[width=0.395\textwidth]{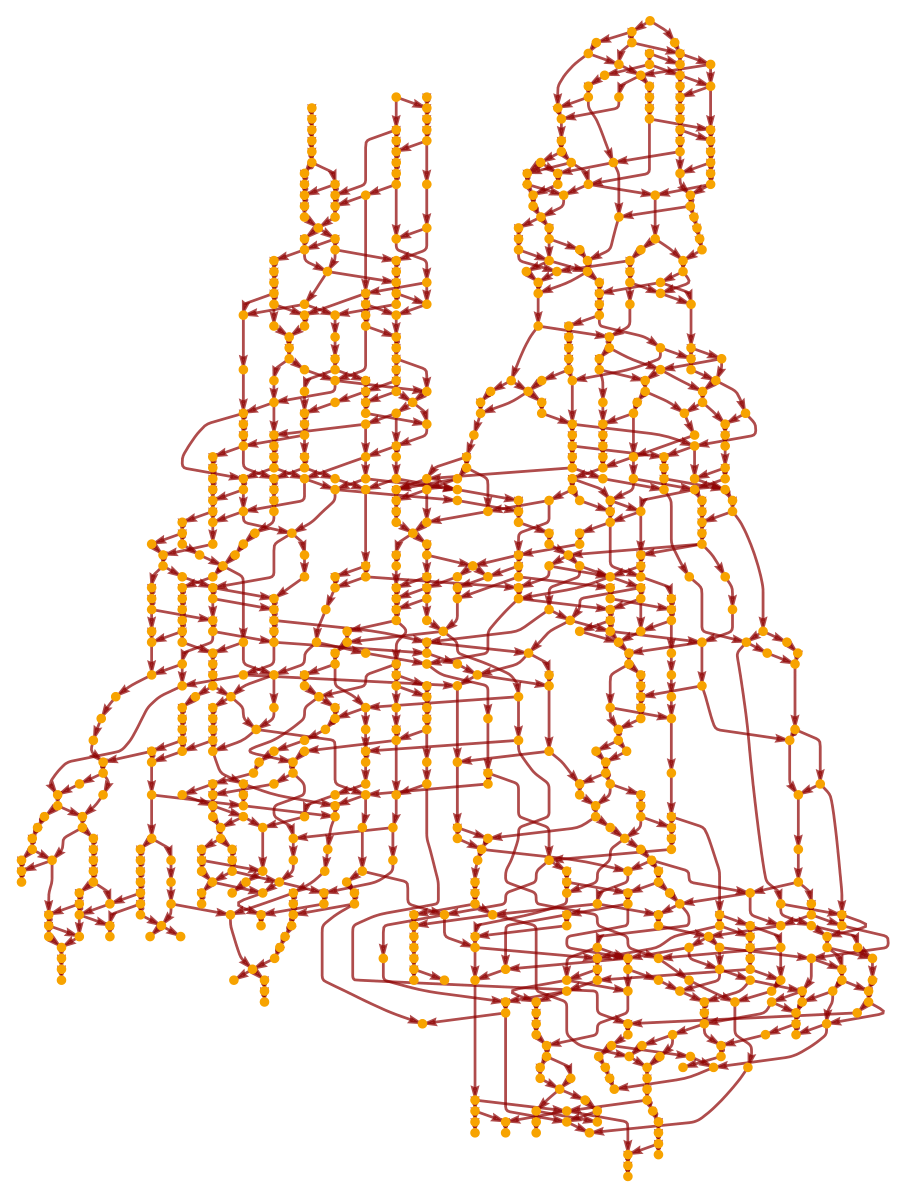}
\caption{Causal graphs corresponding to the discrete characteristic structure of the head-on collision of Schwarzschild black holes (convergence) test at time ${t = 12 M}$, produced via numerical solution of the Einstein field equations and by pure Wolfram model evolution (with set substitution rule ${\left\lbrace \left\lbrace x, y \right\rbrace, \left\lbrace y, z \right\rbrace, \left\lbrace z, w \right\rbrace, \left\lbrace w, v \right\rbrace \right\rbrace \to \left\lbrace \left\lbrace y, u \right\rbrace, \left\lbrace u, v \right\rbrace, \left\lbrace w, x \right\rbrace, \left\lbrace x, u \right\rbrace \right\rbrace}$), respectively, with a resolution of 800 hypergraph vertices.}
\label{fig:Figure76}
\end{figure}

\begin{figure}[ht]
\centering
\includegraphics[width=0.395\textwidth]{NumericalRelativity100}\hspace{0.1\textwidth}
\includegraphics[width=0.395\textwidth]{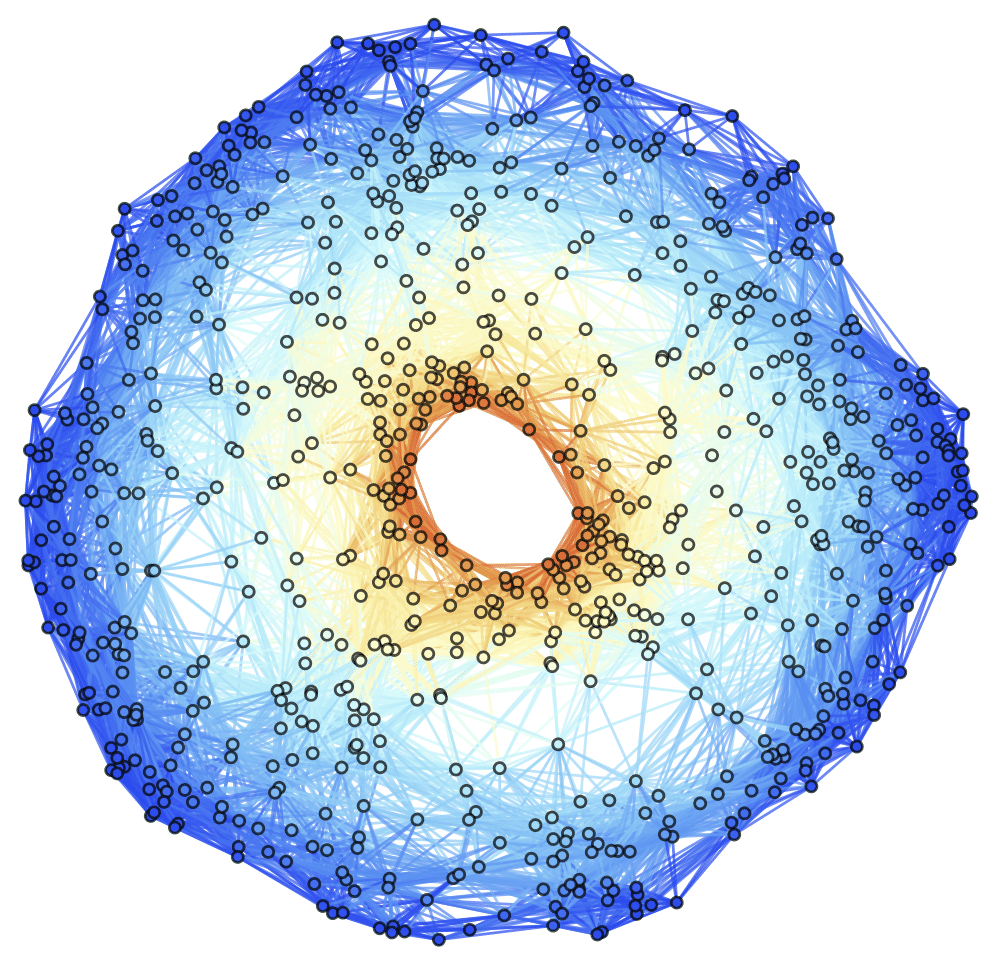}
\caption{Spatial hypergraphs corresponding to the post-ringdown hypersurface configuration of the head-on collision of Schwarzschild black holes (convergence) test at time ${t = 24 M}$, produced via numerical solution of the Einstein field equations and by pure Wolfram model evolution (with set substitution rule ${\left\lbrace \left\lbrace x, y \right\rbrace, \left\lbrace y, z \right\rbrace, \left\lbrace z, w \right\rbrace, \left\lbrace w, v \right\rbrace \right\rbrace \to \left\lbrace \left\lbrace y, u \right\rbrace, \left\lbrace u, v \right\rbrace, \left\lbrace w, x \right\rbrace, \left\lbrace x, u \right\rbrace \right\rbrace}$), respectively, with a resolution of 800 vertices. The hypergraphs have been colored using the local curvature in the Schwarzschild conformal factor ${\psi}$.}
\label{fig:Figure77}
\end{figure}

Now that we have the ability to extract arbitrary curvature information from the combinatorial structure of a given hypergraph and/or causal graph, and we have techniques for determining when a given Wolfram model evolution rule satisfies the Einstein field equations (through extremization of the Einstein-Hilbert action) in the continuum limit, we can proceed to define the initial data for our relativistic Cauchy problem by constructing a hypergraph ${\mathcal{H}}$ that is guaranteed to be faithfully embeddable into a given Lorentzian manifold ${\left( \mathcal{M}, g \right)}$, with embedding map ${\Phi}$, by means of a Poisson point process. More precisely, we construct the elements of ${\Phi \left( \mathcal{H} \right)}$ using a Poisson distribution, in which the probability of $n$ vertices of ${\mathcal{H}}$ lying within a spatial region of volume $v$ is given by:

\begin{equation}
P_{v} \left( n \right) = \frac{\left( \rho_c v \right)n}{n!} \exp \left( - \rho_c v \right),
\end{equation}
such that the expectation value for the number of vertices ${\hat{n}}$ lying with a spatial region of volume $v$ in the random hypergraph ${\mathcal{H}}$ in ${\Phi \left( \mathcal{H} \right)}$ is therefore identically:

\begin{equation}
\left\langle \hat{n} \right\rangle = \rho_c v.
\end{equation}
A pair of vertices in ${\mathcal{H}}$ can then be linked by a hyperedge if and only if the corresponding points lie within a certain cut-off distance (defined relative to the discretization cut-off ${V_c = \rho_{c}^{-1}}$) in ${\Phi \left( \mathcal{H} \right)}$. We shall use this procedure first to set up the standard Brill-Lindquist binary black hole initial data for a pair of static Schwarzschild black holes, each of mass ${0.5 M}$, and with initial separation of ${1 M}$, such that the center of mass of the binary is at the center of the initial hypergraph. The standard expression for the Schwarzschild conformal factor ${\psi}$:

\begin{equation}
\psi = \left( 1 + \frac{M}{2r} \right)^{-2},
\end{equation}
is used for coloring the intermediate hypergraphs, and the solution is evolved in the first instance until the final time of ${t = 12 M}$, with an intermediate check at time ${t = 6 M}$, after which the two black holes merge and the merged black hole enters its gravitational ringdown phase; the initial, intermediate and final hypersurface configurations are shown in Figures \ref{fig:Figure73}, \ref{fig:Figure74} and \ref{fig:Figure75}, respectively, showing comparison between the direct numerical solution of the Einstein field equations and the pure Wolfram model evolution (with set substitution rule ${\left\lbrace \left\lbrace x, y \right\rbrace, \left\lbrace y, z \right\rbrace, \left\lbrace z, w \right\rbrace, \left\lbrace w, v \right\rbrace \right\rbrace \to \left\lbrace \left\lbrace y, u \right\rbrace, \left\lbrace u, v \right\rbrace, \left\lbrace w, x \right\rbrace, \left\lbrace x, u \right\rbrace \right\rbrace}$), with a resolution of 800 vertices. Figure \ref{fig:Figure76} shows the discrete characteristic structure of the solutions after time ${t = 12 M}$ (as represented using causal graphs). The hypersurface configuration following gravitational ringdown at time ${t = 24 M}$ is shown in Figure \ref{fig:Figure77}.

\begin{figure}[ht]
\centering
\includegraphics[width=0.895\textwidth]{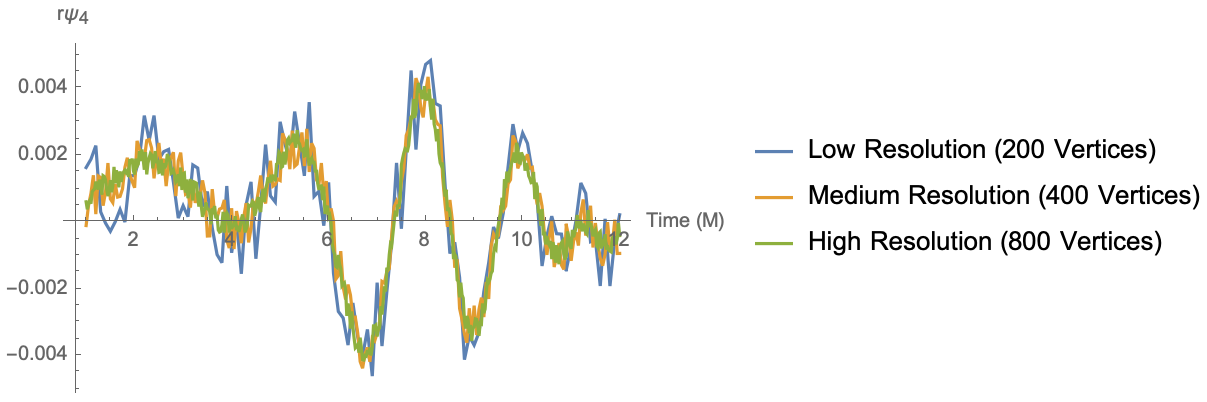}
\caption{Convergence test of the head-on collision of Schwarzschild black holes after time ${t = 12 M}$, showing the real part of the ${\ell = 2}$, ${m = 0}$ mode of the radial Weyl scalar ${r \Psi_4}$ extrapolated on a sphere of radius ${R = 6 M}$, with resolutions of 200, 400 and 800 vertices, respectively.}
\label{fig:Figure78}
\end{figure}

As a means of demonstrating convergence and the successful extraction of gravitational wave data in the pure Wolfram model case, we show the real part of the ${\ell = 2}$, ${m = 0}$ mode of the radial Weyl scalar ${r \Psi_4}$ at time ${t = 12 M}$, extrapolated over a sphere of radius ${R = 6 M}$ using the usual fourth-order interpolation techniques across both the polar and the azimuthal angular directions, in Figure \ref{fig:Figure78}, with resolutions of 200, 400 and 800 vertices. We confirm, as previously, that the ADM mass of the final Schwarzschild black hole configuration (computed as usual by integrating over a single surface surrounding the boundary of asymptotic flatness) is approximately equal to the sum of the ADM masses of the two initial Schwarzschild black holes (computed as usual by integrating over a pair of surfaces surrounding the two boundaries of asymptotic flatness), and also that the linear and angular momenta converge to become approximately zero in the post-ringdown phase, as required. The convergence rates for the Hamiltonian constraint after time ${t = 24 M}$, with respect to the ${L_1}$, ${L_2}$ and ${L_{\infty}}$ norms, illustrating approximately second-order convergence of the pure Wolfram model evolution scheme, are shown in Table \ref{tab:Table6}.

\begin{table}
\centering
\begin{tabular}{|c|c|c|c|c|c|c|}
\hline
Vertices & ${\epsilon \left( L_1 \right)}$ & ${\epsilon \left( L_2 \right)}$ & ${\epsilon \left( L_{\infty} \right)}$ & ${\mathcal{O} \left( L_1 \right)}$ & ${\mathcal{O} \left( L_2 \right)}$ & ${\mathcal{O} \left( L_{\infty} \right)}$\\
\hline\hline
100 & ${6.80 \times 10^{-1}}$ & ${3.15 \times 10^{-1}}$ & ${8.97 \times 10^{-1}}$ & - & - & -\\
\hline
200 & ${1.45 \times 10^{-1}}$ & ${6.76 \times 10^{-2}}$ & ${2.74 \times 10^{-1}}$ & 2.23 & 2.22 & 1.71\\
\hline
400 & ${4.29 \times 10^{-2}}$ & ${1.30 \times 10^{-2}}$ & ${4.22 \times 10^{-2}}$ & 1.76 & 2.38 & 2.70\\
\hline
800 & ${1.44 \times 10^{-2}}$ & ${2.65 \times 10^{-3}}$ & ${1.96 \times 10^{-2}}$ & 1.58 & 2.29 & 1.11\\
\hline
1600 & ${5.29 \times 10^{-3}}$ & ${1.15 \times 10^{-3}}$ & ${2.76 \times 10^{-3}}$ & 1.44 & 1.20 & 2.83\\
\hline
\end{tabular}
\caption{Convergence rates for the head-on collision of Schwarzschild black holes test with respect to the ${L_1}$, ${L_2}$ and ${L_{\infty}}$ norms for the Hamiltonian constraint $H$ after time ${t = 24 M}$, showing approximately second-order convergence.}
\label{tab:Table6}
\end{table}

\begin{figure}[ht]
\centering
\includegraphics[width=0.395\textwidth]{NumericalRelativity107}\hspace{0.1\textwidth}
\includegraphics[width=0.395\textwidth]{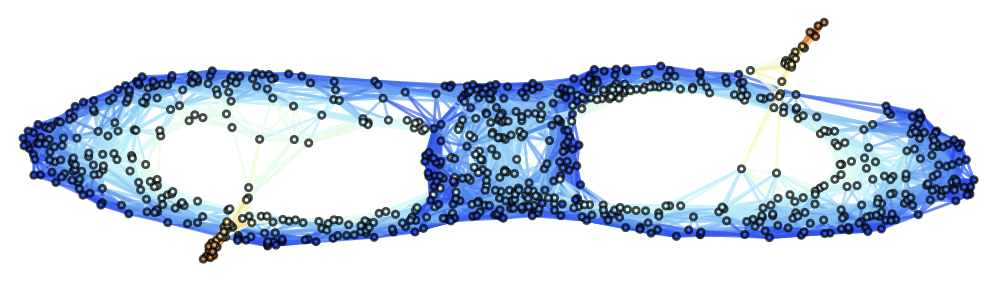}
\caption{Spatial hypergraphs corresponding to the initial hypersurface configuration of the head-on collision of rapidly rotating Kerr black holes (angular momentum) test with ${a = 0.6}$ at time ${t = 0 M}$, produced via numerical solution of the Einstein field equations and by pure Wolfram model evolution (with set substitution rule ${\left\lbrace \left\lbrace x, y \right\rbrace, \left\lbrace y, z \right\rbrace, \left\lbrace z, w \right\rbrace, \left\lbrace w, v \right\rbrace \right\rbrace \to \left\lbrace \left\lbrace y, u \right\rbrace, \left\lbrace u, v \right\rbrace, \left\lbrace w, x \right\rbrace, \left\lbrace x, u \right\rbrace \right\rbrace}$), respectively, with a resolution of 800 vertices. The hypergraphs have been colored using the local curvature in the Boyer-Lindquist conformal factor ${\psi}$.}
\label{fig:Figure79}
\end{figure}

\begin{figure}[ht]
\centering
\includegraphics[width=0.395\textwidth]{NumericalRelativity110}\hspace{0.1\textwidth}
\includegraphics[width=0.395\textwidth]{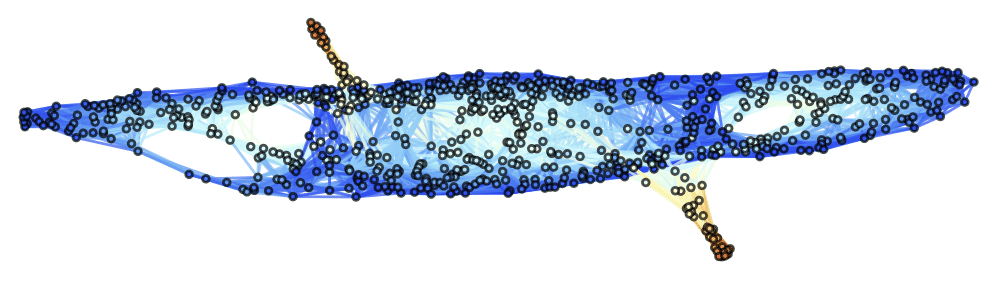}
\caption{Spatial hypergraphs corresponding to the intermediate hypersurface configuration of the head-on collision of rapidly rotating Kerr black holes (angular momentum) test with ${a = 0.6}$ at time ${t = 6 M}$, produced via numerical solution of the Einstein field equations and by pure Wolfram model evolution (with set substitution rule ${\left\lbrace \left\lbrace x, y \right\rbrace, \left\lbrace y, z \right\rbrace, \left\lbrace z, w \right\rbrace, \left\lbrace w, v \right\rbrace \right\rbrace \to \left\lbrace \left\lbrace y, u \right\rbrace, \left\lbrace u, v \right\rbrace, \left\lbrace w, x \right\rbrace, \left\lbrace x, u \right\rbrace \right\rbrace}$), respectively, with a resolution of 800 vertices. The hypergraphs have been colored using the local curvature in the Boyer-Lindquist conformal factor ${\psi}$.}
\label{fig:Figure80}
\end{figure}

\begin{figure}[ht]
\centering
\includegraphics[width=0.395\textwidth]{NumericalRelativity113}\hspace{0.1\textwidth}
\includegraphics[width=0.395\textwidth]{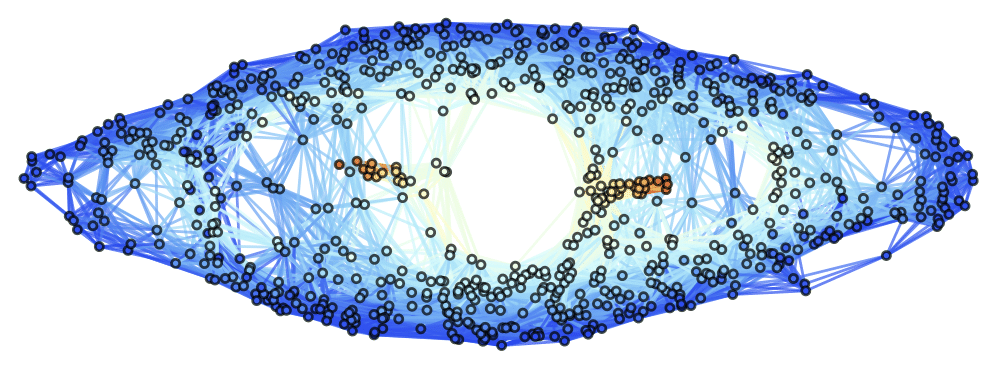}
\caption{Spatial hypergraphs corresponding to the final hypersurface configuration of the head-on collision of rapidly rotating Kerr black holes (angular momentum) test with ${a = 0.6}$ at time ${t = 12 M}$, produced via numerical solution of the Einstein field equations and by pure Wolfram model evolution (with set substitution rule ${\left\lbrace \left\lbrace x, y \right\rbrace, \left\lbrace y, z \right\rbrace, \left\lbrace z, w \right\rbrace, \left\lbrace w, v \right\rbrace \right\rbrace \to \left\lbrace \left\lbrace y, u \right\rbrace, \left\lbrace u, v \right\rbrace, \left\lbrace w, x \right\rbrace, \left\lbrace x, u \right\rbrace \right\rbrace}$), respectively, with a resolution of 800 vertices. The hypergraphs have been colored using the local curvature in the Boyer-Lindquist conformal factor ${\psi}$.}
\label{fig:Figure81}
\end{figure}

\begin{figure}[ht]
\centering
\includegraphics[width=0.395\textwidth]{NumericalRelativity116}\hspace{0.1\textwidth}
\includegraphics[width=0.395\textwidth]{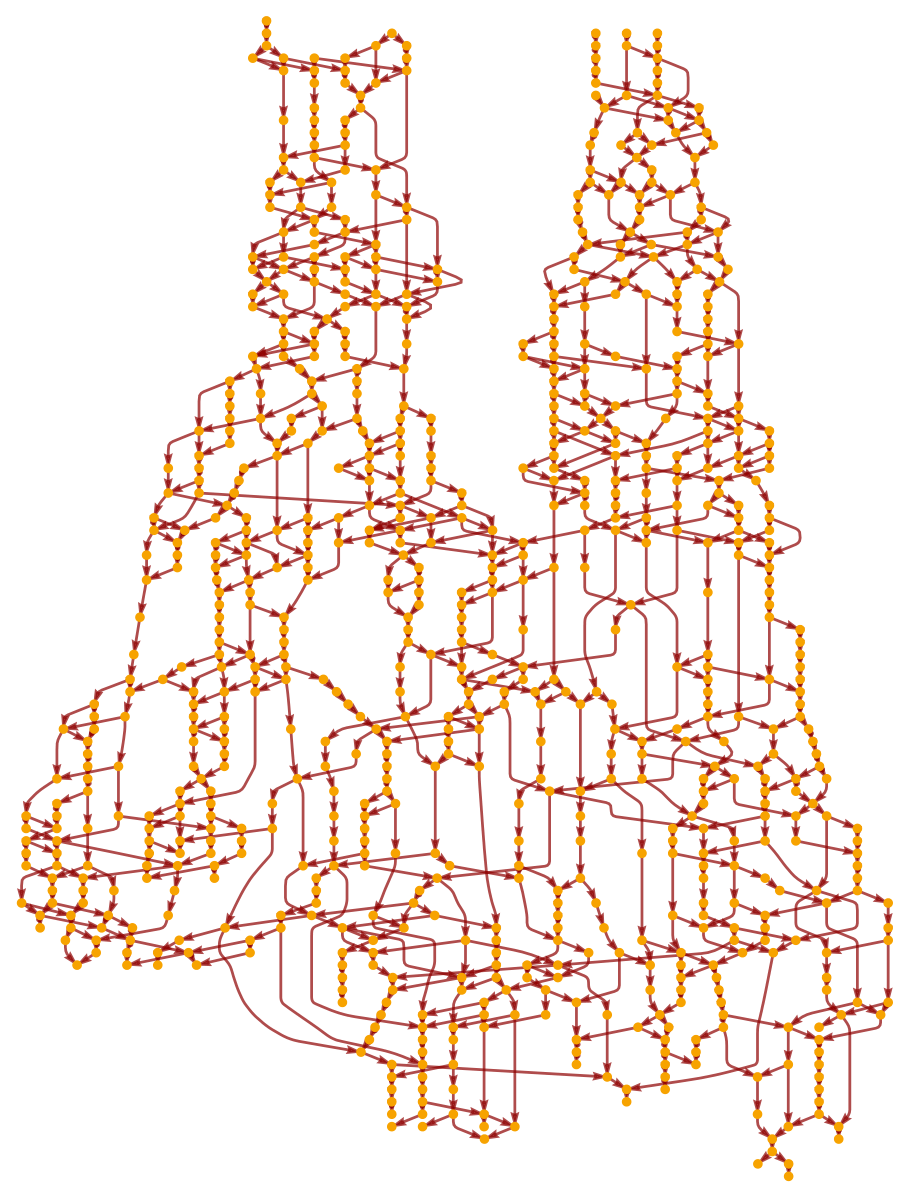}
\caption{Causal graphs corresponding to the discrete characteristic structure of the head-on collision of rapidly rotating Kerr black holes (angular momentum) test with ${a = 0.6}$ at time ${t = 12 M}$, produced via numerical solution of the Einstein field equations and by pure Wolfram model evolution (with set substitution rule ${\left\lbrace \left\lbrace x, y \right\rbrace, \left\lbrace y, z \right\rbrace, \left\lbrace z, w \right\rbrace, \left\lbrace w, v \right\rbrace \right\rbrace \to \left\lbrace \left\lbrace y, u \right\rbrace, \left\lbrace u, v \right\rbrace, \left\lbrace w, x \right\rbrace, \left\lbrace x, u \right\rbrace \right\rbrace}$), respectively, with a resolution of 800 hypergraph vertices.}
\label{fig:Figure82}
\end{figure}

\begin{figure}[ht]
\centering
\includegraphics[width=0.395\textwidth]{NumericalRelativity128}\hspace{0.1\textwidth}
\includegraphics[width=0.395\textwidth]{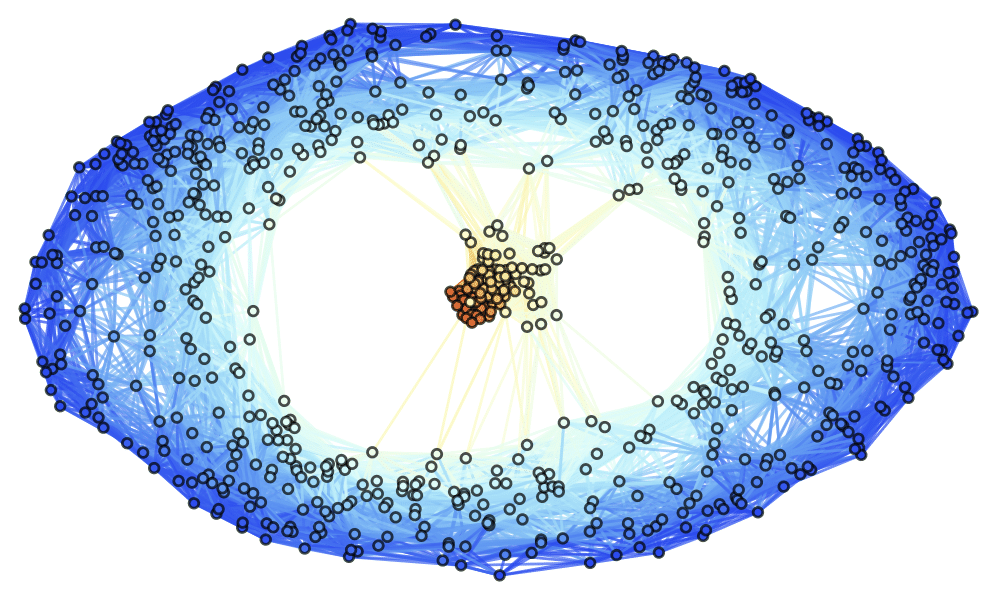}
\caption{Spatial hypergraphs corresponding to the post-ringdown hypersurface configuration of the head-on collision of rapidly rotating Kerr black holes (angular momentum) test with ${a = 0.6}$ at time ${t = 24 M}$, produced via numerical solution of the Einstein field equations and by pure Wolfram model evolution (with set substitution rule ${\left\lbrace \left\lbrace x, y \right\rbrace, \left\lbrace y, z \right\rbrace, \left\lbrace z, w \right\rbrace, \left\lbrace w, v \right\rbrace \right\rbrace \to \left\lbrace \left\lbrace y, u \right\rbrace, \left\lbrace u, v \right\rbrace, \left\lbrace w, x \right\rbrace, \left\lbrace x, u \right\rbrace \right\rbrace}$), respectively, with a resolution of 800 vertices. The hypergraphs have been colored using the local curvature in the Boyer-Lindquist conformal factor ${\psi}$.}
\label{fig:Figure83}
\end{figure}

Finally, we shall use this Poisson point process methodology to set up the Brill-Lindquist binary black hole initial data for a pair of rapidly rotating Kerr black holes, each of mass ${0.5 M}$, and with initial separation of ${1 M}$, such that the center of mass of the binary is at the center of the initial hypergraph. Each black hole is initialized with its angular momentum parameter ${a = \frac{J}{M} = 0.6}$. The standard expression for the Boyer-Lindquist conformal factor ${\psi}$:

\begin{equation}
\psi = \frac{\Sigma^{\frac{1}{4}}}{\sqrt{\sin \left( \lambda \right)}},
\end{equation}
where ${\Sigma}$ denotes the usual Boyer-Lindquist quantity:

\begin{equation}
\Sigma = r^2 + a^2 \cos \left( \theta \right),
\end{equation}
and where ${\lambda}$ is a parameter used in the definition of a new radial coordinate ${\tilde{r}}$:

\begin{equation}
\tilde{r} = \frac{\alpha^2 \cos^2 \left( \frac{\lambda}{2} \right) + \delta^2 \sin^2 \left( \frac{\lambda}{2} \right)}{\alpha \sin \left( \lambda \right)} + M,
\end{equation}
with the parameters ${\delta}$, ${\lambda}$ and ${\alpha}$ satisfying:

\begin{equation}
\delta = \sqrt{M^2 - a^2}, \qquad 0 \leq \lambda \leq \pi, \qquad \alpha > 0,
\end{equation}
is used for coloring the intermediate hypergraphs, and the solution is evolved in the first instance until the final time of ${t = 12 M}$, with an intermediate check at time ${t = 6 M}$, after which the two black holes merge and the merged black hole enters its gravitational ringdown phase, as above; the initial, intermediate and final hypersurface configurations are shown in Figures \ref{fig:Figure79}, \ref{fig:Figure80} and \ref{fig:Figure81}, respectively, showing comparison between the direct numerical solutions of the Einstein field equations and the pure Wolfram model evolution (with set substitution system ${\left\lbrace \left\lbrace x, y \right\rbrace, \left\lbrace y, z \right\rbrace, \left\lbrace z, w \right\rbrace, \left\lbrace w, v \right\rbrace \right\rbrace \to \left\lbrace \left\lbrace y, u \right\rbrace, \left\lbrace u, v \right\rbrace, \left\lbrace w, x \right\rbrace, \left\lbrace x, u \right\rbrace \right\rbrace}$), with a resolution of 800 vertices. Figure \ref{fig:Figure82} shows the discrete characteristic structure of the solutions after time ${t = 12 M}$ (as represented using causal graphs). The hypersurface configuration following gravitational ringdown at time ${t = 24 M}$ is shown in Figure \ref{fig:Figure83}.

\begin{figure}[ht]
\centering
\includegraphics[width=0.895\textwidth]{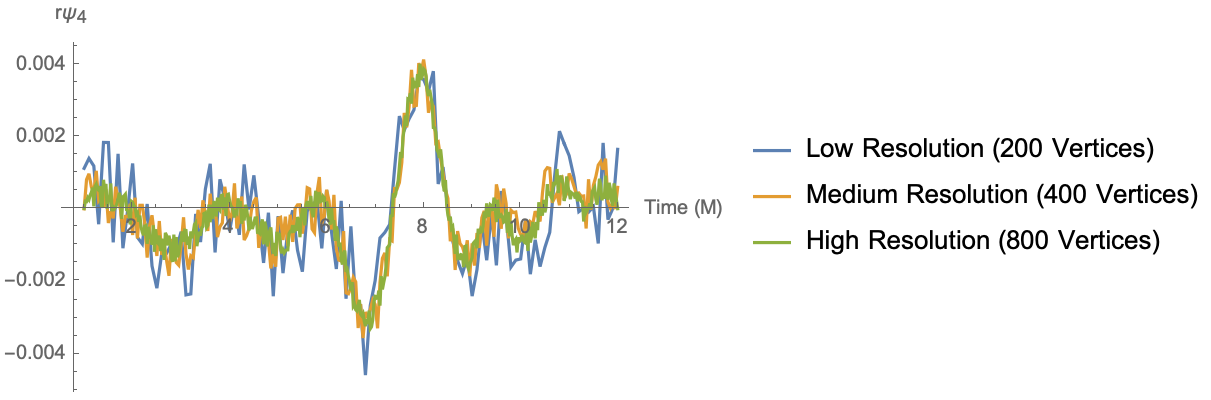}
\caption{Convergence test of the head-on collision of rapidly rotating Kerr black holes (angular momentum) test after time ${t = 12 M}$, showing the real part of the ${\ell = 2}$, ${m = 0}$ mode of the radial Weyl scalar ${r \Psi_4}$ extrapolated on a sphere of radius ${R = 60 M}$, with resolutions of 200, 400 and 800 vertices, respectively.}
\label{fig:Figure84}
\end{figure}

As with the static Schwarzschild case, we show the real part of the ${\ell = 2}$, ${m = 0}$ mode of the radial Weyl scalar ${r \Psi_4}$ at time ${t = 12 M}$ as means of demonstrating convergence and the successful extraction of gravitational wave data in the pure Wolfram model case, extrapolated over a sphere of radius ${R = 6 M}$ using the usual fourth-order interpolation techniques across both the polar and azimuthal angular directions, in Figure \ref{fig:Figure84}, with resolutions of 200, 400 and 800 vertices. We confirm, as above, that the ADM mass of the final Kerr black hole configuration (computed as usual by integrating over a single surface surrounding the boundary of asymptotic flatness) is approximately equal to the sum of ADM masses of the two initial Kerr black holes (computed as usual by integrating over a pair of surfaces surrounding the two boundaries of asymptotic flatness), as well as that the ADM linear momentum converges to be approximately zero in the post-ringdown phase. The angular momentum parameter ${a = \frac{J}{M}}$ of the final Kerr black hole configuration converges to be approximately equal to the angular momentum parameters of the two initial Kerr black holes, as expected. The convergence rates for the Hamiltonian constraint after time ${t = 24 M}$, with respect to the ${L_1}$, ${L_2}$ and ${L_{\infty}}$ norms, illustrating approximately second-order convergence of the pure Wolfram model evolution scheme, are shown in Table \ref{tab:Table7}.

\begin{table}
\centering
\begin{tabular}{|c|c|c|c|c|c|c|}
\hline
Vertices & ${\epsilon \left( L_1 \right)}$ & ${\epsilon \left( L_2 \right)}$ & ${\epsilon \left( L_{\infty} \right)}$ & ${\mathcal{O} \left( L_1 \right)}$ & ${\mathcal{O} \left( L_2 \right)}$ & ${\mathcal{O} \left( L_{\infty} \right)}$\\
\hline\hline
100 & ${7.49 \times 10^{-1}}$ & ${9.38 \times 10^{-1}}$ & ${2.28 \times 10^{-1}}$ & - & - & -\\
\hline
200 & ${2.26 \times 10^{-1}}$ & ${3.47 \times 10^{-1}}$ & ${5.25 \times 10^{-2}}$ & 1.73 & 1.44 & 2.12\\
\hline
400 & ${5.69 \times 10^{-2}}$ & ${9.07 \times 10^{-2}}$ & ${1.02 \times 10^{-2}}$ & 1.99 & 1.93 & 2.36\\
\hline
800 & ${1.38 \times 10^{-2}}$ & ${2.30 \times 10^{-2}}$ & ${1.63 \times 10^{-3}}$ & 2.05 & 1.98 & 2.64\\
\hline
1600 & ${4.77 \times 10^{-3}}$ & ${3.16 \times 10^{-3}}$ & ${3.37 \times 10^{-4}}$ & 1.53 & 2.86 & 2.28\\
\hline
\end{tabular}
\caption{Convergence rates for the head-on collision of rapidly rotating Kerr black holes (angular momentum) test with respect to the ${L_1}$, ${L_2}$ and ${L_{\infty}}$ norms for the Hamiltonian constraint $H$ after time ${t = 24 M}$, showing approximately second-order convergence.}
\label{tab:Table7}
\end{table}

\clearpage

\section{Concluding Remarks}
\label{sec:Section6}

This article has presented a novel numerical general relativity code based on the CCZ4/BSSN formulation of the Einstein field equations with constraint violation damping, solved using an unstructured generalization of a standard fourth-order finite difference scheme, in which the Cauchy initial data is defined over a hypergraph of arbitrary topology. The hypergraph topology is then refined and coarsened dynamically as the data is evolved forwards in time, in accordance with  another unstructured generalization of the local AMR technique of Berger and Colella, such that the topology is effectively coupled to projections of the local conformal curvature tensor. This new code was then validated against a variety of standard spacetimes, including static Schwarzschild black holes, rapidly rotating Kerr black holes, maximally extended Schwarzschild black holes, and binary black hole mergers in both rotating and non-rotating cases. Convergence and stability analyses were performed, demonstrating the desired fourth-order convergence properties of the numerical scheme with respect to the ${L_1}$, ${L_2}$ and ${L_{\infty}}$ norms, as well as the desired stability of the ADM Hamiltonian, linear momentum and angular momentum constraints. Finally, the results of this code were compared against hypergraphs obtained by a pure Wolfram model evolution (without any underlying PDE system, but where the hypergraph substitution rules were known a priori to satisfy the Einstein field equations in the continuum limit), and we performed a similar convergence and stability analysis to illustrate that the two sets of discrete spacetimes were indeed converging to equivalent limiting geometries.

We emphasize that the present article represents a highly preliminary investigation of this area, and we make no claim of completeness; for instance, although approximately second-order convergence of the pure Wolfram model evolution scheme was observed, this analysis was only performed for a single example of a hypergraph substitution rule that was known to obey the Einstein field equations in the continuum limit, and it remains highly likely that there exist other substitution rules with similar algorithmic complexity but with considerably more desirable convergence and stability properties which we have not yet found. A systematic investigation of such rules, along with a rigorous comparison of their convergence and stability properties, remains a topic for further investigation - as, indeed, does the derivation of a robust technique for determining (either heuristically or exactly) when a given Wolfram model rule will satisfy the Einstein field equations, in which limit, and with what convergence behavior. Moreover, it is likely that our numerical implementation of a CCZ4/BSSN finite difference code with dynamic topology offers various advantages over more conventional numerical general relativity schemes implemented over meshes with fixed topologies; it would also be extremely interesting to compare the relative advantages and disadvantages of the dynamic vs. fixed topology approaches in some appropriately methodical way in the future.

We also emphasize that this article has thus far considered only vacuum solutions of the Einstein field equations; the next obvious case to consider would be simple non-vacuum solutions in which the energy-momentum tensor is specified purely in terms of a single massive (and minimally coupled) scalar field term, obeying a discrete Klein-Gordon equation defined in terms of the discrete ${\left( d - 1 \right) + 1}$-dimensional d'Alembertian operator ${B_{k}^{\left( d \right)}}$\cite{x}. Even this very simple formulation would suffice to allow one to investigate minimal models of gravitational collapse using (for instance) Gaussian scalar field bubbles evolved under gravity, which would in turn provide a new means of analyzing relativistic critical phenomena, as first described by Choptuik\cite{choptuik}\cite{gundlach2}, and perhaps also permit the extraction of precise numerical values for critical points and the derivation of appropriate supercritical scaling relations. In principle, such a scalar field could then be quantized by means of the covariant Peierls bracket, as first proposed by Johnston\cite{johnston}\cite{johnston2}\cite{johnston3}, which would then permit the definition of a Pauli-Jordan operator (and from it, a Wightman function), and hence the construction of a full non-interacting quantum field theory in curved spacetime with a Gaussian state. Such a formulation would then allow for the numerical study of phenomena such as Hawking radiation, spacetime entanglement entropies\cite{sorkin2}\cite{shah}, and potentially even the ER=EPR hypothesis.

\section*{Acknowledgments}

The author would like to thank Stephen Wolfram for his continual encouragement in the pursuit of the present project, as well as for useful conversations and suggestions.

\end{document}